\documentclass[
12pt,
letterpaper,
reprint,
prx,
aps,
floatfix,
showkeys,
superscriptaddress,
longbibliography,
colorlinks
]{revtex4-2}

\usepackage[utf8]{inputenc}
\usepackage{amsmath,amsthm,amssymb,amsfonts}
\theoremstyle{remark}
\newtheorem{claim}{\textbf{Claim}}[section]
\usepackage{mathtools}

\usepackage{graphicx}
\graphicspath{{./figures/}}
\usepackage{booktabs}
\usepackage{xcolor}
\usepackage{rotating}
\usepackage{orcidlink}
\usepackage{adjustbox}
\usepackage{array}
\usepackage[mathlines]{lineno}
\usepackage{appendix}
\usepackage{aligned-overset}
\usepackage{algorithm}
\usepackage{algpseudocode}
\usepackage{bbm}

\AtBeginDocument{
  \heavyrulewidth=.08em
  \lightrulewidth=.05em
  \cmidrulewidth=.03em
  \belowrulesep=.65ex
  \belowbottomsep=0pt
  \aboverulesep=.4ex
  \abovetopsep=0pt
  \cmidrulesep=\doublerulesep
  \cmidrulekern=.5em
  \defaultaddspace=.5em
}
\newcommand{\todo}[1]{}

\usepackage{hyperref}

\let\originalleft\left
\let\originalright\right
\renewcommand{\left}{\mathopen{}\mathclose\bgroup\originalleft}
\renewcommand{\right}{\aftergroup\egroup\originalright}
\newcommand{\abs}[1]{\left\lvert#1\right\rvert} 
\newcommand{\absSmall}[1]{\lvert#1\rvert}

\newcommand{\bra}[1]{\langle#1\rvert} 
\newcommand{\ket}[1]{\lvert#1\rangle} 
\newcommand{\proj}[1]{\ket{#1}\bra{#1}} 
\newcommand{\braket}[2]{ \langle #1 | #2 \rangle} 
\newcommand{\braopket}[3]{\langle #1 | #2 | #3\rangle} 
\newcommand{\ev}[1]{\left\langle #1 \right\rangle} 
\newcommand{\evSmall}[1]{\langle #1 \rangle} 
\newcommand{\tr}[1]{\text{Tr}\left[#1\right]}

\newcommand{\trSmall}[1]{\text{Tr}[#1]}  
 
\newcommand{\re}[1]{\text{Re}\left[#1\right]}  
\newcommand{\reSmall}[1]{\text{Re}[#1]} 
\newcommand{\im}[1]{\text{Im}\left[#1\right]} 
\newcommand{\imSmall}[1]{\text{Im}[#1]} 
\newcommand{\evText}[1]{\text{E}\left[#1\right]} 
\newcommand{\var}[1]{\text{Var}\left[#1\right]} 
\newcommand{\varSmall}[1]{\text{Var}[#1]} 
\newcommand{\varSub}[2]{\text{Var}_{#1}\left[#2\right]}
\newcommand{\varSubSmall}[2]{\text{Var}_{#1}[#2]}

\newcommand{\bmatrixByJames}[1]{\left[\;\begin{matrix}#1\end{matrix}\;\right]}

\newcommand{\intg}[3]{\int_{#1}^{#2} \text{d}#3\;}

\newcommand{\intginf}[1]{\intg{-\infty}{\infty}{#1}}

\newcommand{\iintginfTwoArg}[2]{\iint_{-\infty}^{\infty} \text{d}#1\text{d}#2\;}
\newcommand{\deriv}[2]{\frac{\text{d}#1}{\text{d}#2}}

\newcommand{\h}[1]{\hat{#1}}

\newcommand{\R}{\mathbb{R}} 
\newcommand{\C}{\mathbb{C}}

\newcommand{\IQ}{\mathcal{I}_Q}
\newcommand{\IC}{\mathcal{I}_C}

\newcommand{\om}{\omega}
\newcommand{\Om}{\Omega}
\newcommand{\si}{\sigma}
\newcommand{\Si}{\Sigma}

\newcommand{\T}{\text{T}}

\newcommand{\diag}[1]{\text{diag}\left(#1\right)}
\newcommand{\diagSmall}[1]{\text{diag}(#1)}
\newcommand{\order}[1]{\mathcal{O}(#1)}
\newcommand{\orderten}[1]{\mathcal{O}(10^{#1})}

\newcommand{\where}{\intertext{where}}

\begin{document}

\title{Stochastic waveform estimation at the fundamental quantum limit}

\author{James~W.~Gardner\,\orcidlink{0000-0002-8592-1452}}
\email{james.gardner@anu.edu.au}
\affiliation{OzGrav-ANU, Centre for Gravitational Astrophysics, Research Schools of Physics, and of Astronomy and Astrophysics, The Australian National University, Canberra, ACT 2601, Australia}
\affiliation{Walter Burke Institute for Theoretical Physics, California Institute of Technology, Pasadena, California 91125, USA} 
\author{Tuvia Gefen\,\orcidlink{0000-0002-3235-4917}}
\email{tgefen@caltech.edu}
\affiliation{Institute for Quantum Information and Matter, California Institute of Technology, Pasadena, California 91125, USA}
\author{Simon A. Haine\,\orcidlink{0000-0003-1534-1492}}
\affiliation{Department of Quantum Science and Technology and Department of Fundamental and Theoretical Physics, Research School of Physics, The Australian National University, Canberra, ACT 0200, Australia}
\author{Joseph~J.~Hope\,\orcidlink{0000-0002-5260-1380}}
\affiliation{Department of Quantum Science and Technology and Department of Fundamental and Theoretical Physics, Research School of Physics, The Australian National University, Canberra, ACT 0200, Australia}
\author{John Preskill\,\orcidlink{0000-0002-2421-4762}}
\affiliation{Walter Burke Institute for Theoretical Physics, California Institute of Technology, Pasadena, California 91125, USA}
\affiliation{Institute for Quantum Information and Matter, California Institute of Technology, Pasadena, California 91125, USA}
\author{Yanbei Chen\,\orcidlink{0000-0002-9730-9463}\,}
\affiliation{Walter Burke Institute for Theoretical Physics, California Institute of Technology, Pasadena, California 91125, USA}
\author{Lee McCuller\,\orcidlink{0000-0003-0851-0593}}
\affiliation{Institute for Quantum Information and Matter, California Institute of Technology, Pasadena, California 91125, USA} 
\affiliation{Division of Physics, Mathematics and Astronomy, California Institute of Technology, Pasadena, California 91125, USA}

\date{\today}
\begin{abstract}
    Although measuring the deterministic waveform of a weak classical force is a well-studied problem, estimating a random waveform, such as the spectral density of a stochastic signal field, is much less well-understood despite it being a widespread task at the frontier of experimental physics. State-of-the-art precision sensors of random forces must account for the underlying quantum nature of the measurement, but the optimal quantum protocol for interrogating such linear sensors is not known. We derive the fundamental precision limit, the extended channel quantum Cram\'er-Rao bound, and the optimal protocol that attains it. In the experimentally relevant regime where losses dominate, we prove that non-Gaussian state preparation and measurements are required for optimality. We discuss how this non-Gaussian protocol could improve searches for signatures of quantum gravity, stochastic gravitational waves, and axionic dark matter.
\end{abstract}
\maketitle
\allowdisplaybreaks

The estimation of the spectral density of a classical process is a ubiquitous task in continuous-variable quantum systems. 
Examples include searching for excess noise in optical interferometers due to quantum gravity~\cite{VerlindePLB21ObservationalSignatures,LiPRD23InterferometerResponse,McCuller22SinglePhotonSignal,ChouCQG17HolometerInstrument,VermeulenCQG21ExperimentObserving}, probing stochastic gravitational waves with the Laser Interferometric Gravitational-wave Observatory (LIGO)~\cite{AasiCQG15AdvancedLIGO,BuikemaPRD20SensitivityPerformance,RomanoLRR17DetectionMethods}, and hunting for axionic dark matter with microwave cavities~\cite{kim2010axions,choi2021recent,rosenberg2000searches,graham2015experimental}.
The remarkable sensitivity of contemporary devices demands that we contend with quantum noise, the fundamental uncertainty in the state of the device arising from the Heisenberg Uncertainty Principle. The stochastic signal of interest must be distinguished against the natural fluctuations arising from the measurement of the device. 

\begin{figure}
    \centering
    \includegraphics[width=\columnwidth]{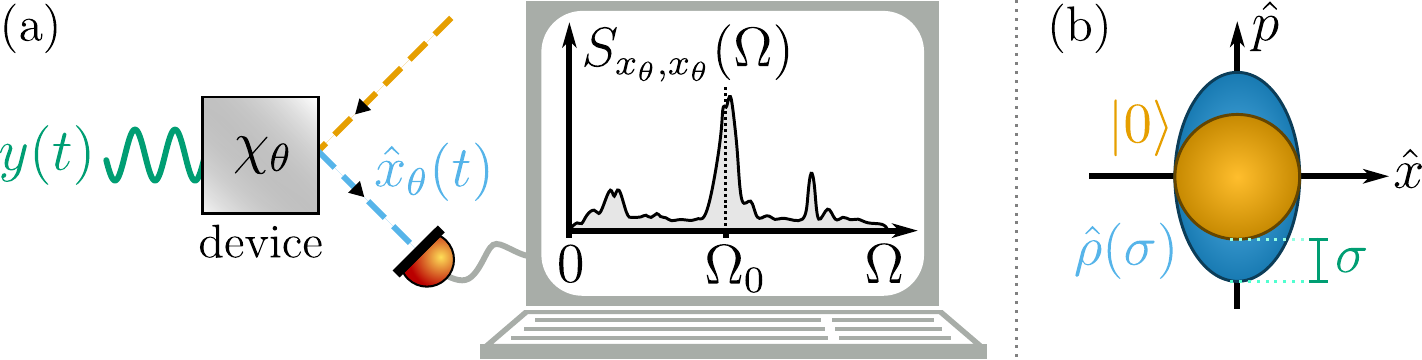}
    \caption{(a) Spectral density of the outgoing bosonic mode from a linear quantum device coupled to a classical random process $y(t)$, e.g.\ LIGO coupled to a stochastic gravitational wave. (b) Phase plane representation of the analogous single-parameter problem at a particular frequency $\Omega_0$. Given vacuum input $\ket0$ with covariance matrix $\diag{\frac{1}{2},\frac{1}{2}}$, then the final quantum state $\h\rho(\si)$ is a squeezed thermal state with covariance matrix $\diag{\frac{1}{2},\frac{1}{2}+\si^2}$.} 
    \label{fig:device}
\end{figure}

This task, sometimes called ``noise spectroscopy'', can be formally expressed in the language of quantum metrology,  the study of estimating parameters encoded in quantum states. As shown in Fig.~\hyperref[fig:device]{\ref*{fig:device}a}, we consider a linear quantum device coupled to a stochastic signal, a classical continuous random variable $y(t)$ at times $t$. We assume this to be a stationary random process with (one-sided) power spectral density $S_{yy}(\Om)\propto\abs{y(\Om)}^2$ in terms of its Fourier transform $y(\Om):=\intginf{t} e^{i\Om t} y(t)$ at the positive frequency $\Om$. In the input/output formalism~\cite{GardinerPRA85InputOutput}, information about $y(t)$ is encoded in the state of an outgoing bosonic mode of the environment. The canonical quadratures $\h x_\theta(t):=\cos(\theta)\h x(t) + \sin(\theta)\h p(t)$ of this outgoing mode obey
\begin{align}\label{eq:TD IO}
    \h x_\theta(t) &= \h x_\theta^{(0)}(t) + \intginf{t'} \chi_\theta(t-t')y(t')
    \intertext{where $\h x_\theta^{(0)}(t)$ is the free solution and $\chi_\theta$ is the device's susceptibility~\cite{KuboRPP66FluctuationdissipationTheorem,BuonannoPRD02SignalRecycled}. In the Fourier domain, Eq.~\ref{eq:TD IO} becomes}
    \label{eq:IOrelation}
    \h x_\theta(\Om) &= \h x_\theta^{(0)}(\Om) + \chi_\theta(\Om)y(\Om)
    \intertext{such that the spectral density measured from $\h x_\theta(t)$ is }
    \label{eq:IO-SD}
    S_{x_\theta x_\theta}(\Om) &= S^{(0)}_{x_\theta x_\theta}(\Om)+ \abs{\chi_\theta(\Om)}^2S_{yy}(\Om)
\end{align}
where $S^{(0)}_{x_\theta x_\theta}$, which includes the quantum noise. Here, the one-sided power spectral density between two stationary operators $\h z_1$ and $\h z_2$ is defined from the anticommutator $\{\cdot,\cdot\}$ as
\begin{align*}    
    2\pi\delta(\Omega-\Omega')S_{z_1z_2}(\Omega) := 
    \langle\{\h z_1(\Om),\h z_2^\dag(\Om')\}\rangle.
\end{align*}

Formally, our goal is the optimal unbiased estimation of the continuum $S_{yy}(\Om)$ from measurements of the outgoing mode such that our estimate has the minimal mean square error (MSE) at each $\Om$. We note that this resembles the problem of estimating the uncertainty, $\sigma$, of a single bosonic mode, and so treat that case extensively. 

Much attention has been previously dedicated to studying the fundamental precision limits of estimating the mean values of the signal, $\evText{y(\Om)}$, for various protocols and quantum devices, e.g.\ see Refs.~\cite{TsangPRL11FundamentalQuantum,MiaoPRL17FundamentalQuantum,gardner2024achieving}. Here, we focus instead on estimating the spectral density $S_{yy}(\Om)$ of the signal rather than its mean value. This is a much less studied problem. In fact, it was recently shown that this problem is fundamentally different from mean value estimation in that non-Gaussian measurement techniques are required to obtain optimal estimates~\cite{NgPRA16SpectrumAnalysis}. There are still many critical open questions, however, such as the effects of imperfections and the preparation of different initial quantum states.

To guide the reader, we summarise the main results of our work as follows:
\begin{itemize}
    \item In Sec.~\ref{sec:single-parameter problem}, we reduce the estimation problem at a fixed frequency to the study of a harmonic oscillator undergoing a random displacement channel. The task is to estimate the standard deviation $\si$ of the Gaussian distribution of random displacements and, in some cases, to simultaneously estimate the mean displacement $\mu$.
    \item In Sec.~\ref{sec:lossless review}, we review the prior literature on estimating $\si$ in the ideal lossless limit using an initial vacuum state. A number-resolving measurement, e.g.\ photon counting, is optimal~\cite{NgPRA16SpectrumAnalysis,PhysRevA.107.012611}. In contrast, Gaussian measurements, such as homodyne detection, suffer the ``Rayleigh curse'' when the signal $\sigma$ is small.    
    The goal of our work is to find the optimal initial state and measurement scheme that achieves the ultimate limit on precision, the extended channel quantum Fisher information (ECQFI), in the presence of loss. We also consider the simultaneous estimation of $\mu$ and $\si$.

   \item In Sec.~\ref{sec:loss}, we discuss imperfections. We focus on the case of a loss $\eta$ occurring before the signal is encoded by the random displacement channel. We assume negligible loss after the encoding and negligible additive classical noise, $\sigma_C$. Experimentally, the relevant regime is that of small signals $\sigma^2 \ll 1/2$ and high loss $\eta\gg\si^2$. 
    \item In Sec.~\ref{sec:state}, we find the optimal protocol for estimating $\si$. In the lossless case, where the initial state obeys the energy constraint $\ev{\h n} = N$, 
    we show that the ECQFI is saturated by preparing an initial single-mode squeezed vacuum (SMSV) and then, after the encoding, anti-squeezing and performing a number measurement. In the lossy case, we show that preparing a two-mode squeezed vacuum (TMSV) is optimal, but only if negligible loss occurs on the ancilla mode. In the experimentally relevant regime of high loss on all modes, numerics indicate that the ECQFI is attainable, without an ancilla, using highly non-Gaussian states and measurements.     
    \item In Sec.~\ref{sec:simultaneous estimation}, we discuss how to simultaneous estimate $\mu$ and $\si$. We assume that the initial state is vacuum (which may not be optimal). For separable measurements on $M$ copies of the final state, we construct an adaptive measurement scheme in which we learn about $\mu$ via quadrature measurement and then use that information to displace back to the origin and learn about $\si$ via number measurement. To saturate the quantum Fisher information for the fixed vacuum input state, however, a collective measurement on the $M$ modes is required.
    \item Finally, in Sec.~\ref{sec:stochastic waveform estimation}, we discuss how our results concerning single-mode channel estimation can be leveraged for estimating the power spectral density of a continuously varying signal, we propose experimental implementations to realise the optimal estimation protocols, and we apply our results to searches for signatures of quantum gravity, stochastic gravitational waves, and axionic dark matter.
\end{itemize}

\section{Canonical noise estimation problem}
\label{sec:single-parameter problem}

The continuous-variable problem of estimating $S_{yy}(\Om)$ at each frequency $\Om$ can be simplified to a continuum of independent single-variable estimation problems by linearity. We use the technique of decomposing the outgoing mode into its cosine and sine phases~\cite{gardner2024achieving}. This is equivalent to decomposing the Fourier transform of the quadratures in Eq.~\ref{eq:IOrelation} into its real and imaginary parts as
\begin{align}\label{eq:spec_components}
\vec{\h q}  
&:=
\frac{1}{\sqrt{\pi T}}
\bmatrixByJames{
        \re{\h x(\Omega)} \\
        \re{\h p(\Omega)} \\
        \im{\h x(\Omega)} \\
        \im{\h p(\Omega)}
    }
    \\&\approx 
    \frac{1}{\sqrt{\pi T}}
    \intg{0}{T}{t} \bmatrixByJames{
        \cos(\Om t) \h x(t) \\
        \cos(\Om t) \h p(t) \\
        \sin(\Om t) \h x(t) \\
        \sin(\Om t) \h p(t) 
    }\nonumber
\end{align}
where $T$ is the long integration time. 
The four elements of $\vec{\h q}$ are nonstationary Hermitian observables which resemble the quadratures of two real harmonic oscillators at each frequency since $[{\vec{\h q}}_1, {\vec{\h q}}_2]=[{\vec{\h q}}_3, {\vec{\h q}}_4]=i$ (where $\hbar=1$) and all other commutators are zero. These observables contain the same information as the two stationary but complex quadratures $\h x(\Om)$ and $\h p(\Om)$. We emphasise that there are always four independent real degrees of freedom associated with each positive frequency, regardless of the formalism.

Although the variance of each of the components of $\vec{\h q}$ is defined, the spectral density is not, since they are nonstationary. Since the signal process $y(t)$ is stationary, however, the excess noise in the real and imaginary parts of the output are each equal to half of the total variance associated with $S_{yy}(\Om)$. If the device is also stationary, then the free covariance matrix is block diagonal and the two harmonic oscillators $({\vec{\h q}}_1, {\vec{\h q}}_2)$ and $({\vec{\h q}}_3, {\vec{\h q}}_4)$ are independent.

We assume that $\chi_0=0$ so that the signal appears in only the $\h p(t)$ quadrature of the outgoing mode. In the context of an optical interferometer like LIGO, this corresponds to assuming that the resonant optical cavities are all tuned and that the Michelson interferometer is held at total destructive interference at the output port such that signal appears in only one quadrature of the outgoing optical mode~\cite{bond2016interferometer}. Thus, for a Gaussian initial state, the encoding adds $\diag{0, \si^2, 0, \si^2}$ to the 4-by-4 covariance matrix of the parts $\frac{1}{2}\langle\{\vec{\h q}^{\,(0)}_j, \vec{\h q}^{\,(0)}_k\}\rangle$ where the signal $\si$ with gain $G$ is
\begin{align}\label{eq:sigma}
    \si^2:=G(\Om)S_{yy}(\Om), \quad G(\Om):=\frac{1}{2}\lvert\chi_{\frac{\pi}{2}}(\Om)\rvert^2
    .
\end{align}

Estimating the continuum $S_{yy}(\Om)$ is, therefore, equivalent to a continuum of independent copies of the following canonical single-parameter estimation problem (with two copies for each positive $\Om$): Given a single canonical mode $\vec{\h x}=(\h x, \h p)^\T$ and a Gaussian initial state with 2-by-1 mean vector $\vec{\mu}=\langle\vec{\h x}\rangle$ and 2-by-2 covariance matrix $\Si_{jk}=\frac{1}{2}\langle\{\vec{\h x}^{\,(0)}_j, \vec{\h x}^{\,(0)}_k\}\rangle$, estimate $\si^2$ by measuring the quantum state after the Gaussian encoding channel $\Lambda_\si$ is applied. (A quantum channel is a completely positive, trace-preserving linear map between density matrices.) Here, $\Lambda_\si$ leaves $\vec{\mu}$ invariant and sends $\Si\mapsto\Si+\diag{0, \si^2}$. For example, the vacuum $\ket0\bra0$ with covariance matrix $\Si_0=\diag{\frac{1}{2}, \frac{1}{2}}$ becomes the squeezed thermal state $\h\rho(\si)$ with covariance matrix $\diag{\frac{1}{2}, \frac{1}{2}+\si^2}$ as shown in Fig.~\hyperref[fig:device]{\ref*{fig:device}b} (which depicts one of the two oscillators at $\Omega_0$).

This non-unitary channel $\Lambda_\si$ is equivalent to a random displacement along the $\hat{p}$ quadrature with the Kraus representation
\begin{align}\label{eq:encoding}
    \Lambda_\si(\h \rho) = \intginf{\alpha} p(\alpha) \h U_\alpha\h\rho\h U_\alpha^\dag
\end{align}
where $p(\alpha)\sim \mathcal{N}(0, \si^2)$ is the weighting of the different unitaries
and $\h U_\alpha = \exp(i\alpha \h x)$ displaces $\h p$ to $\h p + \alpha$. 
When $\si \h x$ is small, we may expand Eq.~\ref{eq:encoding} to obtain the channel 
\begin{align}\label{eq:encoding_expansion}
     \Lambda_\si(\h\rho) &= \h\rho + \si^2\left( \h x\h\rho\h x - \frac{1}{2}\{\h x^2, \h\rho\}\right) + \order{\si^4\h x^4},      
\end{align}
where the odd terms vanish because $p(\alpha)$ is an even function in $\alpha$. 
In the position basis of the canonical mode, $\Lambda_\si$ is the following decoherence channel~\cite{zurek2003decoherence,Zurek2007}:
\begin{align}
    \label{eq:decoherence channel in position basis}
    \braopket{x}{\Lambda_\si(\h \rho)}{x'} &= 
    e^{-\frac{1}{2} \si^2 (x-x')^2} \braopket{x}{\h \rho}{x'}.
\end{align}
Our goal now is to understand the optimal protocol for estimating $\si^2$ at a given frequency.

\subsection{Review of Fisher Information}
\label{sec:FI review 1}
We now review our main tools for addressing this single-parameter estimation problem: the concepts of Classical and Quantum Fisher Information (FI)~\cite{BraunsteinPRL94StatisticalDistance,Wiseman09QuantumMeasurement,PezzeRMP18QuantumMetrology}.

Suppose that a real parameter of interest $\theta$ is encoded in a quantum state $\h\rho(\theta)$ and that we estimate $\theta$ by performing a given measurement [i.e.\ a positive operator-valued measure (POVM)] with associated probability distribution $p(x|\theta)$. The minimal MSE $\Delta^2\theta$ of unbiased estimation of $\theta$ from $M$ measurement results satisfies the \textit{Classical Cram\'er-Rao Bound (CCRB)}, $\Delta^2\theta\geq\frac{1}{M}[\IC(\theta)]^{-1}$, where $\IC(\theta)$ is the \textit{Classical Fisher Information (CFI)} given by 
\begin{align}\label{eq:CFI}
    \IC(\si)=\intginf{x}\frac{[\partial_\theta p(x|\theta)]^2}{p(x|\theta)}.
\end{align}
This bound can be attained by maximal likelihood estimation in the asymptotic limit such that the Central Limit Theorem applies. 
In general, saturating the CFI may require a parameter that is sufficiently small or has a narrow enough prior, as well as the ability to perform a large number of independent measurements.
A useful property of the CFI is that if $p(x|\theta)$ and $p(y|\theta)$ are independent distributions, e.g.\ describing two separate measurements, then the total CFI from observing one outcome from each distribution is simply the sum of the individual CFIs.

This may not be the optimal measurement, however, for extracting the maximal information about $\theta$ from $\h\rho(\theta)$. For single-parameter estimation, the \textit{Quantum Fisher Information (QFI)} is the CFI maximised over all possible measurements (POVMs), $\IQ(\theta)=\sup\IC(\theta)$. In terms of the eigenvalues $p_j$ and eigenvectors $\ket{\phi_j}$ of the final state $\h\rho(\theta)=\sum_j p_j \ket{\phi_j}\bra{\phi_j}$, the QFI can be shown to be
\begin{align*}
    \IQ(\theta) = \sum_{j,k}\frac{2}{p_j+p_k}\abs{\braopket{\phi_j}{\partial_\theta\h \rho(\theta)}{\phi_k}}^2
\end{align*}
where the sum runs over only $j,k$ such that $p_j+p_k>0$. 
For example, if a parameter $\theta$ is encoded by a unitary $\exp(-i\theta\h H)$ applied to a pure state $\ket\psi$, then the QFI is $\IQ(\theta) = 4 \varSubSmall{\ket\psi}{\h H}$ independent of $\theta$.

The \textit{Quantum Cram\'er-Rao Bound (QCRB)}, $\Delta^2\theta\geq\frac{1}{M}[\IQ(\theta)]^{-1}$, provides the fundamental minimal MSE that can be achieved by maximum likelihood estimation from the $M$ outcomes of the optimal measurement scheme. Similarly, the minimal fractional MSE with respect to a parameter $\theta$ is bounded too, $\frac{\Delta^2\theta}{\theta^2}\geq\frac{1}{M}[\theta^2\IQ(\theta)]^{-1}$. The analogue of the additivity of the CFI for independent distributions is that the QFI for a product state $\h\rho_1(\theta)\otimes\h\rho_2(\theta)$ is simply the sum of the individual QFIs.

\subsection{Standard deviation versus variance}
\label{sec:stdev-to-var}

For our single-parameter problem, we choose to estimate the standard deviation $\si$. This is equivalent to estimating the variance $\si^2$ since $\si\geq0$. The QFIs with respect to $\si$ and $\si^2$ are related by $\IQ(\si) = 4\si^2\IQ(\si^2)$ through the chain rule. Although $\IQ(\si^2)$ may diverge and $\IQ(\si)$ stay finite as $\si\rightarrow0$, the limiting behaviour of the fractional MSEs is consistent since, in that event, $[\si^4\IQ(\si^2)]^{-1}\propto[\si^2\IQ(\si)]^{-1}\rightarrow\infty$.

\section{Lossless vacuum limit}
\label{sec:lossless review}

\begin{figure}
    \centering
    \includegraphics[width=\columnwidth]{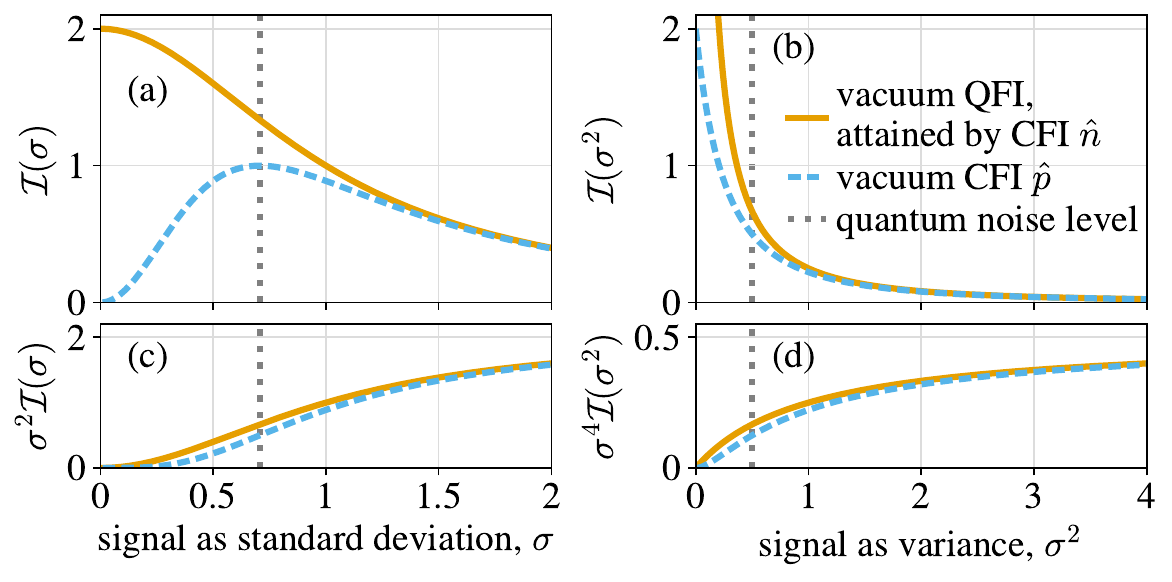}
    \caption{Fisher information with respect to (a) standard deviation $\si$ or (b) variance $\si^2$ for the single-parameter estimation problem in the lossless vacuum limit. The level of quantum noise for the initial vacuum state, shown by the dotted grey vertical line, is equal to $\varSmall{\h p} = \frac{1}{2}$. A quadrature measurement $\h p$, with CFI shown in the dashed blue curve, is optimal for signals well above this level. A number-resolving measurement $\h n$, however, is optimal---its CFI attains the QFI shown in the solid orange curve---for all $\si$ including as $\si\rightarrow0$. In this limit, the fractional MSE with respect to (c) $\si$ or (d) $\si^2$ diverges regardless of the measurement scheme since $\theta^2\IQ(\theta)$ converges to zero.} 
    \label{fig:losslessFI}
\end{figure} 

The task of estimating $\si$ from $\Lambda_\si$ in Eq.~\ref{eq:encoding}, in the absence of loss when the input state is the vacuum, was previously studied in Refs.~\cite{NgPRA16SpectrumAnalysis,PhysRevA.107.012611,takeoka2003unambiguous}. We briefly discuss this ideal lossless regime here for completeness.

Suppose that we apply $\Lambda_\si$ to $\h\rho=\ket0\bra0$ such that the final covariance matrix is $\Si=\diag{\frac{1}{2},\frac{1}{2}+\si^2}$. The QFI for a signal encoded solely in the covariance matrix of a single-mode Gaussian state is~\cite{monras2013phase}
\begin{align}\label{eq:Gaussian QFI CM}
    \IQ(\si) &=  \frac{\tr{(\Si^{-1}\partial_\si\Si)^2}}{2(1+\gamma^2)}+\frac{2(\partial_\si \gamma)^2}{1-\gamma^4}
\end{align}
where $\gamma = \det(2\Si)^{-\frac{1}{2}}$ is the purity of the Gaussian state. For the vacuum state, Eq.~\ref{eq:Gaussian QFI CM} implies that the QFI is 
\begin{align}\label{eq:ideal vacuum case QFI}
    \IQ(\si)=\frac{2}{1+\si^2} \xrightarrow[\si\rightarrow0]{} 2.
\end{align}
We will write the important limit of $\lim_{\si\rightarrow0}\IQ(\si)$ as $\IQ(\si=0)$ henceforth. Note that it is nonzero for the QFI.

We want to know what measurement will saturate the QFI in Eq.~\ref{eq:ideal vacuum case QFI}. A homodyne measurement of $\h p$ is a Gaussian measurement for which the CFI with respect to the total standard deviation $\varsigma=\sqrt{\frac{1}{2}+\si^2}$ is $\IC^{\h p}(\varsigma)=\frac{2}{\varsigma^2}$. This implies that $\IC^{\h p}(\si)=\frac{2\si^2}{\left(\frac{1}{2}+\si^2\right)^2}$ such that the CFI for quadrature measurement saturates the QFI in Eq.~\ref{eq:ideal vacuum case QFI} in the classical regime of $\si\gg\frac{1}{\sqrt2}$, but vanishes in the quantum regime of $\sigma\rightarrow0$ where $\IC^{\h p}(\si)\approx 8\si^2$ as shown in Fig.~\hyperref[fig:losslessFI]{\ref*{fig:losslessFI}a}. If we estimate $\si^2$ instead of $\si$, as shown in Fig.~\hyperref[fig:losslessFI]{\ref*{fig:losslessFI}b}, then the behaviour of the fractional MSE is the same as discussed in Sec.~\ref{sec:stdev-to-var} and shown in Fig.~\hyperref[fig:losslessFI]{\ref*{fig:losslessFI}c}. The conventional quadrature measurement, therefore, is highly inefficient in the relevant limit of $\sigma \rightarrow 0$. Intuitively, the small change in the variance induced by the weak signal $\sigma^{2}\ll\frac{1}{2}$ is masked by the vacuum quantum noise, leading to this vanishing sensitivity.

In comparison, a number-resolving measurement of $\h n = \h a^\dag \h a$ can be shown to saturate the QFI for all $\si$. Since $\h U_\alpha\ket0=\ket{\alpha'}$ is a coherent state with amplitude $\alpha'=i\alpha/\sqrt{2}$, we can compute the probabilities in the number basis after the encoding in Eq.~\ref{eq:encoding} as 
\begin{align}\label{eq:number prob dist}
    p(n) &= \intginf{\alpha} \abs{\braket{n}{\alpha'}}^2p(\alpha)    
    \\&= \frac{(2n)!}{2^{2n} (n!)^2} \frac{\si^{2n}}{\left(\si^2+1\right)^{n+\frac{1}{2}}}\nonumber
    .
\end{align}
Then, the CFI for number measurement, calculated as $\IC^{\h n}(\si)=\sum_{n=0}^\infty \frac{[\partial_\si p(n)]^2}{p(n)}$ by Eq.~\ref{eq:CFI}, equals the QFI in Eq.~\ref{eq:ideal vacuum case QFI} as shown in Fig.~\ref{fig:losslessFI}.

This estimation problem is analogous to quantum super-resolution~\cite{PhysRevA.107.012611,tsang2016quantum} where the CFI of the na\"ive Gaussian measurement converges to zero at zero signal---called the ``Rayleigh curse''---but the CFI of the optimal non-Gaussian measurement is positive and attains the QFI. Analogously, here, we define the \textit{Rayleigh curse} to refer to the FI converging to zero at zero signal. We have seen above that the Rayleigh curse can be avoided in the lossless vacuum limit by performing a number measurement.

\section{Loss and classical noise}
\label{sec:loss}

Realistically, the quantum state will experience noise channels before and after the encoding such that the noiseless channel $\Lambda_{\sigma}$ becomes the noisy channel $\Lambda_{\sigma}^\text{noisy}$. We restrict our attention to Gaussian noise channels. We want to understand how these imperfections limit our ability to estimate $\si$ and whether the Rayleigh curse can still be avoided to increase QFI in the small signal limit.

\subsection{Gaussian states}
We first consider preparing a Gaussian initial state in the limit of $\si \rightarrow 0$. For Gaussian noise channels, $\partial_{\sigma}\Lambda_{\sigma}^\text{noisy}\rightarrow0$ as $\si \rightarrow 0$ and information about $\sigma$ is encoded only in the covariance matrix of the output Gaussian state. Whether the Rayleigh curse reappears is addressed by the following claim:
\begin{claim}
\label{claim:simplectic eigenvalues}
For any Gaussian state such that a parameter $\sigma$ is encoded only on its covariance matrix $\Sigma\left(\sigma\right)$ and $\lim_{\si\rightarrow0}\partial_{\sigma}\Sigma=0$, then $\IQ(\si=0)\neq0$ (i.e.\ the state does not exhibit the Rayleigh curse) if and only if there exists a symplectic eigenvalue of $\Sigma$ equal to $\frac{1}{2}+ k \sigma^{2}$ for some constant $k>0$.
\end{claim}
The proof of this claim is given in Appendix~\ref{sec:simplectic eigenvalue proof}. For a single-mode Gaussian state, this claim implies that overcoming the Rayleigh curse is possible only if $\Lambda_{\sigma=0}^\text{noisy}\left(\rho\right)$ is pure. In that case, the QFI comes only from the purity (the second term in Eq.~\ref{eq:Gaussian QFI CM}): the QFI is non-vanishing in the limit $\sigma\rightarrow 0$, and equal to $\IQ(\si=0)=2k'$, if and only if the purity is $\gamma=1-k'\sigma^{2}$ for some constant $k'$. For a multi-mode Gaussian state in the limit $\sigma\rightarrow0$, at least one of the modes needs to be pure for the QFI not to vanish. 

\subsection{Loss channel}
\label{sec:Loss channel}
\begin{figure}
    \centering
    \includegraphics[width=\columnwidth]{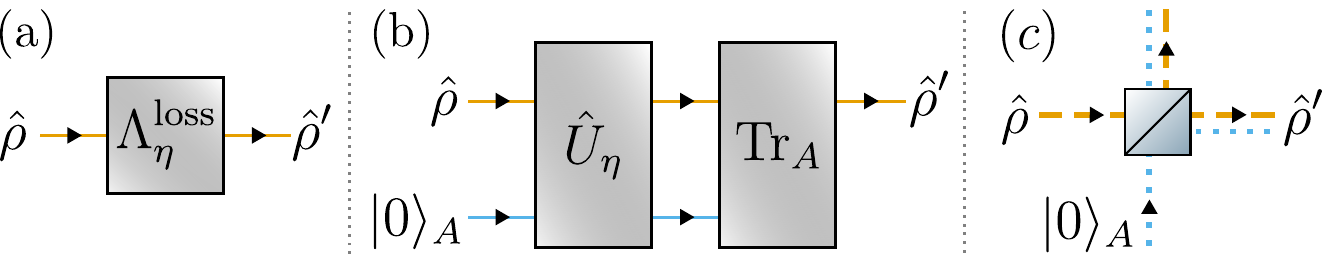}
    \caption{The quantum state $\h\rho'$ after a loss can be thought of as (a) the result, $\h\rho'=\Lambda^\text{loss}_\eta(\h\rho)$, of a non-unitary quantum channel $\Lambda^\text{loss}_\eta$; (b) the result of a beamsplitter unitary $\h U_\eta$ with an ancillary vacuum mode $\ket{0}_A$ which is then traced out; or (c) for an optical system, the state at one of the output ports of a fictitious beamsplitter with a vacuum mode.}
    \label{fig:loss_diagram}
\end{figure}

We apply Claim~\ref{claim:simplectic eigenvalues} to a couple of examples of common Gaussian noise channels, assuming an initial vacuum state. Firstly, consider the ``pure'' or ``cold'' loss channel $\Lambda^\text{loss}_\eta$ which models, for example, losses due to coupling with a zero temperature bath. This can be modelled as a beamsplitter operation with an ancillary vacuum mode that is then traced out as shown in Fig.~\ref{fig:loss_diagram}. For a loss $\eta\in(0, 1)$, this channel has the following Kraus representation~\cite{liu2004kraus} 
\begin{align*}
    \Lambda^\text{loss}_\eta(\h\rho) &= \sum_{n=0}^\infty \h K_n \h\rho \h K_n^\dag
    \intertext{where the Kraus operators are}
    \h K_n &=
    \frac{1}{\sqrt{n!}}\left(\sqrt{\eta}\h a\right)^n
    (1-\eta)^{\frac{1}{2}(\h a^\dag \h a - n)}    
    .
\end{align*}
Note that here $\eta$ is the loss and $1 - \eta$ is the efficiency. The action of this channel on a single-mode Gaussian state is to send $\vec{\mu}\mapsto\sqrt{1-\eta}\vec{\mu}$ and $\Si\mapsto(1-\eta)\Si+\eta\Si_0$ where $\Si_0=\diag{\frac{1}{2},\frac{1}{2}}$. If the initial state is a vacuum state, then the final state in the limit of $\si\rightarrow0$ is pure and there is no Rayleigh curse. The precise nonzero value of $\IQ$ depends on whether losses occur before or after the encoding: if before the encoding, then the QFI is unchanged from $\frac{2}{1+\si^2}$ and if after the encoding, then the QFI is $\frac{2(1-\eta)}{1+(1-\eta)\si^2}$. 

In this work, we will focus on the impact of a known loss $\eta$ occurring before the encoding but no loss occurring after the encoding. We address the effect of noise channels occurring after the encoding in Appendix~\ref{sec:measurement noise}, where we show that measurement noise can, in theory, be overcome by using a suitable control unitary. 
We analyze the implications of an unknown loss in Appendix~\ref{sec:loss indeterminacy}, where we justify neglecting this effect given a vacuum input state.

\subsection{Classical noise channel}
\label{sec:classical noise}
We also consider a Gaussian noise channel $\Lambda^\text{noise}_{\Si_C}$ which models a source of uncorrelated classical noise, e.g.\ thermal fluctuation processes in optics or a non-zero temperature of microwave resonators. For $M$ modes, this channel is a random displacement channel
\begin{align}\label{eq:classical noise channel Kraus representation}
    \Lambda^\text{noise}_{\Si_C}(\h \rho) &= \int_{\C^{M}}\text{d}\vec{\alpha}\; p(\vec{\alpha}) \h D(\vec{\alpha})\h\rho\h D(\vec{\alpha})^\dag 
\end{align}
where $p(\alpha)\sim \mathcal{N}(\mathbf{0}, \frac{1}{2}\Si_C)$ for a positive semi-definite $2M$-by-$2M$ matrix $\Si_C$ and the $M$-mode displacement operator is $\h D(\alpha) = \prod_{i=1}^M \exp(\alpha_i \h a_i^\dag - \alpha_i^*\h a_i)$ where $\h a_i$ is the annihilation operator for the $i$th mode. This classical noise channel acts on Gaussian states as an additive noise source: $\Si\mapsto\Si+\Si_C$. (Note that $\Lambda_\si=\Lambda^\text{noise}_{\diag{0,\si^2}}$ in Eq.~\ref{eq:encoding}.) For a given single-mode Gaussian state with fixed $\ev{\h n}=N$, a nonzero $\Si_C$ will make the final state mixed and always lead to a Rayleigh curse as $\si\rightarrow0$ by Claim~\ref{claim:simplectic eigenvalues}. For example, if $\Si_C=\diag{\si_x^2,\si_p^2}$ with vacuum input, then 
\begin{align} \label{eq:classical noise QFI vacuum}
    \IQ(\sigma) &= \frac{2 \sigma^2\left(1 + 2 \si_x^2 \right)^2}{\kappa \left(\kappa+1\right)},
    \where
    \kappa &= \sigma ^2+\si_x^2+\si_p^2+2  \si_x^2\left(\sigma ^2+\si_p^2\right)
    .
\end{align}
As long as one of $\si_x$ and $\si_p$ is nonzero, then $\IQ(\si=0)=0$ here. In particular, for the isotropic case $\si_C:=\si_x=\si_p$, the Rayleigh curse arises when the signal is dominated by the classical noise, $\si\ll\si_C$.

Classical noise changes the optimal measurement for the vacuum input-state case, such that number measurement alone does not saturate the QFI anymore. Instead, the optimal measurement is to squeeze $\Si\mapsto\diag{e^{-2r},e^{2r}}\Si$ with $e^{2r}=1+2\si_x^2$ immediately after the encoding process. The squeezing level is chosen such that $\Si=\diag{\frac{1}{2}+\si_x^2,\frac{1}{2}+\si^2+\si_p^2}$ becomes $\diag{\frac{1}{2},\frac{1}{2}+\kappa}$. After this additional squeezing operation, we then perform a number measurement. The gain in the CFI from this additional squeezing operation in the limit of small classical noise compared to direct number measurement is $1+4\sigma_{x}^{2}$, hence it is marginal given small $\sigma_{x}$, but it is still highly favourable compared to quadrature measurement.

In summary, classical noise limits the QFI for small signals when preparing the vacuum state. Unless stated otherwise, we assume henceforth that the classical noise is negligible, i.e.\ that the signal is dominant, $\si\gg\si_C$. This is motivated by certain applications, discussed later in Sec.~\ref{sec:applications}, for which the search for stochastic signals is limited by imperfections from decoherence---quantum backgrounds---rather than classical backgrounds. We will revisit nonzero classical noise and whether its impact can be avoided by preparing different initial states in Sec.~\ref{sec:Classical noise case}.

\section{Optimal initial state}
\label{sec:state}

We now consider the initial state and measurement scheme that comprise the optimal protocol for sensing the signal $\si$ encoded by $\Lambda_\si$ with and without losses. In particular, we want to know whether entangled resources and collective measurements are necessary.

\subsection{Review of channel Quantum Fisher Information}
\label{sec:FI review 2}

Building on Sec.~\ref{sec:FI review 1}, here, we introduce our tools for determining the optimal initial state: the channel QFI and extended channel QFI~\cite{EscherNP11GeneralFramework, KolodynskiNJP13EfficientTools}. 

Given an initial state $\ket{\psi}$ and a channel $\Lambda_\theta$ that encodes a parameter $\theta$ in the final state $\h\rho(\theta)=\Lambda_\theta(\ket{\psi}\bra{\psi})$, then let the QFI with respect to $\theta$ be denoted as $\IQ^{\Lambda_\theta(\ket{\psi}\bra{\psi})}(\theta)$. The \textit{channel QFI (CQFI)} of $\Lambda_\theta$, the QFI from the optimal initial state $\ket\psi$, then is
\begin{align}\label{eq:channel QFI definition}
    \IQ^{\Lambda_\theta, \text{no ancilla}}(\theta) = \sup_{\ket\psi}\IQ^{\Lambda_\theta(\ket{\psi}\bra{\psi})}(\theta),
\end{align}
where we emphasise that no ancilla is allowed when calculating the CQFI. From the convexity of the QFI, it suffices to optimize over pure initial states.

If we prepare an initial state $\ket\Psi$ that is also entangled with some ancilla, e.g.\ two-mode squeezed vacuum (TMSV), then we might improve the estimation of $\theta$. Let the joint channel be $\Lambda_\si\otimes \Lambda_A$ where $\Lambda_A$ is some channel that acts on the ancilla. In the noiseless ancilla or ``perfect storage'' case, $\Lambda_A=1_A$ is the identity. In this ideal case, the CQFI of the joint channel $\Lambda_\si\otimes 1_A$ is called the \textit{extended channel QFI (ECQFI)} of $\Lambda_\si$ and is given by
\begin{align}\label{eq:extended channel QFI definition}
    \IQ^{\Lambda_\theta}(\theta) = \sup_{\ket\Psi}\IQ^{(\Lambda_\theta\otimes 1_A)(\ket{\Psi}\bra{\Psi})}(\theta)
    .
\end{align}
The ECQFI is the maximum amount of information about the parameter $\theta$ that can be extracted after the channel $\Lambda_\theta$ acts on the quantum device. Geometrically, this represents a fundamental speed limit on the statistical change of the final quantum state $\h\rho(\theta)$ with respect to local changes in $\theta$. 
Note that the inequality $\IQ^{\Lambda_\theta}(\theta) \geq \IQ^{\Lambda_\theta, \text{no ancilla}}(\theta)$ between the ECQFI and CQFI always holds. If, as in some cases, $\IQ^{\Lambda_\theta}(\theta)$ is strictly larger than $\IQ^{\Lambda_\theta, \text{no ancilla}}(\theta)$, then entanglement with an ancilla is a required resource for optimal signal extraction~\cite{Demkowicz-DobrzanskiPRL14UsingEntanglement}.

Realistically, however, we expect that $\Lambda_A=\Lambda^\text{loss}_{\eta_A}$ where $\eta_A$ is some ancilla loss from imperfect storage of the ancilla. The CQFI of this joint channel $\Lambda_\si\otimes \Lambda^\text{loss}_{\eta_A}$ then represents the maximum information feasibly available. The key question to ask is then whether this joint channel CQFI can be saturated without using an ancilla. 

When exploring these limits, we can also impose other physical restrictions. For example, since the unconstrained ECQFI $\IQ^{\Lambda_\theta}(\theta)$ might be unbounded at $\theta=0$ in the lossless case, we can constrain the average energy $\langle{\h H}\rangle \propto \ev{\h n} + \frac{1}{2}$ of the initial state of the harmonic oscillator. Let the ECQFI in Eq.~\ref{eq:extended channel QFI definition} constrained to initial states $\ket\Psi$ with $\ev{\h n}=N$ average occupation number per mode be denoted $\IQ^{\Lambda_\theta, N}(\theta)$. (Note that while $\ev{\h n}\leq N$ may be a more natural constraint, the bounds that we find are always non-decreasing in $N$ such that it suffices to consider $\ev{\h n}=N$.) 

We need to calculate the ECQFI of our channel $\Lambda_\si$ with respect to $\si$ to determine the optimal initial state and whether entanglement is a required resource, with and without noise channels on the system and ancilla.

\subsection{Review of deterministic case}
\label{sec:Review of deterministic case}
Before proceeding to our case of a random displacement channel, we briefly review the case of a deterministic displacement channel to establish the similarities and differences between the two cases.

Consider a deterministic displacement of the state $\h\rho$ by $\mu$ along $\h p$ such that we want to estimate $\mu$ from measurements of $\h U_\mu\h\rho\h U_\mu^\dag$ where $\h U_\mu=\exp(i\mu \h x)$. In the lossless case, if $\h\rho=\proj{\psi}$ is pure, then the QFI is $\IQ(\mu) = 4\varSubSmall{\ket{\psi}}{\h x}$. Since $\varSubSmall{\ket{\psi}}{\h x}$ can be made arbitrarily large, the unconstrained ECQFI is unbounded. As such, we constrain the initial state $\ket\psi$ to have $N$ average occupation number per mode, $\ev{\h n}=N$. The maximum value of $\varSub{\ket\psi}{\h x}$ given the constraint of $\ev{\h n}=N$ is attained by an SMSV state, in which case $\varSub{\ket\psi}{\h x}$ is equal to $\xi_N:=N+\frac{1}{2}+\sqrt{N(N+1)}$. Note that $\xi_N\rightarrow2 N$ as $N\rightarrow\infty$. The ECQFI is then
\begin{align}\label{eq:deterministic displacement, lossless ECQFI}
    \IQ^{\h U_\mu, N}(\mu) &= 4\xi_N
    \xrightarrow[N\rightarrow\infty]{} 8 N
\end{align}
which is achieved by preparing an SMSV state and, e.g., measuring the quadrature $\h p$. Another optimal measurement is to instead anti-squeeze after the encoding and then perform a number measurement $\h n$. We emphasise that this asymptotic scaling with $N$ is the fundamental limit for lossless deterministic displacement sensing. The Heisenberg limit of $N^2$ for phase estimation is not possible to achieve for displacements.

Realistically, however, the state will experience some losses such that the total channel is not unitary. Suppose that the state encounters a loss $\eta$ before the encoding such that the total channel is $\Lambda_{\mu}^\text{noisy}(\rho)=\h U_\mu\Lambda^\text{loss}_\eta(\h\rho)\h U_\mu^\dag$. In this lossy case, Ref.~\cite{LatunePRA13QuantumLimit} showed that SMSV is still the optimal initial state and attains the following ECQFI given the constraint of $\ev{\h n}=N$:
\begin{align}\label{eq:deterministic displacement, lossy ECQFI}
    \IQ^{\Lambda_{\mu}^\text{noisy}, N}(\mu) = 
    \frac{4 \xi_N}{\eta  \left(2\xi_N-1\right)+1}
    \xrightarrow[N\rightarrow\infty]{} 
    \frac{2}{\eta}
    .
\end{align}
The high energy limit of the deterministic ECQFI is then bounded by $\frac{2}{\eta}$ which cannot be surpassed using any initial state. For example, this result is known in the LIGO context where, to sense deterministic gravitational waves in the presence of optical losses, it is optimal to inject a squeezed optical state into the ``dark port'' of the Michelson interferometer~\cite{Demkowicz-DobrzanskiPRA13FundamentalQuantum}, as is presently done~\cite{AasiNP13EnhancedSensitivity,TsePRL19QuantumEnhancedAdvanced,McCullerPRL20FrequencyDependentSqueezing}.

\subsection{Lossless case}
\begin{figure}
    \centering
    \includegraphics[width=\columnwidth]{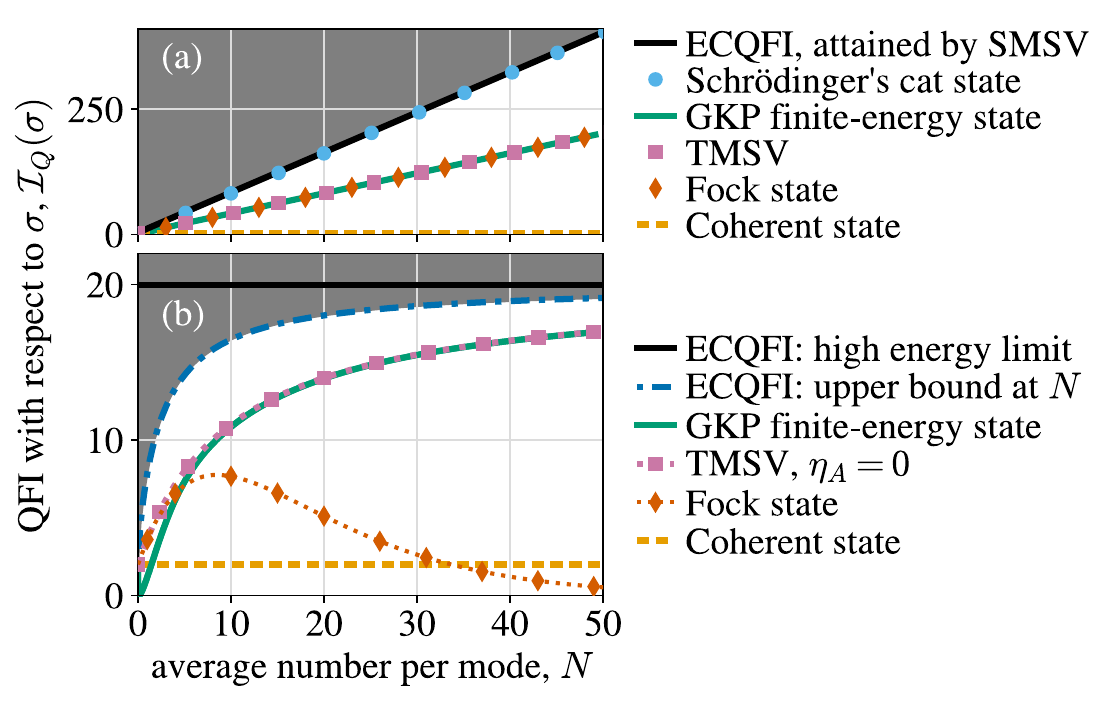}
    \caption{QFI from preparing different initial states, indicated in the legend, versus the initial average occupation number per mode in the (a) lossless case and (b) case of a loss of $\eta=0.1$ occurring before the encoding with $\si=10^{-3}$. We compare the QFI from preparing each initial state to the ultimate precision limit, the extended channel QFI (ECQFI). The shaded grey region is thus inaccessible [but the upper bound at $N$ is loose in (b)].
    In the lossy case, SMSV and TMSV with $\eta_A>0$ exhibit the Rayleigh curse, i.e.\ $\text{QFI}\rightarrow0$ as $\si\rightarrow0$, and the Schr\"odinger's cat state has the same QFI as a coherent state for approximately $N>3$.}
    \label{fig:channelFI}
\end{figure}

We now return to studying the ECQFI of the random displacement channel $\Lambda_\si$ with respect to $\si$. We prove the following claim about the optimal protocol: 
\begin{claim}
\label{claim:lossless QFI UB}

In the lossless case, the ECQFI constrained to initial states with $\ev{\h n}=N$ per mode is
\begin{align}\label{eq:lossless QFI UB}
    \IQ^{\Lambda_\si, N}(\si)=
\frac{4}{2\sigma^{2}+\xi_N^{-1}}
\end{align}
which is saturated by preparing an initial single-mode squeezed vacuum (SMSV) state and performing a measurement such that anti-squeezing is followed by number counting. Entanglement with an ancilla is not a required resource.
\end{claim}
The proof of Claim~\ref{claim:lossless QFI UB} is given in Appendix~\ref{sec:proof of lossless channel QFI}. We use an established technique of bounding the ECQFI using a sequence of purifications of $\Lambda_{\sigma}$ by Uhlmann's Theorem~\cite{EscherNP11GeneralFramework,KolodynskiNJP13EfficientTools,LatunePRA13QuantumLimit}. We find the following upper bound on the ECQFI
\begin{align}
\label{eq:lossless UB}
    \IQ^{\Lambda_\si, N}(\si)\leq\frac{4}{2\sigma^{2}+\var{\h x}^{-1}}
\end{align}
where $\var{\h x}$ is calculated with respect to the initial state. Minimising Eq.~\ref{eq:lossless UB} over the initial state is equivalent to optimizing $\var{\h x}$ given the constraint of $\ev{\h n}=N$, which is attained by an SMSV state. 
By calculating the QFI using Eq.~\ref{eq:Gaussian QFI CM}, we observe that SMSV is optimal for any $\sigma$ and saturates the ECQFI in Eq~\ref{eq:lossless QFI UB}.

Claim~\ref{claim:lossless QFI UB} implies that the behavior of the ECQFI depends on the ratio between $N$ and $\frac{1}{4 \sigma^2}$. For $N\ll\frac{1}{4 \sigma^2}$, $\IQ^{\Lambda_\si, N}(\si)$ grows linearly as $8N$ similarly to Eq.~\ref{eq:deterministic displacement, lossless ECQFI}, but, if $N\gg\frac{1}{4 \sigma^2}$, then $\IQ^{\Lambda_\si, N}(\si)$ converges to $\frac{2}{\sigma^2}$.

We comment on the limit of vanishingly small signals $\si\rightarrow0$ such that the upper bound in Eq.~\ref{eq:lossless UB} becomes $4\var{\h x}$ and is tight for any initial state in this limit~\cite{gorecki2022quantum}. (Curiously, this equals the QFI for deterministic displacements in Sec.~\ref{sec:Review of deterministic case}.) In Fig.~\hyperref[fig:channelFI]{\ref*{fig:channelFI}a}, for this small signal limit of $\si\rightarrow0$, we compare the ECQFI attained by SMSV versus average number $\ev{\h n}$ to the QFI of other states such as coherent, TMSV, Fock, Schr\"odinger's cat, and finite-energy Gottesman-Kitaev-Preskill (GKP) grid states~\cite{GottesmanPRA01EncodingQubit}. (We review GKP finite-energy states in Appendix~\ref{sec:GKP review}.) In the high energy limit of $\ev{\h n}\rightarrow\infty$; cat states also attain the ECQFI of $8 N$; TMSV, GKP finite-energy states, and Fock only grow as $4 N$; and coherent states remain at the vacuum level of $2$ since displacements commute with the encoding channel $\Lambda_\si$. In this limit of $\si\rightarrow0$, an optimal measurement of the final state $\h\rho$ is to project it onto the initial state $\proj{\Psi}$, which can be implemented by an ``echo protocol''~\cite{gilmore2021quantum}. For example, if we prepare SMSV (TMSV), then this measurement means to first perform one-mode (two-mode) anti-squeezing and then a number-resolving measurement.

\subsection{Lossy case}
\begin{figure}
    \centering
    \includegraphics[width=\columnwidth]{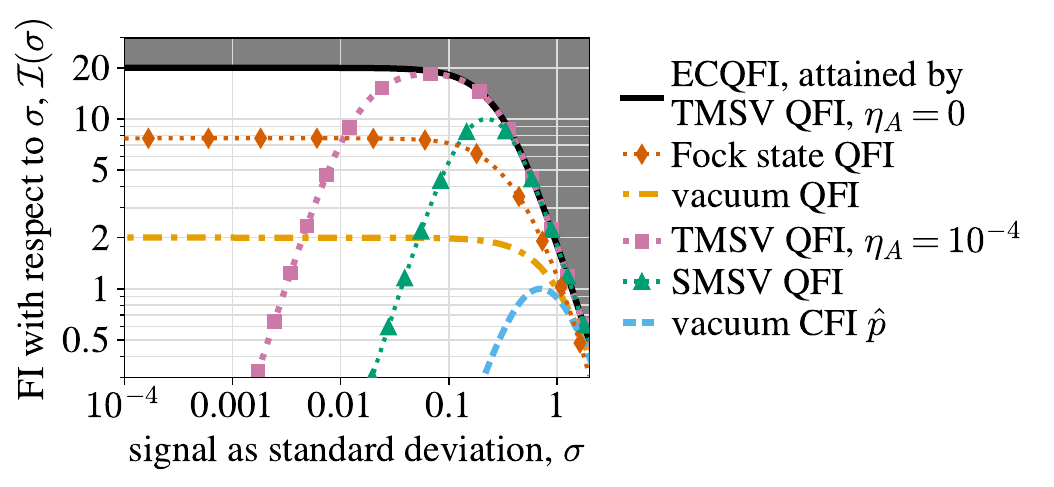}
    \caption{Fisher information (FI) versus the standard deviation $\si$ for different initial states, indicated in the legend, with a loss of $\eta=0.1$ occurring before the encoding. 
    The high energy limit of the ECQFI and the squeezed states is shown. The shaded grey region is inaccessible. For large $\si$, e.g.\ $\si\sim\eta$, we calculate the QFI for the Fock state $\ket 8$ numerically using a truncated Hilbert space of dimension 50.} 
    \label{fig:lossy FI loglog}
\end{figure}

Any actual experiment will experience loss which will dramatically change the ECQFI and optimal initial state compared to the lossless case. Consider using the initial state which is optimal in the lossless case, an SMSV state. Given a loss $\eta$ occurring before the encoding, then the total channel becomes $\Lambda_{\sigma}^\text{noisy}=\Lambda_\si\circ\Lambda^\text{loss}_\eta$ where we assume that the classical noise is negligible. By Eq.~\ref{eq:Gaussian QFI CM}, the QFI for SMSV in the high energy limit is 
\begin{align}\label{eq:SMSV QFI high energy limit}
    \IQ(\si) = \frac{8 \sigma ^2}{\left(\eta +2 \sigma ^2\right)^2}.
\end{align}
In the small loss limit, $\eta \ll \sigma^{2}$, the QFI is approximately $\frac{2}{\eta+\sigma^{2}}$, which will be shown to be optimal at any $\si^2$. 
In the experimentally relevant, loss-dominated regime of $0<\si^2\ll\eta$, SMSV performs poorly. The QFI for SMSV vanishes as $\sigma \rightarrow 0$: it suffers from the Rayleigh curse and performs even worse than vacuum, as shown in Fig.~\ref{fig:lossy FI loglog}. This raises the question: What is the ECQFI in the presence of significant loss, $\eta\gg\si^2$, and which initial states saturate this bound?

We prove the following claim about the optimal protocol in the lossy case:
\begin{claim}
\label{claim:lossy QFI UB}
The ECQFI given a loss $\eta$ before the encoding is given by
\begin{align}\label{eq:lossy ECQFI}
    \IQ^{\Lambda_{\sigma}^\text{noisy}}(\si) =\frac{2}{\eta+\sigma^{2}},
\end{align}
which is attained by preparing a TMSV state with perfect storage $\eta_A=0$ in the high energy limit of $N \rightarrow \infty$.
\end{claim}

The proof of Claim~\ref{claim:lossy QFI UB} is given in Appendix~\ref{sec:proof of lossy UB} where we use the same method as the lossless case but with different purifications, inspired by the deterministic case~\cite{LatunePRA13QuantumLimit}. 

We have also determined the following upper bound on the ECQFI for a given finite $\ev{\h n}=N$:
\begin{align} \label{eq:lossy UB constrained}
    \IQ^{\Lambda_{\sigma}^\text{noisy}, N} &<
    \frac{4}{2(\eta + \si^2) + (1-\eta)\xi_N^{-1}}, 
\end{align}
but this upper bound is not tight for $\eta>0$ and a fixed finite $N$. For example, TMSV with perfect storage $\eta_A=0$, does not saturate this upper bound for finite $N$ as shown in Fig.~\hyperref[fig:channelFI]{\ref*{fig:channelFI}b}.

For finite $\si>0$, if the loss $\eta_A$ on the ancilla mode before the encoding is small $\eta_A\ll\si^2,\eta$, then TMSV saturates the ECQFI in Eq.~\ref{eq:lossy ECQFI} in the high-energy limit of $N\rightarrow\infty$ as shown in Appendix~\ref{sec:TMSV results}. Experimentally, however, this requirement of near-perfect storage of the ancilla mode is likely too stringent to probe the small signals of interest. In the realistic regime of $\si^2\ll\eta,\eta_A$, TMSV does not saturate the ECQFI.

In the $\si\rightarrow0$ limit, all squeezed Gaussian states suffer the Rayleigh curse by Claim~\ref{claim:simplectic eigenvalues}, provided that loss occurs on every mode (i.e.\ with fixed $\eta,\eta_A>0$). The only Gaussian states with non-vanishing QFI are then the coherent states which still have a QFI of $\frac{2}{1+\si^2}$ since they remain coherent after the loss before the encoding. This raises the question of whether it is possible to attain the ECQFI in the realistic regime of $\si^2\ll\eta,\eta_A$ by using non-Gaussian states. We address this question in Sec.~\ref{sec:non-Gaussian states} below and show numerically that indeed it appears to be possible.

\subsection{Limit of small signals}
\label{sec:small signals}
Before moving to discuss our numerical results with non-Gaussian states, let us gain some more understanding of the small signal $\sigma \rightarrow 0$ limit. 
In the lossless case, we already observed that the QFI in this limit is $4\var{\h x}$~\cite{gorecki2022quantum}. Here, we want to understand the lossy case where the state is mixed before the channel is applied. 
In Appendix~\ref{sec:small signal generalisations}, we prove the following general claim
\begin{claim}
\label{claim:small signal limit}
Given an initial state $\h\rho$ and the random unitary channel
\begin{align}\label{eq:random unitary channel}
    \Lambda_{\sigma}(\h\rho)=\intginf{\theta} p\left(\theta\right)\h U_\theta \h\rho \h U_\theta^\dag
\end{align}
where $p(\theta)\sim\mathcal{N}(0,\si^2)$ and $\h U_\theta$ is unitary, then the QFI is
\begin{align*}
    \IQ(\si=0)=4\langle{\h H\h\Pi_\perp\h H}\rangle,
\end{align*}
where $\h H = i\h U^\dag_0 \dot{\h U}_0$ is the Hermitian generator at $\si=0$, $\h\Pi_\perp$ is the projection operator onto the null space of $\h\rho$, and the expectation value is calculated with respect to $\h\rho$, i.e.\ $\evSmall{\h O}=\trSmall{\h\rho\h O}$. An optimal measurement that attains the QFI is measurement of $\h U_0^\dag\h\Pi_\perp\h U_0$.
\end{claim}
This result means that, to obtain information about $\sigma$ in the limit of $\si\rightarrow0$, $\h H$ needs to map some of $\hat \rho$ into its null space. In particular, if $\h\rho$ is full rank, then it suffers the Rayleigh curse. 

For our random displacement channel, given a pure initial state $\ket\psi$, then $\IQ(\si=0)=4\langle{\h x\h\Pi_\perp\h x}\rangle$ where $\h\rho=\Lambda^\text{loss}_\eta(\ket\psi\bra\psi)$ is the state after the loss but before the encoding. This result is useful when $\IQ(\si=0)$ is nonzero. For example, it implies that, for a Fock state $\ket N$, the QFI is $\IQ(\si=0)=2(1-\eta)^N(N+1)$ and is attained by number measurement.

What is the optimal initial state in this limit of $\sigma\rightarrow 0$? The ECQFI is not well defined unless we specify the order of limits since, by Eq.~\ref{eq:extended channel QFI definition},
\begin{align*}
    \lim_{\sigma\rightarrow0} \IQ^{\Lambda_\theta}(\theta)
    \neq \sup_{\ket\Psi}\lim_{\sigma\rightarrow0}\IQ^{(\Lambda_\theta\otimes 1_A)(\ket{\Psi}\bra{\Psi})}(\theta).
\end{align*}
Nevertheless, we claim that $\lim_{\sigma\rightarrow0} \IQ^{\Lambda_\theta}(\theta)$ is the relevant quantity of interest since, in practice, the signal is small but finite $0<\si\ll1$ and we can only search for signals above the classical noise floor. In the following subsection, we discuss a family of initial states that are numerically optimal for a fixed small but finite $0<\si\ll1$.

An open question about this limit of $\si\rightarrow0$ is whether the only single-mode pure initial states that are finite rank after the loss $\Lambda^\text{loss}_\eta$ are either finite superpositions of coherent states or bounded in the Fock basis.

We discuss a generalisation of Claim~\ref{claim:small signal limit} in Appendix~\ref{sec:lindbladian}. Also, in Appendix~\ref{sec:finite dimensions}, we give an example of how the upper bound on the ECQFI which is analogous to Eq.~\ref{eq:lossless QFI UB} can be loose for random unitary channels acting on finite-dimensional systems.

\subsection{Non-Gaussian states}
\label{sec:non-Gaussian states}

We now explore whether there exist non-Gaussian states of the probe that can outperform the Gaussian states and saturate the ECQFI in Eq.~\ref{eq:lossy ECQFI} in the relevant high-loss regime $\si^2\ll\eta,\eta_A$.

We start by analysing the QFI with Fock states. As shown above, the QFI with a Fock state $\ket{N}$ is $\IQ(\si=0)=2(1-\eta)^N(N+1)$. The optimal Fock state $\ket N$ is thus $N\approx-\frac{1}{\log (1-\eta )}-1$, where the Bose enhancement factor $(N+1)$ balances with the loss factor $(1-\eta)^N$ to achieve a QFI of $\IQ(\si=0)\approx\frac{2}{e(1-\eta)\log[(1-\eta)^])}$. In the limit of small $\eta$, $\IQ(\si=0)\approx \frac{2}{e\eta}$, which misses the ECQFI by roughly a factor of $e$, or a penalty of 4.3~dB. 
This is shown in Fig.~\hyperref[fig:channelFI]{\ref*{fig:channelFI}b} for $\eta=0.1$, for which the optimal Fock QFI is $\IQ(\si=0)=7.75$ at $N=8,9$, while the ECQFI is $\frac{2}{\eta}=20$. The key takeaway is that for $\si^2\ll\eta,\eta_A$, Fock states outperform all Gaussian states but do not attain the ECQFI in Eq.~\ref{eq:lossy ECQFI}. 

To attain the ECQFI for $0<\si^2\ll\eta,\eta_A$, we now consider GKP finite-energy states. We review these states in Appendix~\ref{sec:GKP review}, in particular, the $\ket{\text{GKP}_\Delta}$ family of finite-energy states given by a superposition of displaced SMSV states of width $\propto\Delta$ within a Gaussian window of width $\propto\Delta^{-1}$. To calculate the QFI with respect to $\si$ numerically, we model $\ket{\text{GKP}_\Delta}$ in the Fock basis of a truncated Hilbert space. For the lossless case and $\Delta\ll1$, the overall variance of the pure initial state $\ket{\text{GKP}_\Delta}$ is $\var{\h x}=\var{\h p}=\frac{1}{2}\Delta^{-2}$ which equals the average occupation number $\ev{\h n} = N$~\cite{ZhuangNJP20DistributedQuantum} 
such that the QFI for small $\si$ is $4 N$ as shown in Fig.~\hyperref[fig:channelFI]{\ref*{fig:channelFI}a}. A non-Gaussian measurement, which we determine numerically for $\si>0$, is once again required to attain the QFI. (This measurement is more complicated than direct number measurement and it would be interesting to understand what it represents physically.) 
For the lossy case, e.g.\ a loss of $\eta=0.1$ occurring before the encoding, the QFI from preparing GKP finite-energy states $\ket{\text{GKP}_\Delta}$ converges numerically to the QFI of TMSV with perfect storage $\eta_A=0$ for $N>10$, i.e.\ $\Delta<0.2$, as shown in Fig.~\hyperref[fig:channelFI]{\ref*{fig:channelFI}b}. At higher energies, then the QFI increases towards the ECQFI. For example, with $\eta=0.1$ and $\si=10^{-3}$, then $\ket{\text{GKP}_\Delta}$ attains a QFI above 19, within 95\% of the ECQFI of 20, using states with $\ev{\h n}=N>100$ in a truncated Hilbert space of dimension $\order{1000}$. Note that we have only considered the small signal $\si$ behaviour here, since it is the most relevant regime, and not the large $\si$ behaviour shown in Fig.~\ref{fig:lossy FI loglog} for the other initial states. 

We conjecture that preparing GKP finite-energy states of higher average number, which will require more peaks and a larger truncated Hilbert space, can get arbitrarily close to the ECQFI for any fixed $\si^2\ll\eta$. We expect the convergence to be slow given that TMSV with perfect storage $\eta_A=0$ only converges asymptotically. This conjecture is based on the above numerics and, heuristically, the connection between the GKP and TMSV infinite-energy states discussed below. It also would be interesting to understand the performance in the limit of $\si\rightarrow0$.

Moreover, we conjecture that the ECQFI for a fixed finite $\ev{\h n}=N$ is saturated by preparing TMSV with perfect storage for all $N$. Numerically, we have searched for different non-Gaussian single-mode states that perform better than TMSV with perfect storage with the same large $\ev{\h n} = N$ but have not found any. We describe our numerical methods in Appendix~\ref{sec:Biconvex optimisation}. Briefly, we found that sparse superpositions of finitely many Fock states also outperform Fock states and approach the ECQFI at high energies. For example, we found a sparse state $\ket{\psi_\text{num.}}=\sum_{j=0}^{23}c_j\ket{20j}$ with $\ev{\h n}=158.9$ and a QFI of 18.4, within 9\% of the ECQFI of 20 for $\si^2=10^{-6}$ and $\eta=0.1$. Intuitively, e.g., the signal trajectory from $\ket{0}$ to $\ket{1}$ dominates the loss trajectory from $\ket{20}$ to $\ket{1}$ for finite signals $\si^2=10^{-6}$ and $\eta=0.1$. These sparse states are similar to optimised binomial quantum error-correcting codes~\cite{MichaelPRX16NewClass}.

Finally, we remark that the high numerical performance of the GKP finite-energy states for sensing a random displacement in the presence of loss is intriguing, as the GKP infinite-energy state was originally designed for the correction of random displacement noise~\cite{GottesmanPRA01EncodingQubit}. Both the GKP infinite-energy state and the TMSV infinite-energy state can be used to form error correction codes that are sensitive to random displacement signals along $\hat{p}$ yet protected against random displacement noise along $\hat{x}$. (We discuss this further in Appendix~\ref{sec:TMSV results}.) In the next subsection, we will show that this property makes these states resilient to classical noise along $\hat{x}$. It would be interesting to understand how this property is related to their performance in the presence of loss.

\subsection{Classical noise case}
\label{sec:Classical noise case}
We briefly address the case of significant classical noise such that the total channel is $\Lambda_{\sigma}^\text{noisy}=\Lambda_\si\circ\Lambda^\text{noise}_{\Si_C}$ for a given classical noise matrix $\Si_C$ in Eq.~\ref{eq:classical noise channel Kraus representation}. Losses are not present unless otherwise noted. In Appendix~\ref{sec:Optimal initial states with classical noise}, we prove that the ECQFI and optimal initial state depend on $\Si_C$ as follows. 

Firstly, suppose that the classical noise is confined to the same quadrature, $\h p$, as the signal, i.e.\ $\Si_C=\diag{0,\si_p^2}$ with $\si_p>0$. Then, the Rayleigh curse is unavoidable for $\si\ll\si_p$ and the noiseless ECQFI cannot be recovered since there is no way to distinguish the signal and noise. For a fixed finite $\si>0$, the ECQFI is
\begin{align}\label{eq:ECQFI CN parallel}
    \IQ^{\Lambda^\text{noisy}_\si, N}\left(\si\right) = \frac{4\sigma^{2}}{(\sigma^{2}+\sigma_{p}^{2})\left[2(\si^2+\si_p^2) + \xi_N^{-1}\right]},
\end{align}
which reduces to Eq.~\ref{eq:classical noise QFI vacuum} in the vacuum case ($N{=}0$) with $\si_x=0$. The ECQFI in Eq.~\ref{eq:ECQFI CN parallel} is attained for a given $\ev{\h n} = N$ by preparing the appropriate SMSV state and performing the noiseless optimal measurement: anti-squeezing followed by a number measurement. If a loss $\eta$ is also present, then in the high energy limit the ECQFI becomes
\begin{align}\label{eq:clnoise with loss}
    \IQ^{\Lambda^\text{noisy}_\si}\left(\si\right)
    &= \frac{2\si^2}{(\eta + \si^2 + \si^2_p)(\si^2 + \si^2_p)},
\end{align}
such that, in the experimentally relevant regime of $\sigma^2 \ll \sigma_p^2 \ll \eta$, the ECQFI approaches $\frac{2\si^2}{\eta\si_p^2}$ which exhibits the Rayleigh curse. In comparison, the QFI from preparing an SMSV state in the high energy limit is
\begin{align*}
    \IQ(\si) = \frac{8 \sigma ^2}{\left[\eta +2 \left(\sigma ^2+\sigma_p^2\right)\right]^2}.
\end{align*}
The ratio of the ECQFI to the QFI from preparing an SMSV state is thus $\frac{\left(\eta+2\tilde{\sigma}^{2}\right)^2}{4\left(\eta+\tilde{\sigma}^{2}\right)\tilde{\sigma}^{2}}$ which is a function of $\eta/\tilde\si^2$, where $\tilde{\sigma}^{2}:=\sigma^{2}+\sigma_{p}^{2}$. If $\eta \ll \tilde{\sigma}^{2}$, then this ratio approaches one and preparing SMSV is close to optimal. In the above experimentally relevant regime of $\sigma^2 \ll \sigma_p^2 \ll \eta$, however, then this ratio approaches $\frac{\eta}{4\sigma_{p}^{2}}$. If this ratio is large, i.e.\ the classical noise is sufficiently small compared to the loss, then preparing TMSV with perfect storage of the ancilla or preparing non-Gaussian states can outperform SMSV.

Secondly, suppose that the classical noise is confined to the quadrature opposite to the signal, i.e.\ $\Si_C=\diag{\si_x^2,0}$ with $\si_x>0$. The effect of this opposite quadrature noise is fundamentally different from the effect of noise in the same quadrature as the signal; unlike the same quadrature case, noise in the opposite quadrature can be overcome in the limit of large $N$ and does not necessarily lead to a Rayleigh curse. 
For fixed finite values of $\si,\si_x>0$, the noiseless ECQFI in Eq.~\ref{eq:lossless QFI UB} can be recovered by preparing an SMSV state in the high energy limit of $N\rightarrow\infty$. Intuitively, the classical noise can be squeezed after the encoding while simultaneously anti-squeezing the signal. 
A given SMSV state with fixed finite $\ev{\h n} = N$, however, still exhibits the Rayleigh curse as $\si\rightarrow0$ such that SMSV may not be optimal for a given $\ev{\h n} = N$ and $\si$. In comparison, there exist finite-energy states, e.g.\ TMSV with perfect storage, that do not exhibit the Rayleigh curse and which recover their respective noiseless QFI in the high energy limit, e.g.\ $\lim_{N\rightarrow\infty}\lim_{\si\rightarrow0}\IQ(\si)=4 N$ for TMSV.

Finally, suppose that the classical noise appears in both quadratures, i.e.\ $\Si_C=\diag{\si_x^2,\si_p^2}$ with $\si_x,\si_p>0$. For example, suppose that the classical noise is isotropic with $\Si_C=\diag{\si_C^2,\si_C^2}$ and $\si_C>0$. Then, the Rayleigh curse is unavoidable as $\si\rightarrow0$ with $\si_C$ fixed, and the noiseless ECQFI cannot be recovered. The ECQFI with $\si_x=0$ in Eq.~\ref{eq:ECQFI CN parallel}, however, can be recovered for a fixed finite $\si>0$ by preparing SMSV in the high energy limit despite $\si_x>0$. In the isotropic case, this means simply replacing $\si_p$ by $\si_C$ in Eq.~\ref{eq:ECQFI CN parallel}.

\section{Simultaneous estimation of the mean and variance}
\label{sec:simultaneous estimation}
\begin{figure}
    \centering
    \includegraphics[width=\columnwidth]{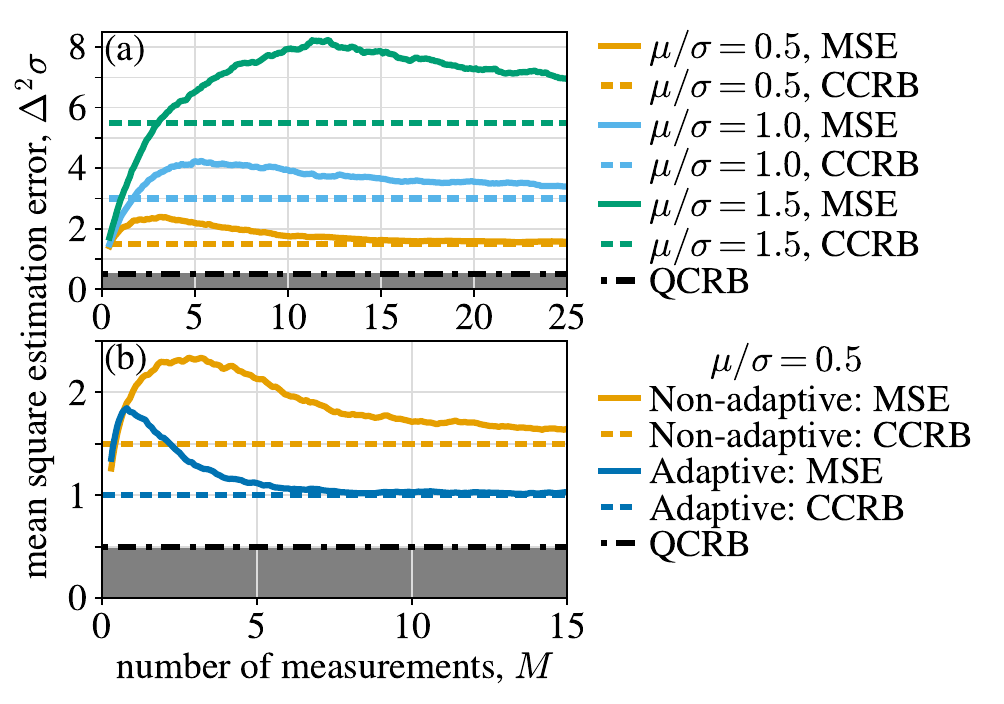} 
    \caption{Numerical mean squared estimation error compared to the Cram\'er-Rao bounds (CRBs) for simultaneous estimation of $\mu$ and $\si$ versus the number of measurements $M$. The shaded grey region is inaccessible with the input state fixed as the vacuum state. (a) Non-adaptive separate measurements scheme with $\sigma=0.05$ and different ratios of $\mu/\sigma$. (b) Adaptive separate measurements are more precise than the non-adaptive strategy with $\sigma=0.05$ and $\mu/\sigma=0.5$, but neither reach the QCRB asymptotically as $M\rightarrow\infty$.}
    \label{fig:simulEst}
\end{figure}

We have focused so far on estimating the variance of a random process. In contrast, most work on quantum metrology up to now has been dedicated to estimating the mean value of a signal rather than its variance. Formally, estimating the deterministic encoding of a signal is equivalent to mean estimation in the zero variance limit. Here, we unify these two efforts as we consider the optimal simultaneous estimation of the mean and the variance. We emphasise that we only consider the case where the initial state is vacuum such that any loss occurring before the encoding can be ignored.

\subsection{Review of quantum multi-parameter estimation}
\label{sec:FI review 3}

We now introduce the tools of quantum multi-parameter estimation, in contrast to single-parameter estimation as reviewed in Sec.~\ref{sec:FI review 1}.
In the multi-parameter case, a vector of real parameters $\vec{\theta}$ is encoded in a quantum state $\h\rho(\vec{\theta})$. Given $M$ measurements of the probability distribution  $p(x|\vec{\theta})$, the CCRB provides a tight lower bound on the covariance matrix, $\Sigma_{\vec{\theta}}$, of the estimators of $\vec{\theta}$ as $\Sigma_{\vec{\theta}}\geq\frac{1}{M}[\IC(\vec{\theta})]^{-1}$, where matrix inequalities $A\geq B$ are interpreted henceforth as $A-B$ being positive semi-definite and the classical Fisher information matrix (CFIM) is 
\begin{align*}
    \IC(\vec{\theta})_{i,j}=\intginf{x}\frac{\partial_{\theta_{i}}p(x|\vec{\theta})\partial_{\theta_{j}}p(x|\vec{\theta})}{p(x|\vec{\theta})}
.
\end{align*}
As in the single parameter case, the CCRB is asymptotically saturable with maximal likelihood estimators and the CFIMs of two independent observations sum: If $p(x|\vec{\theta})$ and $p(y|\vec{\theta})$ are independent distributions, then the total CFIM from observing one outcome from each is the sum of the two CFIMs.

The QCRB provides a further lower bound on the covariance matrix of the estimators given any measurement strategy,
\begin{align*}
    \Sigma_{\vec{\theta}}\geq\frac{1}{M}[\IC(\vec{\theta})]^{-1}\geq\frac{1}{M}[\IQ(\vec{\theta})]^{-1}.
\end{align*}
Here, the quantum Fisher information matrix (QFIM) is defined as
\begin{align*}
    [\IQ(\vec{\theta})]_{ij}=\frac{1}{2}\text{Tr}[\h\rho\{\h L_i,\h L_j\}]
\end{align*}
where the Hermitian operator $\h L_i$ is the symmetric logarithmic derivative (SLD) of $\h \rho$ with respect to $\theta_i$ which is defined implicitly by the equation $\partial_{\theta_i}\h\rho = \frac{1}{2}\{\h L_i,\h \rho\}$. Given the spectral decomposition of $\h\rho=\sum_j p_j \ket{\phi_j}\bra{\phi_j}$, the SLDs are given by
\begin{align*}
   \h L_i&=\sum_{k,l}\frac{2\braopket{\phi_{k}}{\partial_{\theta_{i}}\h\rho}{\phi_{l}}}{p_{k}+p_{l}} \ket{\phi_{k}}\bra{\phi_{l}}   
\end{align*}
where the sum runs over only $k,l$ such that $p_k+p_l>0$. Similarly to the single-parameter case, the QFIM for a product state $\h\rho_1(\theta)\otimes\h\rho_2(\theta)$ is simply the sum of the individual QFIMs. 

Unlike the single parameter case, however, this bound is not saturable a priori. A necessary and sufficient condition for the asymptotic saturability of the QFIM is the weak commutativity condition~\cite{monras2011measurement}
\begin{align}\label{eq:weak commutativity condition}
    \tr{\h \rho [\h L_i, \h L_j]} = 0, \quad \forall i,j.
\end{align}
If this weak commutativity condition holds, then a joint (i.e.\ collective) measurement of many copies of $\hat{\rho}$ may be necessary to asymptotically saturate the QFIM. (If this condition does not hold, then the Holevo CRB instead would be required to find a tight bound~\cite{Holevo2011book,gardner2024achieving}.)

\subsection{QCRB for simultaneous estimation}

We now consider the problem of estimating
$\mu$ and $\si$ from the encoding in Eq.~\ref{eq:encoding} but with $p(\alpha)\sim \mathcal{N}(\mu, \si^2)$. We assume that the initial state is the vacuum state such that it becomes the displaced squeezed thermal state with $\vec{\mu}=(0,\mu)$ and $\Si=\diag{\frac{1}{2}, \frac{1}{2}+\si^2}$. 
The QFIM about parameters $\vec{\theta}$ encoded in a single-mode Gaussian state is given by a generalization of Eq.~\ref{eq:Gaussian QFI CM}~\cite{pinel2013quantum}
\begin{align}\label{eq:Gaussian-QFIM}
    [\IQ(\vec\theta)]_{jk} &= \frac{\tr{\Si^{-1}[\partial_{\theta_j}\Si]\Si^{-1}[\partial_{\theta_k}\Si]}}{2(1+\gamma^2)}+\frac{2[\partial_{\theta_j} \gamma][\partial_{\theta_k} \gamma]}{1-\gamma^4}
    \nonumber
    \\&+ [\partial_{\theta_j}\vec\mu]^\T\Si^{-1}[\partial_{\theta_k}\vec\mu].
\end{align}
Let $\vec{\theta}=(\mu,\si)^\T$, then the QFIM for the vacuum state is
\begin{align}\label{eq:QFIM simul est}
    \IQ(\mu,\si) &= \bmatrixByJames{ \frac{2}{1 + 2 \sigma ^2} & 0 \\
    0 & \frac{2}{1 + \sigma ^2} \\}
\end{align}
such that the QCRB (per measurement) is $\Delta^{2}\mu=\Delta^2\sigma=\frac{1}{2}$ for $\si\ll1$.

This bound is the best we could expect for: the estimation variances of $\sigma$ and $\mu$ given by it are the same as in each respective single parameter case. If the QCRB is tight, therefore, we can estimate the two parameters simultaneously with a precision for each that is identical to its respective single parameter case.

We check the saturability of the QCRB using the weak commutativity condition in Eq.~\ref{eq:weak commutativity condition}. By calculating the SLDs for the Gaussian channel~\cite{monras2013phase,gao2014bounds}, we find that Eq.~\ref{eq:weak commutativity condition} holds for the vacuum case. Our remaining task is to determine the optimal asymptotic measurement scheme that saturates the QCRB and whether it must be a joint measurement acting collectively on multiple copies of the state.

\subsection{Separate measurements}
For the vacuum case, while a number-resolving measurement of $\h n$ is optimal for estimating $\si$ alone (in the $\mu=0$ case) and $\mu$ alone (in the $\si=0$ case), it is inefficient for estimating the two parameters simultaneously. Analogously to Eq.~\ref{eq:number prob dist}, the probability of a number measurement outcome $n$ is
\begin{align*}
    p(n)
    &= \intginf{\alpha} \abs{\braket{n}{\alpha'}}^2p(\alpha)
    \\&=
    \frac{
        e^{
            \frac{-\mu^2}{2(\si^2+1)}
        }
        \si^{2n}
    }{
        n!\left(\si^2+1\right)^{n+\frac{1}{2}}
    }
    \biggl[
    \mu^{2n}
    \frac{
        (\si^{2n}-[4(\si^2+1)]^n)
    }{
        2^{n}\si^{4n}(\si^2+1)^{n}
    }
    \\&+
    n! \;\mathcal{L}^{(-\frac{1}{2})}_n\left(-\frac{2\mu^2}{4 \si^2(\si^2+1)}\right)
    \biggr]
\end{align*}
where $\mathcal{L}^{(\alpha)}_n(x)$ is the generalized Laguerre polynomial. 
In the relevant limit of $\mu,\si \ll 1$, $p(0)\approx 1-a$ and $p(1)\approx a$ where $a=\frac{1}{2}(\mu^2+\si^2)$. 
These probabilities are degenerate with respect to $\mu$ and $\si$ such that only $\sqrt{\mu^{2}+\sigma^{2}}$ can be estimated and $\mu$ can not be distinguished from $\sigma$. 
Indeed, the CFIM for a number measurement is 
\begin{align}\label{eq:simul est number CFIM}
    \IC^{\h n}(\mu,\si) &= \frac{2}{\left(\mu ^2+\si ^2\right)\left(1-\mu ^2+\si ^2\right)}\bmatrixByJames{
    \mu ^2 & \mu  \si  \\
    \mu  \si  & \si ^2 \\}
\end{align}
which is singular, and the only non-zero eigenvalue corresponds to the parameter $\sqrt{\mu^2+\si^2}$.

For simultaneous estimation of $\mu$ and $\si$, therefore, we need to add a measurement that will break this degeneracy. This can be accomplished by adding a quadrature measurement. In the limit of $\mu, \si \ll 1$, quadrature measurement is optimal for estimating $\mu$ yet provides no information about $\si$ since it is Rayleigh cursed, as can be observed from the CFIM 
\begin{align*}
    \IC^{\h p}(\mu,\si) = \bmatrixByJames{\frac{2}{1+2\si^2} & 0 \\ 0 & \frac{8\si^2}{\left(1+2\si^2\right)^2}}
    \xrightarrow[\si\rightarrow0]{}\left(\begin{array}{cc}
2 & 0\\
0 & 0
\end{array}\right).
\end{align*}
Performing many combinations of the two measurements, e.g.\ such that one-half of them are number measurements and the other half are quadrature measurements, should thus allow for simultaneous estimation. For this protocol, the CCRB per measurement is 
\begin{align*}
    \Sigma=\left(\frac{1}{2}\IC^{\h n}+\frac{1}{2}\IC^{\h p}\right)^{-1}=\left(\begin{array}{cc}
    1 & -\frac{\mu}{\sigma}\\
    -\frac{\mu}{2} & 1+\frac{2\mu^{2}}{\sigma^{2}}
    \end{array}\right) 
\end{align*}
such that the minimum mean square errors are $\Delta^{2}\mu=1$ and $\Delta^{2}\sigma=1+\frac{2\mu^{2}}{\sigma^{2}}$.  The QCRB per measurement of $\Delta^{2}\mu=\Delta^{2}\si=\frac{1}{2}$, therefore, is not saturated by this measurement strategy.

Intuitively, $\h p$ provides an optimal measurement of $\mu$ and $\h n$ provides an optimal measurement of $\si$ for a sufficiently small $\mu$. These observables, however, only weakly commute, since $\ev{[\h p, \h n]}=-i\ev{\h x}=0$, such that these measurements cannot be performed simultaneously. Since only one of them is performed each time, this accounts for the factor of two gap from the QCRB. Although $\Delta^2 \si$ suffers an additional uncertainty of $\frac{2\mu^{2}}{\sigma^{2}}$, this can be suppressed asymptotically by reducing $\mu$ adaptively, i.e.\ by estimating $\mu$ and then nulling (displacing the state back towards the vacuum using the estimate of $\mu$) before the number measurement. The adaptive protocol is described further in Appendix~\ref{sec:Adaptive protocol}.

Numerical results of the adaptive and non-adaptive schemes are shown in Fig.~\ref{fig:simulEst}, where we plot the mean squared error in estimating $\sigma$. Denoting the $M$ outcomes of the quadrature measurement and number measurement as $\left\{ p_{i}\right\} _{i=1}^{M}$ and $\left\{ n_{i}\right\} _{i=1}^{M}$ respectively, the estimator of $\mu$ is $\tilde{\mu}=\frac{1}{M}\sum_{i}p_{i}$, and the estimator of $\sigma$ is $\tilde{\sigma}=\sqrt{\frac{2}{M}\sum_{i}n_{i}-\tilde{\mu}^{2}}$. The distribution of $\tilde{\sigma}$ is not Gaussian, and hence it only converges to the CCRB of $\Delta^2\si=1$ asymptotically as shown in Fig.~\ref{fig:simulEst}.

\subsection{Joint measurements}\label{sec:joint measurements}
Suppose that we perform joint measurements on the $M$ copies of the final state $\h \rho=\Lambda_\si(\ket0\bra0)$. This collective state $\h \rho^{\otimes M}$ is an $M$-mode Gaussian state with $2M$-by-1 mean vector $\vec{\mu} = (0, \mu, 0, \mu, \mathellipsis, 0, \mu)^\T$ and $2M$-by-$2M$ covariance matrix $\Si = \bigoplus_{j=1}^M\diag{\frac{1}{2}, \frac{1}{2}+ \si^2}$ in the basis $(\h x_1, \h p_1, \mathellipsis, \h x_M, \h p_M)^\T$. This $M$-mode state can be transformed to the (anti)symmetric basis such that $\vec{\mu} = (\sqrt{M}\mu, 0, 0,0,\mathellipsis, 0, 0)^\T$, where the symmetric mode is listed first, and $\Si$ is unchanged. We emphasise that the elements of this (anti)symmetric basis are orthogonal and commute. This shows that measuring the $\h p$ quadrature of the symmetric mode is a sufficient statistic for $\mu$. The remaining $M-1$ anti-symmetric modes can then be used to measure $\si$ performing a number measurement on each mode. Absorbing the $M$ factors from the CRBs, the CFIM for this joint measurement is $2\,\diag{M, M-1}$ which saturates the QFIM in Eq.~\ref{eq:QFIM simul est} of $2M\diag{1,1}$ in the asymptotic limit of $M\rightarrow\infty$. 

It is therefore possible to saturate the QFIM if we can perform joint measurements on $M$ copies of the final state. These $M$ copies can correspond to different experiments distributed in space or time. For experiments distributed in space, the required transformation to symmetric and anti-symmetric modes could be done using a sequence of beamsplitter unitaries. While, for experiments distributed in time, the transformation may be done using quantum memories, as further discussed in Sec.~\ref{sec:implementations}. 

This protocol is an instance of the following general statement for any quantum system:
\begin{claim}
\label{claim:joint measurement}
Given an initial pure state $\ket{\psi}$ and the random unitary channel
\begin{align}
    \label{eq:random unitary mean and variance}
    \Lambda_{\sigma,\mu}(\ket\psi\bra\psi)=\intginf{\theta} p\left(\theta\right)\h U_\theta\ket\psi\bra\psi \h U_\theta^\dag
\end{align}
where $p(\theta)\sim\mathcal{N}(\mu,\si^2)$ and $\h U_\theta=\exp(-i\theta \h H)$, 
then the QFI with respect to either $\mu\rightarrow0$ or $\sigma\rightarrow0$ equals $4\varSmall{\h H}$ and the QFIM is simultaneously saturable with a joint measurement in the asymptotic limit of $M\rightarrow\infty$. 
\end{claim}

The proof of Claim~\ref{claim:joint measurement} is given in Appendix~\ref{sec:claim_joint_meas}. The key idea is the same as in our particular case above: while the parameter $\mu$ displaces the state in the symmetric subspace of the $M$ copies, $\sigma$ takes the state out of the symmetric subspace. The QFI with respect to $\mu$ is saturated by measuring a suitable basis in the symmetric subspace, while the QFI with respect to $\sigma$ is saturated by measuring the projection onto the anti-symmetric subspace. A similar protocol was found recently in the context of estimating phase diffusion in qubits~\cite{len2022multiparameter}. 

Lastly, we remark that preparing different initial states could improve the joint measurement of the mean and variance, e.g., in the presence of loss the symmetric mode could be prepared in an SMSV state to improve the measurement of the mean $\mu$ and the anti-symmetric modes could be prepared in non-Gaussian states to improve the measurement of the standard deviation $\si$.

\section{Stochastic waveform estimation, implementation, and application}
\label{sec:stochastic waveform estimation}

We now solve the continuous estimation problem of estimating $S_{yy}(\Om)$ at each frequency $\Om$ that we introduced at the start of this work. By the chain rule from $\si$ to $S_{yy}(\Om)$ in Eq.~\ref{eq:sigma}, the QFI with respect to $S_{yy}(\Om)$ is
\begin{align}\label{eq:QFIwrtSss}
    \IQ[S_{yy}(\Om)] &= \frac{G(\Om)^2}{2\si^2}\IQ(\si).
\end{align}
where we include an additional factor of two to account for the independent real and imaginary parts of the mode at $\Om$. For example, if the initial state is the vacuum at each frequency, then the vacuum QFI in Eq.~\ref{eq:ideal vacuum case QFI} implies that
\begin{align}\label{eq:QFIwrtSss vacuum}
    \IQ[S_{yy}(\Om)]=\frac{G(\Om)}{S_{yy}(\Om)[1+G(\Om)S_{yy}(\Om)]}
\end{align}
which is attained by number measurement at each frequency. In the large gain limit of $G(\Om)\rightarrow\infty$, then $\IQ[S_{yy}(\Om)]=S_{yy}(\Om)^{-2}$. This QFI can be improved by preparing non-Gaussian initial states independently in the $\cos(\Om t)$ and $\sin(\Om t)$ signal components at each frequency $\Om$ of Eq.~\ref{eq:spec_components}. 

Although, in principle, there is a continuum of independent parameters $S_{yy}(\Om)$ to estimate, we are often only interested in a finite vector of real parameters, $\vec\theta$, in practice. Additionally, given a limited measurement bandwidth $\Delta \Omega$ and a total measurement interval $\Delta T$, it is only possible to make a finite number of measurements $M = \Delta \Omega \Delta T / \pi$. 
The task then becomes to estimate $\vec\theta$ from the measurement outcomes $\{S_{yy}(\Om_j|\vec\theta)\}_{j=1}^M$ at the frequencies $\{\Om_j\}_{j=1}^M$ such that the QFIM is
\begin{align}\label{eq:stochastic WE QFIM}
    [\IQ(\vec{\theta})]_{ij} = \sum_{k=1}^M \IQ[S_{yy}(\Om_k|\vec\theta)]
    \frac{\partial S_{yy}(\Om_k|\vec\theta)}{\partial\theta_i}\frac{\partial S_{yy}(\Om_k|\vec\theta)}{\partial\theta_j}
\end{align}
where, since the modes at each frequency are independent of each other by linearity, the total QFIM is the sum of the individual QFIMs. Note that the optimal measurement at each frequency to attain the QFIM depends on the initial state.

We may generalize the choice of temporal basis. We have so far assumed that we measure the state of the Fourier component at a given frequency, equivalent to measuring the cosine and sine functions in the time domain as given by Eq.~\ref{eq:spec_components}. We generalize this description, as the Fourier basis is expressed in a formal limit that is not experimentally accessible. Instead, we may measure the state in an arbitrary temporal basis $\{w_j\}^M_{j=1}$ of orthogonal functions such that $\intginf{t} w_j(t) w^*_k(t) = \delta_{jk}$. The temporal basis can be chosen to fit the given signal model and, in particular, each temporal mode, $w_j$, can be associated with a parameter of interest, $\theta_j$. 
The state of each temporal mode is described by the density matrix $\h \rho_k := \iintginfTwoArg{\tau}{\tau'} w_k(\tau)\h \rho(\tau, \tau')w^*_k(\tau')$ and the operators that act on this state may be similarly constructed, e.g.\ the annihilation operator is $\h a_k := \intginf{\tau} w_k(\tau) \h a(\tau)$~\footnote{If the temporal basis modes $w(\tau)$ are complex, then their Fourier transforms will have independent positive and negative frequency components unlike the real basis of cosine and sine functions in Eq.~\ref{eq:spec_components}. Instead, the real and imaginary parts of each complex temporal mode should be studied, where each part reduces to the canonical noise estimation problem of a single harmonic oscillator.}. 
Here, $\h \rho(\tau, \tau')$ is the density matrix of a single transverse mode of the outgoing bosonic field where the wave parameter $\tau = t - cz$ depends on time $t$ and the distance along the spatial propagation axis $z$, along which the mode propagates at speed $c$. For example, this state could represent a paraxial Gaussian beam of light or a wire acting as a transmission line. Note that we still assume that the noise is stationary.

Using a temporal measurement basis is a significant departure from the deterministic displacement case, where we record the timeseries from quadrature measurement and can then choose any temporal basis post-hoc using classical processing. Here, we instead directly perform, e.g, a number measurement of each temporal mode, which is highly advantageous compared to quadrature measurement for sensing stochastic signals. 
The measurement comes at the cost, however, of not being able to later process the data classically to study a different temporal basis. Implementations of the independent preparation of non-Gaussian states and performance of non-Gaussian measurements of each temporal mode must be developed for optimal stochastic waveform estimation. We discuss two possible pathways below: resonant filters and cooperative quantum memories.

\subsection{Potential experimental implementations}
\label{sec:implementations}

We first consider the case of preparing the vacuum state and measuring in the Fourier basis. The QFI for an input vacuum state at a given frequency is saturated by number measurement at that frequency. This measurement can be achieved using a resonant filter to extract a band of Fourier modes followed by a number-resolving measurement. For example, in the optical domain, this may be done using resonating filter cavities and low-background photodetectors~\cite{Vermeulen24PhotonCounting,danilishin2012quantum}. 
This implementation, however, can demand a high number of resonant filters. While in principle we only require one filter per parameter $\theta$, in practice we may require additional filters, e.g., to remove background classical noise. Optically, implementing narrow-band filter cavities may be difficult because long round-trip lengths and low scattering loss are required. Furthermore, implementing the measurement in an arbitrary temporal basis or, e.g., the non-Gaussian measurement required for a GKP finite-energy input state may be challenging to do all-optically.

A more promising pathway is to instead prepare the initial states in a dedicated ancilla and then couple the states from that ancilla to the appropriate temporal basis of the incoming bosonic mode of the main device. The time reversal of the coupling process can then load the outgoing bosonic mode from the device into an ancilla to be measured. A key example of this protocol would be to use the two-photon Raman transitions of atoms~\cite{GorshkovPRA07PhotonStoragea} to transmit and receive states in a given temporal basis. To accomplish this in the future with atom-based experimental platforms, we outline the following five requirements:
\begin{enumerate}
    \item[(R1)] Preparing non-Gaussian bosonic initial states using the electronic states of atoms or ensembles of atoms.
    \item[(R2)] Implementing the optimal non-Gaussian projective measurements using atomic states. 
    \item[(R3)] Multiplexing such preparation and measurement procedures across many such atomic ensembles.
    \item[(R4)] Achieving high cooperativity couplings that minimize transmission loss into and out of these ancillae. 
    \item[(R5)] Creating long-lived or distantly-distributed memories.
\end{enumerate}
In general, an optimal stochastic waveform search would demand simultaneous implementation of (R1) through (R4), which have some competing requirements. (R1) and (R2) are feasible 
with various platforms and schemes, e.g. preparing Fock states with efficient projective readout~\cite{BrownPRA03DeterministicOptical, GeremiaPRL06DeterministicNondestructively, HastrupPRL22ProtocolGenerating, ChenPRL06DeterministicStorable, ThomasSA24DeterministicStorage, FarreraNC16GenerationSingle, KatzNC18LightStorage} and coherent population trapping~\cite{BergmannRMP98CoherentPopulation, VitanovRMP17StimulatedRaman} of atomic states. 
(R3) is feasible using lattice traps~\cite{BlochNP05UltracoldQuantum} or tweezer arrays~\cite{KaufmanNP21QuantumScience}. (R4) is feasible using the Purcell effect via cavity enhancement or nanophotonics~\cite{ReisererRMP15CavitybasedQuantum}. 
Finally, (R5) is required to realize the optimal joint measurement of the associated mean and variance from Sec.~\ref{sec:joint measurements} using different copies of the state, which may correspond to experiments distributed distantly in space (e.g.\ observatory networks jointly estimating common deterministic and stochastic signals) or in time (e.g.\ rare events with collective deterministic and stochastic properties). (R5) has been demonstrated in many kinds of systems~\cite{HammererRMP10QuantumInterface,heshami2016quantum,wei2022towards}, but its integration with the other requirements is an ongoing effort in the field of emerging quantum technologies.

While, in the above discussion, we emphasize optical sensors and atomic memories, microwave systems are similarly promising. They can leverage superconducting cavities and nonlinear junctions~\cite{AcharyaNE23MultiplexedSuperconducting, BlaisPRA04CavityQuantum, ChennQI23TransmonQubit, DevoretS13SuperconductingCircuits, DiVincenzoFP00PhysicalImplementation} to form qubits that can efficiently produce non-Gaussian states~\cite{Campagne-IbarcqN20QuantumError} across multiplexed devices~\cite{diringer2024conditional} for fundamental physics and sensing applications~\cite{MorettiITAS24DesignSimulation}. The need for low classical noise suggests that the greatest benefit will be for microwave sensors operating in the $\gg 200$~MHz range that can search for signals above the thermal black-body radiation of dilution refrigerators~\cite{LamoreauxPRD13AnalysisSinglephoton}, i.e.\ in the $\si\gg\si_C$ regime~\footnote{If measuring signals below the classical noise floor, then we expect the regime $\eta\gg4\si_C^2$ to be also relevant, see the discussion surrounding Eq.~\ref{eq:clnoise with loss} for the one-dimensional case.}.

\subsection{Applications}
\label{sec:applications}

We now consider fundamental physics applications that highlight the metrological advantages of non-Gaussian state preparation and measurement for sensing stochastic signals.

\subsubsection{Quantum gravity}

The Verlinde-Zurek theory of quantum gravity predicts large-scale ``geontropic'' length fluctuations that scale with the holographic surface area of a given volume of spacetime~\cite{VerlindePLB21ObservationalSignatures,LiPRD23InterferometerResponse}. This signal is predicted to manifest as excess noise in the phase quadrature $\h p$ of an optical Michelson interferometer that scales linearly with the arm length. This scenario is described by Eq.~\ref{eq:IO-SD} with $\theta=\frac{\pi}{2}$ where $S_{yy}(\Om)=\alpha\Phi(\Om)$ for some scale factor $\alpha$ to be estimated and known signal morphology $\Phi(\Om)$ (e.g.\ the fiducial spectrum from Ref.~\cite{LiPRD23InterferometerResponse}) such that $\si^2=\alpha G(\Om)\Phi(\Om)$ by Eq.~\ref{eq:sigma}. Previous observations, e.g.\ by LIGO, have constrained $\alpha$ to be less than one~\cite{ChouCQG17HolometerInstrument}. The future Gravity from the Quantum Entanglement of Space-Time (GQuEST) optical interferometry experiment is proposed to further constrain $\alpha$~\cite{Vermeulen24PhotonCounting, McCuller22SinglePhotonSignal} (see also Ref.~\cite{VermeulenCQG21ExperimentObserving}).

In GQuEST, the classical noise background is projected to be seven orders of magnitude below the quantum shot noise, $\si_C^2 \sim 10^{-7} \ll \frac{1}{2}$, while the optical loss is projected to be $\eta,\eta_A\sim 0.1$, limited by resonating filter cavities outside the interferometer. 
(Contributions to the optical loss from inside the interferometer are projected to be as low as $10^{-4}$.) For signals below the classical noise background, i.e.\ $\si\ll\si_C$, this situation is described by Eq.~\ref{eq:clnoise with loss} which exhibits the Rayleigh curse. We search instead for signals above the classical noise background, i.e.\ $\si_C\ll\si\ll\eta,\eta_A$. In this loss-dominated regime, preparing squeezed states or performing quadrature measurements introduces the Rayleigh curse. By Eq.~\ref{eq:QFIwrtSss} and Eq.~\ref{eq:stochastic WE QFIM}, the integrated QFI with respect to $\alpha$ is
\begin{align*}
    \IQ(\alpha) = \sum_{j=1}^M \frac{G(\Om_j)\Phi(\Om_j)}{2\alpha}\IQ(\si).
\end{align*}
For example, if the initial state is a vacuum state, then, by Eq.~\ref{eq:QFIwrtSss vacuum}, the QFI is 
\begin{align}\label{eq:gquest lossless}
    \IQ(\alpha)= \sum_{j=1}^M \frac{G(\Om_j)\Phi(\Om_j)}{\alpha[1+\alpha G(\Om_j)\Phi(\Om_j)]}
\end{align}
which can be attained by number measurement at each frequency $\Om_j$. For the vacuum case, the results from Sec.~\ref{sec:lossless review} thus indicate that photon counting accelerates the accrual of information by $1/4\sigma^2$, compared to quadrature measurement at each frequency. This acceleration reduces the required observing time or number of independent measurements to reach a given confidence level by the same factor. In comparison, if an SMSV state is prepared, for which quadrature measurement is optimal in the high energy limit, then the acceleration compared to the vacuum case with quadrature measurement is a factor of $1/\eta^2$ by Eq.~\ref{eq:SMSV QFI high energy limit}. Thus, photon counting beats squeezing for this application of stochastic signal estimation since $4\si^2\ll\eta$.

The constraint on $\alpha$ can be further improved by preparing non-Gaussian states. The ECQFI in Eq.~\ref{eq:lossy ECQFI} indicates that, if we prepare the optimal states and perform the optimal measurements, then the accrual of information about $\alpha$ could be accelerated by $1/(4 \sigma^2 \eta)$ compared to the vacuum case with quadrature measurement, or by $1/\eta$ compared to the vacuum case with number measurement. Thus, optimal state preparation and measurement, if implemented through the aforementioned emerging quantum technology, could significantly accelerate this fundamental physics application.

\subsubsection{Stochastic gravitational waves}
Gravitational-wave searches may also benefit from our results. LIGO operates as a Michelson interferometer with two ports: the common and differential ports. The common port corresponds to the common mode of the interferometer, i.e.\ the symmetric mode of the two arms, and a large coherent state is injected into it.
The differential port corresponds to the differential (anti-symmetric) mode of the interferometer. Information about the gravitational wave in encoded in the output of the differential port: a passing gravitational wave causes small displacements along the phase quadrature of the quantum states reflecting from the output differential port. Presently, LIGO is optimized to estimate the mean value (i.e.\ signal amplitude) of the gravitational wave by injecting highly squeezed states into the differential port and performing quadrature measurements of the output mode~\cite{LIGOO4DetectorCollaborationPRX23BroadbandQuantum}. This is the optimal strategy for sensing deterministic displacements as discussed in Sec.~\ref{sec:Review of deterministic case}. 

Here, we consider instead operating a gravitational-wave detector as a variance (i.e.\ signal power) sensor for the stochastic signal from a single stochastic source or an incoherent background of many sources~\cite{RomanoLRR17DetectionMethods}. LIGO's present strategy outlined above, however, suffers the Rayleigh curse when estimating small stochastic signals: LIGO's present loss of $\eta\leq 0.23$~\cite{LIGOO4DetectorCollaborationPRX23BroadbandQuantum} imposes a detection horizon of $\si^2 > \eta/2 \approx 0.12$ by Eq.~\ref{eq:SMSV QFI high energy limit} beyond which it suffers the Rayleigh curse. The possible improvement from using the optimal stochastic sensing protocol, which we discuss below, in sensing astrophysical parameters $\vec\theta$ for a given signal model can be calculated from Eq.~\ref{eq:stochastic WE QFIM}. We focus on astrophysical applications at kilohertz frequencies since there, unlike at low frequencies, LIGO's present classical noise is a factor of four below the vacuum shot noise~\cite{LIGOO4DetectorCollaborationPRX23BroadbandQuantum}, i.e.\ $\si_p^2 \sim 0.13$, and noise models predict that it could be made as low as $\si_p^2 \sim 0.05$ in the future~\cite{BuikemaPRD20SensitivityPerformance}. The resulting classical noise horizon ($\si^2 > \si_p^2$) suffered by the optimal protocol is thus up to 2.3 times further than the loss horizon ($\si^2 > \eta/2$) suffered by preparing SMSV states~\footnote{This argument assumes a uniform source density, which is valid for the current generation of gravitational-wave detectors. Future detectors are proposed to have much further reach~ \cite{ReitzeBAAS19CosmicExplorer, MaggioreJCAP20ScienceCase} such that this assumption is no longer valid and a more detailed astrophysical model is required.}. We focus on detecting stochastic signals within this additional quantum-enhanced range.

The optimal stochastic sensing protocol involves preparing non-Gaussian states at the differential port and performing non-Gaussian measurements. (We assume that we still prepare a large coherent state at the common port.) In particular, we consider photon counting directly in the temporal basis of the gravitational-wave signal templates, $\{w_k\}^{\infty}_{k=1}$, as follows~\cite{McCuller22SinglePhotonSignal}. The compact binary coalescence strain signal $y(t)=h(t)$ that LIGO searches for may be decomposed as
\begin{align*}
    h(t) = h_{\text{det.LF}}(t) + h_{\text{det.HF}}(t) + h_{\text{stoc.}}(t),
\end{align*}
where $h_{\text{det.LF}}(t)$ and $h_{\text{det.HF}}(t)$ are low-frequency (pre-merger) and high-frequency (merger and post-merger) deterministic terms, respectively, and $h_{\text{stoc.}}(t)$ is a stochastic term. This stochastic term has an associated temporal correlation function ${\langle h_{\text{stoc.}}(t) h_{\text{stoc.}}(t')\rangle = H(t, t')}$ which is \textit{a priori} non-stationary. If we assume Gaussianity, then, by the Karhunen-Lo\'eve Theorem, we can further decompose this stochastic process into a sum of orthonormal temporal modes, $h_{\text{stoc.}}(t) = \sum^{\infty}_{k=1} \sigma_k n_k w_k(t)$, where $\left\{n_k \right\}^{\infty}_{k=1}$ are independent unit-normal-distributed random variables and $\{\sigma_k\}^{\infty}_{k=1}$ are the parameters of interest. Preparing non-Gaussian states and photon counting in the basis $\{w_k\}^{\infty}_{k=1}$ could accelerate the search for stochastic sources beyond LIGO's present loss horizon. Note that this is distinct from the ``event-stacking'' technique~\cite{ChatziioannouPRD17InferringPostmerger,CriswellPRD23HierarchicalBayesian, SasliPRD24ExploringPotential} for aggregating information on the \textit{deterministic} component of post-merger signals, $h_{\text{det.HF}}(t)$, from multiple binary neutron star coalescences~\footnote{In the deterministic event-stacking case~\cite{ChatziioannouPRD17InferringPostmerger, CriswellPRD23HierarchicalBayesian, SasliPRD24ExploringPotential}, there are threshold effects where the classical Cramer-Rao bound cannot be saturated~\cite{RifeITIT74SingleTone, SteinhardtI8IICASSP85ThresholdsFrequency}. This behavior arises from the nonlinear process of frequency estimation at a low signal-to-noise ratio, and can be expressed as the discrepancy between the classical Cramer-Rao and Barankin bounds~\cite{KnockaertITSP97BarankinBound, ChaumetteITSP08NewBarankin}. Since finding the frequency of excess power is a stochastic estimation problem, studying the quantum analogues of the Barankin bound~\cite{GessnerPRL23HierarchiesFrequentist} may be necessary to understanding the behaviour here at low signal-to-noise ratios.}.

It is also physically possible to operate a gravitational-wave detector to simultaneously estimate the mean and variance as a test for unmodelled physics. For example, there are degrees of freedom in our models of binary-neutron stars that are unconstrained by current data, and possibly additional unmodelled physics that would make the source either intrinsically stochastic or have unpredictable variations over the astrophysical population. Also, for black hole events, we could accelerate tests for any stochastic departure from general relativity using a large ensemble of observed binary black hole events. In either case, by performing joint measurements to estimate $\si$ independently from the observation of several similar $\mu$, we could perform stronger assumption-free tests for unmodelled physics.

\subsubsection{Axionic dark matter}
\label{sec:axions}
Axions are a hypothetical wavelike dark matter candidate that could also solve the strong $\mathcal{CP}$-problem~\cite{kim2010axions,choi2021recent}. More generally, if axionlike particles exist, then it is predicted that they should weakly interact with photons at the coupling rate $g_{a\gamma\gamma}$. To search for axions, therefore, we want to estimate $g_{a\gamma\gamma}$ at each frequency, where each frequency corresponds to a possible mass for the axion. Many experiments involving a microwave cavity in the presence of a static magnetic field are searching for axions, e.g.\ see Refs.~\cite{rosenberg2000searches,graham2015experimental,cameron1993search,du2018search, DixitPRL21SearchingDark, AgrawalPRL24StimulatedEmission}. 
Since the coherence time of the axion is predicted to be short, the resulting displacement of the microwave cavity mode would be stochastic and without a preferred phase on longer timescales.

At a given frequency, this transformation is canonically the symmetric case of the additive Gaussian noise channel $\Lambda^\text{noise}_{\Si_C}$ in Eq.~\ref{eq:classical noise channel Kraus representation} with $\Si_C=\diag{\si^2,\si^2}$~\cite{ShinQI23UltimatePrecision}. This is a two-dimensional random displacement channel with
\begin{align}\label{eq:encoding 2D}
    \Lambda_\si^\text{2D}(\h \rho)     
    &\approx \h\rho + \si^2\left(\h x\h\rho\h x+\h p\h\rho\h p - \frac{1}{2}\{\h x^2+\h p^2, \h\rho\}\right) 
\end{align} 
We want to estimate $\si$ since, up to the local dark matter density, from it we can infer the coupling constant $g_{a\gamma\gamma}$ at the given frequency (i.e.\ axion mass).

Suppose that the initial pure state encounters loss before the encoding such that the total channel is $\Lambda_{\sigma}^\text{noisy}=\Lambda_\si^\text{2D}\circ\Lambda^\text{loss}_\eta$. The ECQFI of this noisy channel with respect to $\si$, with energy constraint $\langle\hat n\rangle = N$, is known to be~\cite{ShinQI23UltimatePrecision}
\begin{align*}
    \IQ^{\Lambda_{\sigma}^\text{noisy}, N}(\si) &= 
    \frac{4 \left(\eta +N \left(\eta +2 \sigma ^2\right)+\sigma ^2\right)}{\left(\eta +\sigma ^2\right) \left(N \left(\eta +2 \sigma ^2\right)+\sigma ^2+1\right)}
    \\&\xrightarrow[N\rightarrow\infty]{} 
    \frac{4}{\eta +\sigma ^2}
\end{align*}
which is attained by TMSV with perfect storage $\eta_A=0$ in the high energy limit. (This limit is simply twice the one-dimensional case in Claim~\ref{claim:lossy QFI UB}.) If $\eta_A\neq0$, however, then we show in Appendix~\ref{sec:TMSV results} that the TMSV QFI vanishes for $\si^2\ll\eta_A$. Moreover, by Claim~\ref{claim:simplectic eigenvalues}, all squeezed states~\cite{zheng2016accelerating,BackesN21QuantumEnhanced} are Rayleigh cursed, i.e.\ the QFI with respect to $\si$ converges to zero in the limit of $\si\rightarrow0$, provided that loss occurs on every mode.
In this loss-dominated regime, can we prepare non-Gaussian states instead? 

It has previously been suggested to prepare Fock states for axion searches~\cite{AgrawalPRL24StimulatedEmission}. Indeed, by Appendix~\ref{sec:lindblad multiple}, the QFI in the limit of $\si\rightarrow0$ is
\begin{align}\label{eq:2D small signal}
    \IQ(\si=0) = 4\left(\evSmall{\h x\h\Pi_\perp\h x}+\evSmall{\h p\h\Pi_\perp\h p}\right)
\end{align}
where $\h\Pi_\perp$ projects onto the null space of the state before $\Lambda_\si^\text{2D}$. Eq.~\ref{eq:2D small signal} implies that the QFI for $\langle \hat n \rangle =N$ is $\IQ(\si = 0) = 4(1-\eta)^N(N+1)$ which is better than the Gaussian states, but the optimal Fock state is again roughly a factor of $e$ (4.3 dB) away from the ECQFI. For an input vacuum or Fock state, the optimal measurement is photon counting~\cite{LamoreauxPRD13AnalysisSinglephoton,DixitPRL21SearchingDark}. (Note that here, while we include a large loss before the encoding, we assume that no loss occurs after the encoding.)

Numerically, we observe that preparing GKP finite-energy states can achieve a QFI of at least 36 which is at least 90\% of the ECQFI of 40 for $\si=10^{-3}$ and $\eta=0.1$. Similarly to the one-dimensional case, we conjecture that higher energy GKP finite-energy states can converge to the ECQFI and are the optimal single-mode states for a given large $\ev{\h n} = N$ and small $\si\ll\eta$. This means that non-Gaussian states beyond Fock states could further accelerate the search for axionic dark matter.

\section{Conclusions and outlook}
We have found the optimal protocol for stochastic waveform estimation using a linear quantum device. We simplified the problem to the single-variable estimation of the excess noise in each temporal mode. For realistic losses, all Gaussian protocols exhibit the Rayleigh curse and fail to attain the ultimate precision limit. Instead, we have shown numerically that it is optimal to prepare non-Gaussian states such as finite-energy GKP states. For small signals above the classical noise floor, this non-Gaussian protocol outperforms all Gaussian protocols by orders of magnitude. We also showed that a joint non-Gaussian measurement protocol is optimal for simultaneously estimating the mean and variance of a stochastic signal. Finally, we demonstrated how our results may be applied to enhance searches for geontropic fluctuations from quantum gravity, stochastic gravitational-wave signals, and axionic dark matter.

There are many related open questions to consider in the future. While we have considered the asymptotic limit and maximum likelihood estimation, it remains to be understood whether we can obtain faster convergence to the fundamental limit or better performance when restricted to a small number of measurements. 
It would also be interesting to understand the relation with the Bayesian estimation problem in which, using the notation from Eq.~\ref{eq:encoding}, $\alpha$ is instead being estimated and $p(\alpha)$ is its prior distribution. We note that there is no immediate relation between this Bayesian problem and the problem of estimating $\si$ that we consider. For example, in the vacuum case with a Gaussian prior~\cite{gill1995applications,van2002detection}, quadrature measurement $\h p$ remains asymptotically optimal for estimating deterministic displacements, whereas here $\IC^{\h p}(\si=0)=0$. It remains to be understood whether a subtler connection exists between this Bayesian problem and our estimation problem.
Further open questions include the impact of different noise channels, determining the optimal states for fixed finite $\ev{\h n} = N$, and the complete estimation of a non-Gaussian stochastic waveform since, e.g., estimating the 4th-order cumulant of a quantum state is of interest to testing nonclassicality~\cite{bednorz2011fourth} and quantum gravity~\cite{howl2021non,haine2021searching,mehdi2023signatures}.

\begin{acknowledgements}
We thank the following people for their advice provided during this research: Rana Adhikari, Evan Hall, Thakur Giriraj Hiranandani, Konrad Lehnert, Katarzyna Macieszczak, Ian MacMillan, Haixing Miao, Swadha Pandey, John Teufel, Mankei Tsang, Sander Vermeulen, Chris Whittle, and Sisi Zhou. We also thank the Caltech Chen Quantum Group and the ANU CGA Squeezer Group. In Fig.~\ref{fig:device}, we use component graphics from Ref.~\cite{ComponentLibrary} with permission. The computations presented here were conducted in the Resnick High Performance Computing Center, a facility supported by Resnick Sustainability Institute at the California Institute of Technology.
This research is supported by the Australian Research Council Centre of Excellence for Gravitational Wave Discovery (Project No. CE170100004). J.W.G. and this research are supported by an Australian Government Research Training Program (RTP) Scholarship and also partially supported by the US NSF grant PHY-2011968. In addition, Y.C. acknowledges the support by the Simons Foundation (Award Number 568762). T.G. acknowledges funding provided by the Institute for Quantum Information and Matter and the Quantum Science and Technology Scholarship of the Israel Council for Higher Education. S.A.H. acknowledges support through an Australian Research Council Future Fellowship grant FT210100809. J.P. acknowledges support from the U.S. Department of Energy Office of Science, Office of Advanced Scientific Computing Research (DE-NA0003525, DE-SC0020290), the U.S. Department of Energy, Office of Science, National Quantum Information Science Research Centers, Quantum Systems Accelerator, and the National Science Foundation (PHY-1733907). The Institute for Quantum Information and Matter is an NSF Physics Frontiers Center. L.M.'s photon counting effort on the GQuEST project is funded in part by the Heising-Simons Foundation through grant 2022-3341.
This paper has been assigned LIGO Document No.\ P2400069.
\end{acknowledgements}

\appendix
\section{Proof of Claim~\ref{claim:simplectic eigenvalues}}
\label{sec:simplectic eigenvalue proof}
Here, we prove Claim~\ref{claim:simplectic eigenvalues} by combining a known result about the eigenvalues of the density matrix and the Williamson decomposition of a Gaussian state~\cite{williamson1936algebraic}.

The optimal measurement basis in the limit of $\si\rightarrow0$ is known to be the eigenbasis of the density matrix $\h\rho$ such that the QFI is non-vanishing if and only if there exists an eigenvalue of $\h\rho$ proportional to $\sigma^{2}$~\cite{gefen2019overcoming}. For an $M$-mode Gaussian state, the Williamson decomposition has the eigenvalues $\prod_{i=1}^M\lambda_{k_{i}}$ and eigenstates $\{\bigotimes_{i=1}^M \h S\left(r_{i}\right)\ket{k_{i}}\}_{k_{1}\mathellipsis k_{M}}$ where $\h S\left(r\right)$ is the single-mode squeezing operator and $\ket{k_i}$ is a Fock state (potentially of a combination of the original modes, e.g.\ TMSV). The \textit{symplectic eigenvalues} associated with the $i$th mode are $\frac{1}{2}+\bar{n}_{i}$ such that the corresponding eigenvalues are $\lambda_{k_{i}}=\frac{1}{1+\bar{n}_{i}}(\frac{\bar{n}_{i}}{1+\bar{n}_{i}})^{k_{i}}$ with degeneracy two. 
Since the QFI is not Rayleigh cursed if and only if there exists some $k_{i}$ such that $\lambda_{k_{i}}\propto\sigma^{2}$~\cite{gefen2019overcoming}, therefore, the QFI is non-vanishing if and only if there exists some mode $i$ such that $\bar{n}_{i}\propto\sigma^{2}$, or, equivalently, a symplectic eigenvalue equal to $\frac{1}{2}+k\sigma^{2}$ for some constant $k$ (not to be confused with $k_i$). 

We remark that, for a single mode Gaussian state, this result can be seen directly from the QFI expression: if $\partial_{\sigma}\Sigma\rightarrow 0$, then the first term in Eq.~\ref{eq:Gaussian QFI CM} also vanishes. The second term in Eq.~\ref{eq:Gaussian QFI CM}, however, is non-zero if and only if $\gamma=1-\beta\sigma^{2}$ which is equivalent to the above condition on the symplectic eigenvalues. 

Here, we have assumed that the noisy channel is analytic at $\si=0$ such that we may expand it in a Taylor series. This will usually be the case. For non-analytic noisy channels, however, this result can be extended: the QFI is non-vanishing if and only if there exists a symplectic eigenvalue equal to $\frac{1}{2}+k\si^l$ with $1<l\leq2$~\cite{gefen2019overcoming}.

\section{Other noise channels}
We have focused on the case of a noise channel with known parameters occurring before the encoding channel. In this section, we briefly discuss noise channels occurring after the encoding channel and explore an example of what happens when the parameters of the noise channel are unknown.

\subsection{Noise occurring after the encoding}
\label{sec:measurement noise}
Noise occurring after the encoding can be divided into two types: (1) noise associated with or ``baked into'' the acquisition of the signal or (2) noise associated with the later measurement. In the former case, it is not possible to insert an arbitrary parameter-independent ``control'' channel between the encoding and noise channels to mitigate the noise. In the latter case of ``measurement noise'', however, it is possible to entirely overcome the noise by implementing a suitable control channel between the encoding and measurement noise channels.

Let us illustrate this point that measurement noise can be overcome for sensing $\si$ in the case of preparing an SMSV initial state with $\ev{\h n} = N$. Note that in the noiseless case, the optimal measurement after the encoding channel is to first anti-squeeze and then perform a number-resolving measurement. Here, we assume that we still anti-squeeze but then perform a different, fixed POVM which can be decomposed into a measurement noise channel followed by a number resolving projective measurement. 
In an optical system, for example, our measurement apparatus is a photodetector and we consider two relevant noise models: detection loss and dark counts. Photodetection with detection loss is modeled as a loss channel $\Lambda^\text{loss}_{\eta_\text{meas}}$ followed by a projective number measurement.
Photodetection with dark counts is modeled as the noise channel $\Lambda^\text{dark}_{N_{\text{meas}}}$ followed by a projective number measurement, where $\Lambda^\text{dark}_{N_{\text{meas}}}$ is defined as:
\begin{align*}
    \Lambda^\text{dark}_{N_{\text{meas}}}(\h \rho) = \sum_{n=0}^\infty \h K_n \h \rho \h K_n^\dag, \quad
    \h K_n = \sum_{m=0}^\infty \sqrt{p_\text{th}(m)} \ket{n+m}\bra{n}
\end{align*}
where $p_\text{th}(m) = \frac{1}{1 + N_{\text{meas}}}\left(\frac{N_{\text{meas}}}{1 + N_{\text{meas}}}\right)^m$ is the number distribution of a thermal state with average number $N_{\text{meas}}$. The results of measuring $\h n$ on $\Lambda^\text{dark}_{N_{\text{meas}}}(\h \rho)$ are equivalent to convolving the results of measuring $\h n$ on $\h\rho$ with $p_\text{th}$. Intuitively, $\Lambda^\text{dark}_{N_{\text{meas}}}$ adds $m$ particles to the quantum state with probability $p_\text{th}(m)$ independent of the state. In comparison, it is unsuitable to model dark counts as an isotropic classical noise channel $\Lambda^\text{noise}_{\Si_{C, \text{meas}}}$ with $\Si_{C, \text{meas}}=\diag{\si_{C, \text{meas}}, \si_{C, \text{meas}}}$ since $\Lambda^\text{noise}_{\Si_{C, \text{meas}}}$ both adds and subtracts particles at a rate which depends on the quantum state due to Bose enhancement.

We first consider the impact of measurement noise in the absence of any additional control channels. In the noiseless case, given an initial SMSV state, the QFI is $\IQ(\si)=4\xi_{N}$ in the limit of $\sigma \ll 1$. In this limit, almost all of the CFI from the projective number measurement after anti-squeezing comes from the $\h n = 1$ single-particle detection probability, $p\left(1\right) \approx \xi_{N}\sigma^{2}$.
In the case of measurement loss $\eta_\text{meas}$, e.g.\ detection loss for an optical system, the total channel before the projective number measurement is $\Lambda_{\sigma}^\text{noisy}=\Lambda^\text{loss}_{\eta_\text{meas}}\circ\Lambda_\text{anti-sqz}\circ\Lambda_\si$ where $\Lambda_\text{anti-sqz}$ is the anti-squeezing unitary channel for the given $N$. The single-particle detection probability is now $p\left(1\right)\approx\left(1-\eta_\text{meas}\right)\xi_{N}\sigma^{2}$ such that the CFI from the subsequent projective number measurement falls to $\IC^{\h n}(\si)=\left(1-\eta_\text{meas}\right)4\xi_{N}$ but the Rayleigh curse is not introduced. This is unlike loss occurring before the anti-squeezing operation which does introduce the Rayleigh curse.
In contrast, the Rayleigh curse will arise in the case of the dark count measurement noise channel. 
In this case, the total channel before the projective number measurement is $\Lambda_{\sigma}^\text{noisy}=\Lambda^\text{dark}_{N_{\text{meas}}}\circ\Lambda_\text{anti-sqz}\circ\Lambda_\si$. The convolved probability distribution of number measurements is then $p(0) = \frac{1 - \xi_N\si^2}{1+N_{\text{meas}}}$ and $p(n) = \frac{N_{\text{meas}}^{n-1}(N_{\text{meas}} + \xi_N\si^2)}{(1+N_{\text{meas}})^{n+1}}$ for $n\geq 1$ such that the CFI is $\IC^{\h n}(\si) = \frac{4 \xi_N ^2 \sigma ^2}{\left(1 - \xi_N  \sigma ^2\right) \left(\xi_N  \sigma ^2+N_{\text{meas}}\right)}$ which vanishes in the limit of $\si\rightarrow0$ for fixed $\xi_N, N_{\text{meas}}>0$.

We now show that there exists, in theory, a unitary that allows us to recover the noiseless QFI. After the anti-squeezing operation, applying any unitary channel $\Lambda_\text{swap}$ which swaps the Fock states $\ket{1}$ and $\ket{k}$ for some $k \gg 1$ and stabilizes the vacuum $\ket0$ (up to a phase) will recover the noiseless QFI in the limit of large enough $k$ as proved below. Intuitively, this control unitary channel $\Lambda_\text{swap}$ amplifies the signal to make it more tolerant to loss and distinguish it from dark counts. 
In the case of measurement loss, e.g.\ detection loss for an optical system, the total channel before the projective number measurement is now $\Lambda_{\sigma}^\text{noisy}=\Lambda^\text{loss}_{\eta_\text{meas}}\circ\Lambda_\text{swap}\circ\Lambda_\text{anti-sqz}\circ\Lambda_\si$. The probability of not detecting zero particles is now $\left(1-\eta_\text{meas}^{k}\right)\xi_{N}\sigma^{2}$, keeping only the relevant $\sigma^{2}$ terms in the limit of $\si\ll1$, such that the CFI from the subsequent projective number measurement is $\IC^{\h n}(\si)=4\xi_{N}\left(1-\eta_\text{meas}^{k}\right)$ which recovers the noiseless QFI of $4\xi_{N}$ in the limit of $k\rightarrow\infty$. 
In the case of the dark count measurement noise, the total channel before the projective number measurement is now $\Lambda_{\sigma}^\text{noisy}=\Lambda^\text{dark}_{N_{\text{meas}}}\circ\Lambda_\text{swap}\circ\Lambda_\text{anti-sqz}\circ\Lambda_\si$. The probability of detecting $k$ or more particles after this channel is
\begin{align*}
    p(\h n\geq k) = \xi_N \sigma ^2 \left[1-\left(\frac{N_{\text{meas}}}{N_{\text{meas}}+1}\right)^k\right]+\left(\frac{N_{\text{meas}}}{N_{\text{meas}}+1}\right)^k
\end{align*}
in the limit of $\xi_{N}\sigma^{2}\ll1$. The CFI from detecting $k$ or more particles is then
\begin{align*}
    \IC^{\h n} &= \frac{4 \xi_N^2 \sigma^2 \left[1-\left(\frac{N_{\text{meas}}}{N_{\text{meas}}+1}\right)^k\right]^2}{\xi_N \sigma ^2 \left[1-\left(\frac{N_{\text{meas}}}{N_{\text{meas}}+1}\right)^k\right]+\left(\frac{N_{\text{meas}}}{N_{\text{meas}}+1}\right)^k}    
\end{align*}
for fixed $\si, N, \bar n > 0$. Since, for fixed $\xi_N, \si, N_{\text{meas}}$ and $\varepsilon>0$, $\exists K$ such that $\left(\frac{N_{\text{meas}}}{N_{\text{meas}}+1}\right)^k<\varepsilon$ for $k>K$, the CFI from number measurement recovers the noiseless QFI of $4\xi_{N}$ in the limit of $k\rightarrow\infty$.
We defer to future work finding possible ways to implement suitable control unitary channels such as $\Lambda_\text{swap}$.

While we assumed a specific fixed POVM above, e.g.\ a noisy photodetector, the results hold more generally~\cite{haine2018using, len2022quantum, zhou2023optimal}. Any fixed POVM that is modelled as a loss or dark counts channel followed by a projective measurement will have the above limitations. For any such POVM, there exists an analogous control unitary channel that can overcome the measurement noise, where the particular control unitary channel needed may depend on the POVM~\cite{len2022quantum, zhou2023optimal}.

In the case of a deterministic displacement channel, it is known that a phase-sensitive amplifier after the encoding can mitigate subsequent measurement loss in the high-gain limit. This is sometimes called a ``Caves's amplifier'' in the context of LIGO~\cite{CavesPRD81QuantummechanicalNoise}. In our case, however, a phase-sensitive amplifier after the encoding would introduce the Rayleigh curse and should be avoided.

\subsection{Unknown loss}
\label{sec:loss indeterminacy}
Here, we discuss the implications of not precisely knowing the loss $\eta$ that the state $\h\rho$ experiences via the loss channel $\Lambda^\text{loss}_\eta$ introduced in Sec.~\ref{sec:Loss channel}. We illustrate this for the input vacuum case and an indeterminate loss occurring after the encoding but before the measurement.

Suppose that the loss $\eta$ follows some probability distribution $p(\eta)$. By the Central Limit Theorem, asymptotically, this distribution approaches $p(\eta)\sim \mathcal{N}(\mu_\eta,\si_\eta^2)$ with some mean $\mu_\eta$ and standard deviation $\si_\eta$. The state after this indeterminate loss is  
\begin{align*}
    \Lambda^\text{indet.}_{\mu_\eta,\si_\eta}(\h\rho)
    &= \intg{0}{1}{\eta} \Lambda^\text{loss}_\eta(\h\rho) p(\eta)    
    .
\end{align*} 
We want to know whether this will introduce the Rayleigh curse in the relevant regime of $\si_\eta\ll1$ where the loss is not precisely known but is well-constrained.

The $n$th moment of a weighted average of distributions is the weighted average of their $n$th moments. The first moment is zero for the vacuum case. The second moment after $\Lambda^\text{loss}_\eta$ is linear in $\eta$ such that the second moment after $\Lambda^\text{indet.}_{\mu_\eta,\si_\eta}$ equals that after $\Lambda^\text{loss}_{\mu_\eta}$, e.g.\ the signal term is $(1-\mu_\eta)\si^2$.
The loss indeterminacy $\si_\eta$ does not affect the first two moments, so, it must be a non-Gaussian perturbation of the 3rd, 4th, and higher-order moments.

For the vacuum case, the final state after $\Lambda^\text{indet.}_{\mu_\eta,\si_\eta}\circ\Lambda_\si$ has the second moment $\evSmall{\h p^2}=\varsigma^2$ and fourth moment $\evSmall{\h p^4}=3\varsigma^4+3\si^2\si_\eta^2$ where $\varsigma^2:=\frac{1}{2}+(1-\mu_\eta)\si^2$. The 4th-order cumulant $3\si^2\si_\eta^2$ indicates that the distribution has slightly fatter tails than a Gaussian. 
Numerically, the CFI from number measurement shows that this perturbation is negligible when $\mu_\eta$ is known and $\si_\eta$ is small. For example, an indeterminacy of $\frac{\si_\eta}{\mu_\eta}=10\%$ leads to an $\orderten{-4}$ fractional change in the CFI with respect to $\si$. By convexity, this implies that the QFI also changes negligibly. 
If we also do not know $\mu_\eta$, then the CFIM with respect to $\mu_\eta$ and $\si$ is not singular and thus they can be estimated simultaneously.

\section{Purifications and the ECQFI}
To prove Claims~\ref{claim:lossless QFI UB} and~\ref{claim:lossy QFI UB} about the ECQFI, we first review the established method that we use to optimise over the initial states.

\subsection{Review of purifications} 
Consider an initial pure state $\ket{\psi}$ and a non-unitary channel $\Lambda_\theta$ such that the final state is $\h\rho(\theta)=\Lambda_\theta(\ket{\psi}\bra\psi)$. The channel $\Lambda_\theta$ can be purified (also called dilated) to a unitary process $\h U_\theta$ acting on $\ket\psi\otimes\ket{\varphi}$ such that the final state is $\h\rho(\theta)=\text{Tr}_{A}[\h U_\theta\left(\ket\psi\otimes\ket{\varphi}\right)]$ for all $\ket\psi$ and $\theta$. Note that the ancilla $\ket{\varphi}$ is independent of $\ket\psi$ and $\theta$ but depends on the purification $\h U_\theta$ chosen. The purification of $\Lambda_\theta$ to $\h U_\theta$ should not be confused with the purification of a mixed state or with the extended channel QFI discussed below. 

The choice of purification, however, is not unique. By Uhlmann's Theorem for the quantum fidelity, the QFI for a fixed initial state $\ket\psi$ is the infimum of the QFI over all possible purifications $\h U_\theta$ of $\Lambda_\theta$~\cite{uhlmann1976transition, EscherNP11GeneralFramework,KolodynskiNJP13EfficientTools,ShinQI23UltimatePrecision} 
\begin{align}\label{eq:Uhlmann theorem}
    \IQ^{\Lambda_\theta(\ket\psi\bra\psi)}(\theta) = \inf_{\h U_\theta} \IQ^{\h U_\theta\left(\ket\psi\otimes\ket{\varphi}\right)}(\theta) 
    .
\end{align}
Here, the QFI of the unitary process $\h U_\theta$ is $4 \varSmall{\h H}$ for all $\theta$, where $\h H=-i\h U_\theta^{\dagger}\dot{\h U}_\theta$ is the Hermitian generator of local displacements in $\theta$ and the variance is calculated with respect to the pure initial state $\ket\psi\otimes\ket{\varphi}$~\cite{boixo2007generalized}. Using this fact, then Eq.~\ref{eq:Uhlmann theorem} becomes
\begin{align*}
    \IQ^{\Lambda_\theta(\ket\psi\bra\psi)}(\theta) = 4 \inf_{\h U_\theta} \varSubSmall{\ket\psi\otimes\ket\varphi}{\h H}
\end{align*}
where $\ket\varphi$ and $\h H$ are determined by the purification $\h U_\theta$ of $\Lambda_\theta$.

The CQFI of $\Lambda_\theta$ in Eq.~\ref{eq:channel QFI definition}, which optimises over the initial state, is then given by
\begin{align*}
    \IQ^{\Lambda_\theta, \text{no ancilla}}(\theta) = 4 \sup_{\ket\psi}\inf_{\h U_\theta} \varSubSmall{\ket\psi\otimes\ket\varphi}{\h H}
    .
\end{align*}
Exchanging the order of maximization and minimisation above results in an upper bound on the CQFI of $\Lambda_\theta$. Remarkably, Ref.~\cite{fujiwara2008fibre} showed that this upper bound is exactly the ECQFI of $\Lambda_\theta$ in Eq.~\ref{eq:extended channel QFI definition}, hence 
\begin{align}\label{eq:extended channel QFI by Uhlmann's}
    \IQ^{\Lambda_\theta}(\theta) = 4 \inf_{\h U_\theta}\sup_{\ket\psi} \varSubSmall{\ket\psi\otimes\ket\varphi}{\h H}
    .
\end{align} 
Thus, to attain the ECQFI, it suffices to find a purification $\h U_\theta$ for which the following upper bound is tight
\begin{align}\label{eq:extended channel QFI by Uhlmann's UB}
    \IQ^{\Lambda_\theta}(\theta)\leq 4 \sup_{\ket\psi} \varSubSmall{\ket\psi\otimes\ket\varphi}{\h H}.
\end{align}
To find such a purification, we will use the following technique of introducing a ``hiding'' unitary $\h U_\text{hide}$~\cite{LatunePRA13QuantumLimit}. Suppose that we have a purification $\h U_\text{enc.}$ of $\Lambda_\theta$. We can generate a family of related purifications $\h U_\theta=\h U_\text{hide}\h U_\text{enc.}$ by performing any $\theta$-dependent unitary transformation $\h U_\text{hide}$ on the ancilla, since the ancilla is traced out to recover $\h\rho(\theta)$. Intuitively, this hiding unitary $\h U_\text{hide}$ is meant to remove (or hide) the excess information about $\theta$ present in the ancilla after $\h U_\text{enc.}$ to make Eq.~\ref{eq:extended channel QFI by Uhlmann's UB} tight.

\subsection{Proof of Claim~\ref{claim:lossless QFI UB}}
\label{sec:proof of lossless channel QFI}
Here, we prove Claim~\ref{claim:lossless QFI UB} about the lossless case by using the above purification method with a hiding unitary.

We purify the channel $\Lambda_{\sigma}$ by introducing an ancillary mode 2 prepared in a vacuum state, such that the resulting encoding unitary is a conditional displacement $\h U_\text{enc.}=\exp(-i\sigma \h H_\text{enc.})$ with the generator $\h H_\text{enc.}=\sqrt{2} \hat{x}_{1}\hat{x}_{2}$. 
We also include a hiding unitary $\h U_\text{hide}=\exp(-i\si \h H_\text{hide})$ such that the overall purification is $\h U_\si=\h U_\text{hide}\h U_\text{enc.}$. Here, $\h H_\text{hide}=\frac{1}{2}g\left(\hat{x}_{2}\hat{p}_{2}+\hat{p}_{2}\hat{x}_{2}\right)$ is a squeezing Hamiltonian and $g$ is a real parameter over which to minimize to obtain a tight upper bound in Eq.~\ref{eq:extended channel QFI by Uhlmann's UB}. 

The Hermitian generator for the overall purification $\h U_\si$ is
\begin{align*}
    \h H&= 
    \h H_\text{enc.}+\h U^{\dagger}_\text{enc.}\h H_\text{hide}\h U_\text{enc.}.    
    \\&=\left(\sqrt{2}+\sqrt{2}g\sigma\right)\hat{x}_{1}\hat{x}_{2}+\frac{g}{2}\left(\hat{x}_{2}\hat{p}_{2}+\hat{p}_{2}\hat{x}_{2}\right)   .
\end{align*}
The QFI of $\h U_\si\ket\psi\otimes\ket{\varphi}$ is therefore
\begin{align}
4\varSmall{\h H}=2\left(\sqrt{2}+\sqrt{2}g\sigma\right)^{2}\ev{\hat{x}_{1}^{2}}+2g^{2}
\label{eq:purified_qfi_1}
\end{align}
where we have assumed that $\ev{\h x_1}=0$ without loss of generality. Here, since mode 2 is in a vacuum state, we also used the fact that
\begin{align}\label{eq:vacuum state variance of xp+px}
\var{\hat{x}_{2}\hat{p}_{2}+\hat{p}_{2}\hat{x}_{2}} 
=2.
\end{align}
Minimizing the QFI in Eq.~\ref{eq:purified_qfi_1} over $g$, we derive the upper bound to the ECQFI in Eq.~\ref{eq:lossless UB}. This equals the QFI from preparing SMSV by Eq.~\ref{eq:Gaussian QFI CM}.

\subsection{Proof of Claim~\ref{claim:lossy QFI UB}}
\label{sec:proof of lossy UB}
Here, we prove Claim~\ref{claim:lossy QFI UB} about the lossy case by using a similar purification method to the lossless case above. Our choice of hiding unitary to handle the loss channel was inspired by the proof of Eq.~\ref{eq:deterministic displacement, lossy ECQFI} for the deterministic case~\cite{LatunePRA13QuantumLimit}.

We purify the total noisy channel $\Lambda_{\sigma}^\text{noisy}=\Lambda_\si\circ\Lambda^\text{loss}_\eta$ as follows. The loss channel $\Lambda^\text{loss}_\eta$ is purified to the following beamsplitter unitary between mode 1 and some ancillary mode 3 which starts in vacuum:
\begin{align*} 
\h U_\eta&= \exp[i\arcsin(\sqrt{\eta})(\h a_1^\dag \h a_3+\h a_3^\dag \h a_1)] \\&= \exp[i\arcsin(\sqrt{\eta})(\h x_1 \h x_3+\h p_1 \h p_3)]
\end{align*}
The encoding channel $\Lambda_\si$ is again purified to $\h U_{\text{enc.}}=\exp(-i\sigma \h  H_\text{enc.})$ with $\h H_{\text{enc.}} =\sqrt{2}\hat{x}_{1}\hat{x}_{2}$ and mode 2 starting in vacuum. 
We also include two hiding unitaries: $\hat{U}_{\text{hide},1}=\exp(-i\sigma \h H_{\text{hide},1})$ with $\h H_{\text{hide},1}=g_1\hat{x}_{2}\hat{x}_{3}$, which removes the excess information due to the purification of the loss, and $\hat{U}_{\text{hide},2}=\exp(-i\sigma \h H_{\text{hide},2})$ with $\h H_{\text{hide},2}=\frac{1}{2}g_2(\hat{x}_{2}\hat{p}_{2}+\hat{p}_{2}\hat{x}_{2})$, which again removes the excess information due to the purification of the random displacement. Here, $g_1$ and $g_2$ are parameters to be minimized over again to obtain a tight bound. 

The overall unitary is thus
\begin{align*}
    \h U_\si&=\hat{U}_{\text{hide},2} \hat{U}_{\text{hide},1} \hat{U}_{\text{enc.}}\hat{U}_{\eta}
    \intertext{the Hermitian generator of which is}
    \h {H}    
    &=\hat{U}_{\eta}^{\dagger}\hat{U}_{\text{enc.}}^{\dagger}\hat{U}_{\text{hide},1}^{\dagger}\h H_{\text{hide},2}\hat{U}_{\text{hide},1}\hat{U}_{\text{enc.}}\hat{U}_{\eta}\\
    &+\hat{U}_{\eta}^{\dagger}\hat{U}_{\text{enc.}}^{\dagger}\h H_{\text{hide},1}\hat{U}_{\text{enc.}}\hat{U}_{\eta}+\hat{U}_{\eta}^{\dagger}\h H_{\text{enc.}}\hat{U}_{\eta},
    \intertext{where there is no term associated with $\h H_\eta$ since $\h U_\eta$ is independent of $\si$. This expression then simplifies to}
    \h {H}&=\frac{g_2}{2}\left(\hat{x}_{2}\hat{p}_{2}+\hat{p}_{2}\hat{x}_{2}\right)\\&+\left(g_2\sigma+1\right)\left(\sqrt{2\eta}+g_1\sqrt{1-\eta}\right)\hat{x}_{3}\hat{x}_{2}\\
    &+\left(g_2\sigma+1\right)\left(\sqrt{2(1-\eta)}-g_1\sqrt{\eta}\right)\hat{x}_{1}\hat{x}_{2}.    
\end{align*}
The QFI of $\h U_\si\ket\psi\otimes\ket{\varphi}$ is therefore 
\begin{align*}
    4\varSmall{\h H}
    &=2g_2^{2}+f\left(g_1\right)\left(g_2\sigma+1\right)^{2}
    \intertext{where, by  Eq.~\ref{eq:vacuum state variance of xp+px},}
    f\left(g_1\right)&=\left(\sqrt{2\eta}+g_1\sqrt{1-\eta}\right)^{2}\\&+2\langle\hat{x}_{1}^{2}\rangle\left(\sqrt{2(1-\eta)}-g_1\sqrt{\eta}\right)^{2}.
\end{align*}
Minimising over the free parameters of the hiding unitaries, $g_1$ and $g_2$, we find that 
\begin{align*}
4\varSmall{\h H} &= \min_{g_1,g_2}\left[ 2g_2^{2}+f\left(g_1\right)\left(g_2\sigma+1\right)^{2}\right]\\
&=\min_{g_2}\left[  2g_2^{2}+\frac{4\ev{\h x_1^2}}{\left(1-\eta\right)+2\eta\ev{\h x_1^2}}\left(g_2\sigma+1\right)^{2}\right]\\
&=\frac{4}{2(\eta+\sigma^{2})+\left(1-\eta\right)\ev{\h x_1^2}^{-1}}.
\end{align*}
Using the optimal value of $\evSmall{\h x_1^2}$ for a given constraint $\ev{\h n}=N$ leads to the upper bound of Eq.~\ref{eq:lossy UB constrained}. This optimal value of $\evSmall{\h x_1^2}$ for a fixed $N$ would correspond to the state after the loss but before the encoding being SMSV, but there is no such state that becomes SMSV after a loss. Note that while, e.g., Schr\"odinger's cat states approach $\evSmall{\h x_1^2}=2N$ for large $N$ too, the relation $\evSmall{\h x_1^2}<\xi_N$ holds for any fixed $N$ and any non-SMSV state. This bound, therefore, is not tight for a fixed finite $N$ as preparing SMSV does not saturate it. 
In the limit of $N\rightarrow\infty$, however, then this bound proves the ECQFI in Eq.~\ref{eq:lossy ECQFI} which is saturated asymptotically by TMSV (see Appendix~\ref{sec:TMSV results}) and thus is tight.

\section{Two-mode squeezed vacuum}
\label{sec:TMSV results}
Here, we analyze the QFI from preparing a two-mode squeezed vacuum (TMSV) state given different noise models (loss and classical noise) and different encoding channels (one-dimensional and two-dimensional random displacement channels).

Ref.~\cite{monras2013phase} showed that the QFI of an $M$-mode Gaussian state with the parameter encoded in the covariance matrix $\Si$ is given by the following formula:
\begin{align}\label{eq:Gaussian QFI CM multi-mode}
    \IQ(\si) &= 2\tr{(\partial_\si \Si) (4\Si\otimes \Si - \omega\otimes\omega)^{-1}(\partial_\si \Si)}
\end{align}
where, if $\Si$ is written in the quadrature basis of $(\h x_1, \mathellipsis, \h x_M, \h p_1, \mathellipsis, \h p_M)^\T$, then the commutator matrix is given by $\om = \bmatrixByJames{0 & \mathbbm{1} \\ - \mathbbm{1} & 0}\nonumber$ where $\mathbbm{1}$ is the $M$-by-$M$ identity matrix. This result generalises Eq.~\ref{eq:Gaussian QFI CM} for single-mode Gaussian states.

For the one-dimensional random displacement channel in Eq.~\ref{eq:encoding} preceded by a loss channel, the QFI of TMSV with imperfect storage $\eta_A>0$ in the high energy limit $N\rightarrow\infty$ and with $\si\ll1$ is given by Eq.~\ref{eq:Gaussian QFI CM multi-mode} as
\begin{align}\label{eq:TMSV HE 1D}
    \IQ(\sigma) &= \frac{2 (1-\eta_A) \sigma ^2 (\eta +\eta_A-2 \eta  \eta_A)^2}{\xi + \sigma ^2 (\eta +\eta_A-2 \eta  \eta_A)^3}
    \where
    \xi &= 2 (1-\eta) \eta  \eta_A \left(\eta ^2 (2 (\eta_A-1) \eta_A+1)-2 \eta  \eta_A^2+\eta_A^2\right)\nonumber.    
\end{align}
Attaining this QFI requires a joint measurement scheme of squeezing followed by a number measurement. For any $\si>0$ and with perfect storage $\eta_A=0$, then the QFI in Eq.~\ref{eq:TMSV HE 1D} converges to the ECQFI in Eq.~\ref{eq:lossy ECQFI} in the high energy limit. 
In Fig.~\ref{fig:channelFI}, for example, we use the full form of the QFI for any $\si$, $\eta$, $\eta_A$, and $N$ which is too verbose to provide here but is found with the same method.

In comparison, for the two-dimensional random displacement channel in Eq.~\ref{eq:encoding 2D} preceded by a loss channel, the high energy QFI for TMSV with imperfect storage is given by Eq.~\ref{eq:Gaussian QFI CM multi-mode} as 
\begin{align}\label{eq:TMSV HE 2D}
    \IQ(\si) = \frac{4 (1-\eta_A) \sigma ^2}{\left(\eta +\sigma ^2\right) \left[(1-\eta) \eta_A+(1-\eta_A) \sigma ^2\right]}.
\end{align}

For $0<\eta_A,\si^2\ll\eta$, the two-dimensional encoding TMSV QFI in Eq.~\ref{eq:TMSV HE 2D} with ancilla loss $\eta_A$ is approximately twice the one-dimensional encoding TMSV QFI in Eq.~\ref{eq:TMSV HE 1D} with ancilla loss $\frac{\eta_A}{2}$. In either case, the TMSV QFI attains the ECQFI at high energy if $\eta_A\ll\si^2,\eta$ but vanishes if $\si^2\ll\eta,\eta_A$.

We now briefly analyze the perpendicular classical noise case discussed in Appendix~\ref{sec:Optimal initial states with classical noise}. Consider the one-dimensional random displacement channel preceded by a perpendicular classical noise channel which adds $\si_x$ to the system and $\si_{x,A}$ to the ancilla. 
Let us first consider the case of perfect storage, i.e.\ $\si_{x,A}=0$. Interestingly, for any $\ev{\h n}=N$ per mode and any amount of perpendicular noise $\sigma_{x} > 0$, the QFI is
\begin{align}\label{eq:TMSV perp CN_0}
\IQ(\si = 0) = \frac{8 N (N+1)}{2 N+1} \xrightarrow[N\rightarrow\infty]{} 4 N
\end{align}
which does not exhibit the Rayleigh curse. Intuitively, for any $N$, there exists a symplectic transformation of $\h\rho$, the state after the encoding channel, to a pair of modes such that one of them is completely decoupled from the perpendicular noise but nevertheless contains some information about the signal. 
An optimal measurement, therefore, would be to apply this symplectic transformation to $\h\rho$ and then perform a number measurement of the perpendicular noise--free mode. 
In the quadrature basis of $(\h x_S, \h x_A, \h p_S, \h p_A)^\T$, where the system $S$ and ancilla $A$ are entangled, the covariance matrix of $\h\rho$ is
\begin{align*}
    \Sigma=\frac{1}{2}\mathbbm{1}
    &+
    \bmatrixByJames{
    \sigma_{x}^{2}+N & \sqrt{N\left(N+1\right)}\\
    \sqrt{N\left(N+1\right)} & N
    }\\
    &\oplus
    \bmatrixByJames{
    \sigma^{2}+N & -\sqrt{N\left(N+1\right)}\\
    -\sqrt{N\left(N+1\right)} & N
    }.
\end{align*} 
After performing a suitable anti-squeezing operation, the inverse of the two-mode squeezing unitary that prepares the TMSV state from vacuum, the covariance matrix becomes
\begin{align*}
    \Si = \frac{1}{2}\mathbbm{1}
    &+\frac{1}{2}\sigma_{x}^{2}\bmatrixByJames{
    N+1 & -\sqrt{N\left(N+1\right)}\\
    -\sqrt{N\left(N+1\right)} & N
    }\\
    &\oplus\frac{1}{2}\sigma^{2}\bmatrixByJames{
    N+1 & \sqrt{N\left(N+1\right)}\\
    \sqrt{N\left(N+1\right)} & N
    }.
\end{align*}
The collective mode with annihilation operator $\hat{a}_{\text{dec}} = \frac{1}{\sqrt{2N+1}} \left(\sqrt{N}\hat{a}_{S}+\sqrt{N+1}\hat{a}_{A}\right)$, therefore, is completely decoupled from the perpendicular noise ($\sigma_{x}^{2}$) yet remains coupled to the signal ($\sigma^{2}$). The covariance matrix of this perpendicular noise--free mode is
\begin{align*}
\Sigma_{\text{dec}}= \frac{1}{2}\mathbbm{1} + \diag{0,\; 2\sigma^{2}\frac{N\left(N+1\right)}{2N+1}}
\end{align*}
such that a number measurement of this mode attains the QFI in Eq.~\ref{eq:TMSV perp CN_0}. This protocol is reminiscent of the idea of displacement noise--free interferometry in which there exist modes unaffected by displacement noise but that nevertheless contain information about the signal~\cite{kawamura2004displacement, chen2006interferometers, gefen2024quantum}.
 
We now consider the case of imperfect storage, i.e.\ $\si_{x,A}>0$. Here, however, the QFI as $\si\rightarrow0$ exhibits the Rayleigh curse by Claim~\ref{claim:simplectic eigenvalues} for a fixed value of $N$, since, for all $i$, $\bar n_i = c_i(\si_x,\si_{x,A}) + \order{\si^2}$ for some $c_i(\si_x,\si_{x,A})>0$ (see Appendix~\ref{sec:simplectic eigenvalue proof}). For a fixed finite signal $\si>0$ and any fixed $\si_x$ and $\si_{x,A}$, the QFI in Eq.~\ref{eq:Gaussian QFI CM multi-mode} attains the noiseless ECQFI of $\frac{2}{\si^2}$ in the high energy limit of $N\rightarrow\infty$ particles per mode.

Finally, we elaborate on the relation between TMSV with perfect storage and GKP states in the high energy limit. The infinite-energy TMSV pure state is the following unnormalized state
\begin{align*}    
    \intginf{x}
    \ket{\h x_S = x}\ket{\h x_A = x}
    = \intginf{p} \ket{\h p_S = p}\ket{\h p_A = -p}
\end{align*}
where $|\h x_j = x_0\rangle$ ($|\h p_j = p_0\rangle$) is the position (momentum) eigenstate of mode $j$ at $x_0$ ($p_0$). We observe that deterministic or random displacements along $\hat{p}_{S}$ keep the state inside the code space of $\left\{\ket{\h x_S = x}\ket{\h x_A = x}\right\}_{x}$, where deterministic (random) displacement induces rotation (dephasing) inside this subspace. In comparison, deterministic or random displacements along $\hat{x}_S$ take the state outside of this code space and can be detected by measuring $\hat{x}_{S}-\hat{x}_{A}$.
This is similar to the GKP infinite-energy state $\ket{\text{GKP}_{\text{ideal}}}$ from Appendix~\ref{sec:GKP review}, for which the code space is $\left\{ \sum_{j=-\infty}^{\infty}\ket{\h p=2j\sqrt{\pi}+p_{0}}\right\} _{p_{0}}$. Deterministic (random) displacements along $\hat{p}$ induce rotation (dephasing) inside this subspace, while displacements along $\hat{x}$ take the state outside of this code space and can be detected by the relevant ``syndrome'' measurement of $\hat{x}\;\text{mod }2 \sqrt{\pi}$~\cite{GottesmanPRA01EncodingQubit}.

\section{Proof of Claim~\ref{claim:small signal limit}}
\label{sec:small signal generalisations}
Here, we prove Claim~\ref{claim:small signal limit} by a Taylor expansion and again using the fact that, in the limit of small signals $\si\rightarrow0$, an optimal measurement is to project onto the eigenbasis.

In general, a unitary $\h U_\theta$ may be expanded around $\theta=0$ to obtain the following approximation up to $\order{\theta^3}$ 
\begin{align}\label{eq:unitary expansion}
    \h U_\theta &\approx \h U_0 + \theta \dot{\h U}_0 + \frac{1}{2} \theta^2 \ddot{\h U}_0
    \intertext{where the unitarity condition $\h U_\theta^\dag \h U_\theta=1$ implies that}
    0 &\approx \dot{\h U}_0^\dag\h U_0+\h U_0^\dag\dot{\h U}_0 \\
    0 &\approx \ddot{\h U}_0^\dag\h U_0+\h U_0^\dag\ddot{\h U}_0+2\dot{\h U}_0^\dag\dot{\h U}_0
    .
\end{align}
Let $\h H := i\h U^\dag_0 \dot{\h U}_0$ and $\h Z:=-\h U_0^\dag\ddot{\h U}_0$ such that $\h Z = \h H^2+2 i \h B$ where $\h H^2 =\reSmall{\h Z}= \dot{\h U}^\dag_0 \dot{\h U}_0$ and $\h B = \frac{1}{2} \imSmall{\h Z}=\frac{i}{4}(\h U_0^\dag\ddot{\h U}_0-\ddot{\h U}_0^\dag\h U_0)$.

Using this expansion around $\theta=0$, then the random unitary channel $\Lambda_\si$ in Eq.~\ref{eq:random unitary channel} is approximately
\begin{align*}
    \Lambda_\si(\h \rho) &\approx \h U_0\h \rho \h U^\dag_0 + \si^2\left( \dot{\h U}_0\h\rho \dot{\h U}^\dag_0+ \frac{1}{2}(\Ddot{\h U}_0\h \rho \h U^\dag_0 +\h U_0\h \rho \Ddot{\h U}^\dag_0)\right) \nonumber.
\end{align*}
Without loss of generality, we can study $\h U_0^\dag\Lambda_\si(\h \rho)\h U_0$ since $\h U_0^\dag$ does not depend on the parameter $\theta$ and thus will not affect the QFI if applied to the state after the channel. The final state then becomes
\begin{align}
    \h U_0^\dag\Lambda_\si(\h \rho)\h U_0
    &\approx \h \rho + \si^2\left(\h H\h\rho \h H - \frac{1}{2}(\h Z\h \rho +\h \rho \h Z^\dag)\right)
    \label{eq:general random unitary channel}
    \\&= \h \rho - i[\si^2\h B,\h\rho] + \si^2\left(\h H\h\rho \h H - \frac{1}{2}\{\h H^2,\h \rho\}\right)
    \nonumber.
\end{align}
This is an approximation to the master equation evolution of $\h\rho$ by the Hamiltonian $\si^2\h B$ and the Lindbladian jump operator $\h H$ with decay rate $\si^2$ in the limit of $\si^2 T \ll 1$ where $T$ is the total evolution time. In the additive case of $\h U_\theta = \exp(-i\theta\h H)$, $\h B=0$ such that $\h Z = \h H^2$ and there is no Hamiltonian evolution. Moreover, $\Lambda_\si$ is then a decoherence channel in the eigenbasis of $\h H$ similarly to Eq.~\ref{eq:decoherence channel in position basis} for $\h H = \h x$.

Since $\partial_\si(\h U_0^\dag\Lambda_\si(\h \rho)\h U_0) \rightarrow0$ as $\si\rightarrow0$, the optimal measurement (after $\h U_0^\dag$) is projection $\h\Pi$ onto the support of $\h\rho$, i.e.\ $\h \Pi = \sum_j \ket{\psi_j}\bra{\psi_j}$ in the eigenbasis $\{\ket{\psi_j}\}_j$ of $\h\rho$~\cite{gefen2019overcoming,gorecki2022quantum}. In particular, since $\trSmall{\h \Pi[\si^2\h B,\h\rho]}=0$, we may ignore the Hamiltonian evolution in the limit of $\si\rightarrow0$.
By Eq.~\ref{eq:general random unitary channel}, the probability to remain in the support of $\h\rho$ is
\begin{align*}
    p=\trSmall{\h \Pi\h U_0^\dag\Lambda_\si(\h \rho)\h U_0}=1+\si^2\left(\evSmall{\h H\h\Pi\h H}-\evSmall{\h H^2}\right)
\end{align*}
and the probability to leave is $1-p=\si^2\evSmall{\h H\h\Pi_\perp\h H}$ where $\h\Pi_\perp=1-\h\Pi$ is the projection onto the null space of $\h\rho$. Here, the expectation values are calculated with respect to $\h\rho$. Since the QFI equals the CFI of the optimal measurement, 
\begin{align}\label{eq:lossy QFI zero signal}
    \IQ(\si=0)&=4\evSmall{\h H\h\Pi_\perp\h H}
    \intertext{and, if $\h\rho$ is pure, then the QFI is}
    \IQ(\si=0)&=4\varSmall{\h H}
\end{align}
which is equal to the QFI for the deterministic unitary case $\h U_\si\h \rho \h U_\si^\dag$. This correspondence between the deterministic and random cases does not hold for $\si>0$. A similar observation for non-unitary channels in the limit of $\si\rightarrow0$ was made previously in Ref.~\cite{takeoka2003unambiguous}.

\section{Estimation of a weak decay rate} 
\label{sec:lindbladian}
Here, we consider the general scenario of estimating a weak decay rate. This is inspired by the fact that Eq.~\ref{eq:general random unitary channel} is analogous to a short-time solution of a master equation, where the jump operator is $\h H$ and $\sigma^{2}$ is equivalent to the decay rate multiplied by the evolution time. We consider the cases of a single jump operator and multiple jump operators sharing the same weak decay rate. Ref.~\cite{takeoka2003unambiguous} previously studied this problem in the case of an initial pure state.

Suppose that a quantum state $\h\rho(t)$ evolves in time $t$ under the following Lindbladian master equation:
\begin{align*}    
    \partial_t \h\rho(t) = \gamma\mathcal{L}_{\h Y}[\h \rho(t)], \quad \mathcal{L}_{\h Y}[\h \rho(t)] = \h Y\h\rho(t)\h Y^\dag - \frac{1}{2}\{\h Y^\dag\h Y, \h\rho(t)\}
\end{align*}
where $\gamma\geq0$ is the decay rate, $\h Y$ is the (potentially non-Hermitian) jump operator, and we seek to estimate $\sqrt{\gamma}$. 
In the short-time or weak-decay limit of $\gamma T c\ll1$, where $T$ is the total evolution time and $c=\trSmall{\h Y\h \rho\h  Y^{\dagger}}$, then the evolution of an initial state $\h\rho$ is approximated by the channel $\Lambda_\gamma$ with the following Kraus representation
\begin{align*}
    \Lambda_\gamma(\h \rho)
    &= \h K_0 \h\rho \h K_0^\dag + \h K_1 \h\rho \h K_1^\dag
    \intertext{where the Kraus operators are}
    \h K_0 &:= 1-\frac{\gamma T}{2}\h Y^\dag\h Y \approx \cos\left(\sqrt{\gamma T\h Y^\dag\h Y}\right) 
    \\ \h K_1 &:=  \sqrt{\gamma T}\h Y \approx \sin\left(\sqrt{\gamma T}\h Y\right)
    \intertext{such that}
    1 &= \h K_0^\dag\h K_0 + \h K_1^\dag\h K_1 + \mathcal{O}\left({(\gamma T)^2}\right)
    .
\end{align*}
Since the total evolution time $T$ is known, we can measure the final state $\Lambda_\gamma(\h \rho)$ to estimate the weak decay rate $\sqrt{\gamma}$. Since $\partial_{\sqrt{\gamma}}\Lambda_\gamma\rightarrow0$ as $\sqrt{\gamma}\rightarrow0$, then the QFI with respect to $\sqrt{\gamma}$ is~\cite{gefen2019overcoming,gorecki2022quantum}
\begin{align}\label{eq:Lindbladian QFI zero signal}
    \IQ(\sqrt{\gamma}=0)=4T\evSmall{\h Y^\dag\h\Pi_\perp\h Y}
\end{align}
where $\h \Pi_\perp$ projects onto the null space of the state $\h\rho=\sum_j p_j \ket{\phi_j}\bra{\phi_j}$ before $\Lambda_\gamma$. 
If $\h\rho=\ket\phi\bra\phi$ is pure, then Eq.~\ref{eq:Lindbladian QFI zero signal} becomes
\begin{align}\label{eq:Lindbladian QFI zero signal, pure state}
    \IQ(\sqrt{\gamma}=0)=4T\left(\evSmall{\h Y^\dag\h Y}-\absSmall{\evSmall{\h Y}}^2\right)
\end{align}
which resembles the formula for the variance $\var{Z}=\evText{\absSmall{Z}^2}-\abs{\evText{Z}}^2$ of a complex random variable $Z$. If $\h Y$ is Hermitian, then $\IQ(\sqrt{\gamma}=0)=4T\varSmall{\h Y}$, which is equivalent to Eq.~\ref{eq:lossy QFI zero signal} for $\gamma T=\si^2$ and $\h Y = \h H$.
If instead $\h\rho$ is mixed, then Eq.~\ref{eq:Lindbladian QFI zero signal} implies that the Rayleigh curse will be avoided if and only if $\h\rho$ is partial rank and the jump operator $\h Y$ transitions some of $\h \rho$ into its null space. $\h\rho$ may be mixed, e.g., because $\Lambda_\gamma$ is preceded by another noise channel $\Lambda_{\gamma'}$ with its own decay rate $\gamma'$, which might not be small, and jump operator $\h Y'$ such that the total channel is $\Lambda_{\gamma}\circ\Lambda_{\gamma'}$. It is perhaps more natural, however, to consider these two processes occurring simultaneously within the following Lindbladian master equation:
\begin{align}
    \label{eq:simultaneous noise processes}
    \partial_t \h\rho(t) = \gamma\mathcal{L}_{\h Y}[\h \rho(t)] + \gamma'\mathcal{L}_{\h Y'}[\h \rho(t)].
\end{align}
We consider an example of this scenario below.

\subsection{Example: qubit noise channels}
Consider a qubit simultaneously undergoing amplitude damping (loss) and dephasing such that the Lindbladian master equation is thus
\begin{align*}
    \partial_t \h\rho(t) = \gamma_{\text{deph}}\mathcal{L}\left(\h \sigma_{z}\right)+\gamma_{\text{loss}}\mathcal{L}\left(\h \sigma_{-}\right).
\end{align*} 
After evolving for a time $T$, the individual solutions to the amplitude damping and dephasing channels are, respectively, 
\begin{align*}
    \Lambda_{\text{deph}}\left(\h \rho\right)&=\bmatrixByJames{
    \rho_{1,1} & \rho_{1,0}e^{-\gamma_{\text{deph}}T}\\
    \rho_{1,0}^{*}e^{-\gamma_{\text{deph}}T} & \rho_{0,0}}
    \\\Lambda_{\text{loss}}\left(\h \rho\right)&=\bmatrixByJames{
    \rho_{1,1}e^{-\gamma_{\text{loss}}T} & \rho_{1,0}e^{-\gamma_{\text{loss}}T/2}\\
    \rho_{1,0}^{*}e^{-\gamma_{\text{loss}}T/2} & \rho_{0,0}+\rho_{1,1}\left(1-e^{-\gamma_{\text{loss}}T}\right)},
\end{align*}
where $\rho_{j,k} = \braopket{j}{\h\rho}{k}$. 
Even though $[\h \sigma_{z}, \h \sigma_{-}]\neq0$, these channels commute such that the solution to the Lindbladian master equation is given by: $\Lambda_{\text{deph}}\circ\Lambda_{\text{loss}}\left(\h \rho\right)=\Lambda_{\text{loss}}\circ\Lambda_{\text{deph}}\left(\h \rho\right)$.

This means that estimating a weak loss $\sqrt{\gamma_{\text{loss}}}\ll1$ is unaffected by dephasing since the optimal initial state for sensing $\Lambda_{\text{loss}}$ is $\ket{1}\bra{1}$ which is stabilized by $\Lambda_{\text{deph}}$. By Eq.~\ref{eq:Lindbladian QFI zero signal, pure state}, the QFI with respect to $\sqrt{\gamma_{\text{loss}}}$ is thus $\IQ(\sqrt{\gamma_{\text{loss}}})=4T$.

On the other hand, estimating a weak dephasing $\sqrt{\gamma_{\text{deph}}}\ll1$ is strongly affected by loss. (We assume that $\gamma_{\text{loss}}>0$ is fixed and known.) Eq.~\ref{eq:Lindbladian QFI zero signal} implies that the QFI for any initial state is Rayleigh cursed since the state after $\Lambda_{\text{loss}}$ is either full rank or $\proj{0}$ which has zero QFI. The QFI with respect to $\sqrt{\gamma_{\text{loss}}}$ therefore vanishes in the limit of $\gamma_{\text{loss}}\rightarrow0$ for any single-qubit initial state.

\subsection{Multiple jump operators}
\label{sec:lindblad multiple}
Suppose instead that the same weak decay rate $\gamma$ is common to $m$ jump operators $\{\h Y_j\}_{j=1}^m$ such that the master equation is
\begin{align*}
    \partial_t \h\rho(t) = \gamma\sum_{j=1}^m \mathcal{L}_{\h Y_j}[\h\rho(t)]    
    .
\end{align*} 
Then, the evolution of $\h \rho$ is approximated in the limit of $\gamma T m c \ll 1$, where $c=\max_{j}\trSmall{\h Y_j \h \rho \h Y_j^{\dagger}}$, by the following channel 
\begin{align*}
    \Lambda_\gamma(\h \rho)
    &= \sum_{j=1}^m \left(\h K_{0,j} \h\rho \h K_{0,j}^\dag + \h K_{1,j} \h\rho \h K_{1,j}^\dag\right) 
    \intertext{where the Kraus operators are}
    \h K_{0,j} &:= \frac{1}{\sqrt{m}}-\frac{\gamma T\sqrt{m}}{2}\h Y_j^\dag\h Y_j 
    \\ \h K_{1,j} &:= \sqrt{\gamma T}\h Y_j. 
\end{align*}

By a similar argument to the $m=1$ case above, then 
\begin{align}\label{eq:lindblad multiple jumps small signal}
    \IQ(\sqrt{\gamma}=0)=4T\sum_{j=1}^m\evSmall{\h Y_j^\dag\h\Pi_\perp\h Y_j}
    .
\end{align}
For example, in Eq.~\ref{eq:encoding 2D}, $\gamma T = \si^2$, $\h Y_1 = \h x$, and $\h Y_2 = \h p$ such that Eq.~\ref{eq:lindblad multiple jumps small signal} implies Eq.~\ref{eq:2D small signal}.

\section{Finite-dimensional systems}
\label{sec:finite dimensions}
Here, we consider a random unitary channel acting on a finite-dimensional system. Given a $d$-dimensional quantum system, the random unitary channel $\Lambda_{\si}$ is defined analogously to Eq.~\ref{eq:random unitary channel} as $\Lambda_{\si}(\h \rho)=\intginf{\theta}p(\theta)\h U_{\theta}\h \rho\h U_{\theta}^{\dag}$, where $p(\theta)\sim\mathcal{N}(0,\si^{2})$ and $\h U_{\theta}$ is a $\theta$-dependent unitary that acts on this system. We discuss the differences between finite-dimensional systems and the infinite-dimensional harmonic oscillator system that we study. In particular, we give an example where the analogous upper bound on the ECQFI to Eq.~\ref{eq:lossless QFI UB} can be loose for $\si>0$.

For finite-dimensional systems, the optimal initial pure state in the limit of $\si\rightarrow0$ is an equal superposition of the states corresponding to the smallest and largest eigenvalues of $\h H$ to maximise $4\varSmall{\h H}$. If a loss precedes the channel, then the mixed state $\h\rho$ before the encoding needs to instead optimise $\langle{\h H\h\Pi_\perp\h H}\rangle$. 

For $\si>0$, however, the same state may no longer be optimal, even in the lossless case. For example, consider sensing the phase diffusion of a qubit, modelled as a random rotation channel $\Lambda_\si$ with $\h U_\theta=\exp(-i\theta \h H)$ and $\h H = \frac{1}{2}\h\si_z$. The optimal initial state for $\si>0$, even allowing for entangled resources, is, e.g., $\ket{\uparrow_x}$ with a QFI of $\frac{\si^2}{e^{\si^2}-1}$ as this saturates the ECQFI as proved below. In comparison, the na\"ive upper bound on the ECQFI, analogous to Claim~\ref{claim:lossless QFI UB}, is $\frac{2}{\si^2+2}$ which is loose for $\si>0$, unlike for the random bosonic displacement channel.

The ECQFI for the random rotation channel $\Lambda_\si$ can be found as follows. The channel has the following Kraus operators
\begin{align*}
    \h K_{1}=\sqrt{\frac{1}{2}\left(1+e^{-\frac{1}{2}\si^2}\right)}\h 1,\quad
    \h K_{2}=\sqrt{\frac{1}{2}\left(1-e^{-\frac{1}{2}\si^2}\right)}\hat{\sigma}_{z}
\end{align*}
such that an upper bound on the ECQFI is~\cite{fujiwara2008fibre}
\begin{align*}
    \IQ^{\Lambda_\si}(\si) &\leq 4 \biggl\|\sum_j \dot{\h K}_j^\dag\dot{\h K}_j\biggr\|
    \\&= \frac{\si^2}{e^{\si^2}-1}
    .
\end{align*}
This implies that $\ket{\uparrow_x}$ is optimal and attains the ECQFI.

\section{Review of GKP finite-energy states}
\label{sec:GKP review}

\begin{figure}
    \centering
    \includegraphics[width=0.85\columnwidth]{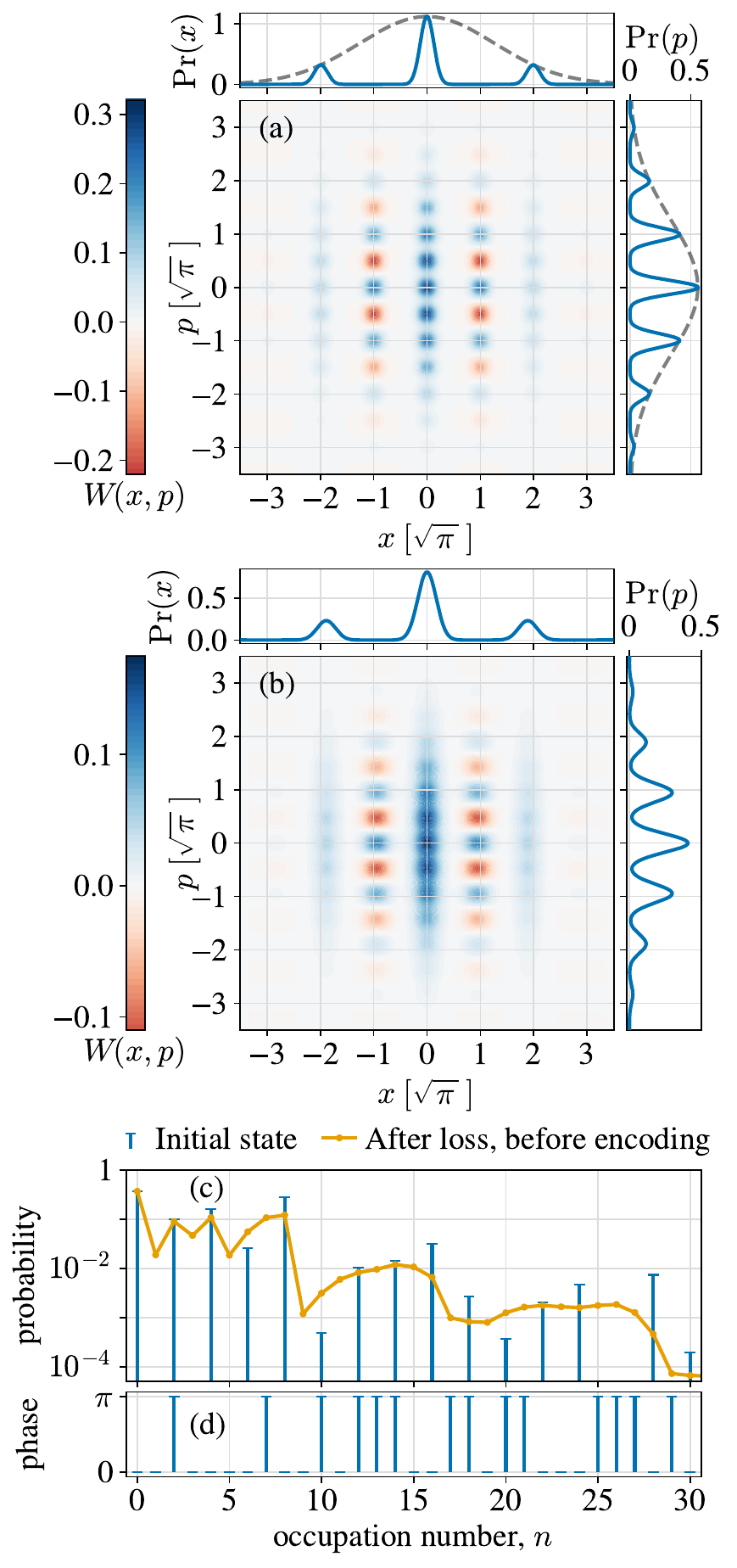}
    \caption{GKP finite-energy state for $\Delta^2\approx0.1$. Wigner function and quadrature probability distributions of (a) the pure initial state and (b) the state after a loss of $\eta=0.1$ before the encoding. The Gaussian envelope of the marginals for the pure initial state is shown. Visually, the loss channel enlarges each sub-vacuum peak and pulls it toward the origin. (c) Fock basis probabilities on a logarithmic scale shown up to $n = 30$ before and after the loss, in a truncated Hilbert space of dimension 100. (d) Fock basis complex phases of the pure initial state. The coefficients are all real.}
    \label{fig:GKP_finite-energy_state}
\end{figure}

Here, we review the families of finite-energy states that we study which approximate the ideal GKP state in the high energy limit. 

The ideal GKP grid state is an unnormalised, infinite-energy single-mode state consisting of an equal superposition of infinitely many, evenly spaced position eigenstates, e.g.\ 
\begin{align*}
    \ket{\text{GKP}_\text{ideal}} &\propto \sum_{j=-\infty}^\infty \ket{\h x=2j\sqrt{\pi}}
    = \sum_{j=-\infty}^\infty \ket{\h p=2j\sqrt{\pi}}
\end{align*}
where $\ket{\h x=x_0}$ ($\ket{\h p=p_0}$) is the position (momentum) eigenstate at $x_0$ ($p_0$)~\cite{GottesmanPRA01EncodingQubit}. There are many different continuous families of normalised, finite-energy states that approximate the same ideal state in the high-energy limit~\cite{TzitrinPRA20ProgressPractical}. One such family $\ket{\text{GKP}_\Delta}$ is defined as a superposition with a Gaussian window function $\sim\mathcal{N}(0,\frac{1}{2}\Delta^{-2})$ of displaced squeezed single-mode states, with variance $\frac{1}{2}\Delta^2$ and evenly spaced by $2\sqrt{\pi}$, as follows
\begin{align*}
    \ket{\text{GKP}_\Delta} &\propto \sum_{j=-\infty}^\infty \exp\left(-2\pi\Delta^2 j^2\right) \h U_{2j\sqrt{\pi}} \ket{\text{SMSV}_{\frac{1}{2}\Delta}}
    \intertext{where $\h U_{\mu}=\exp(-i\mu \h p)$ translates $\h x$ to $\h x + \mu$ and the position-basis wavefunction of the SMSV state is}
    \ket{\text{SMSV}_{\frac{1}{2}\Delta}} &= \frac{1}{\pi^{\frac{1}{4}}\sqrt\Delta} \intginf{x} \exp\left(-\frac{x^2}{2\Delta^2}\right)\ket{\h x = x}
    .
\end{align*}
This family approximates the ideal GKP state, $\ket{\text{GKP}_\text{ideal}}$, in the limit of $\Delta\rightarrow0$. Numerically, for a fixed but small $\Delta > 0$, $\ket{\text{GKP}_\Delta}$ can be approximated by a finite sum over $j=-J, \mathellipsis, J$, where some large integer $J$ is chosen such that $2\pi\Delta^2 J^2\gg1$.

The Wigner function of the pure initial state $\ket{\text{GKP}_\Delta}$ is shown in Fig.~\hyperref[fig:GKP_finite-energy_state]{\ref*{fig:GKP_finite-energy_state}a} where the marginals fit the Gaussian envelope $\sim\mathcal{N}(0,\frac{1}{2}\Delta^{-2})$. The mixed state after a loss, $\h\rho=\Lambda^\text{loss}_\eta(\ket{\text{GKP}_\Delta}\bra{\text{GKP}_\Delta})$, is shown in Fig.~\hyperref[fig:GKP_finite-energy_state]{\ref*{fig:GKP_finite-energy_state}b}, where, visually, the sub-vacuum peaks grow and move towards the origin. Solely for the purpose of plotting, we calculate these Wigner functions using a different family of finite-energy GKP states $\ket{\text{GKP}_\varepsilon'}$ defined as the normalised result of applying the non-unitary operator $\exp(-\varepsilon \h n)$ to the ideal GKP state~\cite{TzitrinPRA20ProgressPractical}. This approximation is valid since, when $\frac{1}{24}\Delta^3$ is negligible compared to $\frac{1}{2}\Delta$, then $\ket{\text{GKP}_\Delta}\approx\ket{\text{GKP}_{\varepsilon=\Delta^2}'}$, e.g.\ this holds for the value of $\varepsilon=0.01$ used in Fig.~\ref{fig:GKP_finite-energy_state}. We use this family $\ket{\text{GKP}_\varepsilon'}$ here since the Wigner function is more efficient to calculate than in the Fock basis representation shown in Fig.~\hyperref[fig:GKP_finite-energy_state]{\ref*{fig:GKP_finite-energy_state}c--d}~\cite{bourassa2021fast}. While, at low energies, the different families of finite-energy GKP states diverge, e.g.\ the behaviour of $\ket{\text{GKP}_\Delta}$ as $N\rightarrow0$ seen in Fig.~\hyperref[fig:channelFI]{\ref*{fig:channelFI}b} may be different from $\ket{\text{GKP}_{\varepsilon=\Delta^2}'}$, we are principally interested in the high-energy limit where they agree.

\section{Numerical methods}
\label{sec:Biconvex optimisation}
Here, we discuss the numerical methods mentioned in Sec.~\ref{sec:non-Gaussian states} that we use to search for the optimal initial single-mode state in the presence of loss. We first discuss our brute-force approach and then an alternate approach that we experimented with.

Our code uses \textsc{Mathematica}~\cite{mathematica} and \textsc{Python}~\cite{python,ipython,jupyter,numpy,matplotlib,pandas,VirtanenNM20SciPyFundamental,qutip,strawberryfields} and is available online~\cite{repo}.

\subsection{Brute-force approach}

\begin{figure}
    \centering
    \includegraphics[width=\columnwidth]{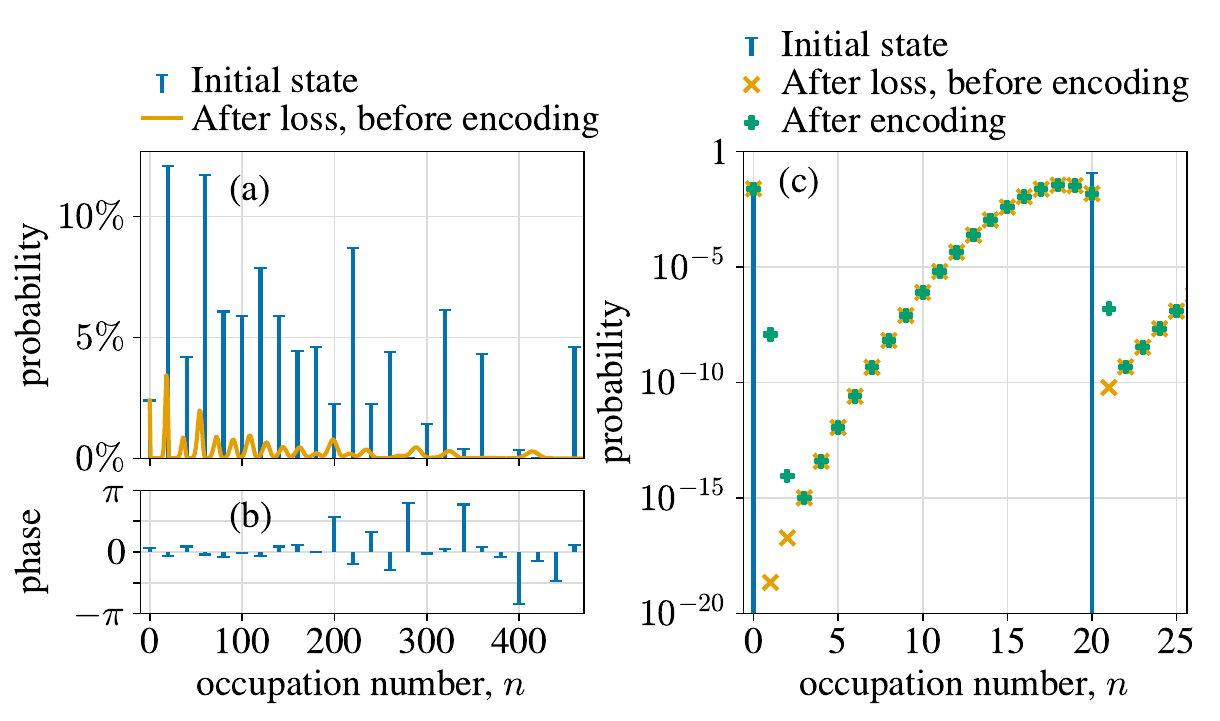}
    \caption{Fock basis (a) probabilities and (b) complex phases of the non-Gaussian initial state $\ket{\psi_\text{num.}}$ found by numerical optimisation. The probabilities after the loss of $\eta=0.1$ but before the encoding are a sum of binomial distributions. (c) An example of two neighbouring components. For $\si=10^{-3}$, the transition to $\ket{1}$ is dominated by the signal trajectory $\ket{0}\mapsto\ket{1}$ rather than the loss trajectory $\ket{20}\mapsto\ket{1}$. But, this is not the case for all neighbouring components of $\ket{\psi_\text{num.}}$.}
    \label{fig:wellseparated_fockstates}
\end{figure}

While we have observed numerically that preparing GKP finite-energy states approaches the ECQFI for $0<\si^2\ll\eta,\eta_A$ in the limit of large $\ev{\h n}=N$, we do not know if this is the only such family of non-Gaussian states. Moreover, while we know that the upper bound in Eq.~\ref{eq:lossy UB constrained} is loose for a fixed finite $\ev{\h n}=N$, we do not know whether TMSV with perfect storage is optimal. We search here numerically for non-Gaussian states that might perform better.

We consider sparse superpositions of finitely many Fock states. For simplicity, we consider superpositions equally-spaced in the number basis, i.e.\ $\ket{\psi}=\sum_{k=0}^{K-1}c_k\ket{mk}$ for some spacing $m$ and number of peaks $K$. For successively larger truncations of the Hilbert space in the number basis, we fix different values of $m$ and $K$ and numerically optimise the basis coefficients $\{c_k\}_{k=0}^{K-1}$ to maximise the QFI using gradient descent and particle swarm optimisation methods. This is a brute-force approach to finding the optimal state.

We show an example of one such state in Fig~\ref{fig:wellseparated_fockstates} which was found numerically within a truncated Hilbert space of dimension 490. This state, $\ket{\psi_\text{num.}}=\sum_{j=0}^{23}c_j\ket{20j}$ where $m=20$ and $K=24$, has a QFI of 18.4 for $\ev{\h n}=158.9$ which is within 9\% of the ECQFI of 20 for $\si^2=10^{-6}$ and $\eta=0.1$. (This value of the QFI appears stable under small perturbations in $\si$ and $\eta$.) It is important here that the signal and loss levels are fixed and finite since the intuition for this sparse state is that, e.g., the transition to $\ket1$ is dominated by the signal trajectory from $\ket0$ rather than the loss trajectory from $\ket{20}$, as shown in Fig.~\hyperref[fig:wellseparated_fockstates]{\ref*{fig:wellseparated_fockstates}c}. If the loss was larger or the signal was smaller, then we expect a larger separation $m$ to be necessary. The optimal measurement for this state is a superposition of number-resolving measurements, i.e.\ projections onto different linear combinations of the Fock states, similar to the generalised parity measurements for binomial codes~\cite{MichaelPRX16NewClass}. We conjecture that optimising over superpositions of unequally-spaced but still sparsely separated Fock states in larger truncations of the Hilbert space can get arbitrarily close to the ECQFI for any $\si^2\ll\eta$.

\subsection{Biconvex optimisation}
Here, we describe this task as a biconvex optimisation problem and discuss how, in principle, it might be solved without using a brute-force approach.

The general problem of finding the optimal protocol, i.e.\ initial state and measurement scheme, for sensing a parameter $\theta$ encoded by a quantum channel $\Lambda_\theta$ may be formulated as follows. This problem reduces to biconvex optimisation of the function $f(\ket{\psi}, X)$ given by~\cite{macieszczak2013quantum}
\begin{align*}
    f(\ket{\psi}, X) := -\braopket{\psi}{\left(-\Lambda_\theta^\dag(X^2)+\dot{\Lambda}_\theta^\dag(X)\right)}{\psi}
\end{align*}
where $\Lambda_\theta^\dag$ is the conjugate channel of $\Lambda_\theta$ in the Heisenberg picture sense, $\dot{\Lambda}_\theta^\dag = \deriv{\Lambda_\theta^\dag}{\theta}$, and $X\in\mathcal{L}^2(\mathcal{H})$ on a Hilbert space $\mathcal{H}$ with finite dimension (e.g.\ on a truncated bosonic Hilbert space). Here, the biconvexity of $f$ means that $f$ is convex with respect to $\ket{\psi}$ if $X$ is held constant and with respect to $X$ if $\ket\psi$ is held constant, but not necessarily with respect to both at once. The global minimum of $f$ exists and is the negative of the CQFI in Eq.~\ref{eq:channel QFI definition}
\begin{align*}
    \IQ^{\Lambda_\theta, \text{no ancilla}}(\theta) = - \inf_X \inf_{\ket\psi} f(\ket{\psi}, X)
\end{align*}
where the order of the infima may be exchanged and the negative sign is included as a convention such that $f$ is biconvex rather than biconcave. If $X$ is the SLD of $\Lambda_\theta(\ket{\psi}\bra{\psi})$ with respect to $\theta$, then $f(\ket\psi, X) = - \IQ^{\Lambda_\theta(\ket{\psi}\bra{\psi})}$ is the negative of the QFI given the initial state $\ket\psi$. The initial state may be restricted to being pure given the convexity of the QFI with respect to the initial state.

In Ref.~\cite{macieszczak2013quantum}, this biconvex optimisation problem is proposed to be solved locally using an alternative convex search (ACS)~\cite{gorski2007biconvex}. The ACS proceeds from a given starting point $\ket{\psi_0}$ by finding the corresponding SLD $X_0$ that solves the convex problem with $\ket{\psi_0}$ held constant, then finds the corresponding $\ket{\psi_1}$ that solves the convex problem with $X_0$ held constant, and so on. Although the ACS is monotonic and converges, it is only a local search and is not guaranteed to find the global minimum although different random starting points $\ket{\psi_0}$ can be used to explore the nonconvex geometry~\cite{gorski2007biconvex}.

In theory, a deterministic global optimisation program (GOP) exists for solving biconvex problems~\cite{floudas1990global}. The GOP, however, only converges in exponential time in the number of confounding variables. Here, this would be proportional to the dimension of the truncated Hilbert space squared in the worst case. Although methods exist to reduce the number of confounding variables in particular cases~\cite{floudas2013deterministic}, the GOP is impractical in general. This limitation has been also discussed in the context of other biconvex optimisation problems in quantum information science, e.g.\ see Refs.~\cite{noh2018quantum,zhou2023optimal}.

While the ACS has proved successful in these other applications, it has not been successful for sensing $\si$ from $\Lambda_\si$. We have observed that the ACS does not converge to or surpass the QFI of the GKP finite-energy states even after 10000 iterations starting from the vacuum or random states, or significantly improve the finite superpositions of Fock states from the brute-force approach when started from them. 
(To generate a random state, we uniformly sample the real and imaginary parts of the coefficients in the number basis in $(-1,1)$ and then normalise.) Determining what aspects of the geometry of our problem limit the ACS is left to future work.

\section{Classical noise and the optimal initial~state}
\label{sec:Optimal initial states with classical noise}
Here, we prove the results discussed in Sec.~\ref{sec:Classical noise case} about the optimal initial state in the presence of significant classical noise but no loss.

Suppose that the encoding channel is preceded by the classical noise channel $\Lambda^\text{noise}_{\Si_C}$ in Eq.~\ref{eq:classical noise channel Kraus representation} for a given classical noise covariance matrix $\Si_C$ such that the total channel is $\Lambda_{\sigma}^\text{noisy}=\Lambda_\si\circ\Lambda^\text{noise}_{\Si_C}$. We consider the following three cases in turn: $\Si_C=\diag{0,\si_p^2}$ where the random displacements from the classical noise are parallel to (i.e.\ in the same quadrature, $\h p$, as) those from the signal; $\Si_C=\diag{\si_x^2,0}$ where the classical noise is perpendicular (i.e.\ in the opposite quadrature, $\h x$) to the signal; and $\Si_C=\diag{\si_x^2,\si_p^2}$ where the classical noise is in both quadratures which includes the isotropic case $\Si_C=\diag{\si_C^2,\si_C^2}$. Whether the Rayleigh curse arises depends on $\Si_C$ as shown below.

\subsection{Parallel classical noise}
For $\Si_C=\diag{0,\si_p^2}$, we prove that the Rayleigh curse is unavoidable and the optimal initial state for a given $\ev{\h n} = N$ is to prepare SMSV.

We first note that the total channel corresponds to a combined classical noise channel:
\begin{align}
\label{eq:conc_rand_disp}
\Lambda_{\sigma}^\text{noisy} = \Lambda_\si\circ\Lambda^\text{noise}_{\diag{0,\si_p^2}} = \Lambda^\text{noise}_{\diag{0,\sigma^{2}+\sigma_{p}^{2}}}
\end{align}
where $\Si_C=\diagSmall{0,\sigma^{2}+\sigma_{p}^{2}}$ adds the signal and noise in quadrature since they are uncorrelated.

Eq.~\ref{eq:conc_rand_disp} holds immediately for Gaussian states, but let us prove it for any initial state $\h \rho$ by direct calculation:
\begin{align*}
\Lambda_{\sigma}^\text{noisy}(\h\rho)
&=\int_{\R^2} \text{d}\alpha_{1}\text{d}\alpha_{2}\; p_{\si^2}\left(\alpha_{2}\right)p_{\si_p^2}\left(\alpha_{1}\right)\h U_{\alpha_{2}}\h U_{\alpha_{1}}\h \rho \h U_{\alpha_{1}}^{\dagger}\h U_{\alpha_{2}}^{\dagger}\\
&=\int_{\R^2} \text{d}\alpha_{1}\text{d}\beta\; p_{\si^2}\left(\beta-\alpha_1\right)p_{\si_p^2}\left(\alpha_{1}\right)\h U_{\beta}\h \rho \h U_{\beta}^{\dagger}\\
\intertext{where $p_{\varsigma^2}(\alpha)\sim\mathcal{N}(0,\varsigma^2)$, $\beta=\alpha_{1}+\alpha_{2}$, and $\h U_\alpha = \exp(i\alpha \h x)$. By recognising the convolution between two Gaussian distributions, then this equals}
&=\int_{\R} \text{d}\beta\; p_{\si^2+\si_p^2}\left(\beta\right)\h U_{\beta}\h \rho \h U_{\beta}^{\dagger}
\\&=\Lambda^\text{noise}_{\diag{0,\sigma^{2}+\sigma_{p}^{2}}}\left(\h\rho\right).
\end{align*}

By the chain rule, Eq.~\ref{eq:conc_rand_disp} implies that the QFI for a given initial state $\h \rho$ is 
\begin{align}
\label{eq:QFI CN parallel}
\IQ^{\Lambda^\text{noisy}_\si(\h\rho)}\left(\si\right)=\frac{\sigma^{2}}{\sigma^{2}+\sigma_{p}^{2}}\; \IQ^{\Lambda_{\si}\left(\h\rho\right)}\left(\sigma=\sqrt{\sigma^{2}+\sigma_{p}^{2}}\right). 
\end{align} 
Hence, the ECQFI $\IQ^{\Lambda^\text{noisy}_\si}\left(\si\right)$ for a given value of $\si_p$ is proportional to the noiseless ECQFI $\IQ^{\Lambda_{\si}}(\si)$ evaluated at the signal with variance $\sigma^2 + \sigma_p^2$ which implies Eq.~\ref{eq:ECQFI CN parallel}. Since preparing SMSV is the optimal initial state that attains the noiseless ECQFI for a fixed $\ev{\h n} = N$, then it is also the optimal initial state in the presence of parallel classical noise. The conversion factor of $\frac{\sigma^{2}}{\sigma^{2}+\sigma_{p}^{2}}$, however, will introduce the Rayleigh curse for $\si\ll\si_p$ and mean that the noiseless ECQFI cannot be recovered regardless of the initial state. 

Before moving on to discuss the other possible cases of classical noise, we analyse what happens when both parallel classical noise and loss are present. Let the total channel be
\begin{align*}
    \Lambda_{\sigma}^\text{noisy} = \Lambda_\si\circ\Lambda^\text{noise}_{\diag{0,\si_p^2}}\circ\Lambda^\text{loss}_\eta.
\end{align*}
By combining Eq.~\ref{eq:lossy ECQFI} and Eq.~\ref{eq:QFI CN parallel}, then the ECQFI is given in Eq.~\ref{eq:clnoise with loss}. By a similar argument to that made below Eq.~\ref{eq:QFI CN parallel}, the optimal initial states are the same as for the case with loss but no classical noise.

\subsection{Perpendicular classical noise}
For $\Si_C=\diag{\si_x,0}$, we prove that, for a fixed finite signal $\si>0$, the noiseless ECQFI can be recovered in the high energy limit, and there exist finite-energy states that do not exhibit the Rayleigh curse as $\si\rightarrow0$.

Suppose that we prepare an SMSV state with $\ev{\h n} = N$, then, given perpendicular classical noise, the QFI calculated from Eq.~\ref{eq:Gaussian QFI CM} is: 
\begin{align*}
\IQ(\si) &= \frac{4}{\left(2\xi_{N}\right)^{-2}\left(\frac{\sigma_{x}}{\sigma}\right)^{2}\frac{\left(\frac{1}{2}\sigma_{x}^{2}+\xi_{N}\right)}{\left(\sigma_{x}^{2}+\xi_{N}\right)^{2}}+\xi_{N}^{-1}+2\sigma{}^{2}}
\\&\xrightarrow[N\rightarrow\infty]{}\frac{4}{\xi_{N}^{-1}+2\sigma{}^{2}}.
\end{align*}
For a fixed finite signal $\si>0$, the above limit as $N\rightarrow\infty$ implies that the noiseless ECQFI in Eq.~\ref{eq:lossless QFI UB} is recovered for large enough $N$ such that $\sigma \xi_{N} \gg \sigma_{x}$. Note that the optimal measurement for SMSV in the noiseless case was to first anti-squeeze and then perform a number-resolving measurement. Here, intuitively, this anti-squeezing operation squeezes the perpendicular noise such that its effect becomes vanishing in the high energy limit.

Note that we assume a particular order of limits above: $\lim_{\si\rightarrow0}\sup_{\ket\Psi}\IQ(\si)$. As discussed in Sec.~\ref{sec:small signals}, this is the relevant order in practice because we are interested in small but finite signals $\si>0$. For completeness, we now briefly discuss the opposite order of limits: $\sup_{\ket\Psi}\lim_{\si\rightarrow0}\IQ(\si)$. While the family of SMSV states is optimal as $N\rightarrow\infty$ for a fixed finite $\si>0$, a given SMSV state with fixed $\ev{\h n}=N$ exhibits the Rayleigh curse as $\si\rightarrow0$ by Claim~\ref{claim:simplectic eigenvalues}.

Unlike the parallel classical noise case, however, the Rayleigh curse can be avoided using different initial states. For example, we show in Appendix~\ref{sec:TMSV results} that preparing TMSV with perfect storage, i.e.\ no noise channels on the ancilla, with a fixed $\ev{\h n}=N$ particles per mode yields a QFI of
\begin{align}\label{eq:TMSV perp CN}
    \IQ(\si = 0) = \frac{8 N (N+1)}{2 N+1} \xrightarrow[N\rightarrow\infty]{} 4 N
\end{align}
which is independent of the amount of perpendicular classical noise. This recovers the noiseless QFI of $4N$ for TMSV in the high energy limit, but not the noiseless ECQFI of $4 \xi_{N} \approx 8N$ at $\si=0$. Numerical analysis suggests that GKP finite-energy states are also resilient to perpendicular classical noise.

We note that the above analysis, while promising, does not include the relevant real-world case of perpendicular noise jointly in the presence of optical loss. There, the degradation of the squeezed state from the loss may prevent the complete removal of perpendicular noise.

\subsection{Classical noise in both quadratures}
For $\Si_C=\diag{\si_x^2,\si_p^2}$ with $\si_x,\si_p>0$, we combine the above results to show that the Rayleigh curse is inevitable, the noiseless ECQFI cannot be recovered, and that SMSV remains the optimal initial state for a fixed $\si>0$ in the high energy limit. These results apply, e.g., to the isotropic case of $\Si_C=\diag{\si_C^2,\si_C^2}$ with $\si_C>0$.

The random displacement channels $\Lambda^\text{noise}_{\diagSmall{\si_x^2,0}}$ and $\Lambda^\text{noise}_{\diagSmall{0,\si_p^2}}$ commute with each other and obey the following relation:
\begin{align*}
    \Lambda^\text{noise}_{\diagSmall{\si_x^2,\si_p^2}}
    &= \Lambda^\text{noise}_{\diagSmall{\si_x^2,0}}\circ\Lambda^\text{noise}_{\diagSmall{0,\si_p^2}}
    \\&= \Lambda^\text{noise}_{\diagSmall{0,\si_p^2}}\circ\Lambda^\text{noise}_{\diagSmall{\si_x^2,0}},
    \intertext{such that, by Eq.~\ref{eq:conc_rand_disp}, the total channel is}
    \Lambda^\text{noisy}_\si &= \Lambda^\text{noise}_{\diagSmall{\si_x^2,\si^2+\si_p^2}}
    .
\end{align*}
Hence, similarly to Eq.~\ref{eq:QFI CN parallel}, the QFI for a given initial state $\h\rho$ is
\begin{align*}
&\IQ^{\Lambda^\text{noisy}_\si(\h\rho)}\left(\si\right)
\\&=\frac{\sigma^{2}}{\sigma^{2}+\sigma_{p}^{2}}\; \IQ^{\left(\Lambda_{\si}\circ\Lambda^\text{noise}_{\diagSmall{\si_x^2,0}}\right)\left(\h\rho\right)}\left(\sigma=\sqrt{\sigma^{2}+\sigma_{p}^{2}}\right).   
\end{align*} 
The optimal initial state here, therefore, is the same as for the perpendicular case where the total channel is $\Lambda_{\si}\circ\Lambda^\text{noise}_{\diagSmall{\si_x^2,0}}$. For a fixed finite $\si>0$, then this implies that preparing SMSV is optimal in the high energy limit but does not recover the noiseless ECQFI due to the $\frac{\sigma^{2}}{\sigma^{2}+\sigma_{p}^{2}}$ factor. In the limit of $\si\ll\si_p$, then the same factor leads to an unavoidable Rayleigh curse.

When we restrict ourselves to study signals above a classical noise floor, e.g.\ in Sec.~\ref{sec:small signals}, we mean signals $\si$ above an isotropic classical noise $\si_C$, i.e\ for $\si\gg\si_C$, since all protocols exhibit the Rayleigh curse below the noise floor, i.e.\ for $\si\ll\si_C$.

\section{Adaptive protocol}
\label{sec:Adaptive protocol}
Here, we detail an adaptive protocol for our separable measurements scheme (quadrature and number measurements) that converges to a factor of two away from the QFIM bound as shown in Fig.~\hyperref[fig:simulEst]{\ref*{fig:simulEst}b}.

Algorithmically, the adaptive measurement protocol proceeds as follows:
\begin{algorithmic}
\State $k \gets 1$
\While{$k \leq M$}
    \State $p_{k} \gets \text{outcome of quadrature measurement $\hat{p}$}$
    \State \text{Displace quadrature $\hat{p}$ by $-\frac{p_{k}}{k}$}
    \State $n_{k} \gets \text{outcome of number measurement $\hat{n}$}$
    \State $k \gets k + 1$
\EndWhile
\end{algorithmic}

Let us denote $\langle \hat{p} \rangle$ after the $k$th step as $\mu_{k}$. 
Since $p_{k}=\mu_{k-1}+n_{k}$, where $n_{k}$ is a Gaussian random variable $n_{k}\sim \mathcal{N}\left(0,\frac{1}{2}\right)$, then, by induction, $\mu_{k}=-\frac{1}{k}\sum_{i=1}^{k}n_{i}$. This implies that $\Delta\mu_{k}=\frac{1}{\sqrt{2k}}$, i.e.\ the ``size'' of $\mu_{k}$ is equal to the statistical uncertainty in estimating $\mu$ after $k$.

After all of the measurements are performed, we then estimate $\mu_{i}$ with $\tilde{\mu}_{i}=\mu_{i}-\mu_{M}$ and estimate $\sigma$ with $\tilde{\sigma}=\sqrt{\frac{2}{M}\sum_{i}n_{i}-\frac{1}{M}\sum_{i}\tilde{\mu}_{i}^{2}}$.

\section{Proof of Claim~\ref{claim:joint measurement}}
\label{sec:claim_joint_meas}

Here, we prove Claim~\ref{claim:joint measurement}. We first calculate the QFI and then explicitly construct the joint measurement which is asymptotically optimal.

In the limit of $\si\rightarrow0$, the QFI with respect to $\si$ for a pure initial state is $4\varSmall{\h H}$ by Claim~\ref{claim:small signal limit}. We can assume without loss of generality that $\mu \ll 1$. Then, the QFI with respect to $\mu$ can be shown to also be $4\varSmall{\h H}$ by expanding the channel $\Lambda_{\sigma,\mu}$ in Eq.~\ref{eq:random unitary mean and variance} to second order in $\mu$ and $\si$ since higher orders will not contribute to the QFI.

We now construct a joint measurement whose CFIM saturates these QFIs simultaneously. Without loss of generality, we assume that $\evSmall{\h H}=0$ and let $V:=\varSmall{\h H}$ for the initial pure state $\ket{\psi}$. Then, $\ket{\phi_1}:=\frac{1}{\sqrt{V}}\h H|\psi\rangle$ and $\ket{\phi_2}:=\frac{1}{\sqrt{l}}(\h H^2 - V)\ket\psi$ are orthogonal to $\ket{\psi}$ where $l$ is a normalization factor.

The final state $\Lambda_{\sigma,\mu}(\ket\psi\bra\psi)$ can be expanded up to second order in $\mu$ and $\sigma$ using this orthonormal set of states $\left\{\ket\psi,\ket{\phi_1},\ket{\phi_2}\right\}$ as
\begin{align*}
\left(\begin{array}{ccc}
1-\left(\mu^{2}+\sigma^{2}\right)V & -i\mu\sqrt{V} & -\frac{1}{2}\left(\mu^{2}+\sigma^{2}\right)l\\
i\mu\sqrt{V} & \left(\mu^{2}+\sigma^{2}\right)V & 0\\
-\frac{1}{2}\left(\mu^{2}+\sigma^{2}\right)l & 0 & 0
\end{array}\right) 
\end{align*}
such that we can restrict attention to the two-dimensional subspace of $\ket\psi$ and $\ket{\phi_1}$.

We want to consider joint measurements on $M$ copies of the final state $\Lambda_{\sigma,\mu}(\ket\psi\bra\psi)$. Consider the following collective states
\begin{align*}
    |e_{i}\rangle:= \ket{\psi}^{\otimes (i-1)}\ket{\phi_1}\ket{\psi}^{\otimes (M-i)}    
    .
\end{align*}
Let the symmetric state be $|e_{s}\rangle:=\frac{1}{\sqrt{M}}\sum_i|e_{i}\rangle$, and let the anti-symmetric states be denoted as $\{ |e_{a,i}\rangle\} _{i=1}^{M-1}$. In this collective basis, the final state is
\begin{align*}
\Lambda_{\sigma,\mu}(\ket\psi\bra\psi)^{\otimes M}
&\approx\left[1-MV(\mu^{2}+\sigma^{2})\right]|{\psi}\rangle^{\otimes M}\langle{\psi}|^{\otimes M}\\&-i\mu\sqrt{MV}\left(|e_{s}\rangle\langle{\psi}|^{\otimes M}-|{\psi}\rangle^{\otimes M}\langle e_{s}|\right)\\&+V\left(\sigma^{2}+\mu^{2}M\right)|e_{s}\rangle\langle e_{s}|\\
&+\sigma^{2}V\sum_{i=1}^{M-1}|e_{a,i}\rangle\langle e_{a,i}|.    
\end{align*}
We now calculate the CFIM obtained with projective measurement onto
$\frac{1}{\sqrt{2}}\left(|{\psi}\rangle^{\otimes M}+ i|e_{s}\rangle\right)$ and $\left\{ |e_{a,i}\rangle\right\} _{i=1}^{M-1},$ and show that it is asymptotically optimal. The probability to be projected onto the anti-symmetric subspace is $\sigma^{2}V\left(M-1\right)$ which provides the information about $\sigma$. While, the probability to be projected onto $\frac{1}{\sqrt{2}}\left(|{\psi}\rangle^{\otimes M}+ i|e_{s}\rangle\right)$ is
\begin{align*}
p=\frac{1}{2}\left(1-\sigma^{2}\left(M-1\right)V\right)+\sqrt{MV}\mu,
\end{align*}
which provides the information about $\mu$ but no information about $\sigma$ in the limit of $\sigma\rightarrow0$. The CFIM with respect to $\mu$ and $\sigma$ in the limit of $\sigma\rightarrow0$ is
\begin{align*}
\IC = 
\left(\begin{array}{cc}
4MV & 0\\
0 & 4\left(M-1\right)V
\end{array}\right),
\end{align*}
which saturates the QFIM and implies that this joint protocol is asymptotically optimal.

\nocite{apsrev42Control}
\bibliographystyle{bibliography/apsrev4-2-trunc.bst}

\renewcommand{\selectlanguage}[1]{}
\bibliography{bibliography/bib_merged}

\begin{thebibliography}{134}%
\makeatletter
\providecommand \@ifxundefined [1]{%
 \@ifx{#1\undefined}
}%
\providecommand \@ifnum [1]{%
 \ifnum #1\expandafter \@firstoftwo
 \else \expandafter \@secondoftwo
 \fi
}%
\providecommand \@ifx [1]{%
 \ifx #1\expandafter \@firstoftwo
 \else \expandafter \@secondoftwo
 \fi
}%
\providecommand \natexlab [1]{#1}%
\providecommand \enquote  [1]{``#1''}%
\providecommand \bibnamefont  [1]{#1}%
\providecommand \bibfnamefont [1]{#1}%
\providecommand \citenamefont [1]{#1}%
\providecommand \href@noop [0]{\@secondoftwo}%
\providecommand \href [0]{\begingroup \@sanitize@url \@href}%
\providecommand \@href[1]{\@@startlink{#1}\@@href}%
\providecommand \@@href[1]{\endgroup#1\@@endlink}%
\providecommand \@sanitize@url [0]{\catcode `\\12\catcode `\$12\catcode `\&12\catcode `\#12\catcode `\^12\catcode `\_12\catcode `\%12\relax}%
\providecommand \@@startlink[1]{}%
\providecommand \@@endlink[0]{}%
\providecommand \url  [0]{\begingroup\@sanitize@url \@url }%
\providecommand \@url [1]{\endgroup\@href {#1}{\urlprefix }}%
\providecommand \urlprefix  [0]{URL }%
\providecommand \Eprint [0]{\href }%
\providecommand \doibase [0]{https://doi.org/}%
\providecommand \selectlanguage [0]{\@gobble}%
\providecommand \bibinfo  [0]{\@secondoftwo}%
\providecommand \bibfield  [0]{\@secondoftwo}%
\providecommand \translation [1]{[#1]}%
\providecommand \BibitemOpen [0]{}%
\providecommand \bibitemStop [0]{}%
\providecommand \bibitemNoStop [0]{.\EOS\space}%
\providecommand \EOS [0]{\spacefactor3000\relax}%
\providecommand \BibitemShut  [1]{\csname bibitem#1\endcsname}%
\let\auto@bib@innerbib\@empty
\bibitem [{\citenamefont {Verlinde}\ and\ \citenamefont {Zurek}(2021)}]{VerlindePLB21ObservationalSignatures}%
  \BibitemOpen
  \bibfield  {author} {\bibinfo {author} {\bibfnamefont {E.~P.}\ \bibnamefont {Verlinde}}\ \bibnamefont {and}\ \bibinfo {author} {\bibfnamefont {K.~M.}\ \bibnamefont {Zurek}},\ }\bibfield  {title} {\bibinfo {title} {Observational {{Signatures}} of {{Quantum Gravity}} in {{Interferometers}}},\ }\href {https://doi.org/10.1016/j.physletb.2021.136663} {\bibfield  {journal} {\bibinfo  {journal} {Phys. Lett. B}\ }\textbf {\bibinfo {volume} {822}},\ \bibinfo {pages} {136663} (\bibinfo {year} {2021})}\BibitemShut {NoStop}%
\bibitem [{\citenamefont {Li}\ \emph {et~al.}(2023)\citenamefont {Li}, \citenamefont {Lee}, \citenamefont {Chen},\ and\ \citenamefont {Zurek}}]{LiPRD23InterferometerResponse}%
  \BibitemOpen
  \bibfield  {author} {\bibinfo {author} {\bibfnamefont {D.}~\bibnamefont {Li}}, \bibinfo {author} {\bibfnamefont {V.~S.~H.}\ \bibnamefont {Lee}}, \bibinfo {author} {\bibfnamefont {Y.}~\bibnamefont {Chen}},\ \bibnamefont {and}\ \bibinfo {author} {\bibfnamefont {K.~M.}\ \bibnamefont {Zurek}},\ }\bibfield  {title} {\bibinfo {title} {Interferometer response to geontropic fluctuations},\ }\href {https://doi.org/10.1103/PhysRevD.107.024002} {\bibfield  {journal} {\bibinfo  {journal} {Phys. Rev. D}\ }\textbf {\bibinfo {volume} {107}},\ \bibinfo {pages} {024002} (\bibinfo {year} {2023})}\BibitemShut {NoStop}%
\bibitem [{\citenamefont {{L. McCuller, Single-{{Photon Signal Sideband Detection}} for {{High-Power Michelson Interferometers}} (2022)}}()}]{McCuller22SinglePhotonSignal}%
  \BibitemOpen
  \bibfield  {author} {\bibinfo {author} {\bibnamefont {{L. McCuller, Single-{{Photon Signal Sideband Detection}} for {{High-Power Michelson Interferometers}} (2022)}}},\ }\href {https://doi.org/10.48550/arXiv.2211.04016} {\bibinfo {title} {arxiv:2211.04016 [hep-ex, physics:physics, physics:quant-ph]}}\BibitemShut {NoStop}%
\bibitem [{\citenamefont {Chou}\ \emph {et~al.}(2017)\citenamefont {Chou}, \citenamefont {Glass}, \citenamefont {Gustafson}, \citenamefont {Hogan}, \citenamefont {Kamai}, \citenamefont {Kwon}, \citenamefont {Lanza}, \citenamefont {McCuller}, \citenamefont {Meyer}, \citenamefont {Richardson}, \citenamefont {Stoughton}, \citenamefont {Tomlin},\ and\ \citenamefont {Weiss}}]{ChouCQG17HolometerInstrument}%
  \BibitemOpen
  \bibfield  {author} {\bibinfo {author} {\bibfnamefont {A.}~\bibnamefont {Chou}}, \bibinfo {author} {\bibfnamefont {H.}~\bibnamefont {Glass}}, \bibinfo {author} {\bibfnamefont {H.~R.}\ \bibnamefont {Gustafson}}, \bibinfo {author} {\bibfnamefont {C.}~\bibnamefont {Hogan}}, \bibinfo {author} {\bibfnamefont {B.~L.}\ \bibnamefont {Kamai}}, \bibinfo {author} {\bibfnamefont {O.}~\bibnamefont {Kwon}}, \bibinfo {author} {\bibfnamefont {R.}~\bibnamefont {Lanza}}, \bibinfo {author} {\bibfnamefont {L.}~\bibnamefont {McCuller}}, \bibinfo {author} {\bibfnamefont {S.~S.}\ \bibnamefont {Meyer}}, \bibinfo {author} {\bibfnamefont {J.}~\bibnamefont {Richardson}}, \bibinfo {author} {\bibfnamefont {C.}~\bibnamefont {Stoughton}}, \bibinfo {author} {\bibfnamefont {R.}~\bibnamefont {Tomlin}},\ \bibnamefont {and}\ \bibinfo {author} {\bibfnamefont {R.}~\bibnamefont {Weiss}},\ }\bibfield  {title} {\bibinfo {title} {The {{Holometer}}: An instrument to probe {{Planckian}} quantum geometry},\ }\href
  {https://doi.org/10.1088/1361-6382/aa5e5c} {\bibfield  {journal} {\bibinfo  {journal} {Class. Quantum Grav.}\ }\textbf {\bibinfo {volume} {34}},\ \bibinfo {pages} {065005} (\bibinfo {year} {2017})}\BibitemShut {NoStop}%
\bibitem [{\citenamefont {Vermeulen}\ \emph {et~al.}(2021)\citenamefont {Vermeulen}, \citenamefont {Aiello}, \citenamefont {Ejlli}, \citenamefont {Griffiths}, \citenamefont {James}, \citenamefont {Dooley},\ and\ \citenamefont {Grote}}]{VermeulenCQG21ExperimentObserving}%
  \BibitemOpen
  \bibfield  {author} {\bibinfo {author} {\bibfnamefont {S.~M.}\ \bibnamefont {Vermeulen}}, \bibinfo {author} {\bibfnamefont {L.}~\bibnamefont {Aiello}}, \bibinfo {author} {\bibfnamefont {A.}~\bibnamefont {Ejlli}}, \bibinfo {author} {\bibfnamefont {W.~L.}\ \bibnamefont {Griffiths}}, \bibinfo {author} {\bibfnamefont {A.~L.}\ \bibnamefont {James}}, \bibinfo {author} {\bibfnamefont {K.~L.}\ \bibnamefont {Dooley}},\ \bibnamefont {and}\ \bibinfo {author} {\bibfnamefont {H.}~\bibnamefont {Grote}},\ }\bibfield  {title} {\bibinfo {title} {An experiment for observing quantum gravity phenomena using twin table-top {{3D}} interferometers},\ }\href {https://doi.org/10.1088/1361-6382/abe757} {\bibfield  {journal} {\bibinfo  {journal} {Class. Quantum Grav.}\ }\textbf {\bibinfo {volume} {38}},\ \bibinfo {pages} {085008} (\bibinfo {year} {2021})}\BibitemShut {NoStop}%
\bibitem [{\citenamefont {Aasi}\ \emph {et~al.}(2015)\citenamefont {Aasi}, \citenamefont {Abbott}, \citenamefont {Abbott}, \citenamefont {Abbott}, \citenamefont {Abernathy}, \citenamefont {Ackley}, \citenamefont {Adams}, \citenamefont {Adams}, \citenamefont {Addesso}, \citenamefont {Adhikari}, \citenamefont {Adya}, \citenamefont {Affeldt}, \citenamefont {Aggarwal}, \citenamefont {Aguiar}, \citenamefont {Ain}, \citenamefont {Ajith}, \citenamefont {Alemic}, \citenamefont {Allen}, \citenamefont {Amariutei}, \citenamefont {Anderson}, \citenamefont {Anderson}, \citenamefont {Arai}, \citenamefont {Araya}, \citenamefont {Arceneaux}, \citenamefont {Areeda}, \citenamefont {Ashton}, \citenamefont {Ast}, \citenamefont {Aston}, \citenamefont {Aufmuth}, \citenamefont {Aulbert}, \citenamefont {Aylott}, \citenamefont {Babak}, \citenamefont {Baker}, \citenamefont {Ballmer}, \citenamefont {Barayoga}, \citenamefont {Barbet}, \citenamefont {Barclay}, \citenamefont {Barish}, \citenamefont {Barker}, \citenamefont {Barr},
  \citenamefont {Barsotti}, \citenamefont {Bartlett}, \citenamefont {Barton}, \citenamefont {Bartos}, \citenamefont {Bassiri}, \citenamefont {Batch}, \citenamefont {Baune}, \citenamefont {Behnke}, \citenamefont {Bell}, \citenamefont {Bell}, \citenamefont {Benacquista}, \citenamefont {Bergman}, \citenamefont {Bergmann}, \citenamefont {Berry}, \citenamefont {Betzwieser}, \citenamefont {Bhagwat}, \citenamefont {Bhandare}, \citenamefont {Bilenko}, \citenamefont {Billingsley}, \citenamefont {Birch}, \citenamefont {Biscans}, \citenamefont {Biwer}, \citenamefont {Blackburn}, \citenamefont {Blackburn}, \citenamefont {Blair}, \citenamefont {Blair}, \citenamefont {Bock}, \citenamefont {Bodiya}, \citenamefont {Bojtos}, \citenamefont {Bond}, \citenamefont {Bork}, \citenamefont {Born}, \citenamefont {Bose}, \citenamefont {Brady}, \citenamefont {Braginsky}, \citenamefont {Brau}, \citenamefont {Bridges}, \citenamefont {Brinkmann}, \citenamefont {Brooks}, \citenamefont {Brown}, \citenamefont {Brown}, \citenamefont {Brown},
  \citenamefont {Buchman}, \citenamefont {Buikema}, \citenamefont {Buonanno}, \citenamefont {Cadonati}, \citenamefont {Bustillo}, \citenamefont {Camp}, \citenamefont {Cannon}, \citenamefont {Cao}, \citenamefont {Capano}, \citenamefont {Caride}, \citenamefont {Caudill}, \citenamefont {Cavagli{\`a}}, \citenamefont {Cepeda}, \citenamefont {Chakraborty}, \citenamefont {Chalermsongsak}, \citenamefont {Chamberlin}, \citenamefont {Chao}, \citenamefont {Charlton}, \citenamefont {Chen}, \citenamefont {Cho}, \citenamefont {Cho}, \citenamefont {Chow}, \citenamefont {Christensen}, \citenamefont {Chu}, \citenamefont {Chung}, \citenamefont {Ciani}, \citenamefont {Clara}, \citenamefont {Clark}, \citenamefont {Collette}, \citenamefont {Cominsky}, \citenamefont {Constancio}, \citenamefont {Cook}, \citenamefont {Corbitt}, \citenamefont {Cornish}, \citenamefont {Corsi}, \citenamefont {Costa}, \citenamefont {Coughlin}, \citenamefont {Countryman}, \citenamefont {Couvares}, \citenamefont {Coward}, \citenamefont {Cowart},
  \citenamefont {Coyne}, \citenamefont {Coyne}, \citenamefont {Craig}, \citenamefont {Creighton}, \citenamefont {Creighton}, \citenamefont {Cripe}, \citenamefont {Crowder}, \citenamefont {Cumming}, \citenamefont {Cunningham}, \citenamefont {Cutler}, \citenamefont {Dahl}, \citenamefont {Canton}, \citenamefont {Damjanic}, \citenamefont {Danilishin}, \citenamefont {Danzmann}, \citenamefont {Dartez}, \citenamefont {Dave}, \citenamefont {Daveloza}, \citenamefont {Davies}, \citenamefont {Daw}, \citenamefont {DeBra}, \citenamefont {Pozzo}, \citenamefont {Denker}, \citenamefont {Dent}, \citenamefont {Dergachev}, \citenamefont {DeRosa}, \citenamefont {DeSalvo}, \citenamefont {Dhurandhar}, \citenamefont {D{\textasciiacute}{\i}az}, \citenamefont {Palma}, \citenamefont {Dojcinoski}, \citenamefont {Dominguez}, \citenamefont {Donovan}, \citenamefont {Dooley}, \citenamefont {Doravari}, \citenamefont {Douglas}, \citenamefont {Downes}, \citenamefont {Driggers}, \citenamefont {Du}, \citenamefont {Dwyer}, \citenamefont
  {Eberle}, \citenamefont {Edo}, \citenamefont {Edwards}, \citenamefont {Edwards}, \citenamefont {Effler}, \citenamefont {Eggenstein}, \citenamefont {Ehrens}, \citenamefont {Eichholz}, \citenamefont {Eikenberry}, \citenamefont {Essick}, \citenamefont {Etzel}, \citenamefont {Evans}, \citenamefont {Evans}, \citenamefont {Factourovich}, \citenamefont {Fairhurst}, \citenamefont {Fan}, \citenamefont {Fang}, \citenamefont {Farr}, \citenamefont {Farr}, \citenamefont {Favata}, \citenamefont {Fays}, \citenamefont {Fehrmann}, \citenamefont {Fejer}, \citenamefont {Feldbaum}, \citenamefont {Ferreira}, \citenamefont {Fisher}, \citenamefont {Frei}, \citenamefont {Freise}, \citenamefont {Frey}, \citenamefont {Fricke}, \citenamefont {Fritschel}, \citenamefont {Frolov}, \citenamefont {{Fuentes-Tapia}}, \citenamefont {Fulda}, \citenamefont {Fyffe}, \citenamefont {Gair}, \citenamefont {Gaonkar}, \citenamefont {Gehrels}, \citenamefont {Gergely{\textasciiacute}}, \citenamefont {Giaime}, \citenamefont {Giardina}, \citenamefont
  {Gleason}, \citenamefont {Goetz}, \citenamefont {Goetz}, \citenamefont {Gondan}, \citenamefont {Gonz{\'a}lez}, \citenamefont {Gordon}, \citenamefont {Gorodetsky}, \citenamefont {Gossan}, \citenamefont {Go{\ss}ler}, \citenamefont {Gr{\"a}f}, \citenamefont {Graff}, \citenamefont {Grant}, \citenamefont {Gras}, \citenamefont {Gray}, \citenamefont {Greenhalgh}, \citenamefont {Gretarsson}, \citenamefont {Grote}, \citenamefont {Grunewald}, \citenamefont {Guido}, \citenamefont {Guo}, \citenamefont {Gushwa}, \citenamefont {Gustafson}, \citenamefont {Gustafson}, \citenamefont {Hacker}, \citenamefont {Hall}, \citenamefont {Hammond}, \citenamefont {Hanke}, \citenamefont {Hanks}, \citenamefont {Hanna}, \citenamefont {Hannam}, \citenamefont {Hanson}, \citenamefont {Hardwick}, \citenamefont {Harry}, \citenamefont {Harry}, \citenamefont {Hart}, \citenamefont {Hartman}, \citenamefont {Haster}, \citenamefont {Haughian}, \citenamefont {Hee}, \citenamefont {Heintze}, \citenamefont {Heinzel}, \citenamefont {Hendry},
  \citenamefont {Heng}, \citenamefont {Heptonstall}, \citenamefont {Heurs}, \citenamefont {Hewitson}, \citenamefont {Hild}, \citenamefont {Hoak}, \citenamefont {Hodge}, \citenamefont {Hollitt}, \citenamefont {Holt}, \citenamefont {Hopkins}, \citenamefont {Hosken}, \citenamefont {Hough}, \citenamefont {Houston}, \citenamefont {Howell}, \citenamefont {Hu}, \citenamefont {Huerta}, \citenamefont {Hughey}, \citenamefont {Husa}, \citenamefont {Huttner}, \citenamefont {Huynh}, \citenamefont {{Huynh-Dinh}}, \citenamefont {Idrisy}, \citenamefont {Indik}, \citenamefont {Ingram}, \citenamefont {Inta}, \citenamefont {Islas}, \citenamefont {Isler}, \citenamefont {Isogai}, \citenamefont {Iyer}, \citenamefont {Izumi}, \citenamefont {Jacobson}, \citenamefont {Jang}, \citenamefont {Jawahar}, \citenamefont {Ji}, \citenamefont {{Jim{\'e}nez-Forteza}}, \citenamefont {Johnson}, \citenamefont {Jones}, \citenamefont {Jones}, \citenamefont {Ju}, \citenamefont {Haris}, \citenamefont {Kalogera}, \citenamefont {Kandhasamy},
  \citenamefont {Kang}, \citenamefont {Kanner}, \citenamefont {Katsavounidis}, \citenamefont {Katzman}, \citenamefont {Kaufer}, \citenamefont {Kaufer}, \citenamefont {Kaur}, \citenamefont {Kawabe}, \citenamefont {Kawazoe}, \citenamefont {Keiser}, \citenamefont {Keitel}, \citenamefont {Kelley}, \citenamefont {Kells}, \citenamefont {Keppel}, \citenamefont {Key}, \citenamefont {Khalaidovski}, \citenamefont {Khalili}, \citenamefont {Khazanov}, \citenamefont {Kim}, \citenamefont {Kim}, \citenamefont {Kim}, \citenamefont {Kim}, \citenamefont {Kim}, \citenamefont {King}, \citenamefont {King}, \citenamefont {Kinzel}, \citenamefont {Kissel}, \citenamefont {Klimenko}, \citenamefont {Kline}, \citenamefont {Koehlenbeck}, \citenamefont {Kokeyama}, \citenamefont {Kondrashov}, \citenamefont {Korobko}, \citenamefont {Korth}, \citenamefont {Kozak}, \citenamefont {Kringel}, \citenamefont {Krishnan}, \citenamefont {Krueger}, \citenamefont {Kuehn}, \citenamefont {Kumar}, \citenamefont {Kumar}, \citenamefont {Kuo}, \citenamefont
  {Landry}, \citenamefont {Lantz}, \citenamefont {Larson}, \citenamefont {Lasky}, \citenamefont {Lazzarini}, \citenamefont {Lazzaro}, \citenamefont {Le}, \citenamefont {Leaci}, \citenamefont {Leavey}, \citenamefont {Lebigot}, \citenamefont {Lee}, \citenamefont {Lee}, \citenamefont {Lee}, \citenamefont {Leong}, \citenamefont {Levin}, \citenamefont {Levine}, \citenamefont {Lewis}, \citenamefont {Li}, \citenamefont {Libbrecht}, \citenamefont {Libson}, \citenamefont {Lin}, \citenamefont {Littenberg}, \citenamefont {Lockerbie}, \citenamefont {Lockett}, \citenamefont {Logue}, \citenamefont {Lombardi}, \citenamefont {Lormand}, \citenamefont {Lough}, \citenamefont {Lubinski}, \citenamefont {L{\"u}ck}, \citenamefont {Lundgren}, \citenamefont {Lynch}, \citenamefont {Ma}, \citenamefont {Macarthur}, \citenamefont {MacDonald}, \citenamefont {Machenschalk}, \citenamefont {MacInnis}, \citenamefont {Macleod}, \citenamefont {{Maga{\~n}a-Sandoval}}, \citenamefont {Magee}, \citenamefont {Mageswaran}, \citenamefont {Maglione},
  \citenamefont {Mailand}, \citenamefont {Mandel}, \citenamefont {Mandic}, \citenamefont {Mangano}, \citenamefont {Mansell}, \citenamefont {M{\'a}rka}, \citenamefont {M{\'a}rka}, \citenamefont {Markosyan}, \citenamefont {Maros}, \citenamefont {Martin}, \citenamefont {Martin}, \citenamefont {Martynov}, \citenamefont {Marx}, \citenamefont {Mason}, \citenamefont {Massinger}, \citenamefont {Matichard}, \citenamefont {Matone}, \citenamefont {Mavalvala}, \citenamefont {Mazumder}, \citenamefont {Mazzolo}, \citenamefont {McCarthy}, \citenamefont {McClelland}, \citenamefont {McCormick}, \citenamefont {McGuire}, \citenamefont {McIntyre}, \citenamefont {McIver}, \citenamefont {McLin}, \citenamefont {McWilliams}, \citenamefont {Meadors}, \citenamefont {Meinders}, \citenamefont {Melatos}, \citenamefont {Mendell}, \citenamefont {Mercer}, \citenamefont {Meshkov}, \citenamefont {Messenger}, \citenamefont {Meyers}, \citenamefont {Miao}, \citenamefont {Middleton}, \citenamefont {Mikhailov}, \citenamefont {Miller},
  \citenamefont {Miller}, \citenamefont {Millhouse}, \citenamefont {Ming}, \citenamefont {Mirshekari}, \citenamefont {Mishra}, \citenamefont {Mitra}, \citenamefont {Mitrofanov}, \citenamefont {Mitselmakher}, \citenamefont {Mittleman}, \citenamefont {Moe}, \citenamefont {Mohanty}, \citenamefont {Mohapatra}, \citenamefont {Moore}, \citenamefont {Moraru}, \citenamefont {Moreno}, \citenamefont {Morriss}, \citenamefont {Mossavi}, \citenamefont {{Mow-Lowry}}, \citenamefont {Mueller}, \citenamefont {Mueller}, \citenamefont {Mukherjee}, \citenamefont {Mullavey}, \citenamefont {Munch}, \citenamefont {Murphy}, \citenamefont {Murray}, \citenamefont {Mytidis}, \citenamefont {Nash}, \citenamefont {Nayak}, \citenamefont {Necula}, \citenamefont {Nedkova}, \citenamefont {Newton}, \citenamefont {Nguyen}, \citenamefont {Nielsen}, \citenamefont {Nissanke}, \citenamefont {Nitz}, \citenamefont {Nolting}, \citenamefont {Normandin}, \citenamefont {Nuttall}, \citenamefont {Ochsner}, \citenamefont {O'Dell}, \citenamefont {Oelker},
  \citenamefont {Ogin}, \citenamefont {Oh}, \citenamefont {Oh}, \citenamefont {Ohme}, \citenamefont {Oppermann}, \citenamefont {Oram}, \citenamefont {O'Reilly}, \citenamefont {Ortega}, \citenamefont {O'Shaughnessy}, \citenamefont {Osthelder}, \citenamefont {Ott}, \citenamefont {Ottaway}, \citenamefont {Ottens}, \citenamefont {Overmier}, \citenamefont {Owen}, \citenamefont {Padilla}, \citenamefont {Pai}, \citenamefont {Pai}, \citenamefont {Palashov}, \citenamefont {{Pal-Singh}}, \citenamefont {Pan}, \citenamefont {Pankow}, \citenamefont {Pannarale}, \citenamefont {Pant}, \citenamefont {Papa}, \citenamefont {Paris}, \citenamefont {Patrick}, \citenamefont {Pedraza}, \citenamefont {Pekowsky}, \citenamefont {Pele}, \citenamefont {Penn}, \citenamefont {Perreca}, \citenamefont {Phelps}, \citenamefont {Pierro}, \citenamefont {Pinto}, \citenamefont {Pitkin}, \citenamefont {Poeld}, \citenamefont {Post}, \citenamefont {Poteomkin}, \citenamefont {Powell}, \citenamefont {Prasad}, \citenamefont {Predoi}, \citenamefont
  {Premachandra}, \citenamefont {Prestegard}, \citenamefont {Price}, \citenamefont {Principe}, \citenamefont {Privitera}, \citenamefont {Prix}, \citenamefont {Prokhorov}, \citenamefont {Puncken}, \citenamefont {P{\"u}rrer}, \citenamefont {Qin}, \citenamefont {Quetschke}, \citenamefont {Quintero}, \citenamefont {Quiroga}, \citenamefont {{Quitzow-James}}, \citenamefont {Raab}, \citenamefont {Rabeling}, \citenamefont {Radkins}, \citenamefont {Raffai}, \citenamefont {Raja}, \citenamefont {Rajalakshmi}, \citenamefont {Rakhmanov}, \citenamefont {Ramirez}, \citenamefont {Raymond}, \citenamefont {Reed}, \citenamefont {Reid}, \citenamefont {Reitze}, \citenamefont {Reula}, \citenamefont {Riles}, \citenamefont {Robertson}, \citenamefont {Robie}, \citenamefont {Rollins}, \citenamefont {Roma}, \citenamefont {Romano}, \citenamefont {Romanov}, \citenamefont {Romie}, \citenamefont {Rowan}, \citenamefont {R{\"u}diger}, \citenamefont {Ryan}, \citenamefont {Sachdev}, \citenamefont {Sadecki}, \citenamefont {Sadeghian},
  \citenamefont {Saleem}, \citenamefont {Salemi}, \citenamefont {Sammut}, \citenamefont {Sandberg}, \citenamefont {Sanders}, \citenamefont {Sannibale}, \citenamefont {{Santiago-Prieto}}, \citenamefont {Sathyaprakash}, \citenamefont {Saulson}, \citenamefont {Savage}, \citenamefont {Sawadsky}, \citenamefont {Scheuer}, \citenamefont {Schilling}, \citenamefont {Schmidt}, \citenamefont {Schnabel}, \citenamefont {Schofield}, \citenamefont {Schreiber}, \citenamefont {Schuette}, \citenamefont {Schutz}, \citenamefont {Scott}, \citenamefont {Scott}, \citenamefont {Sellers}, \citenamefont {Sengupta}, \citenamefont {Sergeev}, \citenamefont {Serna}, \citenamefont {Sevigny}, \citenamefont {Shaddock}, \citenamefont {Shahriar}, \citenamefont {Shaltev}, \citenamefont {Shao}, \citenamefont {Shapiro}, \citenamefont {Shawhan}, \citenamefont {Shoemaker}, \citenamefont {Sidery}, \citenamefont {Siemens}, \citenamefont {Sigg}, \citenamefont {Silva}, \citenamefont {Simakov}, \citenamefont {Singer}, \citenamefont {Singer},
  \citenamefont {Singh}, \citenamefont {Sintes}, \citenamefont {Slagmolen}, \citenamefont {Smith}, \citenamefont {Smith}, \citenamefont {Smith}, \citenamefont {{Smith-Lefebvre}}, \citenamefont {Son}, \citenamefont {Sorazu}, \citenamefont {Souradeep}, \citenamefont {Staley}, \citenamefont {Stebbins}, \citenamefont {Steinke}, \citenamefont {Steinlechner}, \citenamefont {Steinlechner}, \citenamefont {Steinmeyer}, \citenamefont {Stephens}, \citenamefont {Steplewski}, \citenamefont {Stevenson}, \citenamefont {Stone}, \citenamefont {Strain}, \citenamefont {Strigin}, \citenamefont {Sturani}, \citenamefont {Stuver}, \citenamefont {Summerscales}, \citenamefont {Sutton}, \citenamefont {Szczepanczyk}, \citenamefont {Szeifert}, \citenamefont {Talukder}, \citenamefont {Tanner}, \citenamefont {T{\'a}pai}, \citenamefont {Tarabrin}, \citenamefont {Taracchini}, \citenamefont {Taylor}, \citenamefont {Tellez}, \citenamefont {Theeg}, \citenamefont {Thirugnanasambandam}, \citenamefont {Thomas}, \citenamefont {Thomas},
  \citenamefont {Thorne}, \citenamefont {Thorne}, \citenamefont {Thrane}, \citenamefont {Tiwari}, \citenamefont {Tomlinson}, \citenamefont {Torres}, \citenamefont {Torrie}, \citenamefont {Traylor}, \citenamefont {Tse}, \citenamefont {Tshilumba}, \citenamefont {Ugolini}, \citenamefont {Unnikrishnan}, \citenamefont {Urban}, \citenamefont {Usman}, \citenamefont {Vahlbruch}, \citenamefont {Vajente}, \citenamefont {Valdes}, \citenamefont {Vallisneri}, \citenamefont {van Veggel}, \citenamefont {Vass}, \citenamefont {Vaulin}, \citenamefont {Vecchio}, \citenamefont {Veitch}, \citenamefont {Veitch}, \citenamefont {Venkateswara}, \citenamefont {{Vincent-Finley}}, \citenamefont {Vitale}, \citenamefont {Vo}, \citenamefont {Vorvick}, \citenamefont {Vousden}, \citenamefont {Vyatchanin}, \citenamefont {Wade}, \citenamefont {Wade}, \citenamefont {Wade}, \citenamefont {Walker}, \citenamefont {Wallace}, \citenamefont {Walsh}, \citenamefont {Wang}, \citenamefont {Wang}, \citenamefont {Wang}, \citenamefont {Ward}, \citenamefont
  {Warner}, \citenamefont {Was}, \citenamefont {Weaver}, \citenamefont {Weinert}, \citenamefont {Weinstein}, \citenamefont {Weiss}, \citenamefont {Welborn}, \citenamefont {Wen}, \citenamefont {Wessels}, \citenamefont {Westphal}, \citenamefont {Wette}, \citenamefont {Whelan}, \citenamefont {Whitcomb}, \citenamefont {White}, \citenamefont {Whiting}, \citenamefont {Wilkinson}, \citenamefont {Williams}, \citenamefont {Williams}, \citenamefont {Williamson}, \citenamefont {Willis}, \citenamefont {Willke}, \citenamefont {Wimmer}, \citenamefont {Winkler}, \citenamefont {Wipf}, \citenamefont {Wittel}, \citenamefont {Woan}, \citenamefont {Worden}, \citenamefont {Xie}, \citenamefont {Yablon}, \citenamefont {Yakushin}, \citenamefont {Yam}, \citenamefont {Yamamoto}, \citenamefont {Yancey}, \citenamefont {Yang}, \citenamefont {Zanolin}, \citenamefont {Zhang}, \citenamefont {Zhang}, \citenamefont {Zhang}, \citenamefont {Zhang}, \citenamefont {Zhao}, \citenamefont {Zhou}, \citenamefont {Zhu}, \citenamefont {Zucker},
  \citenamefont {Zuraw},\ and\ \citenamefont {Zweizig}}]{AasiCQG15AdvancedLIGO}%
  \BibitemOpen
  \bibfield  {author} {\bibinfo {author} {\bibfnamefont {J.}~\bibnamefont {Aasi}}, \bibinfo {author} {\bibfnamefont {B.~P.}\ \bibnamefont {Abbott}}, \bibinfo {author} {\bibfnamefont {R.}~\bibnamefont {Abbott}}, \bibnamefont {et~al.},\ }\bibfield  {title} {\bibinfo {title} {Advanced {{LIGO}}},\ }\href {https://doi.org/10.1088/0264-9381/32/7/074001} {\bibfield  {journal} {\bibinfo  {journal} {Class. Quantum Grav.}\ }\textbf {\bibinfo {volume} {32}},\ \bibinfo {pages} {074001} (\bibinfo {year} {2015})}\BibitemShut {NoStop}%
\bibitem [{\citenamefont {Buikema}\ \emph {et~al.}(2020)\citenamefont {Buikema}, \citenamefont {Cahillane}, \citenamefont {Mansell}, \citenamefont {Blair}, \citenamefont {Abbott}, \citenamefont {Adams}, \citenamefont {Adhikari}, \citenamefont {Ananyeva}, \citenamefont {Appert}, \citenamefont {Arai}, \citenamefont {Areeda}, \citenamefont {Asali}, \citenamefont {Aston}, \citenamefont {Austin}, \citenamefont {Baer}, \citenamefont {Ball}, \citenamefont {Ballmer}, \citenamefont {Banagiri}, \citenamefont {Barker}, \citenamefont {Barsotti}, \citenamefont {Bartlett}, \citenamefont {Berger}, \citenamefont {Betzwieser}, \citenamefont {Bhattacharjee}, \citenamefont {Billingsley}, \citenamefont {Biscans}, \citenamefont {Blair}, \citenamefont {Bode}, \citenamefont {Booker}, \citenamefont {Bork}, \citenamefont {Bramley}, \citenamefont {Brooks}, \citenamefont {Brown}, \citenamefont {Cannon}, \citenamefont {Chen}, \citenamefont {Ciobanu}, \citenamefont {Clara}, \citenamefont {Cooper}, \citenamefont {Corley}, \citenamefont
  {Countryman}, \citenamefont {Covas}, \citenamefont {Coyne}, \citenamefont {Datrier}, \citenamefont {Davis}, \citenamefont {Di~Fronzo}, \citenamefont {Dooley}, \citenamefont {Driggers}, \citenamefont {Dupej}, \citenamefont {Dwyer}, \citenamefont {Effler}, \citenamefont {Etzel}, \citenamefont {Evans}, \citenamefont {Evans}, \citenamefont {Feicht}, \citenamefont {{Fernandez-Galiana}}, \citenamefont {Fritschel}, \citenamefont {Frolov}, \citenamefont {Fulda}, \citenamefont {Fyffe}, \citenamefont {Giaime}, \citenamefont {Giardina}, \citenamefont {Godwin}, \citenamefont {Goetz}, \citenamefont {Gras}, \citenamefont {Gray}, \citenamefont {Gray}, \citenamefont {Green}, \citenamefont {Gustafson}, \citenamefont {Gustafson}, \citenamefont {Hanks}, \citenamefont {Hanson}, \citenamefont {Hardwick}, \citenamefont {Hasskew}, \citenamefont {Heintze}, \citenamefont {{Helmling-Cornell}}, \citenamefont {Holland}, \citenamefont {Jones}, \citenamefont {Kandhasamy}, \citenamefont {Karki}, \citenamefont {Kasprzack}, \citenamefont
  {Kawabe}, \citenamefont {Kijbunchoo}, \citenamefont {King}, \citenamefont {Kissel}, \citenamefont {Kumar}, \citenamefont {Landry}, \citenamefont {Lane}, \citenamefont {Lantz}, \citenamefont {Laxen}, \citenamefont {Lecoeuche}, \citenamefont {Leviton}, \citenamefont {Liu}, \citenamefont {Lormand}, \citenamefont {Lundgren}, \citenamefont {Macas}, \citenamefont {MacInnis}, \citenamefont {Macleod}, \citenamefont {M{\'a}rka}, \citenamefont {M{\'a}rka}, \citenamefont {Martynov}, \citenamefont {Mason}, \citenamefont {Massinger}, \citenamefont {Matichard}, \citenamefont {Mavalvala}, \citenamefont {McCarthy}, \citenamefont {McClelland}, \citenamefont {McCormick}, \citenamefont {McCuller}, \citenamefont {McIver}, \citenamefont {McRae}, \citenamefont {Mendell}, \citenamefont {Merfeld}, \citenamefont {Merilh}, \citenamefont {Meylahn}, \citenamefont {Mistry}, \citenamefont {Mittleman}, \citenamefont {Moreno}, \citenamefont {{Mow-Lowry}}, \citenamefont {Mozzon}, \citenamefont {Mullavey}, \citenamefont {Nelson},
  \citenamefont {Nguyen}, \citenamefont {Nuttall}, \citenamefont {Oberling}, \citenamefont {Oram}, \citenamefont {O'Reilly}, \citenamefont {Osthelder}, \citenamefont {Ottaway}, \citenamefont {Overmier}, \citenamefont {Palamos}, \citenamefont {Parker}, \citenamefont {Payne}, \citenamefont {Pele}, \citenamefont {Penhorwood}, \citenamefont {Perez}, \citenamefont {Pirello}, \citenamefont {Radkins}, \citenamefont {Ramirez}, \citenamefont {Richardson}, \citenamefont {Riles}, \citenamefont {Robertson}, \citenamefont {Rollins}, \citenamefont {Romel}, \citenamefont {Romie}, \citenamefont {Ross}, \citenamefont {Ryan}, \citenamefont {Sadecki}, \citenamefont {Sanchez}, \citenamefont {Sanchez}, \citenamefont {Saravanan}, \citenamefont {Savage}, \citenamefont {Schaetzl}, \citenamefont {Schnabel}, \citenamefont {Schofield}, \citenamefont {Schwartz}, \citenamefont {Sellers}, \citenamefont {Shaffer}, \citenamefont {Sigg}, \citenamefont {Slagmolen}, \citenamefont {Smith}, \citenamefont {Soni}, \citenamefont {Sorazu},
  \citenamefont {Spencer}, \citenamefont {Strain}, \citenamefont {Sun}, \citenamefont {Szczepa{\'n}czyk}, \citenamefont {Thomas}, \citenamefont {Thomas}, \citenamefont {Thorne}, \citenamefont {Toland}, \citenamefont {Torrie}, \citenamefont {Traylor}, \citenamefont {Tse}, \citenamefont {Urban}, \citenamefont {Vajente}, \citenamefont {Valdes}, \citenamefont {{Vander-Hyde}}, \citenamefont {Veitch}, \citenamefont {Venkateswara}, \citenamefont {Venugopalan}, \citenamefont {Viets}, \citenamefont {Vo}, \citenamefont {Vorvick}, \citenamefont {Wade}, \citenamefont {Ward}, \citenamefont {Warner}, \citenamefont {Weaver}, \citenamefont {Weiss}, \citenamefont {Whittle}, \citenamefont {Willke}, \citenamefont {Wipf}, \citenamefont {Xiao}, \citenamefont {Yamamoto}, \citenamefont {Yu}, \citenamefont {Yu}, \citenamefont {Zhang}, \citenamefont {Zucker},\ and\ \citenamefont {Zweizig}}]{BuikemaPRD20SensitivityPerformance}%
  \BibitemOpen
  \bibfield  {author} {\bibinfo {author} {\bibfnamefont {A.}~\bibnamefont {Buikema}}, \bibinfo {author} {\bibfnamefont {C.}~\bibnamefont {Cahillane}}, \bibinfo {author} {\bibfnamefont {G.~L.}\ \bibnamefont {Mansell}}, \bibnamefont {et~al.},\ }\bibfield  {title} {\bibinfo {title} {Sensitivity and performance of the {{Advanced LIGO}} detectors in the third observing run},\ }\href {https://doi.org/10.1103/PhysRevD.102.062003} {\bibfield  {journal} {\bibinfo  {journal} {Phys. Rev. D}\ }\textbf {\bibinfo {volume} {102}},\ \bibinfo {pages} {062003} (\bibinfo {year} {2020})}\BibitemShut {NoStop}%
\bibitem [{\citenamefont {Romano}\ and\ \citenamefont {Cornish}(2017)}]{RomanoLRR17DetectionMethods}%
  \BibitemOpen
  \bibfield  {author} {\bibinfo {author} {\bibfnamefont {J.~D.}\ \bibnamefont {Romano}}\ \bibnamefont {and}\ \bibinfo {author} {\bibfnamefont {{\relax Neil}.~J.}\ \bibnamefont {Cornish}},\ }\bibfield  {title} {\bibinfo {title} {Detection methods for stochastic gravitational-wave backgrounds: A unified treatment},\ }\href {https://doi.org/10.1007/s41114-017-0004-1} {\bibfield  {journal} {\bibinfo  {journal} {Living. Rev. Relativ.}\ }\textbf {\bibinfo {volume} {20}},\ \bibinfo {pages} {2} (\bibinfo {year} {2017})}\BibitemShut {NoStop}%
\bibitem [{\citenamefont {Kim}\ and\ \citenamefont {Carosi}(2010)}]{kim2010axions}%
  \BibitemOpen
  \bibfield  {author} {\bibinfo {author} {\bibfnamefont {J.~E.}\ \bibnamefont {Kim}}\ \bibnamefont {and}\ \bibinfo {author} {\bibfnamefont {G.}~\bibnamefont {Carosi}},\ }\bibfield  {title} {\bibinfo {title} {Axions and the strong ${{C P}}$ problem},\ }\href {https://doi.org/10.1103/RevModPhys.82.557} {\bibfield  {journal} {\bibinfo  {journal} {Rev. Mod. Phys.}\ }\textbf {\bibinfo {volume} {82}},\ \bibinfo {pages} {557} (\bibinfo {year} {2010})}\BibitemShut {NoStop}%
\bibitem [{\citenamefont {Choi}\ \emph {et~al.}(2021)\citenamefont {Choi}, \citenamefont {Im},\ and\ \citenamefont {Shin}}]{choi2021recent}%
  \BibitemOpen
  \bibfield  {author} {\bibinfo {author} {\bibfnamefont {K.}~\bibnamefont {Choi}}, \bibinfo {author} {\bibfnamefont {S.~H.}\ \bibnamefont {Im}},\ \bibnamefont {and}\ \bibinfo {author} {\bibfnamefont {C.~S.}\ \bibnamefont {Shin}},\ }\bibfield  {title} {\bibinfo {title} {Recent progress in the physics of axions and axion-like particles},\ }\href {https://doi.org/10.1146/annurev-nucl-120720-031147} {\bibfield  {journal} {\bibinfo  {journal} {Annu. Rev. Nucl. Part. Sci.}\ }\textbf {\bibinfo {volume} {71}},\ \bibinfo {pages} {225} (\bibinfo {year} {2021})}\BibitemShut {NoStop}%
\bibitem [{\citenamefont {Rosenberg}\ and\ \citenamefont {Van~Bibber}(2000)}]{rosenberg2000searches}%
  \BibitemOpen
  \bibfield  {author} {\bibinfo {author} {\bibfnamefont {L.~J.}\ \bibnamefont {Rosenberg}}\ \bibnamefont {and}\ \bibinfo {author} {\bibfnamefont {K.~A.}\ \bibnamefont {Van~Bibber}},\ }\bibfield  {title} {\bibinfo {title} {Searches for invisible axions},\ }\href {https://doi.org/10.1016/S0370-1573(99)00045-9} {\bibfield  {journal} {\bibinfo  {journal} {Phys. Rep.}\ }\textbf {\bibinfo {volume} {325}},\ \bibinfo {pages} {1} (\bibinfo {year} {2000})}\BibitemShut {NoStop}%
\bibitem [{\citenamefont {Graham}\ \emph {et~al.}(2015)\citenamefont {Graham}, \citenamefont {Irastorza}, \citenamefont {Lamoreaux}, \citenamefont {Lindner},\ and\ \citenamefont {van Bibber}}]{graham2015experimental}%
  \BibitemOpen
  \bibfield  {author} {\bibinfo {author} {\bibfnamefont {P.~W.}\ \bibnamefont {Graham}}, \bibinfo {author} {\bibfnamefont {I.~G.}\ \bibnamefont {Irastorza}}, \bibinfo {author} {\bibfnamefont {S.~K.}\ \bibnamefont {Lamoreaux}}, \bibinfo {author} {\bibfnamefont {A.}~\bibnamefont {Lindner}},\ \bibnamefont {and}\ \bibinfo {author} {\bibfnamefont {K.~A.}\ \bibnamefont {van Bibber}},\ }\bibfield  {title} {\bibinfo {title} {Experimental searches for the axion and axion-like particles},\ }\href {https://doi.org/10.1146/annurev-nucl-102014-022120} {\bibfield  {journal} {\bibinfo  {journal} {Annu. Rev. Nucl. Part. Sci.}\ }\textbf {\bibinfo {volume} {65}},\ \bibinfo {pages} {485} (\bibinfo {year} {2015})}\BibitemShut {NoStop}%
\bibitem [{\citenamefont {Gardiner}\ and\ \citenamefont {Collett}(1985)}]{GardinerPRA85InputOutput}%
  \BibitemOpen
  \bibfield  {author} {\bibinfo {author} {\bibfnamefont {C.~W.}\ \bibnamefont {Gardiner}}\ \bibnamefont {and}\ \bibinfo {author} {\bibfnamefont {M.~J.}\ \bibnamefont {Collett}},\ }\bibfield  {title} {\bibinfo {title} {Input and output in damped quantum systems: {{Quantum}} stochastic differential equations and the master equation},\ }\href {https://doi.org/10.1103/PhysRevA.31.3761} {\bibfield  {journal} {\bibinfo  {journal} {Phys. Rev. A}\ }\textbf {\bibinfo {volume} {31}},\ \bibinfo {pages} {3761} (\bibinfo {year} {1985})}\BibitemShut {NoStop}%
\bibitem [{\citenamefont {Kubo}(1966)}]{KuboRPP66FluctuationdissipationTheorem}%
  \BibitemOpen
  \bibfield  {author} {\bibinfo {author} {\bibfnamefont {R.}~\bibnamefont {Kubo}},\ }\bibfield  {title} {\bibinfo {title} {The fluctuation-dissipation theorem},\ }\href {https://doi.org/10.1088/0034-4885/29/1/306} {\bibfield  {journal} {\bibinfo  {journal} {Rep. Prog. Phys.}\ }\textbf {\bibinfo {volume} {29}},\ \bibinfo {pages} {255} (\bibinfo {year} {1966})}\BibitemShut {NoStop}%
\bibitem [{\citenamefont {Buonanno}\ and\ \citenamefont {Chen}(2002)}]{BuonannoPRD02SignalRecycled}%
  \BibitemOpen
  \bibfield  {author} {\bibinfo {author} {\bibfnamefont {A.}~\bibnamefont {Buonanno}}\ \bibnamefont {and}\ \bibinfo {author} {\bibfnamefont {Y.}~\bibnamefont {Chen}},\ }\bibfield  {title} {\bibinfo {title} {Signal recycled laser-interferometer gravitational-wave detectors as optical springs},\ }\href {https://doi.org/10.1103/PhysRevD.65.042001} {\bibfield  {journal} {\bibinfo  {journal} {Phys. Rev. D}\ }\textbf {\bibinfo {volume} {65}},\ \bibinfo {pages} {042001} (\bibinfo {year} {2002})}\BibitemShut {NoStop}%
\bibitem [{\citenamefont {Tsang}\ \emph {et~al.}(2011)\citenamefont {Tsang}, \citenamefont {Wiseman},\ and\ \citenamefont {Caves}}]{TsangPRL11FundamentalQuantum}%
  \BibitemOpen
  \bibfield  {author} {\bibinfo {author} {\bibfnamefont {M.}~\bibnamefont {Tsang}}, \bibinfo {author} {\bibfnamefont {H.~M.}\ \bibnamefont {Wiseman}},\ \bibnamefont {and}\ \bibinfo {author} {\bibfnamefont {C.~M.}\ \bibnamefont {Caves}},\ }\bibfield  {title} {\bibinfo {title} {Fundamental {{Quantum Limit}} to {{Waveform Estimation}}},\ }\href {https://doi.org/10.1103/PhysRevLett.106.090401} {\bibfield  {journal} {\bibinfo  {journal} {Phys. Rev. Lett.}\ }\textbf {\bibinfo {volume} {106}},\ \bibinfo {pages} {090401} (\bibinfo {year} {2011})}\BibitemShut {NoStop}%
\bibitem [{\citenamefont {Miao}\ \emph {et~al.}(2017)\citenamefont {Miao}, \citenamefont {Adhikari}, \citenamefont {Ma}, \citenamefont {Pang},\ and\ \citenamefont {Chen}}]{MiaoPRL17FundamentalQuantum}%
  \BibitemOpen
  \bibfield  {author} {\bibinfo {author} {\bibfnamefont {H.}~\bibnamefont {Miao}}, \bibinfo {author} {\bibfnamefont {R.~X.}\ \bibnamefont {Adhikari}}, \bibinfo {author} {\bibfnamefont {Y.}~\bibnamefont {Ma}}, \bibinfo {author} {\bibfnamefont {B.}~\bibnamefont {Pang}},\ \bibnamefont {and}\ \bibinfo {author} {\bibfnamefont {Y.}~\bibnamefont {Chen}},\ }\bibfield  {title} {\bibinfo {title} {Towards the {{Fundamental Quantum Limit}} of {{Linear Measurements}} of {{Classical Signals}}},\ }\href {https://doi.org/10.1103/PhysRevLett.119.050801} {\bibfield  {journal} {\bibinfo  {journal} {Phys. Rev. Lett.}\ }\textbf {\bibinfo {volume} {119}},\ \bibinfo {pages} {050801} (\bibinfo {year} {2017})}\BibitemShut {NoStop}%
\bibitem [{\citenamefont {Gardner}\ \emph {et~al.}(2024)\citenamefont {Gardner}, \citenamefont {Gefen}, \citenamefont {Haine}, \citenamefont {Hope},\ and\ \citenamefont {Chen}}]{gardner2024achieving}%
  \BibitemOpen
  \bibfield  {author} {\bibinfo {author} {\bibfnamefont {J.~W.}\ \bibnamefont {Gardner}}, \bibinfo {author} {\bibfnamefont {T.}~\bibnamefont {Gefen}}, \bibinfo {author} {\bibfnamefont {S.~A.}\ \bibnamefont {Haine}}, \bibinfo {author} {\bibfnamefont {J.~J.}\ \bibnamefont {Hope}},\ \bibnamefont {and}\ \bibinfo {author} {\bibfnamefont {Y.}~\bibnamefont {Chen}},\ }\bibfield  {title} {\bibinfo {title} {Achieving the fundamental quantum limit of linear waveform estimation},\ }\href {https://doi.org/10.1103/PhysRevLett.132.130801} {\bibfield  {journal} {\bibinfo  {journal} {Phys. Rev. Lett.}\ }\textbf {\bibinfo {volume} {132}},\ \bibinfo {pages} {130801} (\bibinfo {year} {2024})}\BibitemShut {NoStop}%
\bibitem [{\citenamefont {Ng}\ \emph {et~al.}(2016)\citenamefont {Ng}, \citenamefont {Ang}, \citenamefont {Wheatley}, \citenamefont {Yonezawa}, \citenamefont {Furusawa}, \citenamefont {Huntington},\ and\ \citenamefont {Tsang}}]{NgPRA16SpectrumAnalysis}%
  \BibitemOpen
  \bibfield  {author} {\bibinfo {author} {\bibfnamefont {S.}~\bibnamefont {Ng}}, \bibinfo {author} {\bibfnamefont {S.~Z.}\ \bibnamefont {Ang}}, \bibinfo {author} {\bibfnamefont {T.~A.}\ \bibnamefont {Wheatley}}, \bibinfo {author} {\bibfnamefont {H.}~\bibnamefont {Yonezawa}}, \bibinfo {author} {\bibfnamefont {A.}~\bibnamefont {Furusawa}}, \bibinfo {author} {\bibfnamefont {E.~H.}\ \bibnamefont {Huntington}},\ \bibnamefont {and}\ \bibinfo {author} {\bibfnamefont {M.}~\bibnamefont {Tsang}},\ }\bibfield  {title} {\bibinfo {title} {Spectrum analysis with quantum dynamical systems},\ }\href {https://doi.org/10.1103/PhysRevA.93.042121} {\bibfield  {journal} {\bibinfo  {journal} {Phys. Rev. A}\ }\textbf {\bibinfo {volume} {93}},\ \bibinfo {pages} {042121} (\bibinfo {year} {2016})}\BibitemShut {NoStop}%
\bibitem [{\citenamefont {Tsang}(2023)}]{PhysRevA.107.012611}%
  \BibitemOpen
  \bibfield  {author} {\bibinfo {author} {\bibfnamefont {M.}~\bibnamefont {Tsang}},\ }\bibfield  {title} {\bibinfo {title} {Quantum noise spectroscopy as an incoherent imaging problem},\ }\href {https://doi.org/10.1103/PhysRevA.107.012611} {\bibfield  {journal} {\bibinfo  {journal} {Phys. Rev. A}\ }\textbf {\bibinfo {volume} {107}},\ \bibinfo {pages} {012611} (\bibinfo {year} {2023})}\BibitemShut {NoStop}%
\bibitem [{\citenamefont {Bond}\ \emph {et~al.}(2016)\citenamefont {Bond}, \citenamefont {Brown}, \citenamefont {Freise},\ and\ \citenamefont {Strain}}]{bond2016interferometer}%
  \BibitemOpen
  \bibfield  {author} {\bibinfo {author} {\bibfnamefont {C.}~\bibnamefont {Bond}}, \bibinfo {author} {\bibfnamefont {D.}~\bibnamefont {Brown}}, \bibinfo {author} {\bibfnamefont {A.}~\bibnamefont {Freise}},\ \bibnamefont {and}\ \bibinfo {author} {\bibfnamefont {K.~A.}\ \bibnamefont {Strain}},\ }\bibfield  {title} {\bibinfo {title} {Interferometer techniques for gravitational-wave detection},\ }\href {https://doi.org/10.1007/s41114-016-0002-8} {\bibfield  {journal} {\bibinfo  {journal} {Living Rev. Relativ.}\ }\textbf {\bibinfo {volume} {19}},\ \bibinfo {pages} {1} (\bibinfo {year} {2016})}\BibitemShut {NoStop}%
\bibitem [{\citenamefont {Zurek}(2003)}]{zurek2003decoherence}%
  \BibitemOpen
  \bibfield  {author} {\bibinfo {author} {\bibfnamefont {W.~H.}\ \bibnamefont {Zurek}},\ }\bibfield  {title} {\bibinfo {title} {Decoherence, einselection, and the quantum origins of the classical},\ }\href {https://doi.org/10.1103/RevModPhys.75.715} {\bibfield  {journal} {\bibinfo  {journal} {Rev. Mod. Phys.}\ }\textbf {\bibinfo {volume} {75}},\ \bibinfo {pages} {715} (\bibinfo {year} {2003})}\BibitemShut {NoStop}%
\bibitem [{\citenamefont {Zurek}(2007)}]{Zurek2007}%
  \BibitemOpen
  \bibfield  {author} {\bibinfo {author} {\bibfnamefont {W.~H.}\ \bibnamefont {Zurek}},\ }\bibinfo {title} {Decoherence and the transition from quantum to classical --- revisited},\ in\ \href {https://doi.org/10.1007/978-3-7643-7808-0_1} {\emph {\bibinfo {booktitle} {Quantum Decoherence: Poincar{\'e} Seminar 2005}}}\ (\bibinfo  {publisher} {Birkh{\"a}user Basel},\ \bibinfo {address} {Basel},\ \bibinfo {year} {2007})\ pp.\ \bibinfo {pages} {1--31}\BibitemShut {NoStop}%
\bibitem [{\citenamefont {Braunstein}\ and\ \citenamefont {Caves}(1994)}]{BraunsteinPRL94StatisticalDistance}%
  \BibitemOpen
  \bibfield  {author} {\bibinfo {author} {\bibfnamefont {S.~L.}\ \bibnamefont {Braunstein}}\ \bibnamefont {and}\ \bibinfo {author} {\bibfnamefont {C.~M.}\ \bibnamefont {Caves}},\ }\bibfield  {title} {\bibinfo {title} {Statistical distance and the geometry of quantum states},\ }\href {https://doi.org/10.1103/PhysRevLett.72.3439} {\bibfield  {journal} {\bibinfo  {journal} {Phys. Rev. Lett.}\ }\textbf {\bibinfo {volume} {72}},\ \bibinfo {pages} {3439} (\bibinfo {year} {1994})}\BibitemShut {NoStop}%
\bibitem [{\citenamefont {Wiseman}\ and\ \citenamefont {Milburn}(2009)}]{Wiseman09QuantumMeasurement}%
  \BibitemOpen
  \bibfield  {author} {\bibinfo {author} {\bibfnamefont {H.~M.}\ \bibnamefont {Wiseman}}\ \bibnamefont {and}\ \bibinfo {author} {\bibfnamefont {G.~J.}\ \bibnamefont {Milburn}},\ }\href {https://doi.org/10.1017/CBO9780511813948} {\emph {\bibinfo {title} {Quantum {{Measurement}} and {{Control}}}}}\ (\bibinfo  {publisher} {Cambridge University Press},\ \bibinfo {address} {Cambridge},\ \bibinfo {year} {2009})\BibitemShut {NoStop}%
\bibitem [{\citenamefont {Pezz{\`e}}\ \emph {et~al.}(2018)\citenamefont {Pezz{\`e}}, \citenamefont {Smerzi}, \citenamefont {Oberthaler}, \citenamefont {Schmied},\ and\ \citenamefont {Treutlein}}]{PezzeRMP18QuantumMetrology}%
  \BibitemOpen
  \bibfield  {author} {\bibinfo {author} {\bibfnamefont {L.}~\bibnamefont {Pezz{\`e}}}, \bibinfo {author} {\bibfnamefont {A.}~\bibnamefont {Smerzi}}, \bibinfo {author} {\bibfnamefont {M.~K.}\ \bibnamefont {Oberthaler}}, \bibinfo {author} {\bibfnamefont {R.}~\bibnamefont {Schmied}},\ \bibnamefont {and}\ \bibinfo {author} {\bibfnamefont {P.}~\bibnamefont {Treutlein}},\ }\bibfield  {title} {\bibinfo {title} {Quantum metrology with nonclassical states of atomic ensembles},\ }\href {https://doi.org/10.1103/RevModPhys.90.035005} {\bibfield  {journal} {\bibinfo  {journal} {Rev. Mod. Phys.}\ }\textbf {\bibinfo {volume} {90}},\ \bibinfo {pages} {035005} (\bibinfo {year} {2018})}\BibitemShut {NoStop}%
\bibitem [{\citenamefont {Takeoka}\ \emph {et~al.}(2003)\citenamefont {Takeoka}, \citenamefont {Ban},\ and\ \citenamefont {Sasaki}}]{takeoka2003unambiguous}%
  \BibitemOpen
  \bibfield  {author} {\bibinfo {author} {\bibfnamefont {M.}~\bibnamefont {Takeoka}}, \bibinfo {author} {\bibfnamefont {M.}~\bibnamefont {Ban}},\ \bibnamefont {and}\ \bibinfo {author} {\bibfnamefont {M.}~\bibnamefont {Sasaki}},\ }\bibfield  {title} {\bibinfo {title} {Unambiguous quantum-state filtering},\ }\href {https://doi.org/10.1103/PhysRevA.68.012307} {\bibfield  {journal} {\bibinfo  {journal} {Phys. Rev. A}\ }\textbf {\bibinfo {volume} {68}},\ \bibinfo {pages} {012307} (\bibinfo {year} {2003})}\BibitemShut {NoStop}%
\bibitem [{\citenamefont {{A. Monras, Phase space formalism for quantum estimation of Gaussian states (2013)}}()}]{monras2013phase}%
  \BibitemOpen
  \bibfield  {author} {\bibinfo {author} {\bibnamefont {{A. Monras, Phase space formalism for quantum estimation of Gaussian states (2013)}}},\ }\href {https://doi.org/10.48550/arXiv.1303.3682} {\bibinfo {title} {arxiv:1303.3682 [quant-ph]}}\BibitemShut {NoStop}%
\bibitem [{\citenamefont {Tsang}\ \emph {et~al.}(2016)\citenamefont {Tsang}, \citenamefont {Nair},\ and\ \citenamefont {Lu}}]{tsang2016quantum}%
  \BibitemOpen
  \bibfield  {author} {\bibinfo {author} {\bibfnamefont {M.}~\bibnamefont {Tsang}}, \bibinfo {author} {\bibfnamefont {R.}~\bibnamefont {Nair}},\ \bibnamefont {and}\ \bibinfo {author} {\bibfnamefont {X.-M.}\ \bibnamefont {Lu}},\ }\bibfield  {title} {\bibinfo {title} {Quantum theory of superresolution for two incoherent optical point sources},\ }\href {https://doi.org/10.1103/PhysRevX.6.031033} {\bibfield  {journal} {\bibinfo  {journal} {Phys. Rev. X}\ }\textbf {\bibinfo {volume} {6}},\ \bibinfo {pages} {031033} (\bibinfo {year} {2016})}\BibitemShut {NoStop}%
\bibitem [{\citenamefont {Liu}\ \emph {et~al.}(2004)\citenamefont {Liu}, \citenamefont {{\"O}zdemir}, \citenamefont {Miranowicz},\ and\ \citenamefont {Imoto}}]{liu2004kraus}%
  \BibitemOpen
  \bibfield  {author} {\bibinfo {author} {\bibfnamefont {Y.-x.}\ \bibnamefont {Liu}}, \bibinfo {author} {\bibfnamefont {{\c{S}}.~K.}\ \bibnamefont {{\"O}zdemir}}, \bibinfo {author} {\bibfnamefont {A.}~\bibnamefont {Miranowicz}},\ \bibnamefont {and}\ \bibinfo {author} {\bibfnamefont {N.}~\bibnamefont {Imoto}},\ }\bibfield  {title} {\bibinfo {title} {Kraus representation of a damped harmonic oscillator and its application},\ }\href {https://doi.org/10.1103/PhysRevA.70.042308} {\bibfield  {journal} {\bibinfo  {journal} {Phys. Rev. A}\ }\textbf {\bibinfo {volume} {70}},\ \bibinfo {pages} {042308} (\bibinfo {year} {2004})}\BibitemShut {NoStop}%
\bibitem [{\citenamefont {Escher}\ \emph {et~al.}(2011)\citenamefont {Escher}, \citenamefont {{de Matos Filho}},\ and\ \citenamefont {Davidovich}}]{EscherNP11GeneralFramework}%
  \BibitemOpen
  \bibfield  {author} {\bibinfo {author} {\bibfnamefont {{\relax BM}.}~\bibnamefont {Escher}}, \bibinfo {author} {\bibfnamefont {R.~L.}\ \bibnamefont {{de Matos Filho}}},\ \bibnamefont {and}\ \bibinfo {author} {\bibfnamefont {L.}~\bibnamefont {Davidovich}},\ }\bibfield  {title} {\bibinfo {title} {General framework for estimating the ultimate precision limit in noisy quantum-enhanced metrology},\ }\href {https://doi.org/10.1038/nphys1958} {\bibfield  {journal} {\bibinfo  {journal} {Nat. Phys.}\ }\textbf {\bibinfo {volume} {7}},\ \bibinfo {pages} {406} (\bibinfo {year} {2011})}\BibitemShut {NoStop}%
\bibitem [{\citenamefont {Kolodynski}\ and\ \citenamefont {{Demkowicz-Dobrzanski}}(2013)}]{KolodynskiNJP13EfficientTools}%
  \BibitemOpen
  \bibfield  {author} {\bibinfo {author} {\bibfnamefont {J.}~\bibnamefont {Kolodynski}}\ \bibnamefont {and}\ \bibinfo {author} {\bibfnamefont {R.}~\bibnamefont {{Demkowicz-Dobrzanski}}},\ }\bibfield  {title} {\bibinfo {title} {Efficient tools for quantum metrology with uncorrelated noise},\ }\href {https://doi.org/10.1088/1367-2630/15/7/073043} {\bibfield  {journal} {\bibinfo  {journal} {New J. Phys.}\ }\textbf {\bibinfo {volume} {15}},\ \bibinfo {pages} {073043} (\bibinfo {year} {2013})}\BibitemShut {NoStop}%
\bibitem [{\citenamefont {{Demkowicz-Dobrza{\'n}ski}}\ and\ \citenamefont {Maccone}(2014)}]{Demkowicz-DobrzanskiPRL14UsingEntanglement}%
  \BibitemOpen
  \bibfield  {author} {\bibinfo {author} {\bibfnamefont {R.}~\bibnamefont {{Demkowicz-Dobrza{\'n}ski}}}\ \bibnamefont {and}\ \bibinfo {author} {\bibfnamefont {L.}~\bibnamefont {Maccone}},\ }\bibfield  {title} {\bibinfo {title} {Using {{Entanglement Against Noise}} in {{Quantum Metrology}}},\ }\href {https://doi.org/10.1103/PhysRevLett.113.250801} {\bibfield  {journal} {\bibinfo  {journal} {Phys. Rev. Lett.}\ }\textbf {\bibinfo {volume} {113}},\ \bibinfo {pages} {250801} (\bibinfo {year} {2014})}\BibitemShut {NoStop}%
\bibitem [{\citenamefont {Latune}\ \emph {et~al.}(2013)\citenamefont {Latune}, \citenamefont {Escher}, \citenamefont {{de Matos Filho}},\ and\ \citenamefont {Davidovich}}]{LatunePRA13QuantumLimit}%
  \BibitemOpen
  \bibfield  {author} {\bibinfo {author} {\bibfnamefont {C.~L.}\ \bibnamefont {Latune}}, \bibinfo {author} {\bibfnamefont {B.~M.}\ \bibnamefont {Escher}}, \bibinfo {author} {\bibfnamefont {R.~L.}\ \bibnamefont {{de Matos Filho}}},\ \bibnamefont {and}\ \bibinfo {author} {\bibfnamefont {L.}~\bibnamefont {Davidovich}},\ }\bibfield  {title} {\bibinfo {title} {Quantum limit for the measurement of a classical force coupled to a noisy quantum-mechanical oscillator},\ }\href {https://doi.org/10.1103/PhysRevA.88.042112} {\bibfield  {journal} {\bibinfo  {journal} {Phys. Rev. A}\ }\textbf {\bibinfo {volume} {88}},\ \bibinfo {pages} {042112} (\bibinfo {year} {2013})}\BibitemShut {NoStop}%
\bibitem [{\citenamefont {{Demkowicz-Dobrza{\'n}ski}}\ \emph {et~al.}(2013)\citenamefont {{Demkowicz-Dobrza{\'n}ski}}, \citenamefont {Banaszek},\ and\ \citenamefont {Schnabel}}]{Demkowicz-DobrzanskiPRA13FundamentalQuantum}%
  \BibitemOpen
  \bibfield  {author} {\bibinfo {author} {\bibfnamefont {R.}~\bibnamefont {{Demkowicz-Dobrza{\'n}ski}}}, \bibinfo {author} {\bibfnamefont {K.}~\bibnamefont {Banaszek}},\ \bibnamefont {and}\ \bibinfo {author} {\bibfnamefont {R.}~\bibnamefont {Schnabel}},\ }\bibfield  {title} {\bibinfo {title} {Fundamental quantum interferometry bound for the squeezed-light-enhanced gravitational wave detector {{GEO}} 600},\ }\href {https://doi.org/10.1103/PhysRevA.88.041802} {\bibfield  {journal} {\bibinfo  {journal} {Phys. Rev. A}\ }\textbf {\bibinfo {volume} {88}},\ \bibinfo {pages} {041802} (\bibinfo {year} {2013})}\BibitemShut {NoStop}%
\bibitem [{\citenamefont {Aasi}\ \emph {et~al.}(2013)\citenamefont {Aasi}, \citenamefont {Abadie}, \citenamefont {Abbott}, \citenamefont {Abbott}, \citenamefont {Abbott}, \citenamefont {Abernathy}, \citenamefont {Adams}, \citenamefont {Adams}, \citenamefont {Addesso}, \citenamefont {Adhikari}, \citenamefont {Affeldt}, \citenamefont {Aguiar}, \citenamefont {Ajith}, \citenamefont {Allen}, \citenamefont {Amador~Ceron}, \citenamefont {Amariutei}, \citenamefont {Anderson}, \citenamefont {Anderson}, \citenamefont {Arai}, \citenamefont {Araya}, \citenamefont {Arceneaux}, \citenamefont {Ast}, \citenamefont {Aston}, \citenamefont {Atkinson}, \citenamefont {Aufmuth}, \citenamefont {Aulbert}, \citenamefont {Austin}, \citenamefont {Aylott}, \citenamefont {Babak}, \citenamefont {Baker}, \citenamefont {Ballmer}, \citenamefont {Bao}, \citenamefont {Barayoga}, \citenamefont {Barker}, \citenamefont {Barr}, \citenamefont {Barsotti}, \citenamefont {Barton}, \citenamefont {Bartos}, \citenamefont {Bassiri}, \citenamefont {Batch},
  \citenamefont {Bauchrowitz}, \citenamefont {Behnke}, \citenamefont {Bell}, \citenamefont {Bell}, \citenamefont {Bergmann}, \citenamefont {Berliner}, \citenamefont {Bertolini}, \citenamefont {Betzwieser}, \citenamefont {Beveridge}, \citenamefont {Beyersdorf}, \citenamefont {Bhadbhade}, \citenamefont {Bilenko}, \citenamefont {Billingsley}, \citenamefont {Birch}, \citenamefont {Biscans}, \citenamefont {Black}, \citenamefont {Blackburn}, \citenamefont {Blackburn}, \citenamefont {Blair}, \citenamefont {Bland}, \citenamefont {Bock}, \citenamefont {Bodiya}, \citenamefont {Bogan}, \citenamefont {Bond}, \citenamefont {Bork}, \citenamefont {Born}, \citenamefont {Bose}, \citenamefont {Bowers}, \citenamefont {Brady}, \citenamefont {Braginsky}, \citenamefont {Brau}, \citenamefont {Breyer}, \citenamefont {Bridges}, \citenamefont {Brinkmann}, \citenamefont {Britzger}, \citenamefont {Brooks}, \citenamefont {Brown}, \citenamefont {Brown}, \citenamefont {Buckland}, \citenamefont {Br{\"u}ckner}, \citenamefont {Buchler},
  \citenamefont {Buonanno}, \citenamefont {{Burguet-Castell}}, \citenamefont {Byer}, \citenamefont {Cadonati}, \citenamefont {Camp}, \citenamefont {Campsie}, \citenamefont {Cannon}, \citenamefont {Cao}, \citenamefont {Capano}, \citenamefont {Carbone}, \citenamefont {Caride}, \citenamefont {Castiglia}, \citenamefont {Caudill}, \citenamefont {Cavagli{\`a}}, \citenamefont {Cepeda}, \citenamefont {Chalermsongsak}, \citenamefont {Chao}, \citenamefont {Charlton}, \citenamefont {Chen}, \citenamefont {Chen}, \citenamefont {Cho}, \citenamefont {Chow}, \citenamefont {Christensen}, \citenamefont {Chu}, \citenamefont {Chua}, \citenamefont {Chung}, \citenamefont {Ciani}, \citenamefont {Clara}, \citenamefont {Clark}, \citenamefont {Clark}, \citenamefont {Constancio~Junior}, \citenamefont {Cook}, \citenamefont {Corbitt}, \citenamefont {Cordier}, \citenamefont {Cornish}, \citenamefont {Corsi}, \citenamefont {Costa}, \citenamefont {Coughlin}, \citenamefont {Countryman}, \citenamefont {Couvares}, \citenamefont {Coward},
  \citenamefont {Cowart}, \citenamefont {Coyne}, \citenamefont {Craig}, \citenamefont {Creighton}, \citenamefont {Creighton}, \citenamefont {Cumming}, \citenamefont {Cunningham}, \citenamefont {Dahl}, \citenamefont {Damjanic}, \citenamefont {Danilishin}, \citenamefont {Danzmann}, \citenamefont {Daudert}, \citenamefont {Daveloza}, \citenamefont {Davies}, \citenamefont {Daw}, \citenamefont {Dayanga}, \citenamefont {Deleeuw}, \citenamefont {Denker}, \citenamefont {Dent}, \citenamefont {Dergachev}, \citenamefont {DeRosa}, \citenamefont {DeSalvo}, \citenamefont {Dhurandhar}, \citenamefont {Di~Palma}, \citenamefont {D{\'i}az}, \citenamefont {Dietz}, \citenamefont {Donovan}, \citenamefont {Dooley}, \citenamefont {Doravari}, \citenamefont {Drasco}, \citenamefont {Drever}, \citenamefont {Driggers}, \citenamefont {Du}, \citenamefont {Dumas}, \citenamefont {Dwyer}, \citenamefont {Eberle}, \citenamefont {Edwards}, \citenamefont {Effler}, \citenamefont {Ehrens}, \citenamefont {Eikenberry}, \citenamefont {Engel},
  \citenamefont {Essick}, \citenamefont {Etzel}, \citenamefont {Evans}, \citenamefont {Evans}, \citenamefont {Evans}, \citenamefont {Factourovich}, \citenamefont {Fairhurst}, \citenamefont {Fang}, \citenamefont {Farr}, \citenamefont {Farr}, \citenamefont {Favata}, \citenamefont {Fazi}, \citenamefont {Fehrmann}, \citenamefont {Feldbaum}, \citenamefont {Finn}, \citenamefont {Fisher}, \citenamefont {Foley}, \citenamefont {Forsi}, \citenamefont {Fotopoulos}, \citenamefont {Frede}, \citenamefont {Frei}, \citenamefont {Frei}, \citenamefont {Freise}, \citenamefont {Frey}, \citenamefont {Fricke}, \citenamefont {Friedrich}, \citenamefont {Fritschel}, \citenamefont {Frolov}, \citenamefont {Fujimoto}, \citenamefont {Fulda}, \citenamefont {Fyffe}, \citenamefont {Gair}, \citenamefont {Garcia}, \citenamefont {Gehrels}, \citenamefont {Gelencser}, \citenamefont {Gergely}, \citenamefont {Ghosh}, \citenamefont {Giaime}, \citenamefont {Giampanis}, \citenamefont {Giardina}, \citenamefont {{Gil-Casanova}}, \citenamefont {Gill},
  \citenamefont {Gleason}, \citenamefont {Goetz}, \citenamefont {Gonz{\'a}lez}, \citenamefont {Gordon}, \citenamefont {Gorodetsky}, \citenamefont {Gossan}, \citenamefont {Go{\ss}ler}, \citenamefont {Graef}, \citenamefont {Graff}, \citenamefont {Grant}, \citenamefont {Gras}, \citenamefont {Gray}, \citenamefont {Greenhalgh}, \citenamefont {Gretarsson}, \citenamefont {Griffo}, \citenamefont {Grote}, \citenamefont {Grover}, \citenamefont {Grunewald}, \citenamefont {Guido}, \citenamefont {Gustafson}, \citenamefont {Gustafson}, \citenamefont {Hammer}, \citenamefont {Hammond}, \citenamefont {Hanks}, \citenamefont {Hanna}, \citenamefont {Hanson}, \citenamefont {Haris}, \citenamefont {Harms}, \citenamefont {Harry}, \citenamefont {Harry}, \citenamefont {Harstad}, \citenamefont {Hartman}, \citenamefont {Haughian}, \citenamefont {Hayama}, \citenamefont {Heefner}, \citenamefont {Heintze}, \citenamefont {Hendry}, \citenamefont {Heng}, \citenamefont {Heptonstall}, \citenamefont {Heurs}, \citenamefont {Hewitson},
  \citenamefont {Hild}, \citenamefont {Hoak}, \citenamefont {Hodge}, \citenamefont {Holt}, \citenamefont {Holtrop}, \citenamefont {Hong}, \citenamefont {Hooper}, \citenamefont {Hough}, \citenamefont {Howell}, \citenamefont {Huang}, \citenamefont {Huerta}, \citenamefont {Hughey}, \citenamefont {Huttner}, \citenamefont {Huynh}, \citenamefont {{Huynh-Dinh}}, \citenamefont {Ingram}, \citenamefont {Inta}, \citenamefont {Isogai}, \citenamefont {Ivanov}, \citenamefont {Iyer}, \citenamefont {Izumi}, \citenamefont {Jacobson}, \citenamefont {James}, \citenamefont {Jang}, \citenamefont {Jang}, \citenamefont {Jesse}, \citenamefont {Johnson}, \citenamefont {Jones}, \citenamefont {Jones}, \citenamefont {Jones}, \citenamefont {Ju}, \citenamefont {Kalmus}, \citenamefont {Kalogera}, \citenamefont {Kandhasamy}, \citenamefont {Kang}, \citenamefont {Kanner}, \citenamefont {Kasturi}, \citenamefont {Katsavounidis}, \citenamefont {Katzman}, \citenamefont {Kaufer}, \citenamefont {Kawabe}, \citenamefont {Kawamura}, \citenamefont
  {Kawazoe}, \citenamefont {Keitel}, \citenamefont {Kelley}, \citenamefont {Kells}, \citenamefont {Keppel}, \citenamefont {Khalaidovski}, \citenamefont {Khalili}, \citenamefont {Khazanov}, \citenamefont {Kim}, \citenamefont {Kim}, \citenamefont {Kim}, \citenamefont {Kim}, \citenamefont {Kim}, \citenamefont {King}, \citenamefont {Kinzel}, \citenamefont {Kissel}, \citenamefont {Klimenko}, \citenamefont {Kline}, \citenamefont {Kokeyama}, \citenamefont {Kondrashov}, \citenamefont {Koranda}, \citenamefont {Korth}, \citenamefont {Kozak}, \citenamefont {Kozameh}, \citenamefont {Kremin}, \citenamefont {Kringel}, \citenamefont {Krishnan}, \citenamefont {Kucharczyk}, \citenamefont {Kuehn}, \citenamefont {Kumar}, \citenamefont {Kumar}, \citenamefont {Kuper}, \citenamefont {Kurdyumov}, \citenamefont {Kwee}, \citenamefont {Lam}, \citenamefont {Landry}, \citenamefont {Lantz}, \citenamefont {Lasky}, \citenamefont {Lawrie}, \citenamefont {Lazzarini}, \citenamefont {Roux}, \citenamefont {Leaci}, \citenamefont {Lee},
  \citenamefont {Lee}, \citenamefont {Lee}, \citenamefont {Lee}, \citenamefont {Leong}, \citenamefont {Levine}, \citenamefont {Lhuillier}, \citenamefont {Lin}, \citenamefont {Litvine}, \citenamefont {Liu}, \citenamefont {Liu}, \citenamefont {Lockerbie}, \citenamefont {Lodhia}, \citenamefont {Loew}, \citenamefont {Logue}, \citenamefont {Lombardi}, \citenamefont {Lormand}, \citenamefont {Lough}, \citenamefont {Lubinski}, \citenamefont {L{\"u}ck}, \citenamefont {Lundgren}, \citenamefont {Macarthur}, \citenamefont {Macdonald}, \citenamefont {Machenschalk}, \citenamefont {MacInnis}, \citenamefont {Macleod}, \citenamefont {{Maga{\~n}a-Sandoval}}, \citenamefont {Mageswaran}, \citenamefont {Mailand}, \citenamefont {Manca}, \citenamefont {Mandel}, \citenamefont {Mandic}, \citenamefont {M{\'a}rka}, \citenamefont {M{\'a}rka}, \citenamefont {Markosyan}, \citenamefont {Maros}, \citenamefont {Martin}, \citenamefont {Martin}, \citenamefont {Martinov}, \citenamefont {Marx}, \citenamefont {Mason}, \citenamefont {Matichard},
  \citenamefont {Matone}, \citenamefont {Matzner}, \citenamefont {Mavalvala}, \citenamefont {May}, \citenamefont {Mazzolo}, \citenamefont {McAuley}, \citenamefont {McCarthy}, \citenamefont {McClelland}, \citenamefont {McGuire}, \citenamefont {McIntyre}, \citenamefont {McIver}, \citenamefont {Meadors}, \citenamefont {Mehmet}, \citenamefont {Meier}, \citenamefont {Melatos}, \citenamefont {Mendell}, \citenamefont {Mercer}, \citenamefont {Meshkov}, \citenamefont {Messenger}, \citenamefont {Meyer}, \citenamefont {Miao}, \citenamefont {Miller}, \citenamefont {Mingarelli}, \citenamefont {Mitra}, \citenamefont {Mitrofanov}, \citenamefont {Mitselmakher}, \citenamefont {Mittleman}, \citenamefont {Moe}, \citenamefont {Mokler}, \citenamefont {Mohapatra}, \citenamefont {Moraru}, \citenamefont {Moreno}, \citenamefont {Mori}, \citenamefont {Morriss}, \citenamefont {Mossavi}, \citenamefont {{Mow-Lowry}}, \citenamefont {Mueller}, \citenamefont {Mueller}, \citenamefont {Mukherjee}, \citenamefont {Mullavey}, \citenamefont
  {Munch}, \citenamefont {Murphy}, \citenamefont {Murray}, \citenamefont {Mytidis}, \citenamefont {Kumar}, \citenamefont {Nash}, \citenamefont {Nayak}, \citenamefont {Necula}, \citenamefont {Newton}, \citenamefont {Nguyen}, \citenamefont {Nishida}, \citenamefont {Nishizawa}, \citenamefont {Nitz}, \citenamefont {Nolting}, \citenamefont {Normandin}, \citenamefont {Nuttall}, \citenamefont {O'Dell}, \citenamefont {O'Reilly}, \citenamefont {O'Shaughnessy}, \citenamefont {Ochsner}, \citenamefont {Oelker}, \citenamefont {Ogin}, \citenamefont {Oh}, \citenamefont {Oh}, \citenamefont {Ohme}, \citenamefont {Oppermann}, \citenamefont {Osthelder}, \citenamefont {Ott}, \citenamefont {Ottaway}, \citenamefont {Ottens}, \citenamefont {Ou}, \citenamefont {Overmier}, \citenamefont {Owen}, \citenamefont {Padilla}, \citenamefont {Pai}, \citenamefont {Pan}, \citenamefont {Pankow}, \citenamefont {Papa}, \citenamefont {Paris}, \citenamefont {Parkinson}, \citenamefont {Pedraza}, \citenamefont {Penn}, \citenamefont {Peralta},
  \citenamefont {Perreca}, \citenamefont {Phelps}, \citenamefont {Pickenpack}, \citenamefont {Pierro}, \citenamefont {Pinto}, \citenamefont {Pitkin}, \citenamefont {Pletsch}, \citenamefont {P{\"o}ld}, \citenamefont {Postiglione}, \citenamefont {Poux}, \citenamefont {Predoi}, \citenamefont {Prestegard}, \citenamefont {Price}, \citenamefont {Prijatelj}, \citenamefont {Privitera}, \citenamefont {Prokhorov}, \citenamefont {Puncken}, \citenamefont {Quetschke}, \citenamefont {Quintero}, \citenamefont {{Quitzow-James}}, \citenamefont {Raab}, \citenamefont {Radkins}, \citenamefont {Raffai}, \citenamefont {Raja}, \citenamefont {Rakhmanov}, \citenamefont {Ramet}, \citenamefont {Raymond}, \citenamefont {Reed}, \citenamefont {Reed}, \citenamefont {Reid}, \citenamefont {Reitze}, \citenamefont {Riesen}, \citenamefont {Riles}, \citenamefont {Roberts}, \citenamefont {Robertson}, \citenamefont {Robinson}, \citenamefont {Roddy}, \citenamefont {Rodriguez}, \citenamefont {Rodriguez}, \citenamefont {Rodruck}, \citenamefont
  {Rollins}, \citenamefont {Romie}, \citenamefont {R{\"o}ver}, \citenamefont {Rowan}, \citenamefont {R{\"u}diger}, \citenamefont {Ryan}, \citenamefont {Salemi}, \citenamefont {Sammut}, \citenamefont {Sandberg}, \citenamefont {Sanders}, \citenamefont {Sankar}, \citenamefont {Sannibale}, \citenamefont {Santamar{\'i}a}, \citenamefont {{Santiago-Prieto}}, \citenamefont {Santostasi}, \citenamefont {Sathyaprakash}, \citenamefont {Saulson}, \citenamefont {Savage}, \citenamefont {Schilling}, \citenamefont {Schnabel}, \citenamefont {Schofield}, \citenamefont {Schuette}, \citenamefont {Schulz}, \citenamefont {Schutz}, \citenamefont {Schwinberg}, \citenamefont {Scott}, \citenamefont {Scott}, \citenamefont {Seifert}, \citenamefont {Sellers}, \citenamefont {Sengupta}, \citenamefont {Sergeev}, \citenamefont {Shaddock}, \citenamefont {Shahriar}, \citenamefont {Shaltev}, \citenamefont {Shao}, \citenamefont {Shapiro}, \citenamefont {Shawhan}, \citenamefont {Shoemaker}, \citenamefont {Sidery}, \citenamefont {Siemens},
  \citenamefont {Sigg}, \citenamefont {Simakov}, \citenamefont {Singer}, \citenamefont {Singer}, \citenamefont {Sintes}, \citenamefont {Skelton}, \citenamefont {Slagmolen}, \citenamefont {Slutsky}, \citenamefont {Smith}, \citenamefont {Smith}, \citenamefont {Smith}, \citenamefont {{Smith-Lefebvre}}, \citenamefont {Son}, \citenamefont {Sorazu}, \citenamefont {Souradeep}, \citenamefont {Stefszky}, \citenamefont {Steinert}, \citenamefont {Steinlechner}, \citenamefont {Steinlechner}, \citenamefont {Steplewski}, \citenamefont {Stevens}, \citenamefont {Stochino}, \citenamefont {Stone}, \citenamefont {Strain}, \citenamefont {Strigin}, \citenamefont {Stroeer}, \citenamefont {Stuver}, \citenamefont {Summerscales}, \citenamefont {Susmithan}, \citenamefont {Sutton}, \citenamefont {Szeifert}, \citenamefont {Talukder}, \citenamefont {Tanner}, \citenamefont {Tarabrin}, \citenamefont {Taylor}, \citenamefont {Thomas}, \citenamefont {Thomas}, \citenamefont {Thorne}, \citenamefont {Thorne}, \citenamefont {Thrane},
  \citenamefont {Tiwari}, \citenamefont {Tokmakov}, \citenamefont {Tomlinson}, \citenamefont {Torres}, \citenamefont {Torrie}, \citenamefont {Traylor}, \citenamefont {Tse}, \citenamefont {Ugolini}, \citenamefont {Unnikrishnan}, \citenamefont {Vahlbruch}, \citenamefont {Vallisneri}, \citenamefont {van~der Sluys}, \citenamefont {van Veggel}, \citenamefont {Vass}, \citenamefont {Vaulin}, \citenamefont {Vecchio}, \citenamefont {Veitch}, \citenamefont {Veitch}, \citenamefont {Venkateswara}, \citenamefont {Verma}, \citenamefont {{Vincent-Finley}}, \citenamefont {Vitale}, \citenamefont {Vo}, \citenamefont {Vorvick}, \citenamefont {Vousden}, \citenamefont {Vyatchanin}, \citenamefont {Wade}, \citenamefont {Wade}, \citenamefont {Wade}, \citenamefont {Waldman}, \citenamefont {Wallace}, \citenamefont {Wan}, \citenamefont {Wang}, \citenamefont {Wang}, \citenamefont {Wang}, \citenamefont {Wanner}, \citenamefont {Ward}, \citenamefont {Was}, \citenamefont {Weinert}, \citenamefont {Weinstein}, \citenamefont {Weiss},
  \citenamefont {Welborn}, \citenamefont {Wen}, \citenamefont {Wessels}, \citenamefont {West}, \citenamefont {Westphal}, \citenamefont {Wette}, \citenamefont {Whelan}, \citenamefont {Whitcomb}, \citenamefont {Wiseman}, \citenamefont {White}, \citenamefont {Whiting}, \citenamefont {Wiesner}, \citenamefont {Wilkinson}, \citenamefont {Willems}, \citenamefont {Williams}, \citenamefont {Williams}, \citenamefont {Williams}, \citenamefont {Willis}, \citenamefont {Willke}, \citenamefont {Wimmer}, \citenamefont {Winkelmann}, \citenamefont {Winkler}, \citenamefont {Wipf}, \citenamefont {Wittel}, \citenamefont {Woan}, \citenamefont {Wooley}, \citenamefont {Worden}, \citenamefont {Yablon}, \citenamefont {Yakushin}, \citenamefont {Yamamoto}, \citenamefont {Yancey}, \citenamefont {Yang}, \citenamefont {{Yeaton-Massey}}, \citenamefont {Yoshida}, \citenamefont {Yum}, \citenamefont {Zanolin}, \citenamefont {Zhang}, \citenamefont {Zhang}, \citenamefont {Zhao}, \citenamefont {Zhu}, \citenamefont {Zhu}, \citenamefont {Zotov},
  \citenamefont {Zucker},\ and\ \citenamefont {Zweizig}}]{AasiNP13EnhancedSensitivity}%
  \BibitemOpen
  \bibfield  {author} {\bibinfo {author} {\bibfnamefont {J.}~\bibnamefont {Aasi}}, \bibinfo {author} {\bibfnamefont {J.}~\bibnamefont {Abadie}}, \bibinfo {author} {\bibfnamefont {B.~P.}\ \bibnamefont {Abbott}}, \bibnamefont {et~al.},\ }\bibfield  {title} {\bibinfo {title} {Enhanced sensitivity of the {{LIGO}} gravitational wave detector by using squeezed states of light},\ }\href {https://doi.org/10.1038/nphoton.2013.177} {\bibfield  {journal} {\bibinfo  {journal} {Nat. Photonics}\ }\textbf {\bibinfo {volume} {7}},\ \bibinfo {pages} {613} (\bibinfo {year} {2013})}\BibitemShut {NoStop}%
\bibitem [{\citenamefont {Tse}\ \emph {et~al.}(2019)\citenamefont {Tse}, \citenamefont {Yu}, \citenamefont {Kijbunchoo}, \citenamefont {{Fernandez-Galiana}}, \citenamefont {Dupej}, \citenamefont {Barsotti}, \citenamefont {Blair}, \citenamefont {Brown}, \citenamefont {Dwyer}, \citenamefont {Effler}, \citenamefont {Evans}, \citenamefont {Fritschel}, \citenamefont {Frolov}, \citenamefont {Green}, \citenamefont {Mansell}, \citenamefont {Matichard}, \citenamefont {Mavalvala}, \citenamefont {McClelland}, \citenamefont {McCuller}, \citenamefont {McRae}, \citenamefont {Miller}, \citenamefont {Mullavey}, \citenamefont {Oelker}, \citenamefont {Phinney}, \citenamefont {Sigg}, \citenamefont {Slagmolen}, \citenamefont {Vo}, \citenamefont {Ward}, \citenamefont {Whittle}, \citenamefont {Abbott}, \citenamefont {Adams}, \citenamefont {Adhikari}, \citenamefont {Ananyeva}, \citenamefont {Appert}, \citenamefont {Arai}, \citenamefont {Areeda}, \citenamefont {Asali}, \citenamefont {Aston}, \citenamefont {Austin}, \citenamefont
  {Baer}, \citenamefont {Ball}, \citenamefont {Ballmer}, \citenamefont {Banagiri}, \citenamefont {Barker}, \citenamefont {Bartlett}, \citenamefont {Berger}, \citenamefont {Betzwieser}, \citenamefont {Bhattacharjee}, \citenamefont {Billingsley}, \citenamefont {Biscans}, \citenamefont {Blair}, \citenamefont {Bode}, \citenamefont {Booker}, \citenamefont {Bork}, \citenamefont {Bramley}, \citenamefont {Brooks}, \citenamefont {Buikema}, \citenamefont {Cahillane}, \citenamefont {Cannon}, \citenamefont {Chen}, \citenamefont {Ciobanu}, \citenamefont {Clara}, \citenamefont {Cooper}, \citenamefont {Corley}, \citenamefont {Countryman}, \citenamefont {Covas}, \citenamefont {Coyne}, \citenamefont {Datrier}, \citenamefont {Davis}, \citenamefont {Di~Fronzo}, \citenamefont {Driggers}, \citenamefont {Etzel}, \citenamefont {Evans}, \citenamefont {Feicht}, \citenamefont {Fulda}, \citenamefont {Fyffe}, \citenamefont {Giaime}, \citenamefont {Giardina}, \citenamefont {Godwin}, \citenamefont {Goetz}, \citenamefont {Gras},
  \citenamefont {Gray}, \citenamefont {Gray}, \citenamefont {Gupta}, \citenamefont {Gustafson}, \citenamefont {Gustafson}, \citenamefont {Hanks}, \citenamefont {Hanson}, \citenamefont {Hardwick}, \citenamefont {Hasskew}, \citenamefont {Heintze}, \citenamefont {{Helmling-Cornell}}, \citenamefont {Holland}, \citenamefont {Jones}, \citenamefont {Kandhasamy}, \citenamefont {Karki}, \citenamefont {Kasprzack}, \citenamefont {Kawabe}, \citenamefont {King}, \citenamefont {Kissel}, \citenamefont {Kumar}, \citenamefont {Landry}, \citenamefont {Lane}, \citenamefont {Lantz}, \citenamefont {Laxen}, \citenamefont {Lecoeuche}, \citenamefont {Leviton}, \citenamefont {Liu}, \citenamefont {Lormand}, \citenamefont {Lundgren}, \citenamefont {Macas}, \citenamefont {MacInnis}, \citenamefont {Macleod}, \citenamefont {M{\'a}rka}, \citenamefont {M{\'a}rka}, \citenamefont {Martynov}, \citenamefont {Mason}, \citenamefont {Massinger}, \citenamefont {McCarthy}, \citenamefont {McCormick}, \citenamefont {McIver}, \citenamefont {Mendell},
  \citenamefont {Merfeld}, \citenamefont {Merilh}, \citenamefont {Meylahn}, \citenamefont {Mistry}, \citenamefont {Mittleman}, \citenamefont {Moreno}, \citenamefont {{Mow-Lowry}}, \citenamefont {Mozzon}, \citenamefont {Nelson}, \citenamefont {Nguyen}, \citenamefont {Nuttall}, \citenamefont {Oberling}, \citenamefont {Oram}, \citenamefont {O'Reilly}, \citenamefont {Osthelder}, \citenamefont {Ottaway}, \citenamefont {Overmier}, \citenamefont {Palamos}, \citenamefont {Parker}, \citenamefont {Payne}, \citenamefont {Pele}, \citenamefont {Perez}, \citenamefont {Pirello}, \citenamefont {Radkins}, \citenamefont {Ramirez}, \citenamefont {Richardson}, \citenamefont {Riles}, \citenamefont {Robertson}, \citenamefont {Rollins}, \citenamefont {Romel}, \citenamefont {Romie}, \citenamefont {Ross}, \citenamefont {Ryan}, \citenamefont {Sadecki}, \citenamefont {Sanchez}, \citenamefont {Sanchez}, \citenamefont {Saravanan}, \citenamefont {Savage}, \citenamefont {Schaetzl}, \citenamefont {Schnabel}, \citenamefont {Schofield},
  \citenamefont {Schwartz}, \citenamefont {Sellers}, \citenamefont {Shaffer}, \citenamefont {Smith}, \citenamefont {Soni}, \citenamefont {Sorazu}, \citenamefont {Spencer}, \citenamefont {Strain}, \citenamefont {Sun}, \citenamefont {Szczepa{\'n}czyk}, \citenamefont {Thomas}, \citenamefont {Thomas}, \citenamefont {Thorne}, \citenamefont {Toland}, \citenamefont {Torrie}, \citenamefont {Traylor}, \citenamefont {Urban}, \citenamefont {Vajente}, \citenamefont {Valdes}, \citenamefont {{Vander-Hyde}}, \citenamefont {Veitch}, \citenamefont {Venkateswara}, \citenamefont {Venugopalan}, \citenamefont {Viets}, \citenamefont {Vorvick}, \citenamefont {Wade}, \citenamefont {Warner}, \citenamefont {Weaver}, \citenamefont {Weiss}, \citenamefont {Willke}, \citenamefont {Wipf}, \citenamefont {Xiao}, \citenamefont {Yamamoto}, \citenamefont {Yap}, \citenamefont {Yu}, \citenamefont {Zhang}, \citenamefont {Zucker},\ and\ \citenamefont {Zweizig}}]{TsePRL19QuantumEnhancedAdvanced}%
  \BibitemOpen
  \bibfield  {author} {\bibinfo {author} {\bibfnamefont {M.}~\bibnamefont {Tse}}, \bibinfo {author} {\bibfnamefont {H.}~\bibnamefont {Yu}}, \bibinfo {author} {\bibfnamefont {N.}~\bibnamefont {Kijbunchoo}}, \bibnamefont {et~al.},\ }\bibfield  {title} {\bibinfo {title} {Quantum-{{Enhanced Advanced LIGO Detectors}} in the {{Era}} of {{Gravitational-Wave Astronomy}}},\ }\href {https://doi.org/10.1103/PhysRevLett.123.231107} {\bibfield  {journal} {\bibinfo  {journal} {Phys. Rev. Lett.}\ }\textbf {\bibinfo {volume} {123}},\ \bibinfo {pages} {231107} (\bibinfo {year} {2019})}\BibitemShut {NoStop}%
\bibitem [{\citenamefont {McCuller}\ \emph {et~al.}(2020)\citenamefont {McCuller}, \citenamefont {Whittle}, \citenamefont {Ganapathy}, \citenamefont {Komori}, \citenamefont {Tse}, \citenamefont {{Fernandez-Galiana}}, \citenamefont {Barsotti}, \citenamefont {Fritschel}, \citenamefont {MacInnis}, \citenamefont {Matichard}, \citenamefont {Mason}, \citenamefont {Mavalvala}, \citenamefont {Mittleman}, \citenamefont {Yu}, \citenamefont {Zucker},\ and\ \citenamefont {Evans}}]{McCullerPRL20FrequencyDependentSqueezing}%
  \BibitemOpen
  \bibfield  {author} {\bibinfo {author} {\bibfnamefont {L.}~\bibnamefont {McCuller}}, \bibinfo {author} {\bibfnamefont {C.}~\bibnamefont {Whittle}}, \bibinfo {author} {\bibfnamefont {D.}~\bibnamefont {Ganapathy}}, \bibnamefont {et~al.},\ }\bibfield  {title} {\bibinfo {title} {Frequency-{{Dependent Squeezing}} for {{Advanced LIGO}}},\ }\href {https://doi.org/10.1103/PhysRevLett.124.171102} {\bibfield  {journal} {\bibinfo  {journal} {Phys. Rev. Lett.}\ }\textbf {\bibinfo {volume} {124}},\ \bibinfo {pages} {171102} (\bibinfo {year} {2020})}\BibitemShut {NoStop}%
\bibitem [{\citenamefont {G{\'o}recki}\ \emph {et~al.}(2022)\citenamefont {G{\'o}recki}, \citenamefont {Riccardi},\ and\ \citenamefont {Maccone}}]{gorecki2022quantum}%
  \BibitemOpen
  \bibfield  {author} {\bibinfo {author} {\bibfnamefont {W.}~\bibnamefont {G{\'o}recki}}, \bibinfo {author} {\bibfnamefont {A.}~\bibnamefont {Riccardi}},\ \bibnamefont {and}\ \bibinfo {author} {\bibfnamefont {L.}~\bibnamefont {Maccone}},\ }\bibfield  {title} {\bibinfo {title} {Quantum metrology of noisy spreading channels},\ }\href {https://doi.org/10.1103/PhysRevLett.129.240503} {\bibfield  {journal} {\bibinfo  {journal} {Phys. Rev. Lett.}\ }\textbf {\bibinfo {volume} {129}},\ \bibinfo {pages} {240503} (\bibinfo {year} {2022})}\BibitemShut {NoStop}%
\bibitem [{\citenamefont {Gottesman}\ \emph {et~al.}(2001)\citenamefont {Gottesman}, \citenamefont {Kitaev},\ and\ \citenamefont {Preskill}}]{GottesmanPRA01EncodingQubit}%
  \BibitemOpen
  \bibfield  {author} {\bibinfo {author} {\bibfnamefont {D.}~\bibnamefont {Gottesman}}, \bibinfo {author} {\bibfnamefont {A.}~\bibnamefont {Kitaev}},\ \bibnamefont {and}\ \bibinfo {author} {\bibfnamefont {J.}~\bibnamefont {Preskill}},\ }\bibfield  {title} {\bibinfo {title} {Encoding a qubit in an oscillator},\ }\href {https://doi.org/10.1103/PhysRevA.64.012310} {\bibfield  {journal} {\bibinfo  {journal} {Phys. Rev. A}\ }\textbf {\bibinfo {volume} {64}},\ \bibinfo {pages} {012310} (\bibinfo {year} {2001})}\BibitemShut {NoStop}%
\bibitem [{\citenamefont {Gilmore}\ \emph {et~al.}(2021)\citenamefont {Gilmore}, \citenamefont {Affolter}, \citenamefont {Lewis-Swan}, \citenamefont {Barberena}, \citenamefont {Jordan}, \citenamefont {Rey},\ and\ \citenamefont {Bollinger}}]{gilmore2021quantum}%
  \BibitemOpen
  \bibfield  {author} {\bibinfo {author} {\bibfnamefont {K.~A.}\ \bibnamefont {Gilmore}}, \bibinfo {author} {\bibfnamefont {M.}~\bibnamefont {Affolter}}, \bibinfo {author} {\bibfnamefont {R.~J.}\ \bibnamefont {Lewis-Swan}}, \bibinfo {author} {\bibfnamefont {D.}~\bibnamefont {Barberena}}, \bibinfo {author} {\bibfnamefont {E.}~\bibnamefont {Jordan}}, \bibinfo {author} {\bibfnamefont {A.~M.}\ \bibnamefont {Rey}},\ \bibnamefont {and}\ \bibinfo {author} {\bibfnamefont {J.~J.}\ \bibnamefont {Bollinger}},\ }\bibfield  {title} {\bibinfo {title} {Quantum-enhanced sensing of displacements and electric fields with two-dimensional trapped-ion crystals},\ }\href {https://doi.org/10.1126/science.abi5226} {\bibfield  {journal} {\bibinfo  {journal} {Science}\ }\textbf {\bibinfo {volume} {373}},\ \bibinfo {pages} {673} (\bibinfo {year} {2021})}\BibitemShut {NoStop}%
\bibitem [{\citenamefont {Zhuang}\ \emph {et~al.}(2020)\citenamefont {Zhuang}, \citenamefont {Preskill},\ and\ \citenamefont {Jiang}}]{ZhuangNJP20DistributedQuantum}%
  \BibitemOpen
  \bibfield  {author} {\bibinfo {author} {\bibfnamefont {Q.}~\bibnamefont {Zhuang}}, \bibinfo {author} {\bibfnamefont {J.}~\bibnamefont {Preskill}},\ \bibnamefont {and}\ \bibinfo {author} {\bibfnamefont {L.}~\bibnamefont {Jiang}},\ }\bibfield  {title} {\bibinfo {title} {Distributed quantum sensing enhanced by continuous-variable error correction},\ }\href {https://doi.org/10.1088/1367-2630/ab7257} {\bibfield  {journal} {\bibinfo  {journal} {New J. Phys.}\ }\textbf {\bibinfo {volume} {22}},\ \bibinfo {pages} {022001} (\bibinfo {year} {2020})}\BibitemShut {NoStop}%
\bibitem [{\citenamefont {Michael}\ \emph {et~al.}(2016)\citenamefont {Michael}, \citenamefont {Silveri}, \citenamefont {Brierley}, \citenamefont {Albert}, \citenamefont {Salmilehto}, \citenamefont {Jiang},\ and\ \citenamefont {Girvin}}]{MichaelPRX16NewClass}%
  \BibitemOpen
  \bibfield  {author} {\bibinfo {author} {\bibfnamefont {M.~H.}\ \bibnamefont {Michael}}, \bibinfo {author} {\bibfnamefont {M.}~\bibnamefont {Silveri}}, \bibinfo {author} {\bibfnamefont {R.~T.}\ \bibnamefont {Brierley}}, \bibinfo {author} {\bibfnamefont {V.~V.}\ \bibnamefont {Albert}}, \bibinfo {author} {\bibfnamefont {J.}~\bibnamefont {Salmilehto}}, \bibinfo {author} {\bibfnamefont {L.}~\bibnamefont {Jiang}},\ \bibnamefont {and}\ \bibinfo {author} {\bibfnamefont {S.~M.}\ \bibnamefont {Girvin}},\ }\bibfield  {title} {\bibinfo {title} {New {{Class}} of {{Quantum Error-Correcting Codes}} for a {{Bosonic Mode}}},\ }\href {https://doi.org/10.1103/PhysRevX.6.031006} {\bibfield  {journal} {\bibinfo  {journal} {Phys. Rev. X}\ }\textbf {\bibinfo {volume} {6}},\ \bibinfo {pages} {031006} (\bibinfo {year} {2016})}\BibitemShut {NoStop}%
\bibitem [{\citenamefont {Monras}\ and\ \citenamefont {Illuminati}(2011)}]{monras2011measurement}%
  \BibitemOpen
  \bibfield  {author} {\bibinfo {author} {\bibfnamefont {A.}~\bibnamefont {Monras}}\ \bibnamefont {and}\ \bibinfo {author} {\bibfnamefont {F.}~\bibnamefont {Illuminati}},\ }\bibfield  {title} {\bibinfo {title} {Measurement of damping and temperature: Precision bounds in gaussian dissipative channels},\ }\href {https://doi.org/10.1103/PhysRevA.83.012315} {\bibfield  {journal} {\bibinfo  {journal} {Phys. Rev. A}\ }\textbf {\bibinfo {volume} {83}},\ \bibinfo {pages} {012315} (\bibinfo {year} {2011})}\BibitemShut {NoStop}%
\bibitem [{\citenamefont {Holevo}(2011)}]{Holevo2011book}%
  \BibitemOpen
  \bibfield  {author} {\bibinfo {author} {\bibfnamefont {A.~S.}\ \bibnamefont {Holevo}},\ }\href {https://doi.org/10.1007/978-88-7642-378-9} {\emph {\bibinfo {title} {Probabilistic and Statistical Aspects of Quantum Theory}}}\ (\bibinfo  {publisher} {Springer Science \& Business Media},\ \bibinfo {year} {2011})\BibitemShut {NoStop}%
\bibitem [{\citenamefont {Pinel}\ \emph {et~al.}(2013)\citenamefont {Pinel}, \citenamefont {Jian}, \citenamefont {Treps}, \citenamefont {Fabre},\ and\ \citenamefont {Braun}}]{pinel2013quantum}%
  \BibitemOpen
  \bibfield  {author} {\bibinfo {author} {\bibfnamefont {O.}~\bibnamefont {Pinel}}, \bibinfo {author} {\bibfnamefont {P.}~\bibnamefont {Jian}}, \bibinfo {author} {\bibfnamefont {N.}~\bibnamefont {Treps}}, \bibinfo {author} {\bibfnamefont {C.}~\bibnamefont {Fabre}},\ \bibnamefont {and}\ \bibinfo {author} {\bibfnamefont {D.}~\bibnamefont {Braun}},\ }\bibfield  {title} {\bibinfo {title} {Quantum parameter estimation using general single-mode gaussian states},\ }\href {https://doi.org/10.1103/PhysRevA.88.040102} {\bibfield  {journal} {\bibinfo  {journal} {Phys. Rev. A}\ }\textbf {\bibinfo {volume} {88}},\ \bibinfo {pages} {040102} (\bibinfo {year} {2013})}\BibitemShut {NoStop}%
\bibitem [{\citenamefont {Gao}\ and\ \citenamefont {Lee}(2014)}]{gao2014bounds}%
  \BibitemOpen
  \bibfield  {author} {\bibinfo {author} {\bibfnamefont {Y.}~\bibnamefont {Gao}}\ \bibnamefont {and}\ \bibinfo {author} {\bibfnamefont {H.}~\bibnamefont {Lee}},\ }\bibfield  {title} {\bibinfo {title} {Bounds on quantum multiple-parameter estimation with gaussian state},\ }\href {https://doi.org/10.1140/epjd/e2014-50560-1} {\bibfield  {journal} {\bibinfo  {journal} {Eur. Phys. J. D}\ }\textbf {\bibinfo {volume} {68}},\ \bibinfo {pages} {1} (\bibinfo {year} {2014})}\BibitemShut {NoStop}%
\bibitem [{\citenamefont {Len}(2022)}]{len2022multiparameter}%
  \BibitemOpen
  \bibfield  {author} {\bibinfo {author} {\bibfnamefont {Y.~L.}\ \bibnamefont {Len}},\ }\bibfield  {title} {\bibinfo {title} {Multiparameter estimation for qubit states with collective measurements: A case study},\ }\href {https://doi.org/10.1088/1367-2630/ac599d} {\bibfield  {journal} {\bibinfo  {journal} {New J. Phys.}\ }\textbf {\bibinfo {volume} {24}},\ \bibinfo {pages} {033037} (\bibinfo {year} {2022})}\BibitemShut {NoStop}%
\bibitem [{Note1()}]{Note1}%
  \BibitemOpen
  \bibinfo {note} {If the temporal basis modes $w(\tau )$ are complex, then their Fourier transforms will have independent positive and negative frequency components unlike the real basis of cosine and sine functions in Eq.~\ref {eq:spec_components}. Instead, the real and imaginary parts of each complex temporal mode should be studied, where each part reduces to the canonical noise estimation problem of a single harmonic oscillator.}\BibitemShut {Stop}%
\bibitem [{\citenamefont {{S. M. Vermeulen, T. Cullen, D. Grass, I. A. O. MacMillan, A. J. Ramirez, J. Wack, B. Ko- rzh, V. S. H. Lee, K. M. Zurek, C. Stoughton, and L. McCuller, Photon Counting Interferome- try to Detect Geontropic Space-Time Fluctuations with GQuEST (2024)}}()}]{Vermeulen24PhotonCounting}%
  \BibitemOpen
  \bibfield  {author} {\bibinfo {author} {\bibnamefont {{S. M. Vermeulen, T. Cullen, D. Grass, I. A. O. MacMillan, A. J. Ramirez, J. Wack, B. Ko- rzh, V. S. H. Lee, K. M. Zurek, C. Stoughton, and L. McCuller, Photon Counting Interferome- try to Detect Geontropic Space-Time Fluctuations with GQuEST (2024)}}},\ }\href {https://doi.org/10.48550/arXiv.2404.07524} {\bibinfo {title} {arxiv:2404.07524 [astro-ph, physics:gr-qc, physics:physics, physics:quant-ph]}}\BibitemShut {NoStop}%
\bibitem [{\citenamefont {Danilishin}\ and\ \citenamefont {Khalili}(2012)}]{danilishin2012quantum}%
  \BibitemOpen
  \bibfield  {author} {\bibinfo {author} {\bibfnamefont {S.~L.}\ \bibnamefont {Danilishin}}\ \bibnamefont {and}\ \bibinfo {author} {\bibfnamefont {F.~Y.}\ \bibnamefont {Khalili}},\ }\bibfield  {title} {\bibinfo {title} {Quantum measurement theory in gravitational-wave detectors},\ }\href {https://doi.org/10.12942/lrr-2012-5} {\bibfield  {journal} {\bibinfo  {journal} {Living Rev. Relativ.}\ }\textbf {\bibinfo {volume} {15}},\ \bibinfo {pages} {1} (\bibinfo {year} {2012})}\BibitemShut {NoStop}%
\bibitem [{\citenamefont {Gorshkov}\ \emph {et~al.}(2007)\citenamefont {Gorshkov}, \citenamefont {Andr{\'e}}, \citenamefont {Lukin},\ and\ \citenamefont {S{\o}rensen}}]{GorshkovPRA07PhotonStoragea}%
  \BibitemOpen
  \bibfield  {author} {\bibinfo {author} {\bibfnamefont {A.~V.}\ \bibnamefont {Gorshkov}}, \bibinfo {author} {\bibfnamefont {A.}~\bibnamefont {Andr{\'e}}}, \bibinfo {author} {\bibfnamefont {M.~D.}\ \bibnamefont {Lukin}},\ \bibnamefont {and}\ \bibinfo {author} {\bibfnamefont {A.~S.}\ \bibnamefont {S{\o}rensen}},\ }\bibfield  {title} {\bibinfo {title} {Photon storage in ${{\Lambda}}$-type optically dense atomic media. {{I}}. {{Cavity}} model},\ }\href {https://doi.org/10.1103/PhysRevA.76.033804} {\bibfield  {journal} {\bibinfo  {journal} {Phys. Rev. A}\ }\textbf {\bibinfo {volume} {76}},\ \bibinfo {pages} {033804} (\bibinfo {year} {2007})}\BibitemShut {NoStop}%
\bibitem [{\citenamefont {Brown}\ \emph {et~al.}(2003)\citenamefont {Brown}, \citenamefont {Dani}, \citenamefont {{Stamper-Kurn}},\ and\ \citenamefont {Whaley}}]{BrownPRA03DeterministicOptical}%
  \BibitemOpen
  \bibfield  {author} {\bibinfo {author} {\bibfnamefont {K.~R.}\ \bibnamefont {Brown}}, \bibinfo {author} {\bibfnamefont {K.~M.}\ \bibnamefont {Dani}}, \bibinfo {author} {\bibfnamefont {D.~M.}\ \bibnamefont {{Stamper-Kurn}}},\ \bibnamefont {and}\ \bibinfo {author} {\bibfnamefont {K.~B.}\ \bibnamefont {Whaley}},\ }\bibfield  {title} {\bibinfo {title} {Deterministic optical {{Fock-state}} generation},\ }\href {https://doi.org/10.1103/PhysRevA.67.043818} {\bibfield  {journal} {\bibinfo  {journal} {Phys. Rev. A}\ }\textbf {\bibinfo {volume} {67}},\ \bibinfo {pages} {043818} (\bibinfo {year} {2003})}\BibitemShut {NoStop}%
\bibitem [{\citenamefont {Geremia}(2006)}]{GeremiaPRL06DeterministicNondestructively}%
  \BibitemOpen
  \bibfield  {author} {\bibinfo {author} {\bibfnamefont {{\relax JM}.}~\bibnamefont {Geremia}},\ }\bibfield  {title} {\bibinfo {title} {Deterministic and {{Nondestructively Verifiable Preparation}} of {{Photon Number States}}},\ }\href {https://doi.org/10.1103/PhysRevLett.97.073601} {\bibfield  {journal} {\bibinfo  {journal} {Phys. Rev. Lett.}\ }\textbf {\bibinfo {volume} {97}},\ \bibinfo {pages} {073601} (\bibinfo {year} {2006})}\BibitemShut {NoStop}%
\bibitem [{\citenamefont {Hastrup}\ and\ \citenamefont {Andersen}(2022)}]{HastrupPRL22ProtocolGenerating}%
  \BibitemOpen
  \bibfield  {author} {\bibinfo {author} {\bibfnamefont {J.}~\bibnamefont {Hastrup}}\ \bibnamefont {and}\ \bibinfo {author} {\bibfnamefont {U.~L.}\ \bibnamefont {Andersen}},\ }\bibfield  {title} {\bibinfo {title} {Protocol for {{Generating Optical Gottesman-Kitaev-Preskill States}} with {{Cavity QED}}},\ }\href {https://doi.org/10.1103/PhysRevLett.128.170503} {\bibfield  {journal} {\bibinfo  {journal} {Phys. Rev. Lett.}\ }\textbf {\bibinfo {volume} {128}},\ \bibinfo {pages} {170503} (\bibinfo {year} {2022})}\BibitemShut {NoStop}%
\bibitem [{\citenamefont {Chen}\ \emph {et~al.}(2006{\natexlab{a}})\citenamefont {Chen}, \citenamefont {Chen}, \citenamefont {Strassel}, \citenamefont {Yuan}, \citenamefont {Zhao}, \citenamefont {Schmiedmayer},\ and\ \citenamefont {Pan}}]{ChenPRL06DeterministicStorable}%
  \BibitemOpen
  \bibfield  {author} {\bibinfo {author} {\bibfnamefont {S.}~\bibnamefont {Chen}}, \bibinfo {author} {\bibfnamefont {Y.-A.}\ \bibnamefont {Chen}}, \bibinfo {author} {\bibfnamefont {T.}~\bibnamefont {Strassel}}, \bibinfo {author} {\bibfnamefont {Z.-S.}\ \bibnamefont {Yuan}}, \bibinfo {author} {\bibfnamefont {B.}~\bibnamefont {Zhao}}, \bibinfo {author} {\bibfnamefont {J.}~\bibnamefont {Schmiedmayer}},\ \bibnamefont {and}\ \bibinfo {author} {\bibfnamefont {J.-W.}\ \bibnamefont {Pan}},\ }\bibfield  {title} {\bibinfo {title} {Deterministic and {{Storable Single-Photon Source Based}} on a {{Quantum Memory}}},\ }\href {https://doi.org/10.1103/PhysRevLett.97.173004} {\bibfield  {journal} {\bibinfo  {journal} {Phys. Rev. Lett.}\ }\textbf {\bibinfo {volume} {97}},\ \bibinfo {pages} {173004} (\bibinfo {year} {2006}{\natexlab{a}})}\BibitemShut {NoStop}%
\bibitem [{\citenamefont {Thomas}\ \emph {et~al.}(2024)\citenamefont {Thomas}, \citenamefont {Wagner}, \citenamefont {Joos}, \citenamefont {Sittig}, \citenamefont {Nawrath}, \citenamefont {Burdekin}, \citenamefont {{de Buy Wenniger}}, \citenamefont {Rasiah}, \citenamefont {{Huber-Loyola}}, \citenamefont {{Sagona-Stophel}}, \citenamefont {H{\"o}fling}, \citenamefont {Jetter}, \citenamefont {Michler}, \citenamefont {Walmsley}, \citenamefont {Portalupi},\ and\ \citenamefont {Ledingham}}]{ThomasSA24DeterministicStorage}%
  \BibitemOpen
  \bibfield  {author} {\bibinfo {author} {\bibfnamefont {S.~E.}\ \bibnamefont {Thomas}}, \bibinfo {author} {\bibfnamefont {L.}~\bibnamefont {Wagner}}, \bibinfo {author} {\bibfnamefont {R.}~\bibnamefont {Joos}}, \bibnamefont {et~al.},\ }\bibfield  {title} {\bibinfo {title} {Deterministic storage and retrieval of telecom light from a quantum dot single-photon source interfaced with an atomic quantum memory},\ }\href {https://doi.org/10.1126/sciadv.adi7346} {\bibfield  {journal} {\bibinfo  {journal} {Sci. Adv.}\ }\textbf {\bibinfo {volume} {10}},\ \bibinfo {pages} {eadi7346} (\bibinfo {year} {2024})}\BibitemShut {NoStop}%
\bibitem [{\citenamefont {Farrera}\ \emph {et~al.}(2016)\citenamefont {Farrera}, \citenamefont {Heinze}, \citenamefont {Albrecht}, \citenamefont {Ho}, \citenamefont {Ch{\'a}vez}, \citenamefont {Teo}, \citenamefont {Sangouard},\ and\ \citenamefont {{de Riedmatten}}}]{FarreraNC16GenerationSingle}%
  \BibitemOpen
  \bibfield  {author} {\bibinfo {author} {\bibfnamefont {P.}~\bibnamefont {Farrera}}, \bibinfo {author} {\bibfnamefont {G.}~\bibnamefont {Heinze}}, \bibinfo {author} {\bibfnamefont {B.}~\bibnamefont {Albrecht}}, \bibinfo {author} {\bibfnamefont {M.}~\bibnamefont {Ho}}, \bibinfo {author} {\bibfnamefont {M.}~\bibnamefont {Ch{\'a}vez}}, \bibinfo {author} {\bibfnamefont {C.}~\bibnamefont {Teo}}, \bibinfo {author} {\bibfnamefont {N.}~\bibnamefont {Sangouard}},\ \bibnamefont {and}\ \bibinfo {author} {\bibfnamefont {H.}~\bibnamefont {{de Riedmatten}}},\ }\bibfield  {title} {\bibinfo {title} {Generation of single photons with highly tunable wave shape from a cold atomic ensemble},\ }\href {https://doi.org/10.1038/ncomms13556} {\bibfield  {journal} {\bibinfo  {journal} {Nat. Commun.}\ }\textbf {\bibinfo {volume} {7}},\ \bibinfo {pages} {13556} (\bibinfo {year} {2016})}\BibitemShut {NoStop}%
\bibitem [{\citenamefont {Katz}\ and\ \citenamefont {Firstenberg}(2018)}]{KatzNC18LightStorage}%
  \BibitemOpen
  \bibfield  {author} {\bibinfo {author} {\bibfnamefont {O.}~\bibnamefont {Katz}}\ \bibnamefont {and}\ \bibinfo {author} {\bibfnamefont {O.}~\bibnamefont {Firstenberg}},\ }\bibfield  {title} {\bibinfo {title} {Light storage for one second in room-temperature alkali vapor},\ }\href {https://doi.org/10.1038/s41467-018-04458-4} {\bibfield  {journal} {\bibinfo  {journal} {Nat. Commun.}\ }\textbf {\bibinfo {volume} {9}},\ \bibinfo {pages} {2074} (\bibinfo {year} {2018})}\BibitemShut {NoStop}%
\bibitem [{\citenamefont {Bergmann}\ \emph {et~al.}(1998)\citenamefont {Bergmann}, \citenamefont {Theuer},\ and\ \citenamefont {Shore}}]{BergmannRMP98CoherentPopulation}%
  \BibitemOpen
  \bibfield  {author} {\bibinfo {author} {\bibfnamefont {K.}~\bibnamefont {Bergmann}}, \bibinfo {author} {\bibfnamefont {H.}~\bibnamefont {Theuer}},\ \bibnamefont {and}\ \bibinfo {author} {\bibfnamefont {B.~W.}\ \bibnamefont {Shore}},\ }\bibfield  {title} {\bibinfo {title} {Coherent population transfer among quantum states of atoms and molecules},\ }\href {https://doi.org/10.1103/RevModPhys.70.1003} {\bibfield  {journal} {\bibinfo  {journal} {Rev. Mod. Phys.}\ }\textbf {\bibinfo {volume} {70}},\ \bibinfo {pages} {1003} (\bibinfo {year} {1998})}\BibitemShut {NoStop}%
\bibitem [{\citenamefont {Vitanov}\ \emph {et~al.}(2017)\citenamefont {Vitanov}, \citenamefont {Rangelov}, \citenamefont {Shore},\ and\ \citenamefont {Bergmann}}]{VitanovRMP17StimulatedRaman}%
  \BibitemOpen
  \bibfield  {author} {\bibinfo {author} {\bibfnamefont {N.~V.}\ \bibnamefont {Vitanov}}, \bibinfo {author} {\bibfnamefont {A.~A.}\ \bibnamefont {Rangelov}}, \bibinfo {author} {\bibfnamefont {B.~W.}\ \bibnamefont {Shore}},\ \bibnamefont {and}\ \bibinfo {author} {\bibfnamefont {K.}~\bibnamefont {Bergmann}},\ }\bibfield  {title} {\bibinfo {title} {Stimulated {{Raman}} adiabatic passage in physics, chemistry, and beyond},\ }\href {https://doi.org/10.1103/RevModPhys.89.015006} {\bibfield  {journal} {\bibinfo  {journal} {Rev. Mod. Phys.}\ }\textbf {\bibinfo {volume} {89}},\ \bibinfo {pages} {015006} (\bibinfo {year} {2017})}\BibitemShut {NoStop}%
\bibitem [{\citenamefont {Bloch}(2005)}]{BlochNP05UltracoldQuantum}%
  \BibitemOpen
  \bibfield  {author} {\bibinfo {author} {\bibfnamefont {I.}~\bibnamefont {Bloch}},\ }\bibfield  {title} {\bibinfo {title} {Ultracold quantum gases in optical lattices},\ }\href {https://doi.org/10.1038/nphys138} {\bibfield  {journal} {\bibinfo  {journal} {Nature Phys}\ }\textbf {\bibinfo {volume} {1}},\ \bibinfo {pages} {23} (\bibinfo {year} {2005})}\BibitemShut {NoStop}%
\bibitem [{\citenamefont {Kaufman}\ and\ \citenamefont {Ni}(2021)}]{KaufmanNP21QuantumScience}%
  \BibitemOpen
  \bibfield  {author} {\bibinfo {author} {\bibfnamefont {A.~M.}\ \bibnamefont {Kaufman}}\ \bibnamefont {and}\ \bibinfo {author} {\bibfnamefont {K.-K.}\ \bibnamefont {Ni}},\ }\bibfield  {title} {\bibinfo {title} {Quantum science with optical tweezer arrays of ultracold atoms and molecules},\ }\href {https://doi.org/10.1038/s41567-021-01357-2} {\bibfield  {journal} {\bibinfo  {journal} {Nat. Phys.}\ }\textbf {\bibinfo {volume} {17}},\ \bibinfo {pages} {1324} (\bibinfo {year} {2021})}\BibitemShut {NoStop}%
\bibitem [{\citenamefont {Reiserer}\ and\ \citenamefont {Rempe}(2015)}]{ReisererRMP15CavitybasedQuantum}%
  \BibitemOpen
  \bibfield  {author} {\bibinfo {author} {\bibfnamefont {A.}~\bibnamefont {Reiserer}}\ \bibnamefont {and}\ \bibinfo {author} {\bibfnamefont {G.}~\bibnamefont {Rempe}},\ }\bibfield  {title} {\bibinfo {title} {Cavity-based quantum networks with single atoms and optical photons},\ }\href {https://doi.org/10.1103/RevModPhys.87.1379} {\bibfield  {journal} {\bibinfo  {journal} {Rev. Mod. Phys.}\ }\textbf {\bibinfo {volume} {87}},\ \bibinfo {pages} {1379} (\bibinfo {year} {2015})}\BibitemShut {NoStop}%
\bibitem [{\citenamefont {Hammerer}\ \emph {et~al.}(2010)\citenamefont {Hammerer}, \citenamefont {S{\o}rensen},\ and\ \citenamefont {Polzik}}]{HammererRMP10QuantumInterface}%
  \BibitemOpen
  \bibfield  {author} {\bibinfo {author} {\bibfnamefont {K.}~\bibnamefont {Hammerer}}, \bibinfo {author} {\bibfnamefont {A.~S.}\ \bibnamefont {S{\o}rensen}},\ \bibnamefont {and}\ \bibinfo {author} {\bibfnamefont {E.~S.}\ \bibnamefont {Polzik}},\ }\bibfield  {title} {\bibinfo {title} {Quantum interface between light and atomic ensembles},\ }\href {https://doi.org/10.1103/RevModPhys.82.1041} {\bibfield  {journal} {\bibinfo  {journal} {Rev. Mod. Phys.}\ }\textbf {\bibinfo {volume} {82}},\ \bibinfo {pages} {1041} (\bibinfo {year} {2010})}\BibitemShut {NoStop}%
\bibitem [{\citenamefont {Heshami}\ \emph {et~al.}(2016)\citenamefont {Heshami}, \citenamefont {England}, \citenamefont {Humphreys}, \citenamefont {Bustard}, \citenamefont {Acosta}, \citenamefont {Nunn},\ and\ \citenamefont {Sussman}}]{heshami2016quantum}%
  \BibitemOpen
  \bibfield  {author} {\bibinfo {author} {\bibfnamefont {K.}~\bibnamefont {Heshami}}, \bibinfo {author} {\bibfnamefont {D.~G.}\ \bibnamefont {England}}, \bibinfo {author} {\bibfnamefont {P.~C.}\ \bibnamefont {Humphreys}}, \bibinfo {author} {\bibfnamefont {P.~J.}\ \bibnamefont {Bustard}}, \bibinfo {author} {\bibfnamefont {V.~M.}\ \bibnamefont {Acosta}}, \bibinfo {author} {\bibfnamefont {J.}~\bibnamefont {Nunn}},\ \bibnamefont {and}\ \bibinfo {author} {\bibfnamefont {B.~J.}\ \bibnamefont {Sussman}},\ }\bibfield  {title} {\bibinfo {title} {Quantum memories: emerging applications and recent advances},\ }\href {https://doi.org/10.1080/09500340.2016.1148212} {\bibfield  {journal} {\bibinfo  {journal} {J. Mod. Opt.}\ }\textbf {\bibinfo {volume} {63}},\ \bibinfo {pages} {2005} (\bibinfo {year} {2016})}\BibitemShut {NoStop}%
\bibitem [{\citenamefont {Wei}\ \emph {et~al.}(2022)\citenamefont {Wei}, \citenamefont {Jing}, \citenamefont {Zhang}, \citenamefont {Liao}, \citenamefont {Yuan}, \citenamefont {Fan}, \citenamefont {Lyu}, \citenamefont {Zhou}, \citenamefont {Wang}, \citenamefont {Deng} \emph {et~al.}}]{wei2022towards}%
  \BibitemOpen
  \bibfield  {author} {\bibinfo {author} {\bibfnamefont {S.-H.}\ \bibnamefont {Wei}}, \bibinfo {author} {\bibfnamefont {B.}~\bibnamefont {Jing}}, \bibinfo {author} {\bibfnamefont {X.-Y.}\ \bibnamefont {Zhang}}, \bibinfo {author} {\bibfnamefont {J.-Y.}\ \bibnamefont {Liao}}, \bibinfo {author} {\bibfnamefont {C.-Z.}\ \bibnamefont {Yuan}}, \bibinfo {author} {\bibfnamefont {B.-Y.}\ \bibnamefont {Fan}}, \bibinfo {author} {\bibfnamefont {C.}~\bibnamefont {Lyu}}, \bibinfo {author} {\bibfnamefont {D.-L.}\ \bibnamefont {Zhou}}, \bibinfo {author} {\bibfnamefont {Y.}~\bibnamefont {Wang}}, \bibinfo {author} {\bibfnamefont {G.-W.}\ \bibnamefont {Deng}}, \bibnamefont {et~al.},\ }\bibfield  {title} {\bibinfo {title} {{Towards Real-World Quantum Networks: A Review}},\ }\href {https://doi.org/10.1002/lpor.202100219} {\bibfield  {journal} {\bibinfo  {journal} {Laser Photonics Rev.}\ }\textbf {\bibinfo {volume} {16}},\ \bibinfo {pages} {2100219} (\bibinfo {year} {2022})}\BibitemShut {NoStop}%
\bibitem [{\citenamefont {Acharya}\ \emph {et~al.}(2023)\citenamefont {Acharya}, \citenamefont {Brebels}, \citenamefont {Grill}, \citenamefont {Verjauw}, \citenamefont {Ivanov}, \citenamefont {Lozano}, \citenamefont {Wan}, \citenamefont {Van~Damme}, \citenamefont {Vadiraj}, \citenamefont {Mongillo}, \citenamefont {Govoreanu}, \citenamefont {Craninckx}, \citenamefont {Radu}, \citenamefont {De~Greve}, \citenamefont {Gielen}, \citenamefont {Catthoor},\ and\ \citenamefont {Poto{\v c}nik}}]{AcharyaNE23MultiplexedSuperconducting}%
  \BibitemOpen
  \bibfield  {author} {\bibinfo {author} {\bibfnamefont {R.}~\bibnamefont {Acharya}}, \bibinfo {author} {\bibfnamefont {S.}~\bibnamefont {Brebels}}, \bibinfo {author} {\bibfnamefont {A.}~\bibnamefont {Grill}}, \bibnamefont {et~al.},\ }\bibfield  {title} {\bibinfo {title} {Multiplexed superconducting qubit control at millikelvin temperatures with a low-power cryo-{{CMOS}} multiplexer},\ }\href {https://doi.org/10.1038/s41928-023-01033-8} {\bibfield  {journal} {\bibinfo  {journal} {Nat. Electron.}\ }\textbf {\bibinfo {volume} {6}},\ \bibinfo {pages} {900} (\bibinfo {year} {2023})}\BibitemShut {NoStop}%
\bibitem [{\citenamefont {Blais}\ \emph {et~al.}(2004)\citenamefont {Blais}, \citenamefont {Huang}, \citenamefont {Wallraff}, \citenamefont {Girvin},\ and\ \citenamefont {Schoelkopf}}]{BlaisPRA04CavityQuantum}%
  \BibitemOpen
  \bibfield  {author} {\bibinfo {author} {\bibfnamefont {A.}~\bibnamefont {Blais}}, \bibinfo {author} {\bibfnamefont {R.-S.}\ \bibnamefont {Huang}}, \bibinfo {author} {\bibfnamefont {A.}~\bibnamefont {Wallraff}}, \bibinfo {author} {\bibfnamefont {S.~M.}\ \bibnamefont {Girvin}},\ \bibnamefont {and}\ \bibinfo {author} {\bibfnamefont {R.~J.}\ \bibnamefont {Schoelkopf}},\ }\bibfield  {title} {\bibinfo {title} {Cavity quantum electrodynamics for superconducting electrical circuits: {{An}} architecture for quantum computation},\ }\href {https://doi.org/10.1103/PhysRevA.69.062320} {\bibfield  {journal} {\bibinfo  {journal} {Phys. Rev. A}\ }\textbf {\bibinfo {volume} {69}},\ \bibinfo {pages} {062320} (\bibinfo {year} {2004})}\BibitemShut {NoStop}%
\bibitem [{\citenamefont {Chen}\ \emph {et~al.}(2023)\citenamefont {Chen}, \citenamefont {Li}, \citenamefont {Lu}, \citenamefont {Warren}, \citenamefont {Kri{\v z}an}, \citenamefont {Kosen}, \citenamefont {Rommel}, \citenamefont {Ahmed}, \citenamefont {Osman}, \citenamefont {Bizn{\'a}rov{\'a}}, \citenamefont {Fadavi~Roudsari}, \citenamefont {Lienhard}, \citenamefont {Caputo}, \citenamefont {Grigoras}, \citenamefont {Gr{\"o}nberg}, \citenamefont {Govenius}, \citenamefont {Kockum}, \citenamefont {Delsing}, \citenamefont {Bylander},\ and\ \citenamefont {Tancredi}}]{ChennQI23TransmonQubit}%
  \BibitemOpen
  \bibfield  {author} {\bibinfo {author} {\bibfnamefont {L.}~\bibnamefont {Chen}}, \bibinfo {author} {\bibfnamefont {H.-X.}\ \bibnamefont {Li}}, \bibinfo {author} {\bibfnamefont {Y.}~\bibnamefont {Lu}}, \bibnamefont {et~al.},\ }\bibfield  {title} {\bibinfo {title} {Transmon qubit readout fidelity at the threshold for quantum error correction without a quantum-limited amplifier},\ }\href {https://doi.org/10.1038/s41534-023-00689-6} {\bibfield  {journal} {\bibinfo  {journal} {npj Quantum Inf}\ }\textbf {\bibinfo {volume} {9}},\ \bibinfo {pages} {1} (\bibinfo {year} {2023})}\BibitemShut {NoStop}%
\bibitem [{\citenamefont {Devoret}\ and\ \citenamefont {Schoelkopf}(2013)}]{DevoretS13SuperconductingCircuits}%
  \BibitemOpen
  \bibfield  {author} {\bibinfo {author} {\bibfnamefont {M.~H.}\ \bibnamefont {Devoret}}\ \bibnamefont {and}\ \bibinfo {author} {\bibfnamefont {R.~J.}\ \bibnamefont {Schoelkopf}},\ }\bibfield  {title} {\bibinfo {title} {Superconducting {{Circuits}} for {{Quantum Information}}: {{An Outlook}}},\ }\href {https://doi.org/10.1126/science.1231930} {\bibfield  {journal} {\bibinfo  {journal} {Science}\ }\textbf {\bibinfo {volume} {339}},\ \bibinfo {pages} {1169} (\bibinfo {year} {2013})}\BibitemShut {NoStop}%
\bibitem [{\citenamefont {DiVincenzo}(2000)}]{DiVincenzoFP00PhysicalImplementation}%
  \BibitemOpen
  \bibfield  {author} {\bibinfo {author} {\bibfnamefont {D.~P.}\ \bibnamefont {DiVincenzo}},\ }\bibfield  {title} {\bibinfo {title} {The {{Physical Implementation}} of {{Quantum Computation}}},\ }\href {https://doi.org/10.1002/1521-3978(200009)48:9/11<771::AID-PROP771>3.0.CO;2-E} {\bibfield  {journal} {\bibinfo  {journal} {Fortschritte Phys.}\ }\textbf {\bibinfo {volume} {48}},\ \bibinfo {pages} {771} (\bibinfo {year} {2000})}\BibitemShut {NoStop}%
\bibitem [{\citenamefont {{Campagne-Ibarcq}}\ \emph {et~al.}(2020)\citenamefont {{Campagne-Ibarcq}}, \citenamefont {Eickbusch}, \citenamefont {Touzard}, \citenamefont {{Zalys-Geller}}, \citenamefont {Frattini}, \citenamefont {Sivak}, \citenamefont {Reinhold}, \citenamefont {Puri}, \citenamefont {Shankar}, \citenamefont {Schoelkopf}, \citenamefont {Frunzio}, \citenamefont {Mirrahimi},\ and\ \citenamefont {Devoret}}]{Campagne-IbarcqN20QuantumError}%
  \BibitemOpen
  \bibfield  {author} {\bibinfo {author} {\bibfnamefont {P.}~\bibnamefont {{Campagne-Ibarcq}}}, \bibinfo {author} {\bibfnamefont {A.}~\bibnamefont {Eickbusch}}, \bibinfo {author} {\bibfnamefont {S.}~\bibnamefont {Touzard}}, \bibinfo {author} {\bibfnamefont {E.}~\bibnamefont {{Zalys-Geller}}}, \bibinfo {author} {\bibfnamefont {N.~E.}\ \bibnamefont {Frattini}}, \bibinfo {author} {\bibfnamefont {V.~V.}\ \bibnamefont {Sivak}}, \bibinfo {author} {\bibfnamefont {P.}~\bibnamefont {Reinhold}}, \bibinfo {author} {\bibfnamefont {S.}~\bibnamefont {Puri}}, \bibinfo {author} {\bibfnamefont {S.}~\bibnamefont {Shankar}}, \bibinfo {author} {\bibfnamefont {R.~J.}\ \bibnamefont {Schoelkopf}}, \bibinfo {author} {\bibfnamefont {L.}~\bibnamefont {Frunzio}}, \bibinfo {author} {\bibfnamefont {M.}~\bibnamefont {Mirrahimi}},\ \bibnamefont {and}\ \bibinfo {author} {\bibfnamefont {M.~H.}\ \bibnamefont {Devoret}},\ }\bibfield  {title} {\bibinfo {title} {Quantum error correction of a qubit encoded in grid states of an oscillator},\
  }\href {https://doi.org/10.1038/s41586-020-2603-3} {\bibfield  {journal} {\bibinfo  {journal} {Nature}\ }\textbf {\bibinfo {volume} {584}},\ \bibinfo {pages} {368} (\bibinfo {year} {2020})}\BibitemShut {NoStop}%
\bibitem [{\citenamefont {Diringer}\ \emph {et~al.}(2024)\citenamefont {Diringer}, \citenamefont {Blumenthal}, \citenamefont {Grinberg}, \citenamefont {Jiang},\ and\ \citenamefont {Hacohen-Gourgy}}]{diringer2024conditional}%
  \BibitemOpen
  \bibfield  {author} {\bibinfo {author} {\bibfnamefont {A.~A.}\ \bibnamefont {Diringer}}, \bibinfo {author} {\bibfnamefont {E.}~\bibnamefont {Blumenthal}}, \bibinfo {author} {\bibfnamefont {A.}~\bibnamefont {Grinberg}}, \bibinfo {author} {\bibfnamefont {L.}~\bibnamefont {Jiang}},\ \bibnamefont {and}\ \bibinfo {author} {\bibfnamefont {S.}~\bibnamefont {Hacohen-Gourgy}},\ }\bibfield  {title} {\bibinfo {title} {Conditional-\textsc{NOT} displacement: Fast multioscillator control with a single qubit},\ }\href {https://doi.org/10.1103/PhysRevX.14.011055} {\bibfield  {journal} {\bibinfo  {journal} {Physical Review X}\ }\textbf {\bibinfo {volume} {14}},\ \bibinfo {pages} {011055} (\bibinfo {year} {2024})}\BibitemShut {NoStop}%
\bibitem [{\citenamefont {Moretti}\ \emph {et~al.}(2024)\citenamefont {Moretti}, \citenamefont {Corti}, \citenamefont {Labranca}, \citenamefont {Ahrens}, \citenamefont {Avallone}, \citenamefont {Babusci}, \citenamefont {Banchi}, \citenamefont {Barone}, \citenamefont {Beretta}, \citenamefont {Borghesi}, \citenamefont {Buonomo}, \citenamefont {Calore}, \citenamefont {Carapella}, \citenamefont {Chiarello}, \citenamefont {Cian}, \citenamefont {Cidronali}, \citenamefont {Costa}, \citenamefont {Cuccoli}, \citenamefont {D'Elia}, \citenamefont {Gioacchino}, \citenamefont {Pascoli}, \citenamefont {Falferi}, \citenamefont {Fanciulli}, \citenamefont {Faverzani}, \citenamefont {Felici}, \citenamefont {Ferri}, \citenamefont {Filatrella}, \citenamefont {Foggetta}, \citenamefont {Gatti}, \citenamefont {Giachero}, \citenamefont {Giazotto}, \citenamefont {Giubertoni}, \citenamefont {Granata}, \citenamefont {Guarcello}, \citenamefont {Lamanna}, \citenamefont {Ligi}, \citenamefont {Maccarrone}, \citenamefont {Macucci},
  \citenamefont {Manara}, \citenamefont {Mantegazzini}, \citenamefont {Marconcini}, \citenamefont {Margesin}, \citenamefont {Mattioli}, \citenamefont {Miola}, \citenamefont {Nucciotti}, \citenamefont {Origo}, \citenamefont {Pagano}, \citenamefont {Paolucci}, \citenamefont {Piersanti}, \citenamefont {Rettaroli}, \citenamefont {Sanguinetti}, \citenamefont {Schifano}, \citenamefont {Spagnolo}, \citenamefont {Tocci}, \citenamefont {Toncelli}, \citenamefont {Torrioli},\ and\ \citenamefont {Vinante}}]{MorettiITAS24DesignSimulation}%
  \BibitemOpen
  \bibfield  {author} {\bibinfo {author} {\bibfnamefont {R.}~\bibnamefont {Moretti}}, \bibinfo {author} {\bibfnamefont {H.~A.}\ \bibnamefont {Corti}}, \bibinfo {author} {\bibfnamefont {D.}~\bibnamefont {Labranca}}, \bibnamefont {et~al.},\ }\bibfield  {title} {\bibinfo {title} {Design and {{Simulation}} of a {{Transmon Qubit Chip}} for {{Axion Detection}}},\ }\href {https://doi.org/10.1109/TASC.2024.3350582} {\bibfield  {journal} {\bibinfo  {journal} {IEEE Trans. Appl. Supercond.}\ }\textbf {\bibinfo {volume} {34}},\ \bibinfo {pages} {1} (\bibinfo {year} {2024})}\BibitemShut {NoStop}%
\bibitem [{\citenamefont {Lamoreaux}\ \emph {et~al.}(2013)\citenamefont {Lamoreaux}, \citenamefont {{van Bibber}}, \citenamefont {Lehnert},\ and\ \citenamefont {Carosi}}]{LamoreauxPRD13AnalysisSinglephoton}%
  \BibitemOpen
  \bibfield  {author} {\bibinfo {author} {\bibfnamefont {S.~K.}\ \bibnamefont {Lamoreaux}}, \bibinfo {author} {\bibfnamefont {K.~A.}\ \bibnamefont {{van Bibber}}}, \bibinfo {author} {\bibfnamefont {K.~W.}\ \bibnamefont {Lehnert}},\ \bibnamefont {and}\ \bibinfo {author} {\bibfnamefont {G.}~\bibnamefont {Carosi}},\ }\bibfield  {title} {\bibinfo {title} {Analysis of single-photon and linear amplifier detectors for microwave cavity dark matter axion searches},\ }\href {https://doi.org/10.1103/PhysRevD.88.035020} {\bibfield  {journal} {\bibinfo  {journal} {Phys. Rev. D}\ }\textbf {\bibinfo {volume} {88}},\ \bibinfo {pages} {035020} (\bibinfo {year} {2013})}\BibitemShut {NoStop}%
\bibitem [{Note2()}]{Note2}%
  \BibitemOpen
  \bibinfo {note} {If measuring signals below the classical noise floor, then we expect the regime $\eta \gg 4\sigma _C^2$ to be also relevant, see the discussion surrounding Eq.~\ref {eq:clnoise with loss} for the one-dimensional case.}\BibitemShut {Stop}%
\bibitem [{\citenamefont {{LIGO O4 Detector Collaboration}}\ \emph {et~al.}(2023)\citenamefont {{LIGO O4 Detector Collaboration}}, \citenamefont {Ganapathy}, \citenamefont {Jia}, \citenamefont {Nakano}, \citenamefont {Xu}, \citenamefont {Aritomi}, \citenamefont {Cullen}, \citenamefont {Kijbunchoo}, \citenamefont {Dwyer}, \citenamefont {Mullavey}, \citenamefont {McCuller}, \citenamefont {Abbott}, \citenamefont {Abouelfettouh}, \citenamefont {Adhikari}, \citenamefont {Ananyeva}, \citenamefont {Appert}, \citenamefont {Arai}, \citenamefont {Aston}, \citenamefont {Ball}, \citenamefont {Ballmer}, \citenamefont {Barker}, \citenamefont {Barsotti}, \citenamefont {Berger}, \citenamefont {Betzwieser}, \citenamefont {Bhattacharjee}, \citenamefont {Billingsley}, \citenamefont {Biscans}, \citenamefont {Bode}, \citenamefont {Bonilla}, \citenamefont {Bossilkov}, \citenamefont {Branch}, \citenamefont {Brooks}, \citenamefont {Brown}, \citenamefont {Bryant}, \citenamefont {Cahillane}, \citenamefont {Cao}, \citenamefont
  {Capote}, \citenamefont {Clara}, \citenamefont {Collins}, \citenamefont {Compton}, \citenamefont {Cottingham}, \citenamefont {Coyne}, \citenamefont {Crouch}, \citenamefont {Csizmazia}, \citenamefont {Dartez}, \citenamefont {Demos}, \citenamefont {Dohmen}, \citenamefont {Driggers}, \citenamefont {Effler}, \citenamefont {Ejlli}, \citenamefont {Etzel}, \citenamefont {Evans}, \citenamefont {Feicht}, \citenamefont {Frey}, \citenamefont {Frischhertz}, \citenamefont {Fritschel}, \citenamefont {Frolov}, \citenamefont {Fulda}, \citenamefont {Fyffe}, \citenamefont {Gateley}, \citenamefont {Giaime}, \citenamefont {Giardina}, \citenamefont {Glanzer}, \citenamefont {Goetz}, \citenamefont {Goetz}, \citenamefont {{Goodwin-Jones}}, \citenamefont {Gras}, \citenamefont {Gray}, \citenamefont {Griffith}, \citenamefont {Grote}, \citenamefont {Guidry}, \citenamefont {Hall}, \citenamefont {Hanks}, \citenamefont {Hanson}, \citenamefont {Heintze}, \citenamefont {{Helmling-Cornell}}, \citenamefont {Holland}, \citenamefont {Hoyland},
  \citenamefont {Huang}, \citenamefont {Inoue}, \citenamefont {James}, \citenamefont {Jennings}, \citenamefont {Karat}, \citenamefont {Karki}, \citenamefont {Kasprzack}, \citenamefont {Kawabe}, \citenamefont {King}, \citenamefont {Kissel}, \citenamefont {Komori}, \citenamefont {Kontos}, \citenamefont {Kumar}, \citenamefont {Kuns}, \citenamefont {Landry}, \citenamefont {Lantz}, \citenamefont {Laxen}, \citenamefont {Lee}, \citenamefont {Lesovsky}, \citenamefont {Llamas}, \citenamefont {Lormand}, \citenamefont {Loughlin}, \citenamefont {Macas}, \citenamefont {MacInnis}, \citenamefont {Makarem}, \citenamefont {Mannix}, \citenamefont {Mansell}, \citenamefont {Martin}, \citenamefont {Mason}, \citenamefont {Matichard}, \citenamefont {Mavalvala}, \citenamefont {Maxwell}, \citenamefont {McCarrol}, \citenamefont {McCarthy}, \citenamefont {McClelland}, \citenamefont {McCormick}, \citenamefont {McRae}, \citenamefont {Mera}, \citenamefont {Merilh}, \citenamefont {Meylahn}, \citenamefont {Mittleman}, \citenamefont
  {Moraru}, \citenamefont {Moreno}, \citenamefont {Nelson}, \citenamefont {Neunzert}, \citenamefont {Notte}, \citenamefont {Oberling}, \citenamefont {O'Hanlon}, \citenamefont {Osthelder}, \citenamefont {Ottaway}, \citenamefont {Overmier}, \citenamefont {Parker}, \citenamefont {Pele}, \citenamefont {Pham}, \citenamefont {Pirello}, \citenamefont {Quetschke}, \citenamefont {Ramirez}, \citenamefont {Reyes}, \citenamefont {Richardson}, \citenamefont {Robinson}, \citenamefont {Rollins}, \citenamefont {Romel}, \citenamefont {Romie}, \citenamefont {Ross}, \citenamefont {Ryan}, \citenamefont {Sadecki}, \citenamefont {Sanchez}, \citenamefont {Sanchez}, \citenamefont {Sanchez}, \citenamefont {Savage}, \citenamefont {Schaetzl}, \citenamefont {Schiworski}, \citenamefont {Schnabel}, \citenamefont {Schofield}, \citenamefont {Schwartz}, \citenamefont {Sellers}, \citenamefont {Shaffer}, \citenamefont {Short}, \citenamefont {Sigg}, \citenamefont {Slagmolen}, \citenamefont {Soike}, \citenamefont {Soni}, \citenamefont
  {Srivastava}, \citenamefont {Sun}, \citenamefont {Tanner}, \citenamefont {Thomas}, \citenamefont {Thomas}, \citenamefont {Thorne}, \citenamefont {Torrie}, \citenamefont {Traylor}, \citenamefont {Ubhi}, \citenamefont {Vajente}, \citenamefont {Vanosky}, \citenamefont {Vecchio}, \citenamefont {Veitch}, \citenamefont {Vibhute}, \citenamefont {{von Reis}}, \citenamefont {Warner}, \citenamefont {Weaver}, \citenamefont {Weiss}, \citenamefont {Whittle}, \citenamefont {Willke}, \citenamefont {Wipf}, \citenamefont {Yamamoto}, \citenamefont {Zhang},\ and\ \citenamefont {Zucker}}]{LIGOO4DetectorCollaborationPRX23BroadbandQuantum}%
  \BibitemOpen
  \bibfield  {author} {\bibinfo {author} {\bibnamefont {{LIGO O4 Detector Collaboration}}}, \bibinfo {author} {\bibfnamefont {D.}~\bibnamefont {Ganapathy}}, \bibinfo {author} {\bibfnamefont {W.}~\bibnamefont {Jia}}, \bibnamefont {et~al.},\ }\bibfield  {title} {\bibinfo {title} {Broadband {{Quantum Enhancement}} of the {{LIGO Detectors}} with {{Frequency-Dependent Squeezing}}},\ }\href {https://doi.org/10.1103/PhysRevX.13.041021} {\bibfield  {journal} {\bibinfo  {journal} {Phys. Rev. X}\ }\textbf {\bibinfo {volume} {13}},\ \bibinfo {pages} {041021} (\bibinfo {year} {2023})}\BibitemShut {NoStop}%
\bibitem [{Note3()}]{Note3}%
  \BibitemOpen
  \bibinfo {note} {This argument assumes a uniform source density, which is valid for the current generation of gravitational-wave detectors. Future detectors are proposed to have much further reach~ \cite {ReitzeBAAS19CosmicExplorer, MaggioreJCAP20ScienceCase} such that this assumption is no longer valid and a more detailed astrophysical model is required.}\BibitemShut {Stop}%
\bibitem [{\citenamefont {Chatziioannou}\ \emph {et~al.}(2017)\citenamefont {Chatziioannou}, \citenamefont {Clark}, \citenamefont {Bauswein}, \citenamefont {Millhouse}, \citenamefont {Littenberg},\ and\ \citenamefont {Cornish}}]{ChatziioannouPRD17InferringPostmerger}%
  \BibitemOpen
  \bibfield  {author} {\bibinfo {author} {\bibfnamefont {K.}~\bibnamefont {Chatziioannou}}, \bibinfo {author} {\bibfnamefont {J.~A.}\ \bibnamefont {Clark}}, \bibinfo {author} {\bibfnamefont {A.}~\bibnamefont {Bauswein}}, \bibinfo {author} {\bibfnamefont {M.}~\bibnamefont {Millhouse}}, \bibinfo {author} {\bibfnamefont {T.~B.}\ \bibnamefont {Littenberg}},\ \bibnamefont {and}\ \bibinfo {author} {\bibfnamefont {N.}~\bibnamefont {Cornish}},\ }\bibfield  {title} {\bibinfo {title} {Inferring the post-merger gravitational wave emission from binary neutron star coalescences},\ }\href {https://doi.org/10.1103/PhysRevD.96.124035} {\bibfield  {journal} {\bibinfo  {journal} {Phys. Rev. D}\ }\textbf {\bibinfo {volume} {96}},\ \bibinfo {pages} {124035} (\bibinfo {year} {2017})}\BibitemShut {NoStop}%
\bibitem [{\citenamefont {Criswell}\ \emph {et~al.}(2023)\citenamefont {Criswell}, \citenamefont {Miller}, \citenamefont {Woldemariam}, \citenamefont {Soultanis}, \citenamefont {Bauswein}, \citenamefont {Chatziioannou}, \citenamefont {Coughlin}, \citenamefont {Jones},\ and\ \citenamefont {Mandic}}]{CriswellPRD23HierarchicalBayesian}%
  \BibitemOpen
  \bibfield  {author} {\bibinfo {author} {\bibfnamefont {A.~W.}\ \bibnamefont {Criswell}}, \bibinfo {author} {\bibfnamefont {J.}~\bibnamefont {Miller}}, \bibinfo {author} {\bibfnamefont {N.}~\bibnamefont {Woldemariam}}, \bibinfo {author} {\bibfnamefont {T.}~\bibnamefont {Soultanis}}, \bibinfo {author} {\bibfnamefont {A.}~\bibnamefont {Bauswein}}, \bibinfo {author} {\bibfnamefont {K.}~\bibnamefont {Chatziioannou}}, \bibinfo {author} {\bibfnamefont {M.~W.}\ \bibnamefont {Coughlin}}, \bibinfo {author} {\bibfnamefont {G.}~\bibnamefont {Jones}},\ \bibnamefont {and}\ \bibinfo {author} {\bibfnamefont {V.}~\bibnamefont {Mandic}},\ }\bibfield  {title} {\bibinfo {title} {Hierarchical {{Bayesian}} method for constraining the neutron star equation of state with an ensemble of binary neutron star postmerger remnants},\ }\href {https://doi.org/10.1103/PhysRevD.107.043021} {\bibfield  {journal} {\bibinfo  {journal} {Phys. Rev. D}\ }\textbf {\bibinfo {volume} {107}},\ \bibinfo {pages} {043021} (\bibinfo {year}
  {2023})}\BibitemShut {NoStop}%
\bibitem [{\citenamefont {Sasli}\ \emph {et~al.}(2024)\citenamefont {Sasli}, \citenamefont {Karnesis},\ and\ \citenamefont {Stergioulas}}]{SasliPRD24ExploringPotential}%
  \BibitemOpen
  \bibfield  {author} {\bibinfo {author} {\bibfnamefont {A.}~\bibnamefont {Sasli}}, \bibinfo {author} {\bibfnamefont {N.}~\bibnamefont {Karnesis}},\ \bibnamefont {and}\ \bibinfo {author} {\bibfnamefont {N.}~\bibnamefont {Stergioulas}},\ }\bibfield  {title} {\bibinfo {title} {Exploring the potential for detecting rotational instabilities in binary neutron star merger remnants with gravitational wave detectors},\ }\href {https://doi.org/10.1103/PhysRevD.109.043045} {\bibfield  {journal} {\bibinfo  {journal} {Phys. Rev. D}\ }\textbf {\bibinfo {volume} {109}},\ \bibinfo {pages} {043045} (\bibinfo {year} {2024})}\BibitemShut {NoStop}%
\bibitem [{Note4()}]{Note4}%
  \BibitemOpen
  \bibinfo {note} {In the deterministic event-stacking case~\cite {ChatziioannouPRD17InferringPostmerger, CriswellPRD23HierarchicalBayesian, SasliPRD24ExploringPotential}, there are threshold effects where the classical Cramer-Rao bound cannot be saturated~\cite {RifeITIT74SingleTone, SteinhardtI8IICASSP85ThresholdsFrequency}. This behavior arises from the nonlinear process of frequency estimation at a low signal-to-noise ratio, and can be expressed as the discrepancy between the classical Cramer-Rao and Barankin bounds~\cite {KnockaertITSP97BarankinBound, ChaumetteITSP08NewBarankin}. Since finding the frequency of excess power is a stochastic estimation problem, studying the quantum analogues of the Barankin bound~\cite {GessnerPRL23HierarchiesFrequentist} may be necessary to understanding the behaviour here at low signal-to-noise ratios.}\BibitemShut {Stop}%
\bibitem [{\citenamefont {Cameron}\ \emph {et~al.}(1993)\citenamefont {Cameron}, \citenamefont {Cantatore}, \citenamefont {Melissinos}, \citenamefont {Ruoso}, \citenamefont {Semertzidis}, \citenamefont {Halama}, \citenamefont {Lazarus}, \citenamefont {Prodell}, \citenamefont {Nezrick}, \citenamefont {Rizzo} \emph {et~al.}}]{cameron1993search}%
  \BibitemOpen
  \bibfield  {author} {\bibinfo {author} {\bibfnamefont {R.}~\bibnamefont {Cameron}}, \bibinfo {author} {\bibfnamefont {G.}~\bibnamefont {Cantatore}}, \bibinfo {author} {\bibfnamefont {A.}~\bibnamefont {Melissinos}}, \bibinfo {author} {\bibfnamefont {G.}~\bibnamefont {Ruoso}}, \bibinfo {author} {\bibfnamefont {Y.}~\bibnamefont {Semertzidis}}, \bibinfo {author} {\bibfnamefont {H.}~\bibnamefont {Halama}}, \bibinfo {author} {\bibfnamefont {D.}~\bibnamefont {Lazarus}}, \bibinfo {author} {\bibfnamefont {A.}~\bibnamefont {Prodell}}, \bibinfo {author} {\bibfnamefont {F.}~\bibnamefont {Nezrick}}, \bibinfo {author} {\bibfnamefont {C.}~\bibnamefont {Rizzo}}, \bibnamefont {et~al.},\ }\bibfield  {title} {\bibinfo {title} {Search for nearly massless, weakly coupled particles by optical techniques},\ }\href {https://doi.org/10.1103/PhysRevD.47.3707} {\bibfield  {journal} {\bibinfo  {journal} {Physical Review D}\ }\textbf {\bibinfo {volume} {47}},\ \bibinfo {pages} {3707} (\bibinfo {year} {1993})}\BibitemShut {NoStop}%
\bibitem [{\citenamefont {Du}\ \emph {et~al.}(2018)\citenamefont {Du}, \citenamefont {Force}, \citenamefont {Khatiwada}, \citenamefont {Lentz}, \citenamefont {Ottens}, \citenamefont {Rosenberg}, \citenamefont {Rybka}, \citenamefont {Carosi}, \citenamefont {Woollett}, \citenamefont {Bowring} \emph {et~al.}}]{du2018search}%
  \BibitemOpen
  \bibfield  {author} {\bibinfo {author} {\bibfnamefont {N.}~\bibnamefont {Du}}, \bibinfo {author} {\bibfnamefont {N.}~\bibnamefont {Force}}, \bibinfo {author} {\bibfnamefont {R.}~\bibnamefont {Khatiwada}}, \bibinfo {author} {\bibfnamefont {E.}~\bibnamefont {Lentz}}, \bibinfo {author} {\bibfnamefont {R.}~\bibnamefont {Ottens}}, \bibinfo {author} {\bibfnamefont {L.}~\bibnamefont {Rosenberg}}, \bibinfo {author} {\bibfnamefont {G.}~\bibnamefont {Rybka}}, \bibinfo {author} {\bibfnamefont {G.}~\bibnamefont {Carosi}}, \bibinfo {author} {\bibfnamefont {N.}~\bibnamefont {Woollett}}, \bibinfo {author} {\bibfnamefont {D.}~\bibnamefont {Bowring}}, \bibnamefont {et~al.},\ }\bibfield  {title} {\bibinfo {title} {Search for invisible axion dark matter with the axion dark matter experiment},\ }\href {https://doi.org/10.1103/PhysRevLett.120.151301} {\bibfield  {journal} {\bibinfo  {journal} {Phys. Rev. Lett.}\ }\textbf {\bibinfo {volume} {120}},\ \bibinfo {pages} {151301} (\bibinfo {year} {2018})}\BibitemShut {NoStop}%
\bibitem [{\citenamefont {Dixit}\ \emph {et~al.}(2021)\citenamefont {Dixit}, \citenamefont {Chakram}, \citenamefont {He}, \citenamefont {Agrawal}, \citenamefont {Naik}, \citenamefont {Schuster},\ and\ \citenamefont {Chou}}]{DixitPRL21SearchingDark}%
  \BibitemOpen
  \bibfield  {author} {\bibinfo {author} {\bibfnamefont {A.~V.}\ \bibnamefont {Dixit}}, \bibinfo {author} {\bibfnamefont {S.}~\bibnamefont {Chakram}}, \bibinfo {author} {\bibfnamefont {K.}~\bibnamefont {He}}, \bibinfo {author} {\bibfnamefont {A.}~\bibnamefont {Agrawal}}, \bibinfo {author} {\bibfnamefont {R.~K.}\ \bibnamefont {Naik}}, \bibinfo {author} {\bibfnamefont {D.~I.}\ \bibnamefont {Schuster}},\ \bibnamefont {and}\ \bibinfo {author} {\bibfnamefont {A.}~\bibnamefont {Chou}},\ }\bibfield  {title} {\bibinfo {title} {Searching for {{Dark Matter}} with a {{Superconducting Qubit}}},\ }\href {https://doi.org/10.1103/PhysRevLett.126.141302} {\bibfield  {journal} {\bibinfo  {journal} {Phys. Rev. Lett.}\ }\textbf {\bibinfo {volume} {126}},\ \bibinfo {pages} {141302} (\bibinfo {year} {2021})}\BibitemShut {NoStop}%
\bibitem [{\citenamefont {Agrawal}\ \emph {et~al.}(2024)\citenamefont {Agrawal}, \citenamefont {Dixit}, \citenamefont {Roy}, \citenamefont {Chakram}, \citenamefont {He}, \citenamefont {Naik}, \citenamefont {Schuster},\ and\ \citenamefont {Chou}}]{AgrawalPRL24StimulatedEmission}%
  \BibitemOpen
  \bibfield  {author} {\bibinfo {author} {\bibfnamefont {A.}~\bibnamefont {Agrawal}}, \bibinfo {author} {\bibfnamefont {A.~V.}\ \bibnamefont {Dixit}}, \bibinfo {author} {\bibfnamefont {T.}~\bibnamefont {Roy}}, \bibinfo {author} {\bibfnamefont {S.}~\bibnamefont {Chakram}}, \bibinfo {author} {\bibfnamefont {K.}~\bibnamefont {He}}, \bibinfo {author} {\bibfnamefont {R.~K.}\ \bibnamefont {Naik}}, \bibinfo {author} {\bibfnamefont {D.~I.}\ \bibnamefont {Schuster}},\ \bibnamefont {and}\ \bibinfo {author} {\bibfnamefont {A.}~\bibnamefont {Chou}},\ }\bibfield  {title} {\bibinfo {title} {Stimulated {{Emission}} of {{Signal Photons}} from {{Dark Matter Waves}}},\ }\href {https://doi.org/10.1103/PhysRevLett.132.140801} {\bibfield  {journal} {\bibinfo  {journal} {Phys. Rev. Lett.}\ }\textbf {\bibinfo {volume} {132}},\ \bibinfo {pages} {140801} (\bibinfo {year} {2024})}\BibitemShut {NoStop}%
\bibitem [{\citenamefont {Shi}\ and\ \citenamefont {Zhuang}(2023)}]{ShinQI23UltimatePrecision}%
  \BibitemOpen
  \bibfield  {author} {\bibinfo {author} {\bibfnamefont {H.}~\bibnamefont {Shi}}\ \bibnamefont {and}\ \bibinfo {author} {\bibfnamefont {Q.}~\bibnamefont {Zhuang}},\ }\bibfield  {title} {\bibinfo {title} {Ultimate precision limit of noise sensing and dark matter search},\ }\href {https://doi.org/10.1038/s41534-023-00693-w} {\bibfield  {journal} {\bibinfo  {journal} {npj Quantum Inf}\ }\textbf {\bibinfo {volume} {9}},\ \bibinfo {pages} {1} (\bibinfo {year} {2023})}\BibitemShut {NoStop}%
\bibitem [{\citenamefont {{H. Zheng, M. Silveri, R. Brierley, S. Girvin, and K. Lehnert, Accelerating dark-matter axion searches with quantum measurement technology (2016)}}()}]{zheng2016accelerating}%
  \BibitemOpen
  \bibfield  {author} {\bibinfo {author} {\bibnamefont {{H. Zheng, M. Silveri, R. Brierley, S. Girvin, and K. Lehnert, Accelerating dark-matter axion searches with quantum measurement technology (2016)}}},\ }\href {https://doi.org/10.48550/arXiv.1607.02529} {\bibinfo {title} {arxiv:1607.02529 [hep-ph]}}\BibitemShut {NoStop}%
\bibitem [{\citenamefont {Backes}\ \emph {et~al.}(2021)\citenamefont {Backes}, \citenamefont {Palken}, \citenamefont {Kenany}, \citenamefont {Brubaker}, \citenamefont {Cahn}, \citenamefont {Droster}, \citenamefont {Hilton}, \citenamefont {Ghosh}, \citenamefont {Jackson}, \citenamefont {Lamoreaux}, \citenamefont {Leder}, \citenamefont {Lehnert}, \citenamefont {Lewis}, \citenamefont {Malnou}, \citenamefont {Maruyama}, \citenamefont {Rapidis}, \citenamefont {Simanovskaia}, \citenamefont {Singh}, \citenamefont {Speller}, \citenamefont {Urdinaran}, \citenamefont {Vale}, \citenamefont {van Assendelft}, \citenamefont {van Bibber},\ and\ \citenamefont {Wang}}]{BackesN21QuantumEnhanced}%
  \BibitemOpen
  \bibfield  {author} {\bibinfo {author} {\bibfnamefont {K.~M.}\ \bibnamefont {Backes}}, \bibinfo {author} {\bibfnamefont {D.~A.}\ \bibnamefont {Palken}}, \bibinfo {author} {\bibfnamefont {S.~A.}\ \bibnamefont {Kenany}}, \bibnamefont {et~al.},\ }\bibfield  {title} {\bibinfo {title} {A quantum enhanced search for dark matter axions},\ }\href {https://doi.org/10.1038/s41586-021-03226-7} {\bibfield  {journal} {\bibinfo  {journal} {Nature}\ }\textbf {\bibinfo {volume} {590}},\ \bibinfo {pages} {238} (\bibinfo {year} {2021})}\BibitemShut {NoStop}%
\bibitem [{\citenamefont {Gill}\ and\ \citenamefont {Levit}(1995)}]{gill1995applications}%
  \BibitemOpen
  \bibfield  {author} {\bibinfo {author} {\bibfnamefont {R.~D.}\ \bibnamefont {Gill}}\ \bibnamefont {and}\ \bibinfo {author} {\bibfnamefont {B.~Y.}\ \bibnamefont {Levit}},\ }\bibfield  {title} {\bibinfo {title} {{Applications of the van Trees inequality: a Bayesian Cram{\'e}r-Rao bound}},\ }\href {https://doi.org/10.2307/3318681} {\bibfield  {journal} {\bibinfo  {journal} {Bernoulli}\ ,\ \bibinfo {pages} {59}} (\bibinfo {year} {1995})}\BibitemShut {NoStop}%
\bibitem [{\citenamefont {Van~Trees}(2002)}]{van2002detection}%
  \BibitemOpen
  \bibfield  {author} {\bibinfo {author} {\bibfnamefont {H.~L.}\ \bibnamefont {Van~Trees}},\ }\href {https://doi.org/10.1002/0471221104} {\emph {\bibinfo {title} {{Detection, Estimation, and Modulation Theory: Part IV}}}}\ (\bibinfo  {publisher} {John Wiley \& Sons, Incorporated},\ \bibinfo {year} {2002})\BibitemShut {NoStop}%
\bibitem [{\citenamefont {Bednorz}\ and\ \citenamefont {Belzig}(2011)}]{bednorz2011fourth}%
  \BibitemOpen
  \bibfield  {author} {\bibinfo {author} {\bibfnamefont {A.}~\bibnamefont {Bednorz}}\ \bibnamefont {and}\ \bibinfo {author} {\bibfnamefont {W.}~\bibnamefont {Belzig}},\ }\bibfield  {title} {\bibinfo {title} {Fourth moments reveal the negativity of the wigner function},\ }\href {https://doi.org/10.1103/PhysRevA.83.052113} {\bibfield  {journal} {\bibinfo  {journal} {Phys. Rev. A}\ }\textbf {\bibinfo {volume} {83}},\ \bibinfo {pages} {052113} (\bibinfo {year} {2011})}\BibitemShut {NoStop}%
\bibitem [{\citenamefont {Howl}\ \emph {et~al.}(2021)\citenamefont {Howl}, \citenamefont {Vedral}, \citenamefont {Naik}, \citenamefont {Christodoulou}, \citenamefont {Rovelli},\ and\ \citenamefont {Iyer}}]{howl2021non}%
  \BibitemOpen
  \bibfield  {author} {\bibinfo {author} {\bibfnamefont {R.}~\bibnamefont {Howl}}, \bibinfo {author} {\bibfnamefont {V.}~\bibnamefont {Vedral}}, \bibinfo {author} {\bibfnamefont {D.}~\bibnamefont {Naik}}, \bibinfo {author} {\bibfnamefont {M.}~\bibnamefont {Christodoulou}}, \bibinfo {author} {\bibfnamefont {C.}~\bibnamefont {Rovelli}},\ \bibnamefont {and}\ \bibinfo {author} {\bibfnamefont {A.}~\bibnamefont {Iyer}},\ }\bibfield  {title} {\bibinfo {title} {Non-gaussianity as a signature of a quantum theory of gravity},\ }\href {https://doi.org/10.1103/PRXQuantum.2.010325} {\bibfield  {journal} {\bibinfo  {journal} {PRX Quantum}\ }\textbf {\bibinfo {volume} {2}},\ \bibinfo {pages} {010325} (\bibinfo {year} {2021})}\BibitemShut {NoStop}%
\bibitem [{\citenamefont {Haine}(2021)}]{haine2021searching}%
  \BibitemOpen
  \bibfield  {author} {\bibinfo {author} {\bibfnamefont {S.~A.}\ \bibnamefont {Haine}},\ }\bibfield  {title} {\bibinfo {title} {Searching for signatures of quantum gravity in quantum gases},\ }\href {https://doi.org/10.1088/1367-2630/abd97d} {\bibfield  {journal} {\bibinfo  {journal} {New J. Phys.}\ }\textbf {\bibinfo {volume} {23}},\ \bibinfo {pages} {033020} (\bibinfo {year} {2021})}\BibitemShut {NoStop}%
\bibitem [{\citenamefont {Mehdi}\ \emph {et~al.}(2023)\citenamefont {Mehdi}, \citenamefont {Hope},\ and\ \citenamefont {Haine}}]{mehdi2023signatures}%
  \BibitemOpen
  \bibfield  {author} {\bibinfo {author} {\bibfnamefont {Z.}~\bibnamefont {Mehdi}}, \bibinfo {author} {\bibfnamefont {J.~J.}\ \bibnamefont {Hope}},\ \bibnamefont {and}\ \bibinfo {author} {\bibfnamefont {S.~A.}\ \bibnamefont {Haine}},\ }\bibfield  {title} {\bibinfo {title} {Signatures of quantum gravity in the gravitational self-interaction of photons},\ }\href {https://doi.org/10.1103/PhysRevLett.130.240203} {\bibfield  {journal} {\bibinfo  {journal} {Phys. Rev. Lett.}\ }\textbf {\bibinfo {volume} {130}},\ \bibinfo {pages} {240203} (\bibinfo {year} {2023})}\BibitemShut {NoStop}%
\bibitem [{\citenamefont {Franzen}()}]{ComponentLibrary}%
  \BibitemOpen
  \bibfield  {author} {\bibinfo {author} {\bibfnamefont {A.}~\bibnamefont {Franzen}},\ }\href@noop {} {\bibinfo {title} {Componentlibrary}},\ \bibinfo {note} {\emph{ComponentLibrary}. 2009. \url{http://www.gwoptics.org/ComponentLibrary/}}\BibitemShut {NoStop}%
\bibitem [{\citenamefont {Williamson}(1936)}]{williamson1936algebraic}%
  \BibitemOpen
  \bibfield  {author} {\bibinfo {author} {\bibfnamefont {J.}~\bibnamefont {Williamson}},\ }\bibfield  {title} {\bibinfo {title} {On the algebraic problem concerning the normal forms of linear dynamical systems},\ }\href {https://doi.org/10.2307/2371062} {\bibfield  {journal} {\bibinfo  {journal} {Am. J. Math.}\ }\textbf {\bibinfo {volume} {58}},\ \bibinfo {pages} {141} (\bibinfo {year} {1936})}\BibitemShut {NoStop}%
\bibitem [{\citenamefont {Gefen}\ \emph {et~al.}(2019)\citenamefont {Gefen}, \citenamefont {Rotem},\ and\ \citenamefont {Retzker}}]{gefen2019overcoming}%
  \BibitemOpen
  \bibfield  {author} {\bibinfo {author} {\bibfnamefont {T.}~\bibnamefont {Gefen}}, \bibinfo {author} {\bibfnamefont {A.}~\bibnamefont {Rotem}},\ \bibnamefont {and}\ \bibinfo {author} {\bibfnamefont {A.}~\bibnamefont {Retzker}},\ }\bibfield  {title} {\bibinfo {title} {Overcoming resolution limits with quantum sensing},\ }\href {https://doi.org/10.1038/s41467-019-12817-y} {\bibfield  {journal} {\bibinfo  {journal} {Nat. Comm.}\ }\textbf {\bibinfo {volume} {10}},\ \bibinfo {pages} {4992} (\bibinfo {year} {2019})}\BibitemShut {NoStop}%
\bibitem [{\citenamefont {Haine}(2018)}]{haine2018using}%
  \BibitemOpen
  \bibfield  {author} {\bibinfo {author} {\bibfnamefont {S.~A.}\ \bibnamefont {Haine}},\ }\bibfield  {title} {\bibinfo {title} {Using interaction-based readouts to approach the ultimate limit of detection-noise robustness for quantum-enhanced metrology in collective spin systems},\ }\href {https://doi.org/10.1103/PhysRevA.98.030303} {\bibfield  {journal} {\bibinfo  {journal} {Phys. Rev. A}\ }\textbf {\bibinfo {volume} {98}},\ \bibinfo {pages} {030303} (\bibinfo {year} {2018})}\BibitemShut {NoStop}%
\bibitem [{\citenamefont {Len}\ \emph {et~al.}(2022)\citenamefont {Len}, \citenamefont {Gefen}, \citenamefont {Retzker},\ and\ \citenamefont {Ko{\l}ody{\'n}ski}}]{len2022quantum}%
  \BibitemOpen
  \bibfield  {author} {\bibinfo {author} {\bibfnamefont {Y.~L.}\ \bibnamefont {Len}}, \bibinfo {author} {\bibfnamefont {T.}~\bibnamefont {Gefen}}, \bibinfo {author} {\bibfnamefont {A.}~\bibnamefont {Retzker}},\ \bibnamefont {and}\ \bibinfo {author} {\bibfnamefont {J.}~\bibnamefont {Ko{\l}ody{\'n}ski}},\ }\bibfield  {title} {\bibinfo {title} {Quantum metrology with imperfect measurements},\ }\href {https://doi.org/10.1038/s41467-022-33563-8} {\bibfield  {journal} {\bibinfo  {journal} {Nat. Comm.}\ }\textbf {\bibinfo {volume} {13}},\ \bibinfo {pages} {6971} (\bibinfo {year} {2022})}\BibitemShut {NoStop}%
\bibitem [{\citenamefont {Zhou}\ \emph {et~al.}(2023)\citenamefont {Zhou}, \citenamefont {Michalakis},\ and\ \citenamefont {Gefen}}]{zhou2023optimal}%
  \BibitemOpen
  \bibfield  {author} {\bibinfo {author} {\bibfnamefont {S.}~\bibnamefont {Zhou}}, \bibinfo {author} {\bibfnamefont {S.}~\bibnamefont {Michalakis}},\ \bibnamefont {and}\ \bibinfo {author} {\bibfnamefont {T.}~\bibnamefont {Gefen}},\ }\bibfield  {title} {\bibinfo {title} {Optimal protocols for quantum metrology with noisy measurements},\ }\href {https://doi.org/10.1103/PRXQuantum.4.040305} {\bibfield  {journal} {\bibinfo  {journal} {PRX Quantum}\ }\textbf {\bibinfo {volume} {4}},\ \bibinfo {pages} {040305} (\bibinfo {year} {2023})}\BibitemShut {NoStop}%
\bibitem [{\citenamefont {Caves}(1981)}]{CavesPRD81QuantummechanicalNoise}%
  \BibitemOpen
  \bibfield  {author} {\bibinfo {author} {\bibfnamefont {C.~M.}\ \bibnamefont {Caves}},\ }\bibfield  {title} {\bibinfo {title} {Quantum-mechanical noise in an interferometer},\ }\href {https://doi.org/10.1103/PhysRevD.23.1693} {\bibfield  {journal} {\bibinfo  {journal} {Phys. Rev. D}\ }\textbf {\bibinfo {volume} {23}},\ \bibinfo {pages} {1693} (\bibinfo {year} {1981})}\BibitemShut {NoStop}%
\bibitem [{\citenamefont {Uhlmann}(1976)}]{uhlmann1976transition}%
  \BibitemOpen
  \bibfield  {author} {\bibinfo {author} {\bibfnamefont {A.}~\bibnamefont {Uhlmann}},\ }\bibfield  {title} {\bibinfo {title} {The ``transition probability'' in the state space of a $\ast$-algebra},\ }\href {https://doi.org/10.1016/0034-4877(76)90060-4} {\bibfield  {journal} {\bibinfo  {journal} {Rep. Math. Phys.}\ }\textbf {\bibinfo {volume} {9}},\ \bibinfo {pages} {273} (\bibinfo {year} {1976})}\BibitemShut {NoStop}%
\bibitem [{\citenamefont {Boixo}\ \emph {et~al.}(2007)\citenamefont {Boixo}, \citenamefont {Flammia}, \citenamefont {Caves},\ and\ \citenamefont {Geremia}}]{boixo2007generalized}%
  \BibitemOpen
  \bibfield  {author} {\bibinfo {author} {\bibfnamefont {S.}~\bibnamefont {Boixo}}, \bibinfo {author} {\bibfnamefont {S.~T.}\ \bibnamefont {Flammia}}, \bibinfo {author} {\bibfnamefont {C.~M.}\ \bibnamefont {Caves}},\ \bibnamefont {and}\ \bibinfo {author} {\bibfnamefont {J.~M.}\ \bibnamefont {Geremia}},\ }\bibfield  {title} {\bibinfo {title} {Generalized limits for single-parameter quantum estimation},\ }\href {https://doi.org/10.1103/PhysRevLett.98.090401} {\bibfield  {journal} {\bibinfo  {journal} {Phys. Rev. Lett.}\ }\textbf {\bibinfo {volume} {98}},\ \bibinfo {pages} {090401} (\bibinfo {year} {2007})}\BibitemShut {NoStop}%
\bibitem [{\citenamefont {Fujiwara}\ and\ \citenamefont {Imai}(2008)}]{fujiwara2008fibre}%
  \BibitemOpen
  \bibfield  {author} {\bibinfo {author} {\bibfnamefont {A.}~\bibnamefont {Fujiwara}}\ \bibnamefont {and}\ \bibinfo {author} {\bibfnamefont {H.}~\bibnamefont {Imai}},\ }\bibfield  {title} {\bibinfo {title} {A fibre bundle over manifolds of quantum channels and its application to quantum statistics},\ }\href {https://doi.org/10.1088/1751-8113/41/25/255304} {\bibfield  {journal} {\bibinfo  {journal} {J. Phys. A}\ }\textbf {\bibinfo {volume} {41}},\ \bibinfo {pages} {255304} (\bibinfo {year} {2008})}\BibitemShut {NoStop}%
\bibitem [{\citenamefont {Kawamura}\ and\ \citenamefont {Chen}(2004)}]{kawamura2004displacement}%
  \BibitemOpen
  \bibfield  {author} {\bibinfo {author} {\bibfnamefont {S.}~\bibnamefont {Kawamura}}\ \bibnamefont {and}\ \bibinfo {author} {\bibfnamefont {Y.}~\bibnamefont {Chen}},\ }\bibfield  {title} {\bibinfo {title} {Displacement-noise-free gravitational-wave detection},\ }\href {https://doi.org/10.1103/PhysRevLett.93.211103} {\bibfield  {journal} {\bibinfo  {journal} {Phys. Rev. Lett.}\ }\textbf {\bibinfo {volume} {93}},\ \bibinfo {pages} {211103} (\bibinfo {year} {2004})}\BibitemShut {NoStop}%
\bibitem [{\citenamefont {Chen}\ \emph {et~al.}(2006{\natexlab{b}})\citenamefont {Chen}, \citenamefont {Pai}, \citenamefont {Somiya}, \citenamefont {Kawamura}, \citenamefont {Sato}, \citenamefont {Kokeyama}, \citenamefont {Ward}, \citenamefont {Goda},\ and\ \citenamefont {Mikhailov}}]{chen2006interferometers}%
  \BibitemOpen
  \bibfield  {author} {\bibinfo {author} {\bibfnamefont {Y.}~\bibnamefont {Chen}}, \bibinfo {author} {\bibfnamefont {A.}~\bibnamefont {Pai}}, \bibinfo {author} {\bibfnamefont {K.}~\bibnamefont {Somiya}}, \bibinfo {author} {\bibfnamefont {S.}~\bibnamefont {Kawamura}}, \bibinfo {author} {\bibfnamefont {S.}~\bibnamefont {Sato}}, \bibinfo {author} {\bibfnamefont {K.}~\bibnamefont {Kokeyama}}, \bibinfo {author} {\bibfnamefont {R.~L.}\ \bibnamefont {Ward}}, \bibinfo {author} {\bibfnamefont {K.}~\bibnamefont {Goda}},\ \bibnamefont {and}\ \bibinfo {author} {\bibfnamefont {E.~E.}\ \bibnamefont {Mikhailov}},\ }\bibfield  {title} {\bibinfo {title} {Interferometers for displacement-noise-free gravitational-wave detection},\ }\href {https://doi.org/10.1103/PhysRevLett.97.151103} {\bibfield  {journal} {\bibinfo  {journal} {Phys. Rev. Lett.}\ }\textbf {\bibinfo {volume} {97}},\ \bibinfo {pages} {151103} (\bibinfo {year} {2006}{\natexlab{b}})}\BibitemShut {NoStop}%
\bibitem [{\citenamefont {Gefen}\ \emph {et~al.}(2024)\citenamefont {Gefen}, \citenamefont {Tarafder}, \citenamefont {Adhikari},\ and\ \citenamefont {Chen}}]{gefen2024quantum}%
  \BibitemOpen
  \bibfield  {author} {\bibinfo {author} {\bibfnamefont {T.}~\bibnamefont {Gefen}}, \bibinfo {author} {\bibfnamefont {R.}~\bibnamefont {Tarafder}}, \bibinfo {author} {\bibfnamefont {R.~X.}\ \bibnamefont {Adhikari}},\ \bibnamefont {and}\ \bibinfo {author} {\bibfnamefont {Y.}~\bibnamefont {Chen}},\ }\bibfield  {title} {\bibinfo {title} {Quantum precision limits of displacement noise-free interferometers},\ }\href {https://doi.org/10.1103/PhysRevLett.132.020801} {\bibfield  {journal} {\bibinfo  {journal} {Phys. Rev. Lett.}\ }\textbf {\bibinfo {volume} {132}},\ \bibinfo {pages} {020801} (\bibinfo {year} {2024})}\BibitemShut {NoStop}%
\bibitem [{\citenamefont {Tzitrin}\ \emph {et~al.}(2020)\citenamefont {Tzitrin}, \citenamefont {Bourassa}, \citenamefont {Menicucci},\ and\ \citenamefont {Sabapathy}}]{TzitrinPRA20ProgressPractical}%
  \BibitemOpen
  \bibfield  {author} {\bibinfo {author} {\bibfnamefont {I.}~\bibnamefont {Tzitrin}}, \bibinfo {author} {\bibfnamefont {J.~E.}\ \bibnamefont {Bourassa}}, \bibinfo {author} {\bibfnamefont {N.~C.}\ \bibnamefont {Menicucci}},\ \bibnamefont {and}\ \bibinfo {author} {\bibfnamefont {K.~K.}\ \bibnamefont {Sabapathy}},\ }\bibfield  {title} {\bibinfo {title} {Progress towards practical qubit computation using approximate {{Gottesman-Kitaev-Preskill}} codes},\ }\href {https://doi.org/10.1103/PhysRevA.101.032315} {\bibfield  {journal} {\bibinfo  {journal} {Phys. Rev. A}\ }\textbf {\bibinfo {volume} {101}},\ \bibinfo {pages} {032315} (\bibinfo {year} {2020})}\BibitemShut {NoStop}%
\bibitem [{\citenamefont {Bourassa}\ \emph {et~al.}(2021)\citenamefont {Bourassa}, \citenamefont {Quesada}, \citenamefont {Tzitrin}, \citenamefont {Sz{\'a}va}, \citenamefont {Isacsson}, \citenamefont {Izaac}, \citenamefont {Sabapathy}, \citenamefont {Dauphinais},\ and\ \citenamefont {Dhand}}]{bourassa2021fast}%
  \BibitemOpen
  \bibfield  {author} {\bibinfo {author} {\bibfnamefont {J.~E.}\ \bibnamefont {Bourassa}}, \bibinfo {author} {\bibfnamefont {N.}~\bibnamefont {Quesada}}, \bibinfo {author} {\bibfnamefont {I.}~\bibnamefont {Tzitrin}}, \bibinfo {author} {\bibfnamefont {A.}~\bibnamefont {Sz{\'a}va}}, \bibinfo {author} {\bibfnamefont {T.}~\bibnamefont {Isacsson}}, \bibinfo {author} {\bibfnamefont {J.}~\bibnamefont {Izaac}}, \bibinfo {author} {\bibfnamefont {K.~K.}\ \bibnamefont {Sabapathy}}, \bibinfo {author} {\bibfnamefont {G.}~\bibnamefont {Dauphinais}},\ \bibnamefont {and}\ \bibinfo {author} {\bibfnamefont {I.}~\bibnamefont {Dhand}},\ }\bibfield  {title} {\bibinfo {title} {Fast simulation of bosonic qubits via gaussian functions in phase space},\ }\href {https://doi.org/10.1103/PRXQuantum.2.040315} {\bibfield  {journal} {\bibinfo  {journal} {PRX Quantum}\ }\textbf {\bibinfo {volume} {2}},\ \bibinfo {pages} {040315} (\bibinfo {year} {2021})}\BibitemShut {NoStop}%
\bibitem [{\citenamefont {{Wolfram Research, Inc.}}(2010)}]{mathematica}%
  \BibitemOpen
  \bibfield  {author} {\bibinfo {author} {\bibnamefont {{Wolfram Research, Inc.}}},\ }\href {https://www.wolfram.com} {\bibinfo {title} {Mathematica 8.0}} (\bibinfo {year} {2010})\BibitemShut {NoStop}%
\bibitem [{\citenamefont {Van~Rossum}\ and\ \citenamefont {Drake}(2009)}]{python}%
  \BibitemOpen
  \bibfield  {author} {\bibinfo {author} {\bibfnamefont {G.}~\bibnamefont {Van~Rossum}}\ \bibnamefont {and}\ \bibinfo {author} {\bibfnamefont {F.~L.}\ \bibnamefont {Drake}},\ }\href {https://doi.org/10.5555/1593511} {\emph {\bibinfo {title} {Python 3 Reference Manual}}}\ (\bibinfo  {publisher} {CreateSpace},\ \bibinfo {address} {Scotts Valley, CA},\ \bibinfo {year} {2009})\BibitemShut {NoStop}%
\bibitem [{\citenamefont {P{\'e}rez}\ and\ \citenamefont {Granger}(2007)}]{ipython}%
  \BibitemOpen
  \bibfield  {author} {\bibinfo {author} {\bibfnamefont {F.}~\bibnamefont {P{\'e}rez}}\ \bibnamefont {and}\ \bibinfo {author} {\bibfnamefont {B.~E.}\ \bibnamefont {Granger}},\ }\bibfield  {title} {\bibinfo {title} {Ipython: a system for interactive scientific computing},\ }\bibfield  {journal} {\bibinfo  {journal} {Comput. Sci. Eng.}\ }\textbf {\bibinfo {volume} {9}},\ \href {https://doi.org/10.1109/MCSE.2007.53} {10.1109/MCSE.2007.53} (\bibinfo {year} {2007})\BibitemShut {NoStop}%
\bibitem [{\citenamefont {Kluyver}\ \emph {et~al.}(2016)\citenamefont {Kluyver}, \citenamefont {Ragan-Kelley}, \citenamefont {P{\'e}rez}, \citenamefont {Granger}, \citenamefont {Bussonnier} \emph {et~al.}}]{jupyter}%
  \BibitemOpen
  \bibfield  {author} {\bibinfo {author} {\bibfnamefont {T.}~\bibnamefont {Kluyver}}, \bibinfo {author} {\bibfnamefont {B.}~\bibnamefont {Ragan-Kelley}}, \bibinfo {author} {\bibfnamefont {F.}~\bibnamefont {P{\'e}rez}}, \bibinfo {author} {\bibfnamefont {B.}~\bibnamefont {Granger}}, \bibinfo {author} {\bibfnamefont {M.}~\bibnamefont {Bussonnier}}, \bibnamefont {et~al.},\ }\bibfield  {title} {\bibinfo {title} {Jupyter notebooks -- a publishing format for reproducible computational workflows},\ }in\ \href {https://doi.org/10.3233/978-1-61499-649-1-87} {\emph {\bibinfo {booktitle} {Positioning and Power in Academic Publishing: Players, Agents and Agendas}}}\ (\bibinfo {year} {2016})\ pp.\ \bibinfo {pages} {87--90}\BibitemShut {NoStop}%
\bibitem [{\citenamefont {Harris}\ \emph {et~al.}(2020)\citenamefont {Harris}, \citenamefont {Millman}, \citenamefont {van~der Walt}, \citenamefont {Gommers}, \citenamefont {Virtanen}, \citenamefont {Cournapeau}, \citenamefont {Wieser}, \citenamefont {Taylor}, \citenamefont {Berg}, \citenamefont {Smith}, \citenamefont {Kern}, \citenamefont {Picus}, \citenamefont {Hoyer}, \citenamefont {van Kerkwijk}, \citenamefont {Brett}, \citenamefont {Haldane}, \citenamefont {del R{\'{i}}o}, \citenamefont {Wiebe}, \citenamefont {Peterson}, \citenamefont {G{\'{e}}rard-Marchant}, \citenamefont {Sheppard}, \citenamefont {Reddy}, \citenamefont {Weckesser}, \citenamefont {Abbasi}, \citenamefont {Gohlke},\ and\ \citenamefont {Oliphant}}]{numpy}%
  \BibitemOpen
  \bibfield  {author} {\bibinfo {author} {\bibfnamefont {C.~R.}\ \bibnamefont {Harris}}, \bibinfo {author} {\bibfnamefont {K.~J.}\ \bibnamefont {Millman}}, \bibinfo {author} {\bibfnamefont {S.~J.}\ \bibnamefont {van~der Walt}}, \bibnamefont {et~al.},\ }\bibfield  {title} {\bibinfo {title} {Array programming with {NumPy}},\ }\href {https://doi.org/10.1038/s41586-020-2649-2} {\bibfield  {journal} {\bibinfo  {journal} {Nature}\ }\textbf {\bibinfo {volume} {585}},\ \bibinfo {pages} {357} (\bibinfo {year} {2020})}\BibitemShut {NoStop}%
\bibitem [{\citenamefont {Hunter}(2007)}]{matplotlib}%
  \BibitemOpen
  \bibfield  {author} {\bibinfo {author} {\bibfnamefont {J.~D.}\ \bibnamefont {Hunter}},\ }\bibfield  {title} {\bibinfo {title} {Matplotlib: A 2d graphics environment},\ }\href {https://doi.org/10.1109/MCSE.2007.55} {\bibfield  {journal} {\bibinfo  {journal} {Comput. Sci. Eng.}\ }\textbf {\bibinfo {volume} {9}},\ \bibinfo {pages} {90} (\bibinfo {year} {2007})}\BibitemShut {NoStop}%
\bibitem [{\citenamefont {{W}es {M}c{K}inney}(2010)}]{pandas}%
  \BibitemOpen
  \bibfield  {author} {\bibinfo {author} {\bibnamefont {{W}es {M}c{K}inney}},\ }\bibfield  {title} {\bibinfo {title} {{D}ata {S}tructures for {S}tatistical {C}omputing in {P}ython},\ }in\ \href {https://doi.org/10.25080/Majora-92bf1922-00a} {\emph {\bibinfo {booktitle} {{P}roceedings of the 9th {P}ython in {S}cience {C}onference}}}\ (\bibinfo {year} {2010})\ pp.\ \bibinfo {pages} {56 -- 61}\BibitemShut {NoStop}%
\bibitem [{\citenamefont {Virtanen}\ \emph {et~al.}(2020)\citenamefont {Virtanen}, \citenamefont {Gommers}, \citenamefont {Oliphant}, \citenamefont {Haberland}, \citenamefont {Reddy}, \citenamefont {Cournapeau}, \citenamefont {Burovski}, \citenamefont {Peterson}, \citenamefont {Weckesser}, \citenamefont {Bright}, \citenamefont {{van der Walt}}, \citenamefont {Brett}, \citenamefont {Wilson}, \citenamefont {Millman}, \citenamefont {Mayorov}, \citenamefont {Nelson}, \citenamefont {Jones}, \citenamefont {Kern}, \citenamefont {Larson}, \citenamefont {Carey}, \citenamefont {Polat}, \citenamefont {Feng}, \citenamefont {Moore}, \citenamefont {VanderPlas}, \citenamefont {Laxalde}, \citenamefont {Perktold}, \citenamefont {Cimrman}, \citenamefont {Henriksen}, \citenamefont {Quintero}, \citenamefont {Harris}, \citenamefont {Archibald}, \citenamefont {Ribeiro}, \citenamefont {Pedregosa},\ and\ \citenamefont {{van Mulbregt}}}]{VirtanenNM20SciPyFundamental}%
  \BibitemOpen
  \bibfield  {author} {\bibinfo {author} {\bibfnamefont {P.}~\bibnamefont {Virtanen}}, \bibinfo {author} {\bibfnamefont {R.}~\bibnamefont {Gommers}}, \bibinfo {author} {\bibfnamefont {T.~E.}\ \bibnamefont {Oliphant}}, \bibnamefont {et~al.},\ }\bibfield  {title} {\bibinfo {title} {{{SciPy}} 1.0: Fundamental algorithms for scientific computing in {{Python}}},\ }\href {https://doi.org/10.1038/s41592-019-0686-2} {\bibfield  {journal} {\bibinfo  {journal} {Nat. Methods}\ }\textbf {\bibinfo {volume} {17}},\ \bibinfo {pages} {261} (\bibinfo {year} {2020})}\BibitemShut {NoStop}%
\bibitem [{\citenamefont {Johansson}\ \emph {et~al.}(2013)\citenamefont {Johansson}, \citenamefont {Nation},\ and\ \citenamefont {Nori}}]{qutip}%
  \BibitemOpen
  \bibfield  {author} {\bibinfo {author} {\bibfnamefont {J.}~\bibnamefont {Johansson}}, \bibinfo {author} {\bibfnamefont {P.}~\bibnamefont {Nation}},\ \bibnamefont {and}\ \bibinfo {author} {\bibfnamefont {F.}~\bibnamefont {Nori}},\ }\bibfield  {title} {\bibinfo {title} {Qutip 2: A python framework for the dynamics of open quantum systems},\ }\href {https://doi.org/10.1016/j.cpc.2012.11.019} {\bibfield  {journal} {\bibinfo  {journal} {Comput. Phys. Commun.}\ }\textbf {\bibinfo {volume} {184}},\ \bibinfo {pages} {1234} (\bibinfo {year} {2013})}\BibitemShut {NoStop}%
\bibitem [{\citenamefont {Killoran}\ \emph {et~al.}(2019)\citenamefont {Killoran}, \citenamefont {Izaac}, \citenamefont {Quesada}, \citenamefont {Bergholm}, \citenamefont {Amy},\ and\ \citenamefont {Weedbrook}}]{strawberryfields}%
  \BibitemOpen
  \bibfield  {author} {\bibinfo {author} {\bibfnamefont {N.}~\bibnamefont {Killoran}}, \bibinfo {author} {\bibfnamefont {J.}~\bibnamefont {Izaac}}, \bibinfo {author} {\bibfnamefont {N.}~\bibnamefont {Quesada}}, \bibinfo {author} {\bibfnamefont {V.}~\bibnamefont {Bergholm}}, \bibinfo {author} {\bibfnamefont {M.}~\bibnamefont {Amy}},\ \bibnamefont {and}\ \bibinfo {author} {\bibfnamefont {C.}~\bibnamefont {Weedbrook}},\ }\bibfield  {title} {\bibinfo {title} {Strawberry fields: A software platform for photonic quantum computing},\ }\href {https://doi.org/10.22331/q-2019-03-11-129} {\bibfield  {journal} {\bibinfo  {journal} {Quantum}\ }\textbf {\bibinfo {volume} {3}},\ \bibinfo {pages} {129} (\bibinfo {year} {2019})}\BibitemShut {NoStop}%
\bibitem [{rep()}]{repo}%
  \BibitemOpen
  \href@noop {} {\bibinfo {title} {pleasantpheasant}},\ \bibinfo {howpublished} {J.~W.~Gardner. \emph{pleasantPheasant}. 2023. \url{https://git.ligo.org/jameswalter.gardner/pleasantpheasant}}\BibitemShut {NoStop}%
\bibitem [{\citenamefont {{K. Macieszczak, Quantum fisher information: Variational principle and simple iterative algorithm for its efficient computation (2013)}}()}]{macieszczak2013quantum}%
  \BibitemOpen
  \bibfield  {author} {\bibinfo {author} {\bibnamefont {{K. Macieszczak, Quantum fisher information: Variational principle and simple iterative algorithm for its efficient computation (2013)}}},\ }\href {https://doi.org/10.48550/arXiv.1312.1356} {\bibinfo {title} {arxiv:1312.1356 [quant-ph]}}\BibitemShut {NoStop}%
\bibitem [{\citenamefont {Gorski}\ \emph {et~al.}(2007)\citenamefont {Gorski}, \citenamefont {Pfeuffer},\ and\ \citenamefont {Klamroth}}]{gorski2007biconvex}%
  \BibitemOpen
  \bibfield  {author} {\bibinfo {author} {\bibfnamefont {J.}~\bibnamefont {Gorski}}, \bibinfo {author} {\bibfnamefont {F.}~\bibnamefont {Pfeuffer}},\ \bibnamefont {and}\ \bibinfo {author} {\bibfnamefont {K.}~\bibnamefont {Klamroth}},\ }\bibfield  {title} {\bibinfo {title} {Biconvex sets and optimization with biconvex functions: a survey and extensions},\ }\href {https://doi.org/10.1007/s00186-007-0161-1} {\bibfield  {journal} {\bibinfo  {journal} {Math. Methods Oper. Res.}\ }\textbf {\bibinfo {volume} {66}},\ \bibinfo {pages} {373} (\bibinfo {year} {2007})}\BibitemShut {NoStop}%
\bibitem [{\citenamefont {Floudas}\ and\ \citenamefont {Visweswaran}(1990)}]{floudas1990global}%
  \BibitemOpen
  \bibfield  {author} {\bibinfo {author} {\bibfnamefont {C.~A.}\ \bibnamefont {Floudas}}\ \bibnamefont {and}\ \bibinfo {author} {\bibfnamefont {V.}~\bibnamefont {Visweswaran}},\ }\bibfield  {title} {\bibinfo {title} {{A global optimization algorithm (GOP) for certain classes of nonconvex NLPs---I. Theory}},\ }\href {https://doi.org/10.1016/0098-1354(90)80020-C} {\bibfield  {journal} {\bibinfo  {journal} {Comput. Chem. Eng.}\ }\textbf {\bibinfo {volume} {14}},\ \bibinfo {pages} {1397} (\bibinfo {year} {1990})}\BibitemShut {NoStop}%
\bibitem [{\citenamefont {Floudas}(2013)}]{floudas2013deterministic}%
  \BibitemOpen
  \bibfield  {author} {\bibinfo {author} {\bibfnamefont {C.~A.}\ \bibnamefont {Floudas}},\ }\href {https://doi.org/10.1007/978-1-4757-4949-6} {\emph {\bibinfo {title} {Deterministic global optimization: theory, methods and applications}}},\ Vol.~\bibinfo {volume} {37}\ (\bibinfo  {publisher} {Springer Science \& Business Media},\ \bibinfo {year} {2013})\BibitemShut {NoStop}%
\bibitem [{\citenamefont {Noh}\ \emph {et~al.}(2018)\citenamefont {Noh}, \citenamefont {Albert},\ and\ \citenamefont {Jiang}}]{noh2018quantum}%
  \BibitemOpen
  \bibfield  {author} {\bibinfo {author} {\bibfnamefont {K.}~\bibnamefont {Noh}}, \bibinfo {author} {\bibfnamefont {V.~V.}\ \bibnamefont {Albert}},\ \bibnamefont {and}\ \bibinfo {author} {\bibfnamefont {L.}~\bibnamefont {Jiang}},\ }\bibfield  {title} {\bibinfo {title} {{Quantum capacity bounds of Gaussian thermal loss channels and achievable rates with Gottesman-Kitaev-Preskill codes}},\ }\href {https://doi.org/10.1109/TIT.2018.2873764} {\bibfield  {journal} {\bibinfo  {journal} {IEEE Transactions on Information Theory}\ }\textbf {\bibinfo {volume} {65}},\ \bibinfo {pages} {2563} (\bibinfo {year} {2018})}\BibitemShut {NoStop}%
\bibitem [{\citenamefont {Reitze}\ \emph {et~al.}(2019)\citenamefont {Reitze} \emph {et~al.}}]{ReitzeBAAS19CosmicExplorer}%
  \BibitemOpen
  \bibfield  {author} {\bibinfo {author} {\bibfnamefont {D.}~\bibnamefont {Reitze}} \bibnamefont {et~al.},\ }\bibfield  {title} {\bibinfo {title} {Cosmic explorer: {{The U}}.{{S}}. contribution to gravitational-wave astronomy beyond {{LIGO}}},\ }in\ \href@noop {} {\emph {\bibinfo {booktitle} {Bull. {{Am}}. {{Astron}}. {{Soc}}.}}},\ Vol.~\bibinfo {volume} {51}\ (\bibinfo {year} {2019})\ p.~\bibinfo {pages} {35},\ \Eprint {https://arxiv.org/abs/1907.04833} {arxiv:1907.04833 [astro-ph.IM]} \BibitemShut {NoStop}%
\bibitem [{\citenamefont {Maggiore}\ \emph {et~al.}(2020)\citenamefont {Maggiore}, \citenamefont {van~den Broeck}, \citenamefont {Bartolo}, \citenamefont {Belgacem}, \citenamefont {Bertacca}, \citenamefont {Bizouard}, \citenamefont {Branchesi}, \citenamefont {Clesse}, \citenamefont {Foffa}, \citenamefont {{Garc{\'i}a-Bellido}}, \citenamefont {Grimm}, \citenamefont {Harms}, \citenamefont {Hinderer}, \citenamefont {Matarrese}, \citenamefont {Palomba}, \citenamefont {Peloso}, \citenamefont {Ricciardone},\ and\ \citenamefont {Sakellariadou}}]{MaggioreJCAP20ScienceCase}%
  \BibitemOpen
  \bibfield  {author} {\bibinfo {author} {\bibfnamefont {M.}~\bibnamefont {Maggiore}}, \bibinfo {author} {\bibfnamefont {C.}~\bibnamefont {van~den Broeck}}, \bibinfo {author} {\bibfnamefont {N.}~\bibnamefont {Bartolo}}, \bibnamefont {et~al.},\ }\bibfield  {title} {\bibinfo {title} {Science {{Case}} for the {{Einstein Telescope}}},\ }\href {https://doi.org/10.1088/1475-7516/2020/03/050} {\bibfield  {journal} {\bibinfo  {journal} {J. Cosmol. Astropart. Phys.}\ }\textbf {\bibinfo {volume} {2020}}\bibinfo  {number} { (03)},\ \bibinfo {pages} {050}}\BibitemShut {NoStop}%
\bibitem [{\citenamefont {Rife}\ and\ \citenamefont {Boorstyn}(1974)}]{RifeITIT74SingleTone}%
  \BibitemOpen
\bibfield  {number} {  }\bibfield  {author} {\bibinfo {author} {\bibfnamefont {D.}~\bibnamefont {Rife}}\ \bibnamefont {and}\ \bibinfo {author} {\bibfnamefont {R.}~\bibnamefont {Boorstyn}},\ }\bibfield  {title} {\bibinfo {title} {Single tone parameter estimation from discrete-time observations},\ }\href {https://doi.org/10.1109/TIT.1974.1055282} {\bibfield  {journal} {\bibinfo  {journal} {IEEE Trans. Inf. Theory}\ }\textbf {\bibinfo {volume} {20}},\ \bibinfo {pages} {591} (\bibinfo {year} {1974})}\BibitemShut {NoStop}%
\bibitem [{\citenamefont {Steinhardt}\ and\ \citenamefont {Bretherton}(1985)}]{SteinhardtI8IICASSP85ThresholdsFrequency}%
  \BibitemOpen
  \bibfield  {author} {\bibinfo {author} {\bibfnamefont {A.}~\bibnamefont {Steinhardt}}\ \bibnamefont {and}\ \bibinfo {author} {\bibfnamefont {C.}~\bibnamefont {Bretherton}},\ }\bibfield  {title} {\bibinfo {title} {Thresholds in frequency estimation},\ }in\ \href {https://doi.org/10.1109/ICASSP.1985.1168170} {\emph {\bibinfo {booktitle} {{{ICASSP}} 85 {{IEEE Int}}. {{Conf}}. {{Acoust}}. {{Speech Signal Process}}.}}},\ Vol.~\bibinfo {volume} {10}\ (\bibinfo {year} {1985})\ pp.\ \bibinfo {pages} {1273--1276}\BibitemShut {NoStop}%
\bibitem [{\citenamefont {Knockaert}(1997)}]{KnockaertITSP97BarankinBound}%
  \BibitemOpen
  \bibfield  {author} {\bibinfo {author} {\bibfnamefont {L.}~\bibnamefont {Knockaert}},\ }\bibfield  {title} {\bibinfo {title} {The {{Barankin}} bound and threshold behavior in frequency estimation},\ }\href {https://doi.org/10.1109/78.622965} {\bibfield  {journal} {\bibinfo  {journal} {IEEE Trans. Signal Process.}\ }\textbf {\bibinfo {volume} {45}},\ \bibinfo {pages} {2398} (\bibinfo {year} {1997})}\BibitemShut {NoStop}%
\bibitem [{\citenamefont {Chaumette}\ \emph {et~al.}(2008)\citenamefont {Chaumette}, \citenamefont {Galy}, \citenamefont {Quinlan},\ and\ \citenamefont {Larzabal}}]{ChaumetteITSP08NewBarankin}%
  \BibitemOpen
  \bibfield  {author} {\bibinfo {author} {\bibfnamefont {E.}~\bibnamefont {Chaumette}}, \bibinfo {author} {\bibfnamefont {{\relax J{\'e}}.}~\bibnamefont {Galy}}, \bibinfo {author} {\bibfnamefont {A.}~\bibnamefont {Quinlan}},\ \bibnamefont {and}\ \bibinfo {author} {\bibfnamefont {P.}~\bibnamefont {Larzabal}},\ }\bibfield  {title} {\bibinfo {title} {A {{New Barankin Bound Approximation}} for the {{Prediction}} of the {{Threshold Region Performance}} of {{Maximum Likelihood Estimators}}},\ }\href {https://doi.org/10.1109/TSP.2008.927805} {\bibfield  {journal} {\bibinfo  {journal} {IEEE Trans. Signal Process.}\ }\textbf {\bibinfo {volume} {56}},\ \bibinfo {pages} {5319} (\bibinfo {year} {2008})}\BibitemShut {NoStop}%
\bibitem [{\citenamefont {Gessner}\ and\ \citenamefont {Smerzi}(2023)}]{GessnerPRL23HierarchiesFrequentist}%
  \BibitemOpen
  \bibfield  {author} {\bibinfo {author} {\bibfnamefont {M.}~\bibnamefont {Gessner}}\ \bibnamefont {and}\ \bibinfo {author} {\bibfnamefont {A.}~\bibnamefont {Smerzi}},\ }\bibfield  {title} {\bibinfo {title} {Hierarchies of {{Frequentist Bounds}} for {{Quantum Metrology}}: {{From Cram}}\'er-{{Rao}} to {{Barankin}}},\ }\href {https://doi.org/10.1103/PhysRevLett.130.260801} {\bibfield  {journal} {\bibinfo  {journal} {Phys. Rev. Lett.}\ }\textbf {\bibinfo {volume} {130}},\ \bibinfo {pages} {260801} (\bibinfo {year} {2023})}\BibitemShut {NoStop}%
\end{thebibliography}%

\end{document}